\documentclass[12pt,oneside,a4paper]{article}

\usepackage{graphicx}
\usepackage{amsmath}
\usepackage{amssymb}
\usepackage{gensymb}
\usepackage[round]{natbib}
\usepackage{hyperref}
\usepackage[top=2cm, bottom=2cm, left=2cm, right=2cm]{geometry}  
\usepackage{xcolor}
\usepackage{subfigure}
\usepackage[multidot]{grffile}
\usepackage{pdflscape}
\usepackage{sectsty}

\graphicspath{{Figures/}}

\begin{document}
\chapterfont{\sffamily}
\allsectionsfont{\sffamily}

\title{{\sffamily{\LARGE \textbf{High-speed ditching of double curvature specimens with cavitation and ventilation}}}}

%
%
%
%

\author{Emanuele Spinosa, Silvano Grizzi, Alessandro Iafrati \\
	CNR-INM (Institute of Marine Engineering), Rome (Italy)  \\
	email address: emanuele.spinosa@cnr.it}

\maketitle

\begin{abstract}
The water entry at high horizontal speed of double-curvature specimens, 
representing the rear part of the fuselage that first contacts the water
during aircraft ditching, is investigated experimentally.
Three shapes, reproducing different aircraft types, are analysed through
pressure and load measurements, supported by underwater high-speed visualizations.
Tests are performed at different horizontal speeds and pitch angles,
while maintaining a constant vertical-to-horizontal velocity ratio.
The longitudinal curvature is found to significantly
affect the hydrodynamics, potentially leading to cavitation
and ventilation at the rear at high speeds. The transverse curvature
influences pressures and loads at both the front and rear,
with higher transverse curvature favouring lateral fluid escape.
Higher pitch angles increase loading in both magnitude and distribution.
Increasing the horizontal speed results in higher loads at the front
and alters cavitation and ventilation modalities at the rear.
Load scaling laws established for flat plates are valid at 
the front but are invalidated at the rear due to cavitation 
and ventilation. The time evolution of the cavitation and wetted
area is analysed through advanced image processing and pressure data.
Comparison with the geometric intersection area between the specimen
and the still water level offers further insight into the water entry hydrodynamics. 
\end{abstract}

\noindent $\copyright$ 2024. This manuscript version is made available
 under the CC-BY-NC-ND~4.0 license \url{http://creativecommons.org/licenses/by-nc-nd/4.0/}

\section{Introduction}
\label{sec:introduction}
%
Aircraft ditching is the manoeuvre performed when the pilots and crew are 
forced to recur to an emergency water landing due to exceptional
circumstances such as engine failure, fuel exhaustion, severe weather
or other critical unexpected situations. 
The goal of ditching is to land the aircraft safely on the water surface,
allowing the occupants to be evacuated and rescued with the lowest 
risk of injury. Although being a rare event, airworthiness regulations
require a precise certification for ditching, including
demonstration tests proving that the aircraft structural 
damage during this eventuality is limited and that a sufficient 
floating time is guaranteed for a safe evacuation of the occupants,
as specified in \citet{easa2020cs25} and \cite{faa2023far25} (see also  
\citet{Desjardins1989}).

The theoretical foundations of ditching
were laid by \citet{von1929impact} and \citet{wagner1932stoss},
which first established a theoretical framework of vertical water entry.
These initial (semi)-analytical models were successively refined
by several researchers 
\citep{logvinovich1969hydrodynamics,scolan2001three,korobkin2006three,korobkin2004analytical,tassin2010assessment}.
forming the basis of \emph{low-fidelity
numerical approaches}, which became popular in the early computational era.
A notable example is the Boundary Element Method (BEM)
\citep{zhao1993water,battistin2003hydrodynamic,iafrati2004initial,iafrati2008hydrodynamic},
based on potential theory. Owing to their simplified discretization 
and rapid computation, these methods are still used today for quick, approximate solutions.

To extend vertical entry solutions to the ditching case, 
where a significant horizontal velocity is also present, 
the multi-section or the 2D+t approach can be used \citep{fontaine1998prediction}, 
which break down the 3D motion into a series of simpler 
2D cross-section analyses \citep{sun2011dynamic}, albeit with some limitations.
The application of a multi-section approach to aircraft ditching is presented
in \citet{del2021water,del5210872potential}. The DITCH code
\citep{bensch2003computational,streckwall2007aircraft} implements a 2D+t method,
complemented by empirical formulations and has also been recently used
with machine learning to predict ditching loads \citep{schwarz2025machine}.

As computing power increased, \textit{high-fidelity} methods
using finer discretization of the Navier–Stokes equations were increasingly employed.
Early high-fidelity tools are reviewed in \cite{seddon2006review} and \citet{hughes2013aerospace}.
These methods often combine fluid dynamics solvers with structural models,
like Finite Element Methods (FEM), for fluid–structure interaction.
%
%
A common approach uses URANS solvers with 
Volume-Of-Fluid (VOF) to capture the free surface evolution. 
This method is applied to aircraft ditching in 
\citep{guo2013effect,qu2015study,qu2016numerical,wang2023dynamic,zheng2021numerical,zheng2025numerical}.
%
%
Free surface evolution can also be captured
using over-set approaches \citep{zha2024numerical,song2023numerical}
and \citet{spinosa2022hydrodynamic}, the latter focusing on
the final phase of ditching after water impact.
%
The ALE method overcomes mesh deformation in Lagrangian methods 
via periodic re-meshing \citep{cheng2011simulation,siemann2017coupled,bisagni2018modelling,feng2020simulation,fan2024water}. 
%
The CEL method uses a fixed Eulerian mesh for fluids and a moving Lagrangian mesh for solids, 
to model fluid–structure interaction \citep{goron2023simulation,goron2025numerical}.
%
Mesh-free methods like Smoothed Particle Hydrodynamics (SPH)
avoid mesh distortion and capture sprays well. However, they require higher computational resources
and still face challenges in modelling cavitation during
ditching \citep{anghileri2011rigid,anghileri2014survey,groenenboom2010hydrodynamics,groenenboom2015fluid,siemann2017coupled,siemann2018advances,bisagni2018modelling,groenenboom2021recent,munoz2025high,wang2024numerical}.
%
Undoubtedly, numerical simulations of ditching have advanced significantly.
However, their use in ditching certification remains limited due
to high computational costs and the need for 
rigorous validation against experimental data.

Experimental investigations enable an accurate replication
of the physical phenomena in a ditching event and are essential
for validating the various computational tools.
However, full-scale tests of an entire aircraft remain impractical and prohibitively expensive.
Attempts to replicate ditching through scaled-model experiments began in the
1950s with pioneering work by \citet{smiley1951experimental}, who tested flat plates entering
water at significant horizontal speeds, obtaining accurate measurements of pressures,
loads, and wetted areas under various conditions. While remarkable for its time,
the study had limitations, particularly a data sampling frequency of around 1 kHz,
insufficient to capture the rapid transients typical of water entry problems.
Another notable experimental ditching campaign from the same
period was conducted by \citet{naca2929}, who performed free-fall tests on various
scaled fuselage models representing different aircraft types. The study primarily examined
the effects of longitudinal and transverse body curvatures on ditching dynamics.
Its main limitation was the inability to control impact conditions.
A review of early experimental ditching activities is provided 
by \citet{smith1957investigations}, while more recent developments 
are covered in \cite{seddon2006review,abrate2011hull}.
More recent scaled model tests of a full aircraft model were presented in 
\citet{climent2006aircraft,zhang2012,song2023numerical,wenli2025experimental}.

The process of scaling aircraft ditching tests from full to model scale 
presents substantial challenges. In ship hydrodynamics, towing tank experiments often
rely on Froude similarity, which ensures correct scaling of the ratio between
inertial and gravitational forces. When Froude scaling is applied, the model test
speed decreases proportionally to the square root of the scale factor. 
In the hydrodynamics of water entry, the Reynolds number, representing
the ratio of inertial to viscous forces, should also be considered, 
however, previous studies have shown that viscous effects are
negligible \citep{facci2015numerical, moghisi1981experimental}.
Aircraft ditching involves high horizontal velocities and large-scale 
impact phenomena, which generate extremely high pressures in some regions
of the wetted surface and very low pressures in others.
Under these conditions, suction, cavitation and ventilation are likely
to occur \citet{hughes2013aerospace,climent2006aircraft,zhang2012,tassin2013two}
Therefore, achieving similarity in terms of cavitation and ventilation numbers
is essential \citep{brennen2014cavitation}. This type of similarity can only be
reproduced through testing in a de-pressurized
environment or by reaching sufficiently high speeds.

To address these challenges, the High-Speed Ditching Facility (HSDF)
at the CNR- Institute of Marine Engineering was designed and built. 
The HSDF allows tests at horizontal speeds of up to 48~m/s, 
with vertical-to-horizontal velocity ratios between 0.03 and 0.05, using models
approximately 0.7~m wide and 1.3~m long. This setup enables near full-scale
hydrodynamic conditions and allows observation of phenomena such as cavitation and ventilation,
which cannot be captured at lower speeds or smaller scales.
Pressures and loads can be acquired at high sampling rates through 
an integrated on-board measurement system, while high-speed underwater
imaging provides complementary insight into the flow phenomena \citep{iafrati2015high}.

Early tests were performed on a flat aluminium plates, 15~mm thick, in 
a configuration similar to
that used in \citet{smiley1951experimental}. Different test conditions
were examined by varying the horizontal speed, the vertical-to-horizontal
velocity ratio and the pitch angle. 
The strong correlation between the spray root propagation velocity and
the intensity of the pressure peak was confirmed and a scaling of the
total loading with respect to the impacting velocity relative to the
body was observed \citep{iafrati2015nav,iafrati2016experimental}. 
The high speeds achievable also allow for a more accurate investigation of 
fluid–structure interaction on thin aluminium plates (3~mm and 0.8~mm thick). 
Experimental results showed that hydrodynamic loads can increase by up to 30\%
in the presence of significant structural deformation. As a result, load estimates
based on tests with rigid specimens may not
always be conservative \citep{spinosa2021experimental}.

Flat plate tests provided validation 
data for ditching models. However, 
the curved geometry of the fuselage lower section required further investigation. 
Transverse curvature effects were examined using 15 mm thick concave and
convex plates \citep{iafrati2018effect}. Furthermore, the longitudinal curvature
of the fuselage rear portion induces suction forces due to negative pressures,
significantly influencing the aircraft ditching dynamics, as 
demonstrated by numerical and experimental studies \citep{climent2006aircraft, zhang2012}.
To elucidate this phenomenon, an experimental campaign was
conducted on rigid specimens exhibiting double curvature.
From a preliminary analysis of the data it was found that cavitation and ventilation
may occur at high speed and they may have important effects on the
magnitude and distribution of the loading
\citep{iafrati2019cavitation,iafrati2020experimental}.

This paper presents additional results from this extensive experimental
campaign. Continuing our past research, we further explore 
the influence of horizontal speed and offer considerations
on cavitation modalities and cavitation number.
A key objective of this paper is also to investigate the effects of 
pitch angle and the longitudinal and transverse curvatures, 
aspects not addressed in previous studies.
Analysis of pressures and total loads on the impacting specimens reveals significant 
effects at the front and rear, linked to cavitation and ventilation phenomena. 
Additionally, the temporal evolution of the wetted area and cavitation zone is
examined through underwater visualizations, with quantitative estimates provided.
The complexity of the underlying physics often results in theoretical
and numerical models exhibiting limited accuracy, especially at high speeds.
This paper offers a physical interpretation alongside a comprehensive
database to support the validation of these models.
%
%
%
\section{Experimental Set-up}
\label{sec:setup}
%
\subsection{The High-Speed Ditching Facility}
%
The experiments are performed in the High Speed Ditching Facility (HSDF) 
of the CNR - Institute of Marine Engineering, described in detail
in \citet{iafrati2015high}.
For the purpose of the present paper, it is worth recalling the
most relevant features.

The facility is composed of a guide along which a properly designed trolley 
can freely move. The guide is about 64~m long and it is suspended by five 
bridges over the CNR-INM towing tank. The guide can be set at different 
inclinations with respect to the still water level of the basin, thus yielding 
different vertical-to-horizontal velocity ratios, in the range of 0.03 to 0.05.

The trolley carries a waterproof box that hosts a ruggedized case, with 
the acquisition system inside. The specimens to be tested are directly 
connected to the the bottom of the trolley. 
The trolley is accelerated by a system of eight bungee cords, set in a
V-shaped configuration, but it is left to move freely along the guide 
shortly before the impact of the specimen with the water. 
Hence, the hydrodynamic loads acting on the model and the constraint
forces that the trolley wheels exert on the guide are the only forces
acting on the system, since gravity can be considered negligible. 
The total mass of the trolley and acquisition box is about 900~kg, which
is large enough to ensure that the horizontal velocity remains nearly
constant during the impact, being the velocity reduction
during the impact phase lower than 5\% of the initial velocity.

A video showing the execution of one of the experiments can be found
at the link \url{https://www.youtube.com/watch?v=1j5-Z96NNyQ}.

\subsection{Specimen shapes}
\label{sec:shapes}
\begin{figure}
\centering
\subfigure[Side View]
{\includegraphics[width=0.5\textwidth]{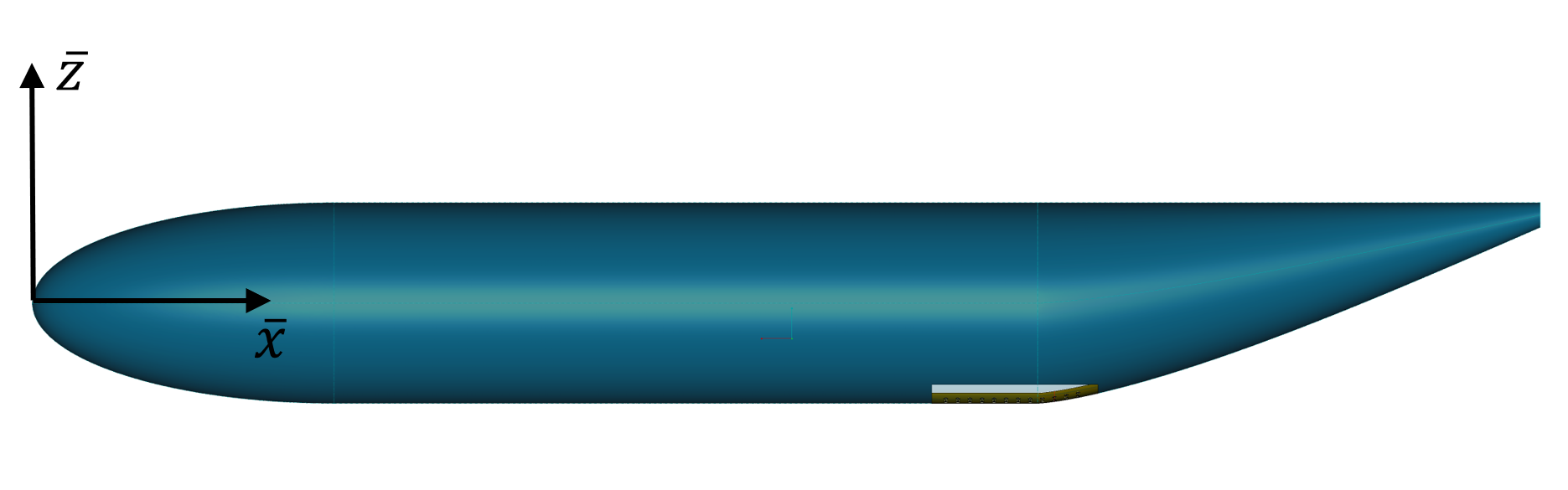}} \\
\subfigure[3D View]
{\includegraphics[width=0.5\textwidth]{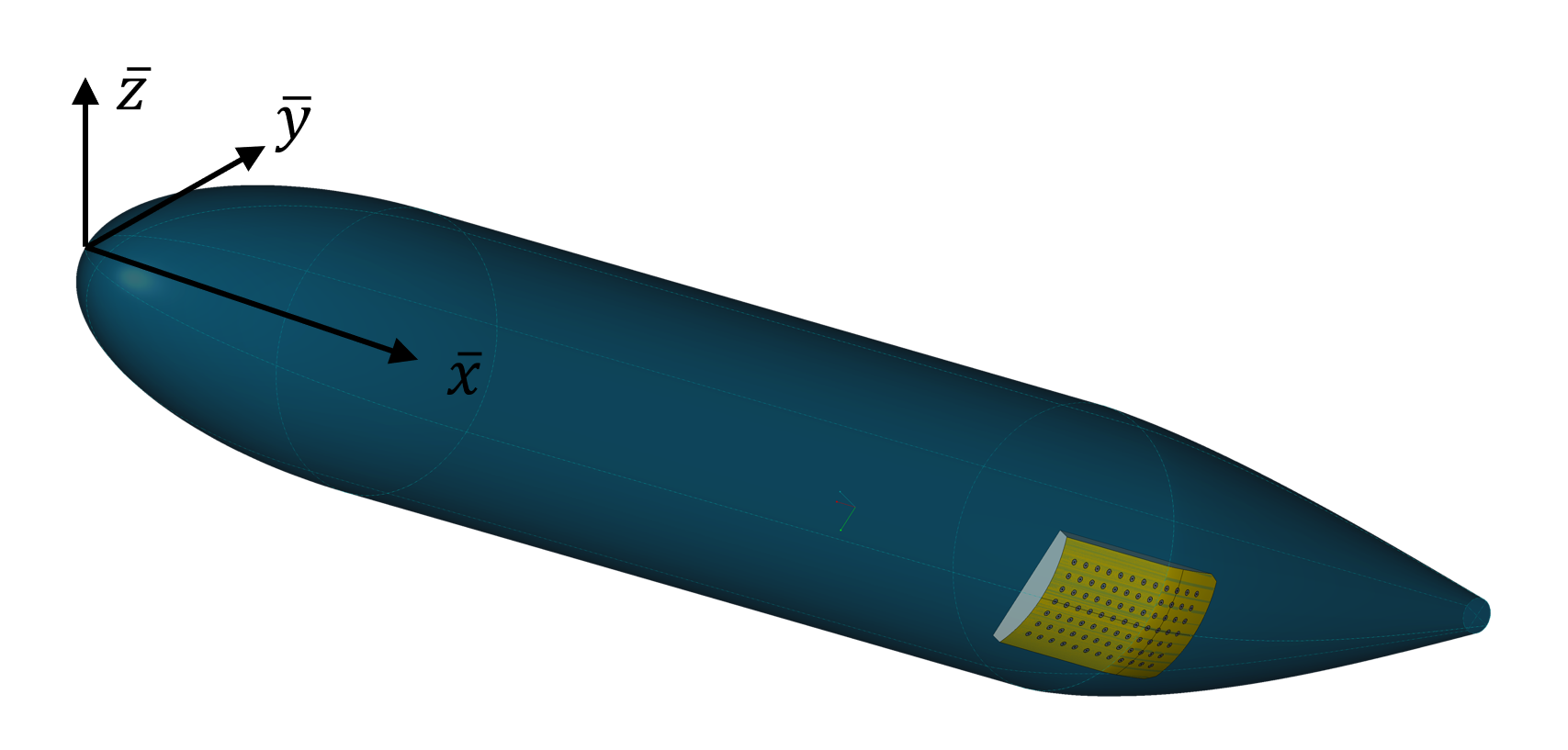}}
\caption{Drawing of the whole fuselage \textbf{S2} (blue) 
and of the extracted specimen part (yellow) in (a) side
view and (b) 3D view. The local coordinate system $(\bar{x},\bar{y},\bar{z})$
is also shown.} 
\label{fig:fuselage_and_specimen}
\end{figure}

%
The tested models are double curvature specimens that reproduce
the part of the fuselage body that first gets in
contact with water during ditching. Just to give an idea, the typical 
shape of the specimen is the yellow portion of the entire fuselage
shown in Figure \ref{fig:fuselage_and_specimen}.
Three fuselage shapes are considered, denoted here as \textbf{S1B}, 
\textbf{S2} and \textbf{S3}, which are described in a fully analytical form
in \citet{iafrati2020experimental}. For the readers' convenience the
details of the analytical formulation are also provided in the following.

\begin{figure}
\centering
\subfigure[Midplane section of \textbf{S1B}]{\includegraphics[width=0.75\textwidth]{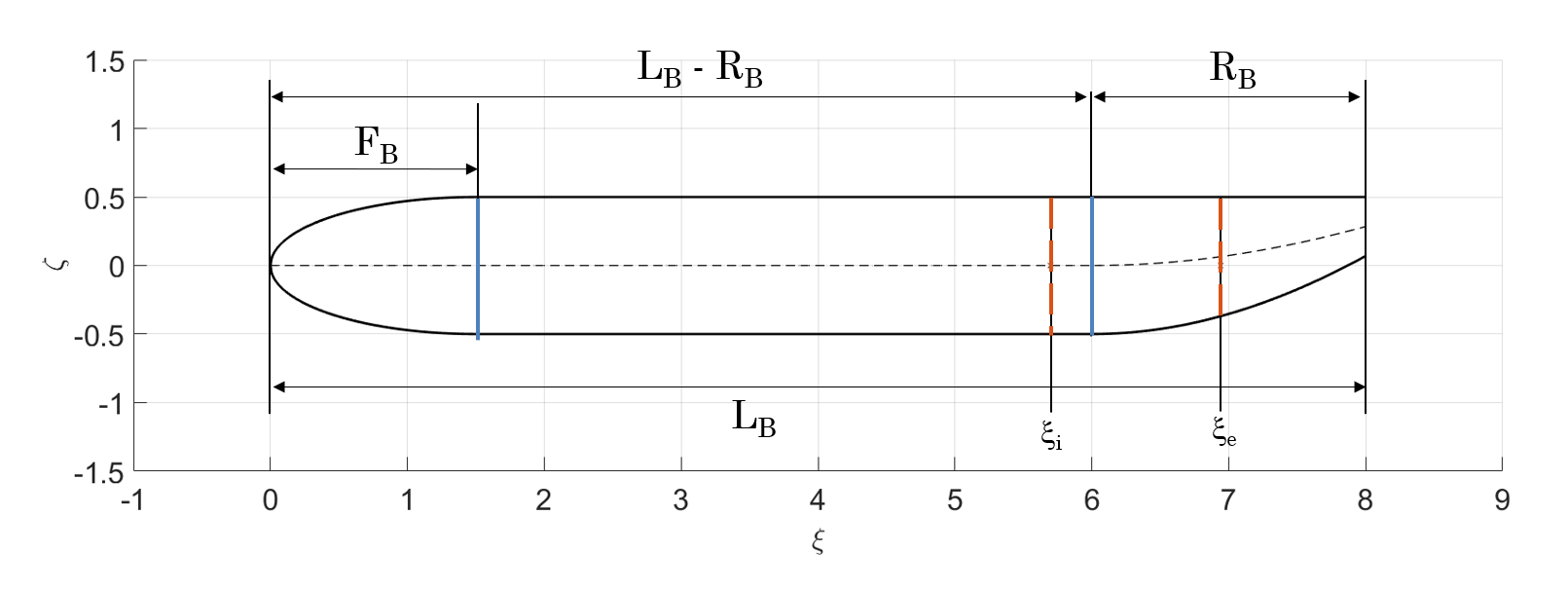}}
\\
\subfigure[Midplane section of \textbf{S2} and \textbf{S3}]{\includegraphics[width=0.75\textwidth]{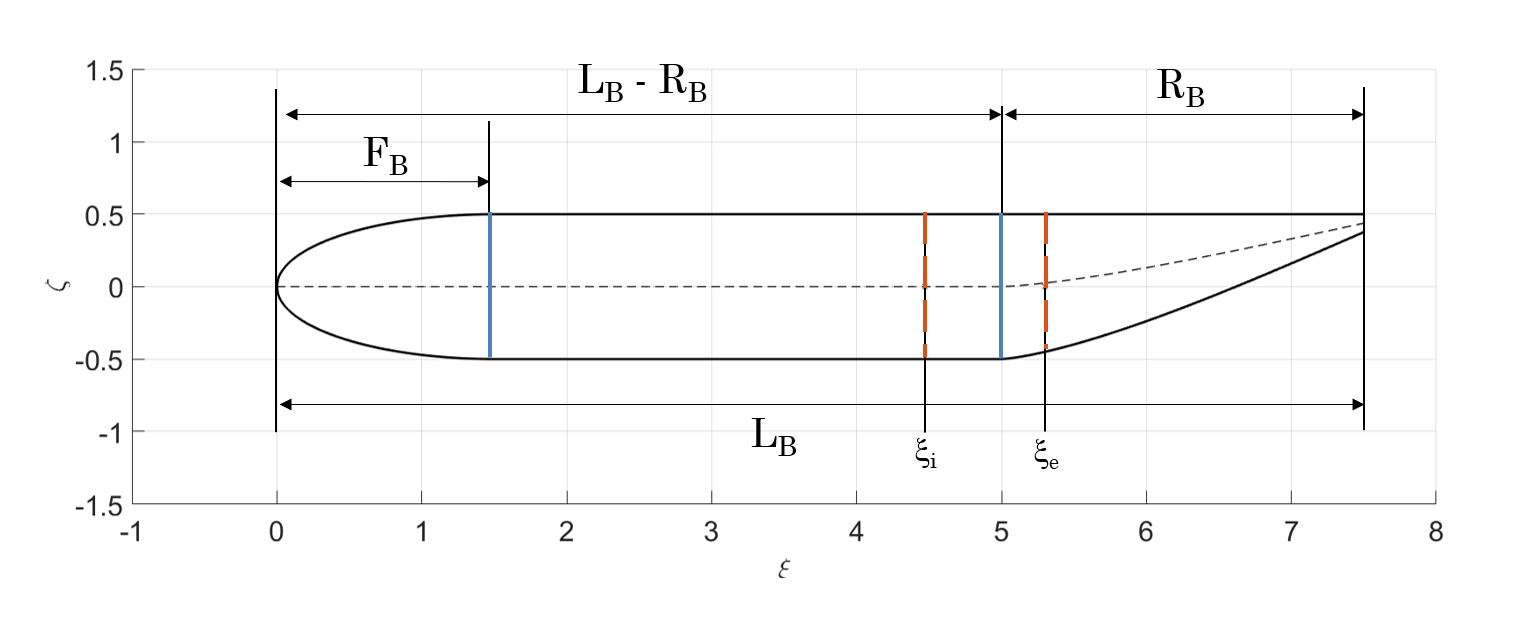}}
\\
\subfigure[Circular-elliptical cross-section of \textbf{S3}]{\includegraphics[width=0.7\textwidth]{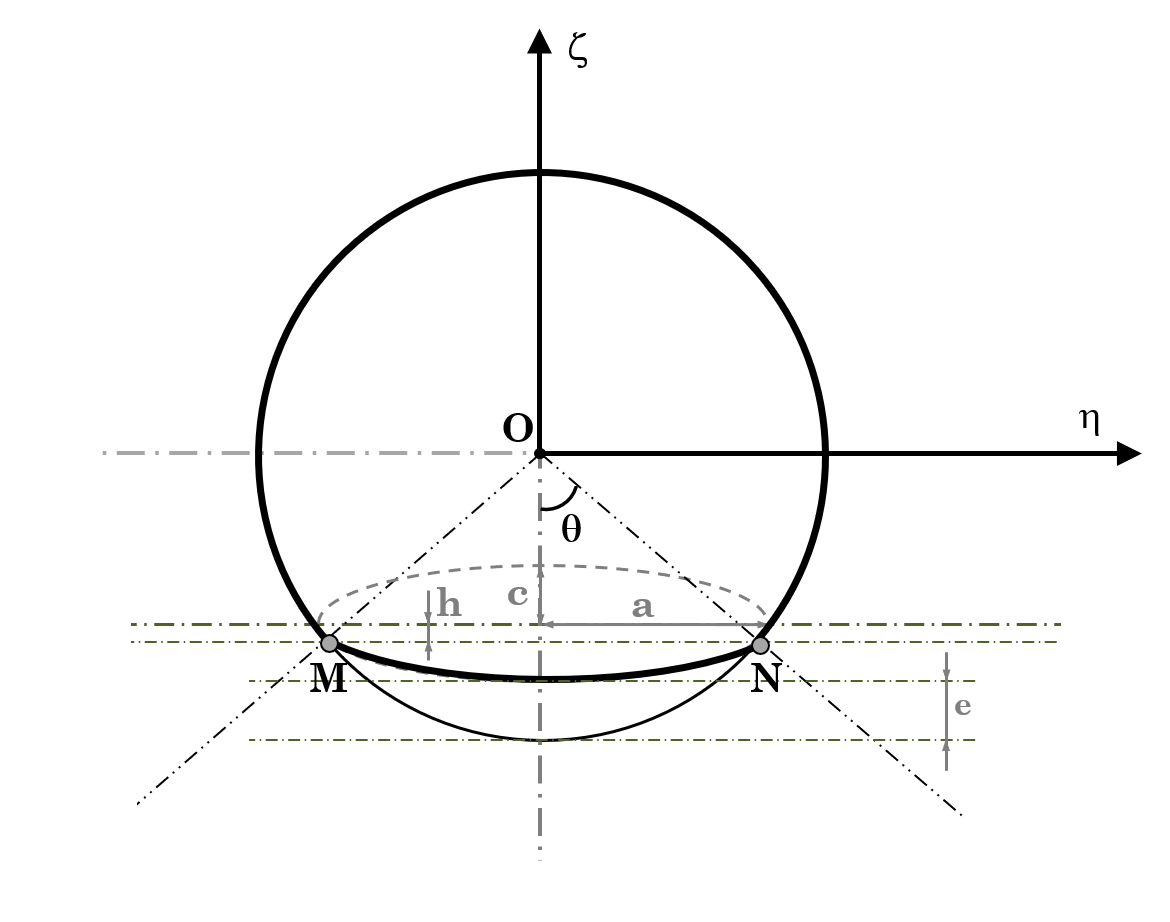}}
\caption{Non-dimensional mid-plane sections of the fuselages
\textbf{S1B} (a) and \textbf{S2}/\textbf{S3} (b), 
where $\xi_i = \bar{x}_i/B$ and $\xi_e=\bar{x}_e/B$, see 
Table \ref{tab:fuselage_shape_parameters}. In (c) the construction
of the semi-elliptical cross-section of the shape \textbf{S3}
is sketched.}
\label{fig:shapes_sideview_with_quotes}
\end{figure}

The shapes are defined in terms of a few non-dimensional parameters,
namely, Length-to-Breadth ratio ($L_B$), Front length-to-Breadth ratio
($F_B$) and Rear length-to-Breadth ratio ($R_B$), where $B$,
the maximum breath of the fuselage, is assumed as the reference length value.
Hence, non-dimensional coordinates are introduced as: $\xi=\bar{x}/B$, 
$\eta=\bar{y}/B$ and $\zeta=\bar{z}/B$ (Figure
\ref{fig:shapes_sideview_with_quotes}).
All shapes are defined by a central region with a constant circular cross-section:
a fore part where the circular cross-sections are uniformly and progressively 
reduced about the centre of the circle until the nose, 
and a rear part where the shape is shrunk, 
with the centre of the circle shifted
in the longitudinal plane in a way that the upper points
are always located at $\zeta = 0.5$.
%
%
The shapes \textbf{S1B} and \textbf{S2} have both a circular
cross-section whose radius $\hat{r}(\xi)$ varies with the 
longitudinal coordinate $\xi$ as follows:
\begin{equation}
\hat{r}(\xi) = \left\{ 
\begin{array}{c}
0.5 \sqrt{1-{\left(\frac{\xi-F_B}{F_B}\right)}^2} \qquad 0 < \xi < F_B \\ \\
0.5  \qquad F_B < \xi < \left(L_B - R_B \right) \\ \\
0.5 + \hat{o}_C(\xi) \qquad \left( L_B - R_B \right) < \xi <L_B
\end{array}
\right. 
\end{equation}
where the offset function $\hat{o}_c(x)$ is defined as
\begin{equation}
\hat{o}_C (\xi) = -0.5 \sin{ \left( \frac{ \xi - (L_B-R_B) }{K \, R_B}
\right) } \cdot \left[ \frac{\xi- (L_B-R_B)}{\sin{ \left(1/i \right) } 
\, R_B} \right]^{1/i}
\label{eq:offset_1}
\end{equation}
$i$ being a non-dimensional parameter. The centre of the cross-section
is located at $\textbf{\textrm{O}} \equiv (\xi,0,0)$ for $\xi \leq
(L_B-R_B)$ and at $\textbf{\textrm{O}} \equiv \left(\xi,0,\hat{o}_C(\xi) 
\right) $ for $\xi > (L_B-R_B)$, thus ensuring that the sections are all
aligned at $\zeta = 0.5$ in the rear fuselage portion.

The shape \textbf{S3} has a circular cross-section blended with an 
elliptical cross-section at the bottom. Starting from the circular
profile, the elliptical portion is defined by two parameters, i.e the 
angle $\theta$ and the ratio $c/h$ of the shorter ellipse semi-axes to 
the offset $h$, see Figure \ref{fig:shapes_sideview_with_quotes}(c).
The contact and the tangency between the circumference and the ellipse 
are enforced at the contact points $M$ and $N$, by choosing the two
semi-axes $a$ and $c$ as:
\begin{equation}
a=\frac{c}{h} \, \frac{ \sin{\theta}}{\sqrt{ {\left(\frac{c}{h} \right)}^2-1}} \qquad c=\frac{a \, {\tan{\theta}}}{\sqrt{{\left(\frac{c}{h}\right)}^2-1}},
\end{equation}
and therefore the bottom part of the section is described by the equation
\begin{equation}
\left( \frac{\eta}{a}\right )^2+\left( \frac{\zeta +1-(e+c)}{c} \right)^2=1,
\end{equation}
where
\begin{equation}
e=1-\cos{\theta}-c \, \left( 1-h/c \right) \;.
\end{equation}

When the circular-elliptical section is used, the equation of the offset
is modified as
\begin{equation}
 \hat{o}_{\textrm{CE}}(\xi) = 2 \hat{o}_C(\xi)/e.
	\label{eq:offset_2}
\end{equation}
which ensure that the bottom profiles of shape \textbf{S2} and
\textbf{S3} are perfectly overlapped.

The parameters that define the three fuselage shapes are listed
in Table \ref{tab:fuselage_shape_parameters}. Therein, the dimensional
parameters $\bar{x}_I$ and $\bar{x}_E$ defining the limits of the 
specimen in the longitudinal direction are also provided. Note that in
all cases the specimens are 1.24~m long and 0.66~m wide.
\begin{table}
\begin{center}
\def~{\hphantom{0}}
\begin{tabular}{ccccc}
Parameter  & \textbf{S1B} & \textbf{S2} & \textbf{S3} \\
$L_B$ [-]     & 8     & 7.5  & 7.5  \\
$F_B$ [-]     & 1.5   & 1.5  & 1.5  \\
$R_B$ [-]     & 2     & 2.5  & 2.5  \\
$K$   [-]     & 2     & 1.55 & 2.55 \\
$i$   [-]     & 1     & 2.6  & 2.6  \\
$C/H$ [-]    & -     & -    & 5    \\
$\theta$ [$^{\circ}$]   & -     & -    & 50   \\
$B$ [m]   & 1  & 1.5 & 1.5 \\
$\bar{x}_I$ [m] & 5.7  & 6.71 & 6.71 \\
$\bar{x}_E$ [m] & 6.94  & 7.95 & 7.95
\end{tabular}
\caption{Parameters defining the whole fuselage shapes and the extracted specimens}
\label{tab:fuselage_shape_parameters}
\end{center}
\end{table}
It is worth noting that the specimen \textbf{S1B} is extracted from a 
fuselage with a breadth (diameter) of 1~m, whereas the fuselage 
\textbf{S2} and \textbf{S3} from a fuselage with a breadth (diameter) of 1.5~m.
The different choice of the was motivated by the need of maximizing, in the 
case of  the specimen \textbf{S1B}, the vertical distance 
between the lowest point, i.e. the first contact point, and the
leading edge of the specimen, which defines the time at which the front of
the specimen gets below the still water level and thus the end of the
impact phase.

\begin{figure}
\centering
\includegraphics[width=0.98\textwidth]{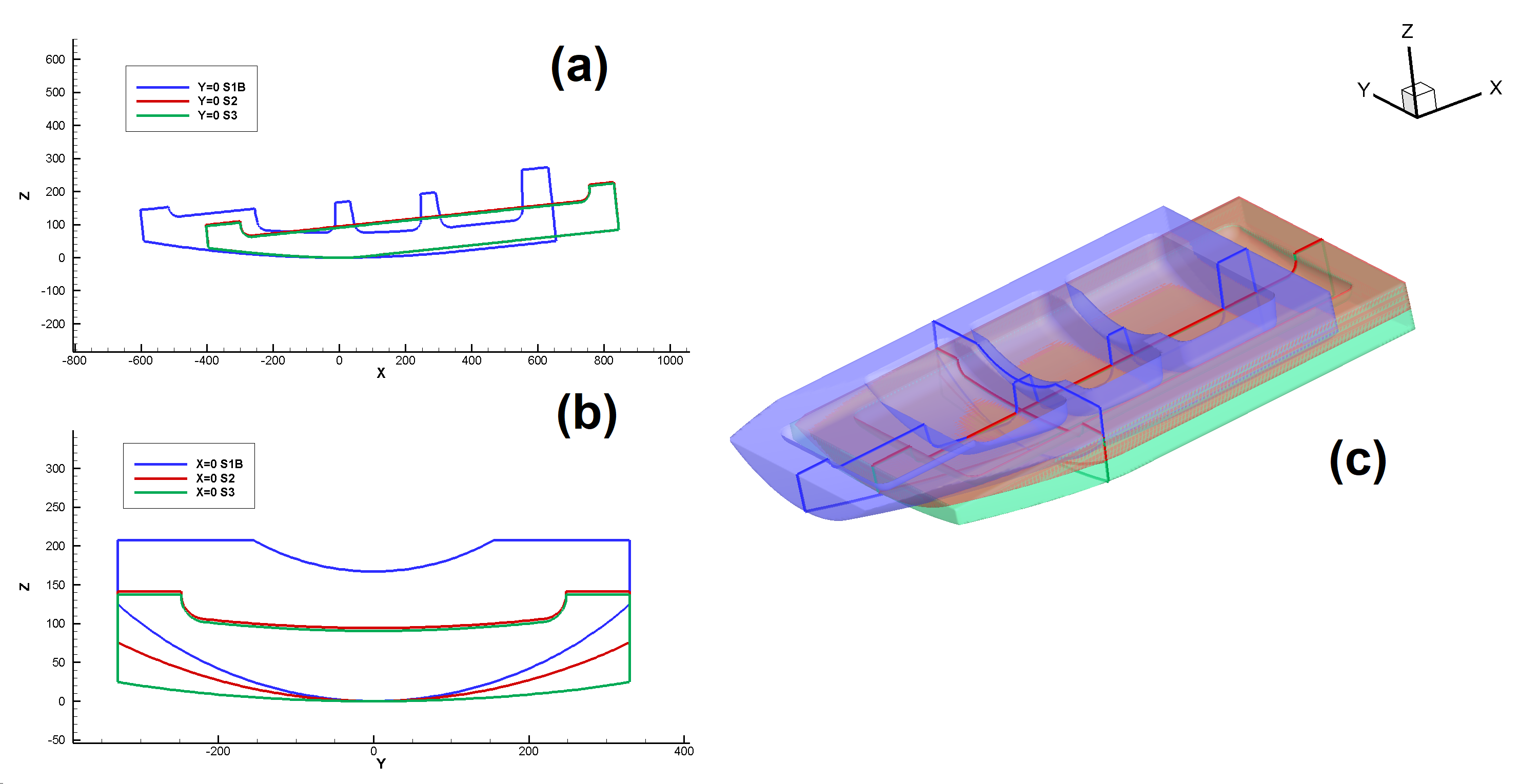}
\caption{Comparison among the geometry of the three specimens aligned with respect to
the lowest Z point, i.e the point of first contact with water:
(a) section on the longitudinal plane 
(b) section on the transverse plane and (c) 3D view}
\label{fig:three_shapes_slices}
\end{figure}

The difference among the three shapes, set at a pitch angle of 6$^{\circ}$, 
is highlighted in \ref{fig:three_shapes_slices}(c), in which a 
three-dimensional view is shown, whereas the longitudinal and transverse 
sections passing through the point of first contact with water, are shown on 
\ref{fig:three_shapes_slices}(a) and (b) respectively.

\begin{figure}
	\centering
		\includegraphics[width=0.98\textwidth]{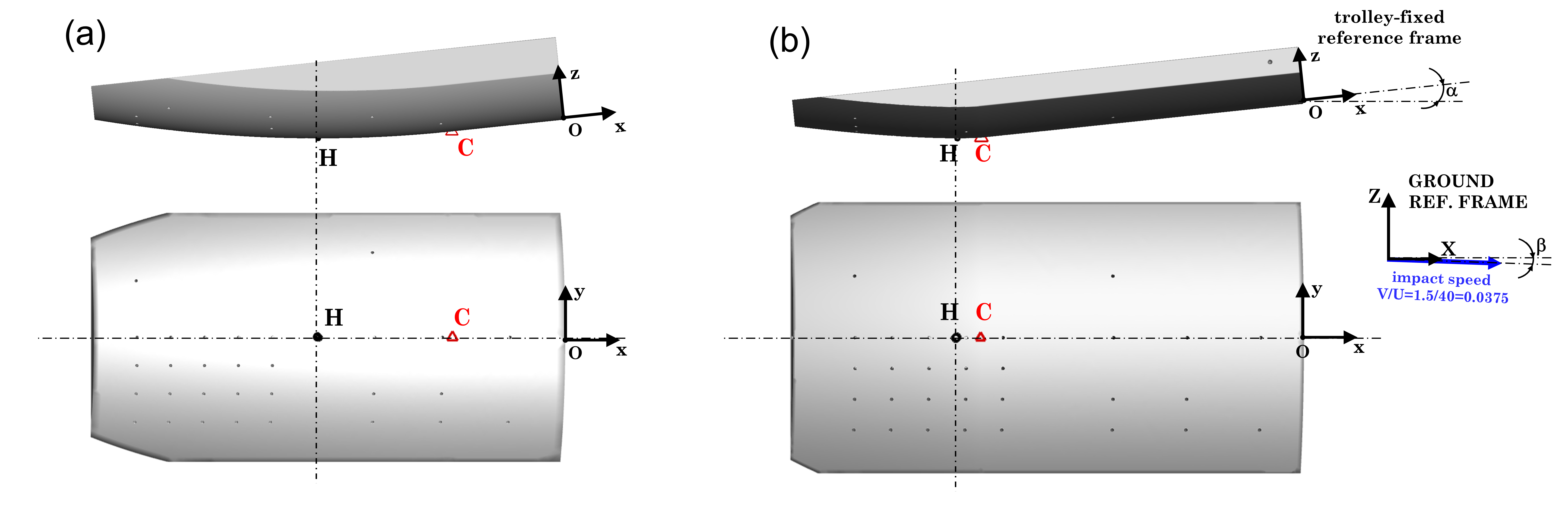}
	\caption{Reference frames, coordinate systems and points of interest 
to study the water entry of the shapes \textbf{S1B} (a) and \textbf{S2}/\textbf{S3} (b) 
at a pitch angle of $\alpha = 6^{\circ}$.}
	\label{fig:ref_frames_points_ALL}
\end{figure}

For the experimental data analysis two reference frames are defined: 
a \emph{ground fixed reference frame} and a \emph{trolley-fixed
(or specimen-fixed) reference frame},
both shown in Figure \ref{fig:ref_frames_points_ALL}.
The coordinates associated with the ground-fixed reference frame are 
indicated with upper-case letters $X$, $Y$ and $Z$. 
In particular, $X$ is parallel to the undisturbed free surface, 
$Z$ is vertical and directed upwards and $Y$ is in the direction
on a right-handed coordinate system. The specimen horizontal
and vertical speed are defined in this reference frame as $U=dX/dt$ and $V=-dZ/dt$,
respectively.
The coordinates associated with the trolley-fixed reference frame are
indicated with lower-case letters $x$, $y$ and $z$, with origin \textbf{O}.
As shown in Figure \ref{fig:ref_frames_points_ALL}
the origin is located at the leading edge of the plate. The
$x$ coordinate is parallel to the longitudinal axis of the
trolley, the $z$-coordinate is normal to it and directed
upwards, whereas the $y$ axis forms and right-handed coordinate system.
The pitch angle $\alpha$ is defined as the angle between the trolley axis $x$ 
and the $X$-direction, whereas the angle $\beta=\arctan{V/U}$ is introduced to
denote the angle between the the impact trajectory of the specimen and the 
undisturbed water level.

For each shape, at given $\alpha$ and $\beta$, it is worth defining the
point \textbf{C} as that where the longitudinal curvature starts changing at 
the rear, which is the point with dimensional $x$-coordinate 
$(L_B-R_B)$ in Figure \ref{fig:shapes_sideview_with_quotes}(a,b), and the point
\textbf{H}, which defines the lowest point of the specimen, and thus the
first getting in contact with water.
As it is possible to see, the two points are very close to each other for the
fuselage shapes \textbf{S2} and \textbf{S3} and relatively far in case
of shape \textbf{S1B}.

\subsection{On-board Measurements}
\label{sec:onboard_measurements}
\begin{figure}
\centering
\subfigure[Shape \textbf{S1B}]
{\includegraphics[width=0.75\textwidth]{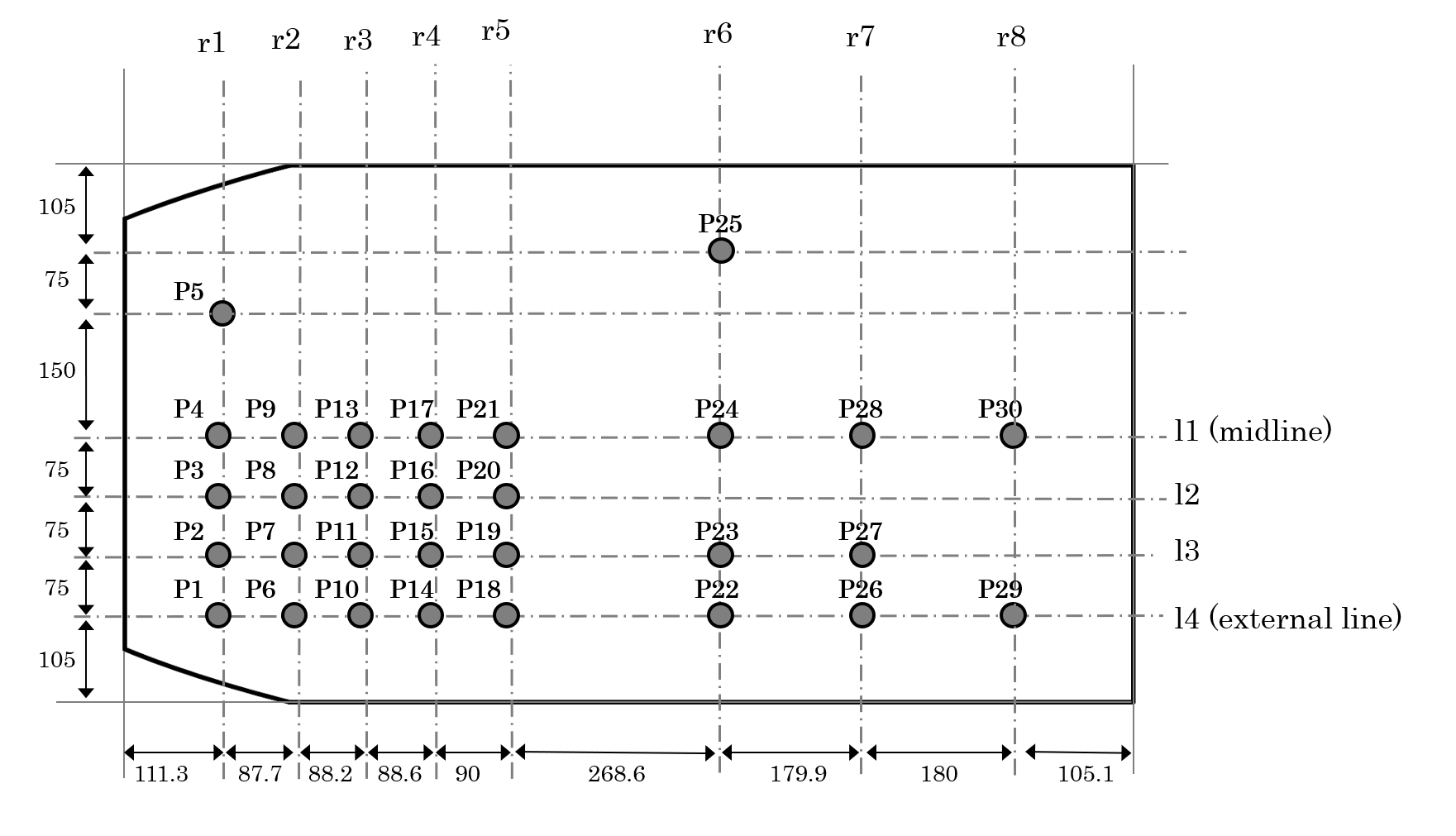}} \\
\subfigure[Shapes \textbf{S2} and \textbf{S3}]
{\includegraphics[width=0.75\textwidth]{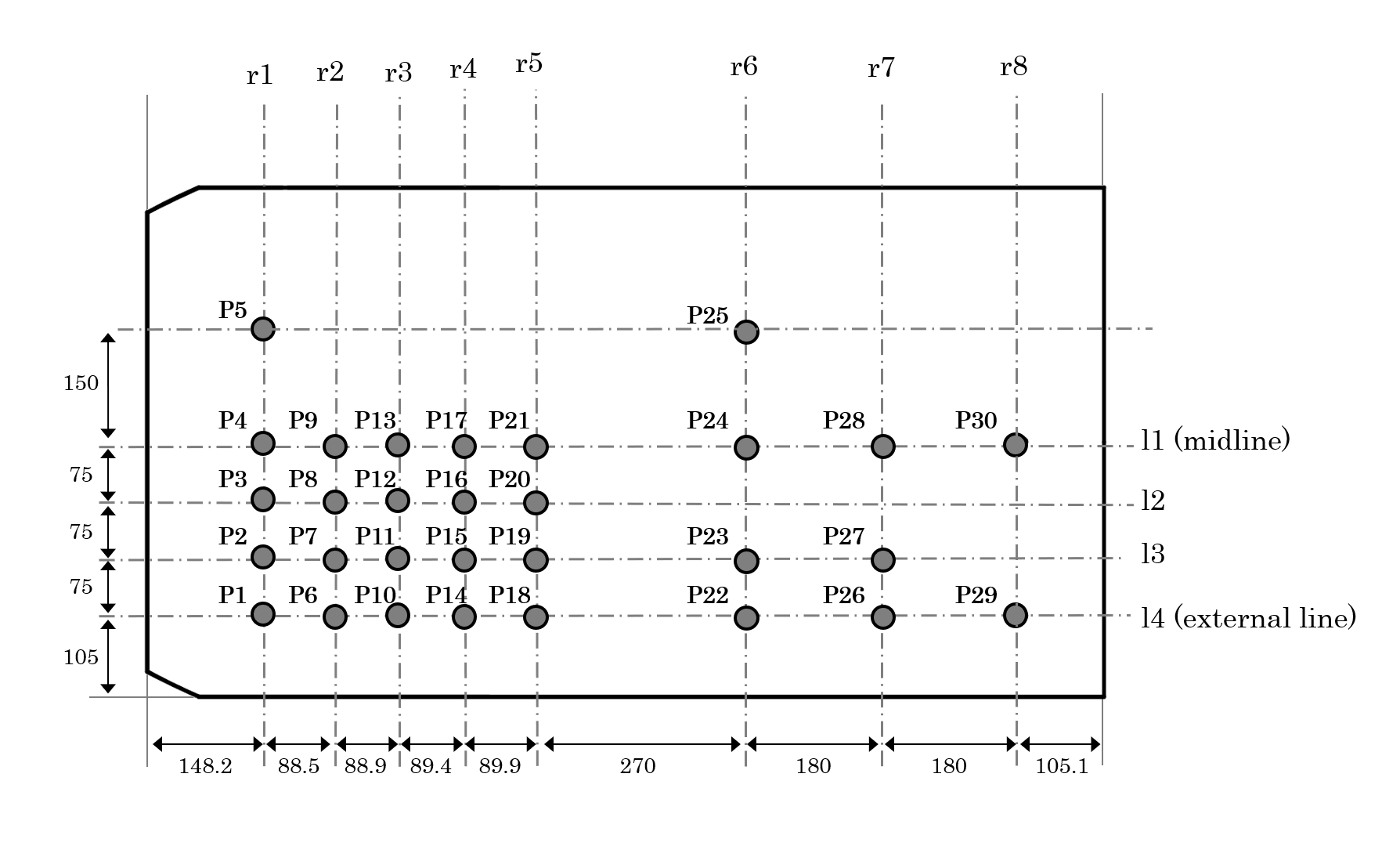}}
\caption{Position of the pressure probes and definition of the row and 
lines used for the discussion. Dimensions are in mm.}
\label{fig:Pressure_Probe_Positions}
\end{figure}

%
Pressures are acquired through thirty pressure probes type Kulite XTL 123B, 
full scale range of 300 psi.
The probes were carefully calibrated 
by the manufacturer prior to delivery and were subsequently re-calibrated
for verification in the metrology laboratory of CNR-INM before installation.
It was found that the sensitivity values were, in all cases,
very close to those provided by the manufacturer, with a discrepancy of
approximately 1\%. Following standard procedures,
the relative expanded uncertainty attributed to random effects was determined
to be 1.7\% of the average measured pressure at a 95\% confidence level
and 0.9\% at a 67\% confidence level. The probe was also tested at a few temperatures 
within its 0–40\,°C compensation range, showing a linear sensitivity of about 0.021\,kPa/°C. 
A typical 5\,°C temperature change, experienced as the probes operate between
approximately 20\,°C and 15\,°C, causes a negligible pressure shift of 0.105\,kPa
(0.0051\% of full scale). This drift is expected and does not require
any further compensation.
The probes are installed flush to the external surface of
the specimen with $\pm $ 0.1~mm accuracy.
The position of the probes is given in Figure 
\ref{fig:Pressure_Probe_Positions}.
In the analysis of the data, beside the specific probe number, 
reference is also made to probe lines ($l1,l2,l3,l4$) and rows
($r1,r2,...,r8$) as indicated in the figures.
The pressure time histories presented in the following
are raw data, unless otherwise specified. 
It is worth noting that sometimes pressure spikes
of non-physical origin may appear, caused, among other reasons,
either by slightly protruding probes generating cavitating vortices
at such speeds \citep{iafrati2015high} or by the temperature shock
experienced by the sensors when they are suddenly immersed in water
at a lower temperature than air \citep{van2013study}. 
It is important to clarify that the presence of these spikes
is not in contradiction with the considerations on temperature sensitivity made above,
since they result from sudden temperature shocks, whereas sensitivity tests are
performed by manually applying temperature variations, which are consequently much more gradual.

\begin{figure}
\centering
\includegraphics[width=0.75\textwidth]{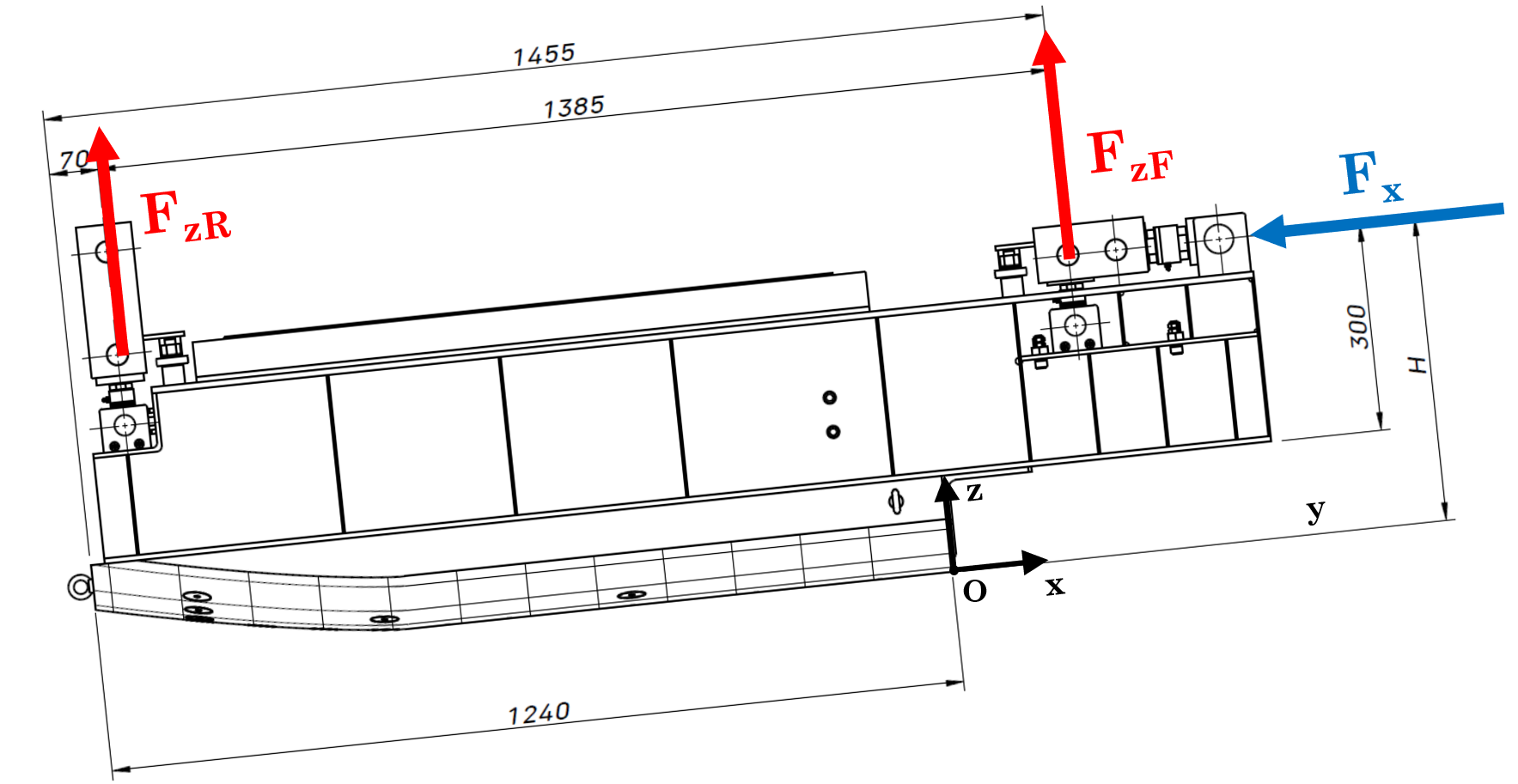}
\caption{Position of the load cells, referring to the specimen \textbf{S2}. 
The total rear and front force direction is indicated, as well as the 
trolley-fixed reference frame. The dimensions are in mm. The value of 
$H$ is 520 mm, 440 mm and 436.2 mm for the shapes \textbf{S1B}, \textbf{S2} 
and \textbf{S3}, respectively.}
\label{fig:LoadCell_Ref_Frame}
\end{figure}

The acquisition box hosting the specimen is connected to the trolley by 
a total of six single-axis load cells, installed between two floating
pins to minimize the loading in other directions, as shown in Figure
\ref{fig:LoadCell_Ref_Frame}.
Four Kistler 9343 cells, with full scale range of 70~kN,
are used in the $z$-direction, whereas two Kistler 9363 cells, 
with full scale range of 120~kN, are used in the $x$-direction.
The load cell signals are conditioned using a charge amplifier type 
Kistler 5073-A421.
The channels are connected so that the loads acting in the $z$-direction
can be separated into the rear $F_{zR}$ and front $F_{zF}$
contributions, whereas $F_{x}$ denotes the load acting in the $x$-direction.
The distance $H$ for the three specimens \textbf{S1B}, \textbf{S2} and 
\textbf{S3} are 520~mm, 440~mm and 436.2~mm respectively. 
Although the single load cell is very accurate when loaded directly
along its axis, the connection to the frame of the acquisition box 
introduces some uncertainties, since the loading is not exactly aligned with
the axes of the cells. Furthermore, due to the inherent free play of the
installations, some errors may be expected for low values of the loading.
For this reason a careful calibration was carried out before
starting the test campaign, and it was found that the recorded data are
always below the actual loading. The underestimate ranges from
about 8\% for a load of order of 1.2~kN and reduces continuously when
increasing the load, being about 4\% for a load of order of 14~kN.
The load was also applied at different locations along the $x$-axis and 
the centre of loads could be retrieved with a satisfactory accuracy by the 
combined data of the rear and the front load cells.
Unless specified differently, also the force time histories 
provided in the following are raw data.

All the data are acquired by a ruggedized acquisition system hosted in
the acquisition box.
The acquisition system is composed of four Dewesoft$^{\copyright}$ Sirius 
modules and one Dewe43 module. All channels are sampled at 200~kHz with a 
double level A/D 24 bit conversion. 

The repeatability of the pressure and load
measurements was thoroughly investigated by \citet{iafrati2015high}
in the context of flat plate tests conducted in the HSDF.
These considerations are directly applicable to the present experiments,
as the same types of pressure probes, load cells
and data acquisition systems were employed. Nevertheless, 
to verify the experimental repeatability under the current conditions,
each test was repeated at least twice. The resulting time
histories showed excellent agreement, with only minor discrepancies.
These were primarily attributed to the inherent difficulty in precisely
reproducing the same target nominal velocity, 
which led to a small variation of approximately 1\%.

%
\subsection{High-Speed Cameras}
%
Two high speed cameras are used to record the tests. 
One camera, type Photron FastCam Mini AX, operating at 5000 fps, is installed at
the side of the facility at the location of the water entry, and it is mainly 
used to retrieve the horizontal velocity at the exact moment 
the specimen touches the water.
A second camera, type Photron FastCam SA-X2, operating at 3000 fps,
is installed on a properly designed waterproof case and positioned 
underwater. This camera records the water entry from below and provides 
important information for a more complete interpretation of the hydrodynamic
phenomena and of the pressures/loads measurements.
Both cameras are synchronized with the on-board acquisition systems.

\subsection{Test Conditions}    
%
The test conditions that are discussed in the present paper are listed in 
Table \ref{tab:test_conditions}. The guide inclination angle is $\beta = \arctan(V/U)$. 
For all tests examined $V/U=1.5/40=0.0375$, so $\beta=2.1476^{\circ}$
(see again Figure \ref{fig:ref_frames_points_ALL}).
It is worth noticing that, due to the use of bungee ropes to accelerate
the trolley, a very precise control of the speed
cannot be achieved, which is the
reason why the actual values of the horizontal speed are
not exactly the nominal expected values, also shown in the table. 
It should be noted that in the discussion from now on,
however, reference will be made to the nominal velocity
values and not to the actual ones reported in the table.

Water temperature was measured on a near-daily basis
using a thermometer. Recorded values ranged from 15 to 18$^{\circ}$ C. 
The air temperature was approximately 20$^{\circ}$ C.

The three groups refer to the test cases examined to investigate the effect 
of the horizontal speed, of the shape and of the pitch angle. 
The test condition highlighted in bold is the one used as a baseline test
condition when examining the effects of the above-mentioned parameters.
\begin{table}
\centering
\begin{tabular}{ccccccc}
\multicolumn{7}{c}{\emph{Effect of speed}} \\
 Shape & Nominal U & U & V & Pitch & V/U & $\left( t_E - t_0 \right)$ \\
       & [m/s]     & [m/s] & [m/s] & [$^{\circ}$] & [-]  & [s] \\
 S2    & 30    & 29.87 & 1.12 & 6     & 0.0375   & 0.0671 \\
 S2    & 35    & 34.5  & 1.29 & 6     & 0.0375   & 0.0515 \\
 \textbf{S2}    & \textbf{40} & \textbf{40.25} &  \textbf{1.51} &  \textbf{6}     &  \textbf{0.0375} & \textbf{0.0491} \\
 S2    & 45 & 46.19 & 1.73 & 6     & 0.0375 & 0.0436  \\
\multicolumn{7}{c}{\emph{Effect of shape}} \\
 Shape & Nominal U & U & V & Pitch & V/U & $\left( t_E - t_0 \right)$ \\
       & [m/s]     & [m/s] & [m/s] & [$^{\circ}$] & [-]  & [s] \\
 S1B   & 40 & 40.45 & 1.52 & 6     & 0.0375   & 0.0260 \\
 \textbf{S2}    & \textbf{40} &  \textbf{40.25} &  \textbf{1.51} &  \textbf{6}     &  \textbf{0.0375} &  \textbf{0.0491} \\
 S3    & 40 & 40.25 & 1.51 & 6     & 0.0375 & 0.0468  \\
\multicolumn{7}{c}{\emph{Effect of pitch angle}} \\
 Shape & Nominal U & U & V & Pitch & V/U & $\left( t_E - t_0 \right)$ \\
       & [m/s]     & [m/s] & [m/s] & [$^{\circ}$] & [-]  & [s] \\
 S2    & 40    & 40.57 & 1.52 & 4     & 0.0375 & 0.0314  \\
 \textbf{S2}    & \textbf{40} &  \textbf{40.25} &  \textbf{1.51} &  \textbf{6}     &  \textbf{0.0375} &  \textbf{0.0491} \\
 S2    & 40 & 40.08 & 1.50 & 8     & 0.0375 & 0.0670
\end{tabular}
\caption{Test Conditions.}
\label{tab:test_conditions}
\end{table}
%
%
%
%
\section{Results and discussion}
\label{sec:results}
%

%
\subsection{Definition of the impact phase}
%
For the purpose of the present study, the data analysis is limited to 
the {\em impact phase}, which lasts from the time of first contact of the 
specimen with the water surface, $t_0$, up to the time $t_E$ at which the 
root of the forward propagating spray reaches the leading edge of the specimen. 
As done in \cite{iafrati2019cavitation}, $t_0$ can be approximated as the time
at which the first pressure probe
exhibits a pressure growth, whereas $t_E$ is
estimated by linear extrapolation starting from the times at which the 
pressure peak passes over the location of the probes P28 
and P30 \citep{iafrati2016experimental}.
As shown in Figure \ref{fig:ref_frames_points_ALL}, for the shapes 
\textbf{S2} and \textbf{S3} at a pitch angle of 6$^{\circ}$ 
the estimation of $t_0$ based on the time history of the pressure measured 
by the probe P17 is expected to be quite accurate, being the probe P17 
located very close to the point of first contact \textbf{H}. 
Instead, for the shape \textbf{S1B} the first contact point is located between 
P21 and P24, therefore the estimate of $t_0$ is more uncertain. The estimate
based on pressure measurements is corrected by computing the time duration 
of the impact phase from the underwater videos and subtracting it from $t_E$. 

It is important to note that tests conducted on specimens can 
effectively capture the hydrodynamic behaviour of the entire fuselage ditching
(with the imposed trajectory, horizontal speed, and constant pitch angle),
as long as the wetted area during water entry remains within the specimen’s boundaries.
This definition of the impact phase ensures that the forward part of the wetted area stays
within the specimen’s area until it reaches the leading edge. Since the front and rear walls
are located above the specimen’s profile, they do not affect the pressures underneath,
at least until the impact phase ends. Minor discrepancies may occur,
after the cavitation region or the wetted area reaches the trailing edge or the lateral edges,
but these differences are believed to be small.

Once $t_0$ and $t_E$ are defined, a non-dimensional time variable, denoted 
as $\tau$ can be defined as
\begin{equation}
	\tau \doteq \frac{t-t_0}{t_E-t_0}\;.
\end{equation}
The use of the non-dimensional time $\tau$ allows a fairer comparison of the
results obtained at different test conditions, and, in particular, at 
different pitch angles.
%
%
\subsection{Hydrodynamics phenomena during the water entry}
\label{hydrodynamics_S2}
\begin{figure}
\centering
\subfigure[$\tau$~=0]
{\includegraphics[width=0.48\textwidth]{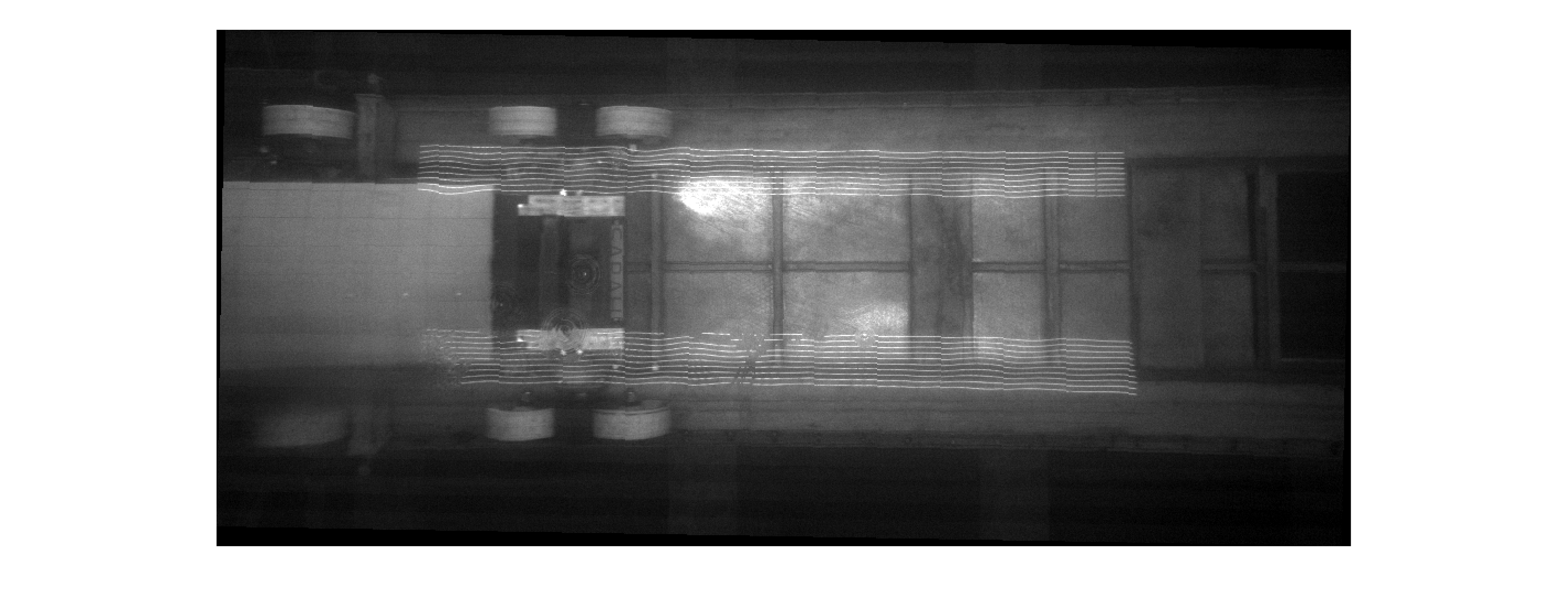}} 
\subfigure[$\tau$~=0.1]
{\includegraphics[width=0.48\textwidth]{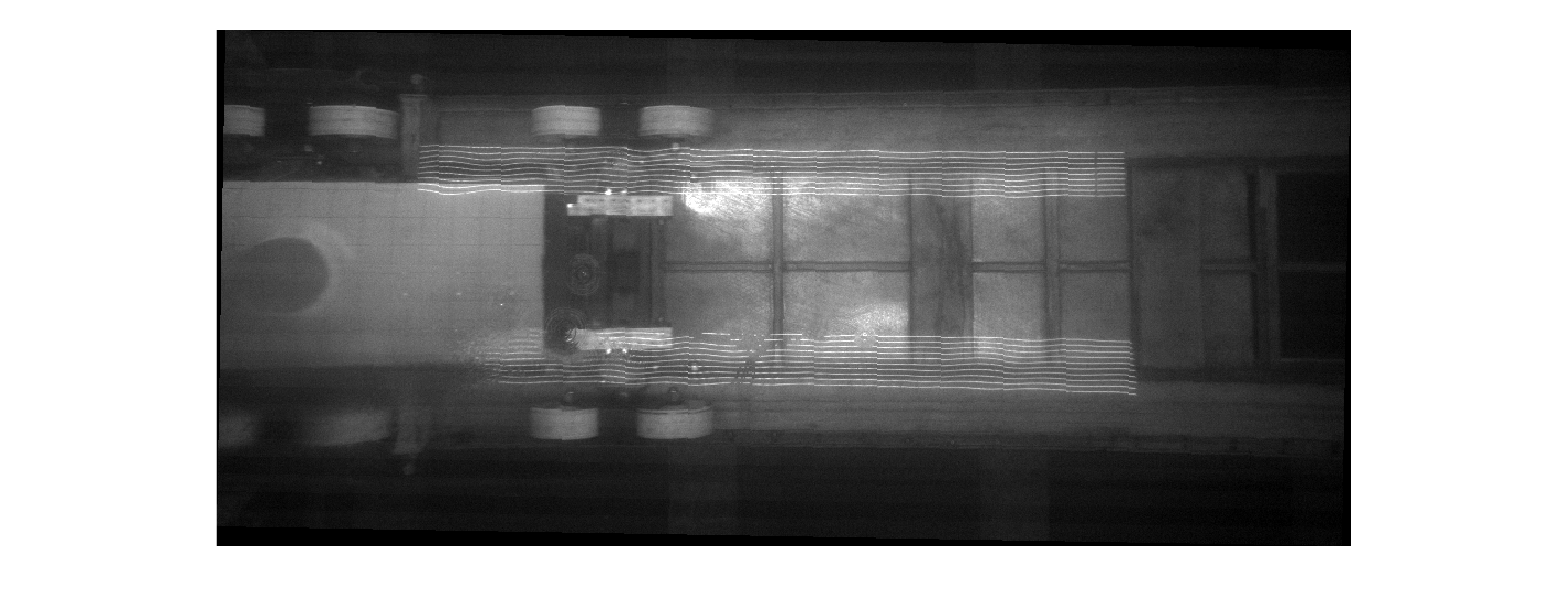}}
\\
\subfigure[$\tau$~=0.2]
{\includegraphics[width=0.48\textwidth]{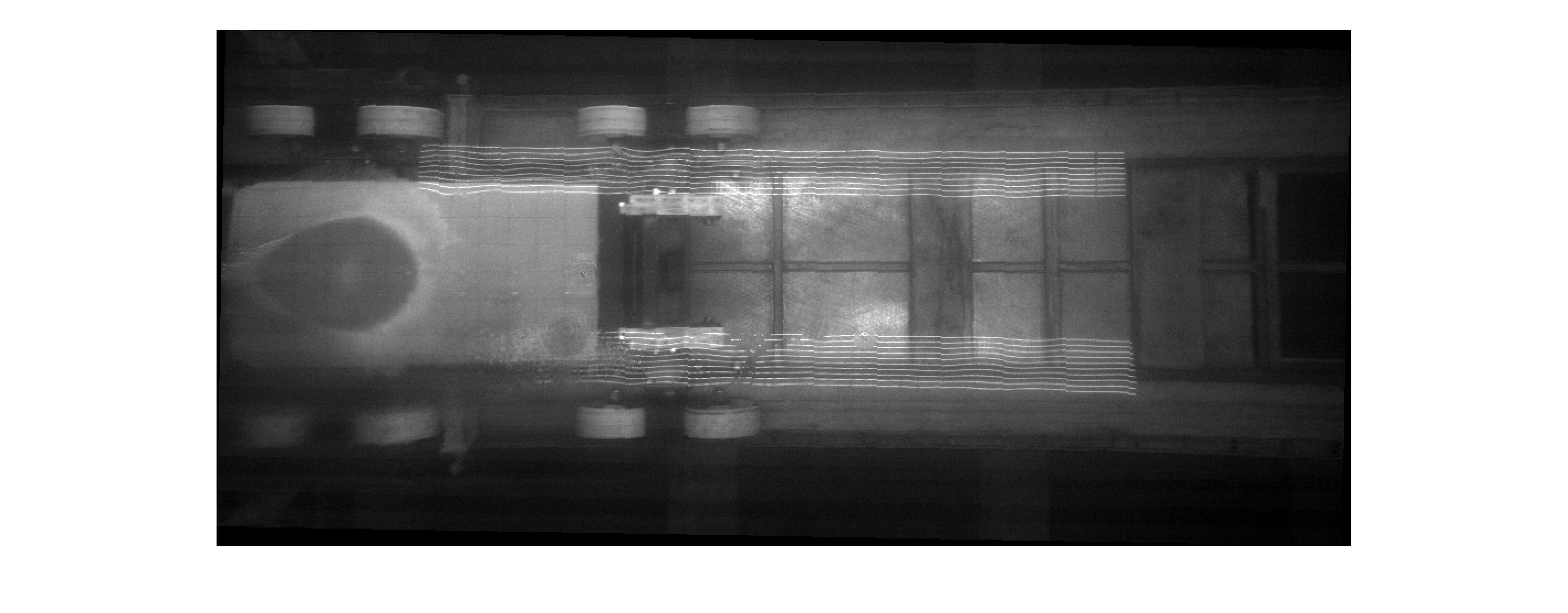}}
\subfigure[$\tau$~=0.3]
{\includegraphics[width=0.48\textwidth]{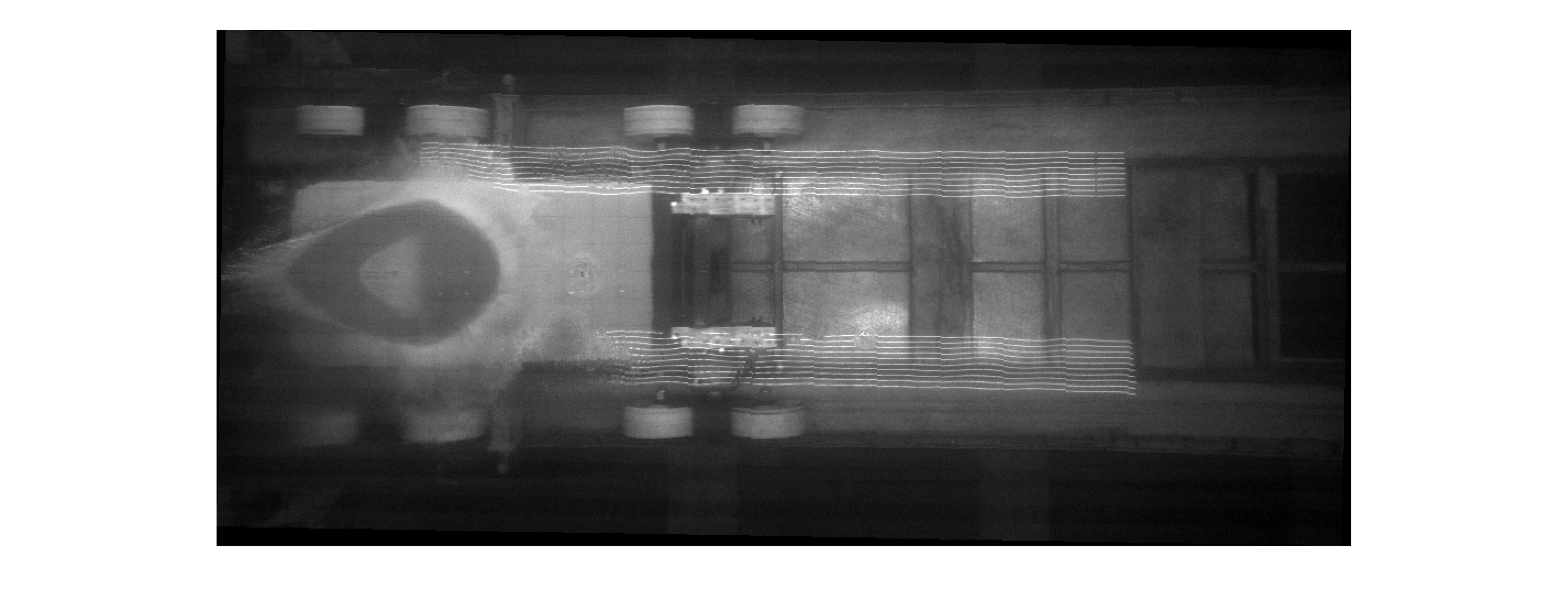}}
\\
\subfigure[$\tau$~=0.4]
{\includegraphics[width=0.48\textwidth]{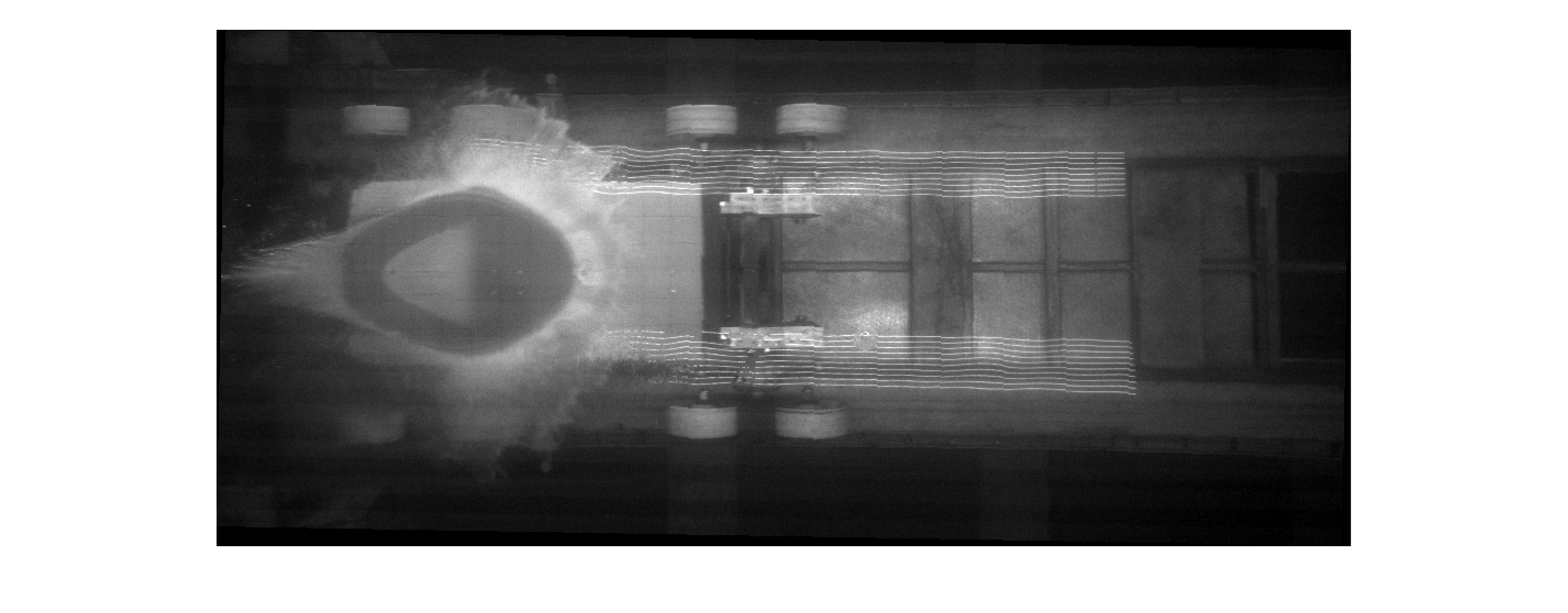}}
\subfigure[$\tau$~=0.5]
{\includegraphics[width=0.48\textwidth]{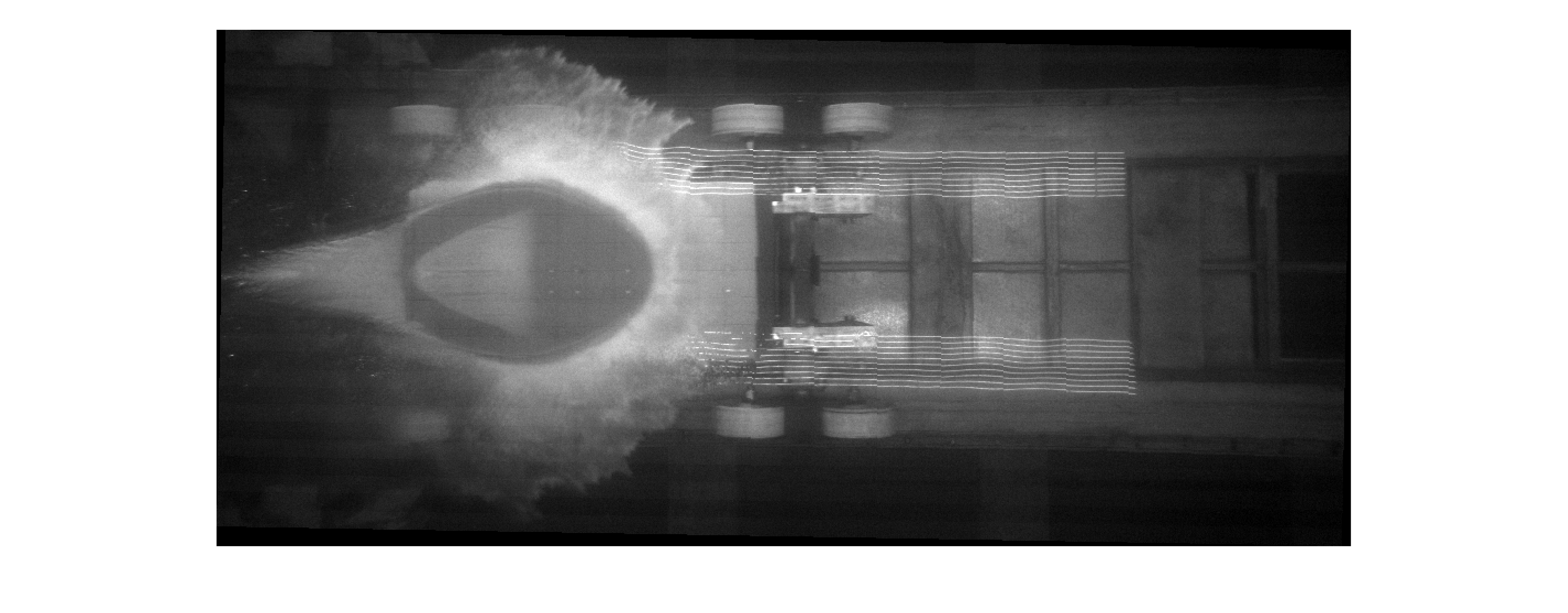}}
\\
\subfigure[$\tau$~=0.6]
{\includegraphics[width=0.48\textwidth]{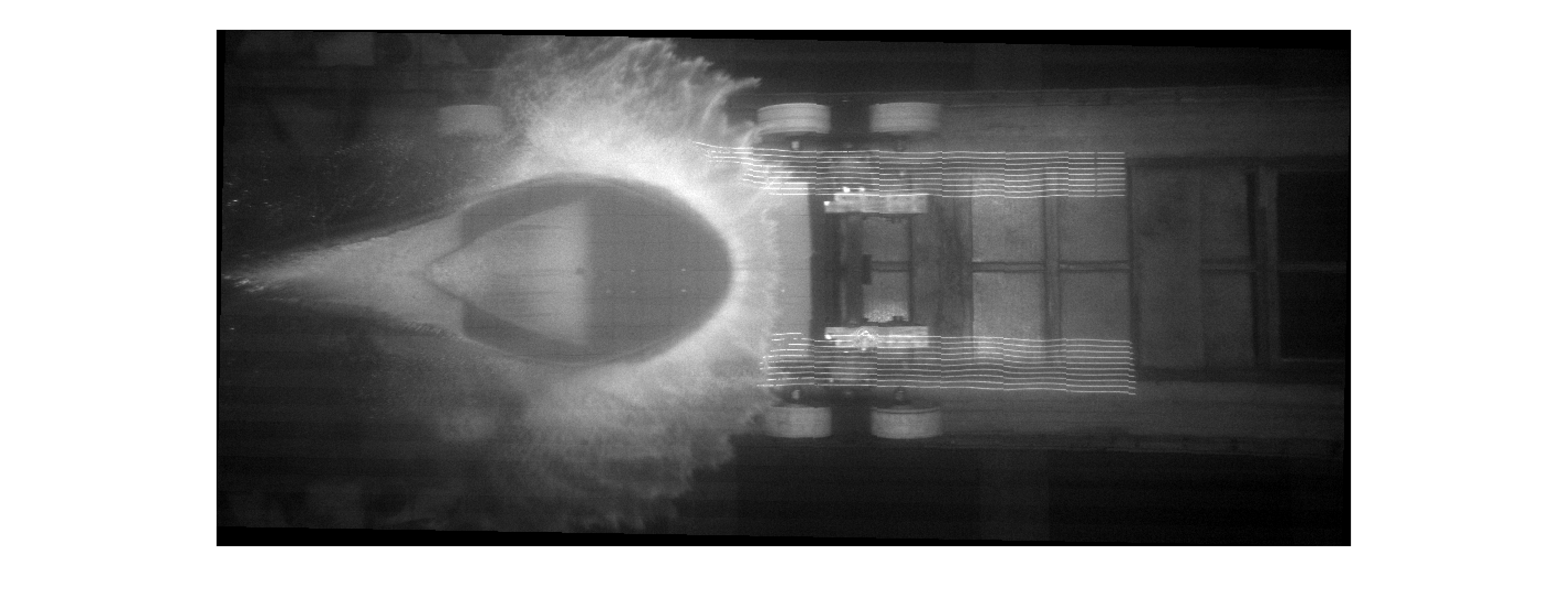}}
\subfigure[$\tau$~=0.7]
{\includegraphics[width=0.48\textwidth]{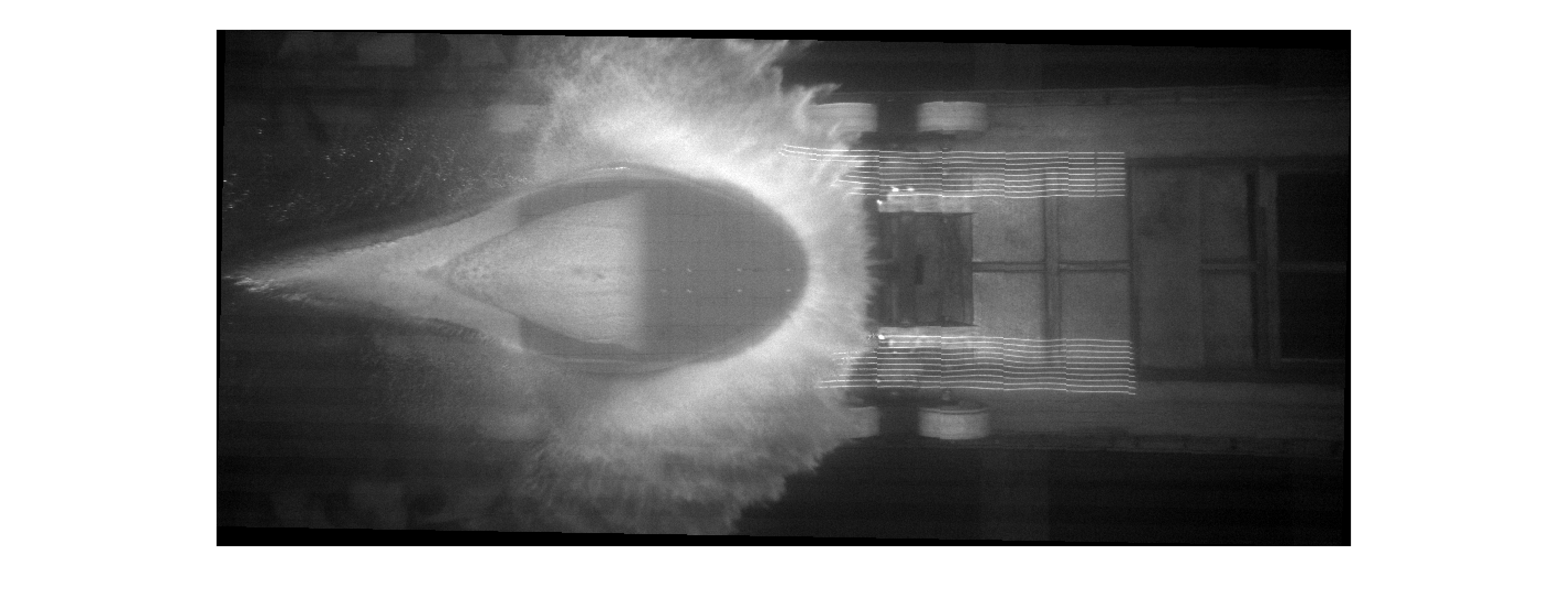}}
\subfigure[$\tau$~=0.8]
{\includegraphics[width=0.48\textwidth]{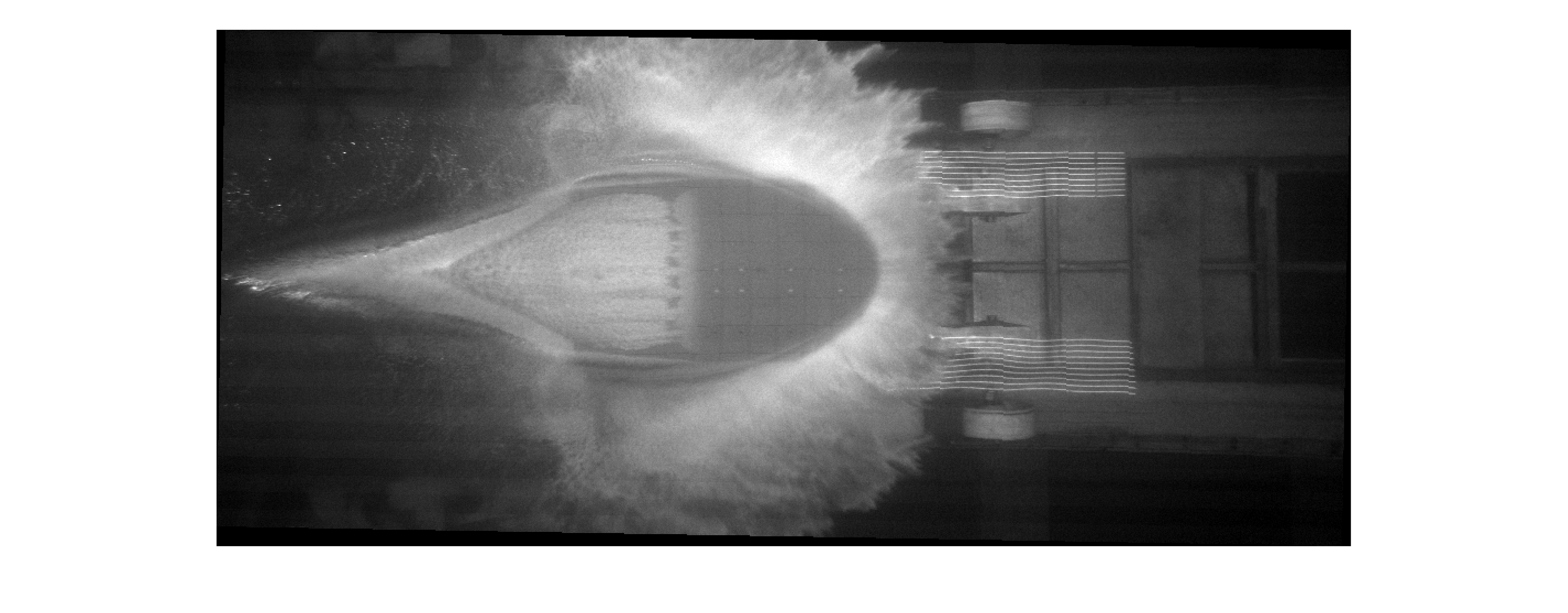}}
\subfigure[$\tau$~=0.9]
{\includegraphics[width=0.48\textwidth]{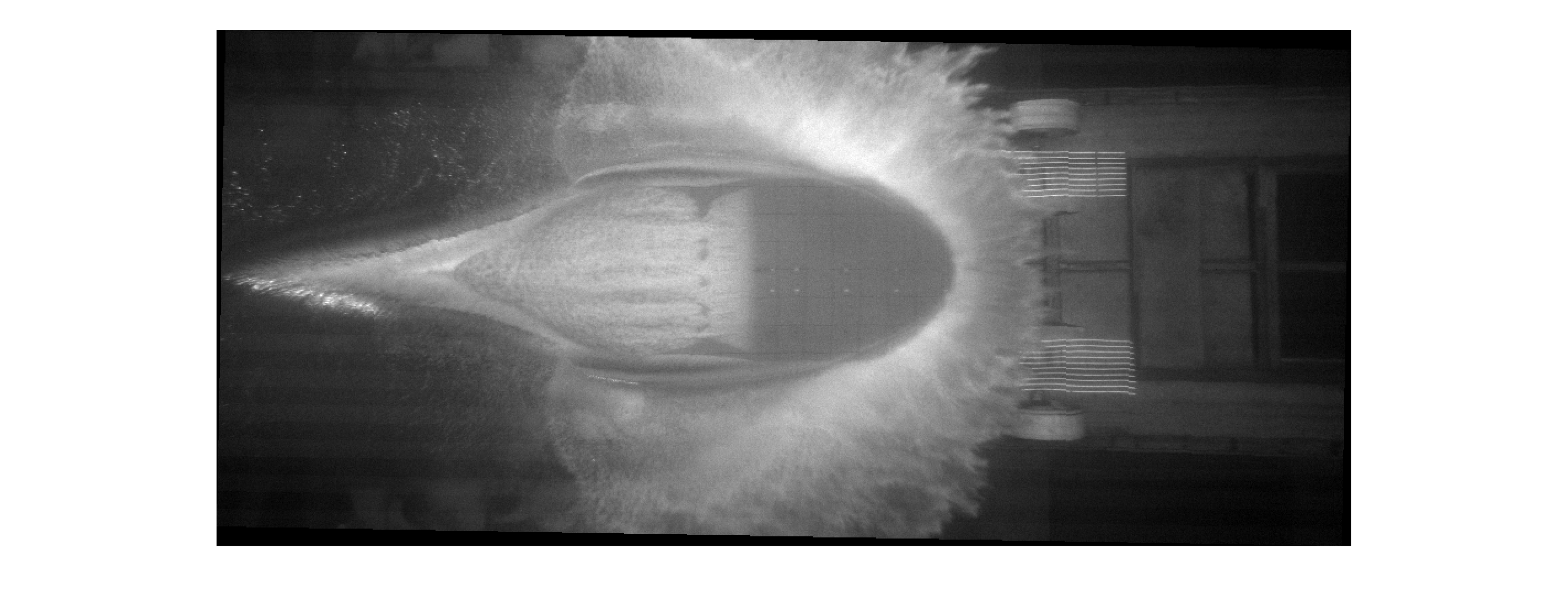}}
\caption{Frames of the underwater video at different $\tau$ during the 
water entry of the shape \textbf{S2} at a nominal horizontal speed of 
$U$=40~m/s and at a pitch angle of $\alpha$=6$^{\circ}$. The specimen moves 
from left to right.}
\label{fig:UWframes_1H2222_17_07_2018_1}
\end{figure}

\begin{figure}%
\centering
\subfigure[Probe positions: \textbf{C} – separation between 
zero-curvature part on the right and finite-curvature part 
on the left; \textbf{H} – first contact point 
at $\alpha$ = 6$^{\circ}$.]
{\includegraphics[width=0.55\textwidth]{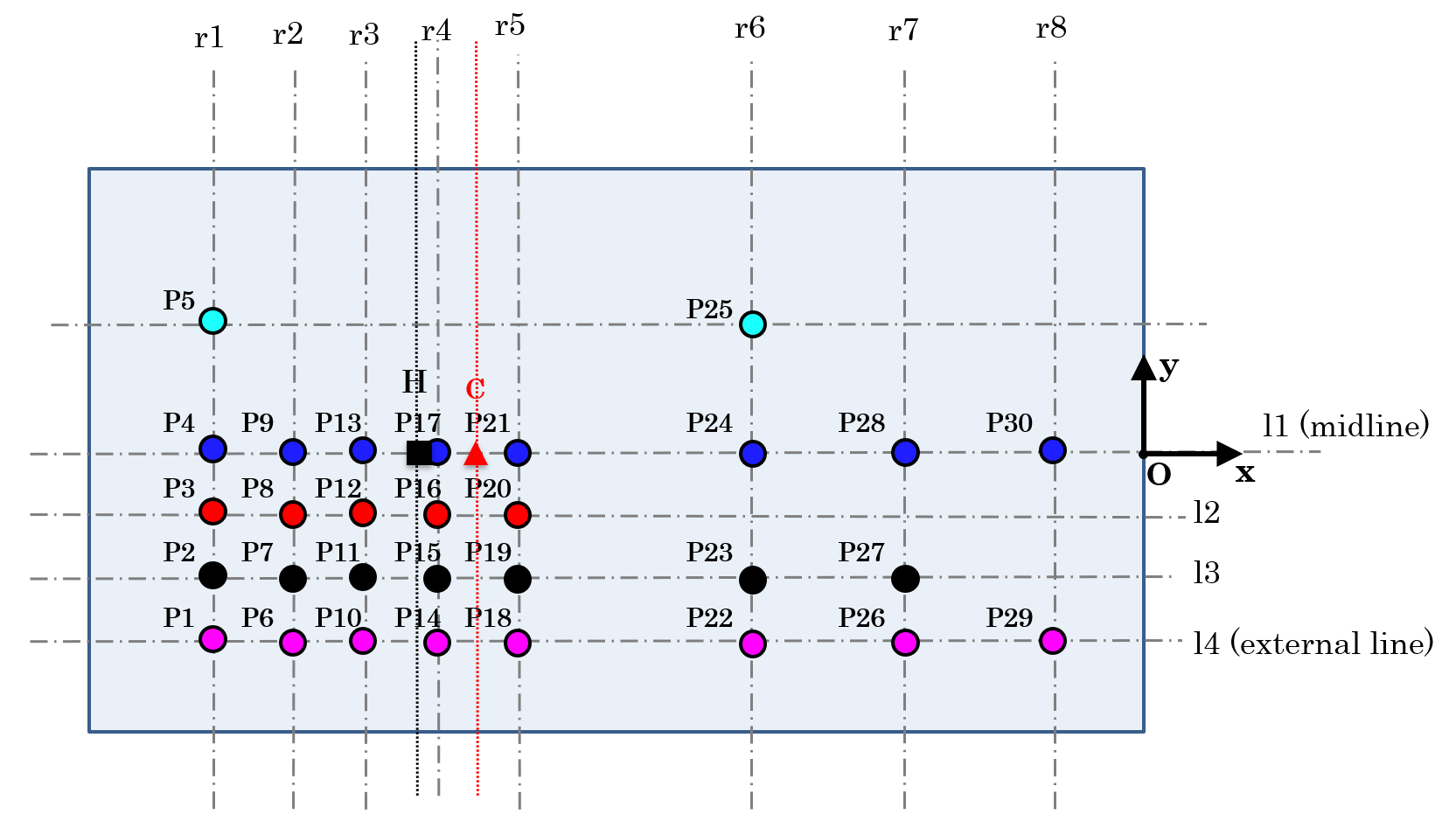}\label{subfig:probes_pos}}
\\
\subfigure[Row $r8$]{\includegraphics[width=0.46\textwidth]{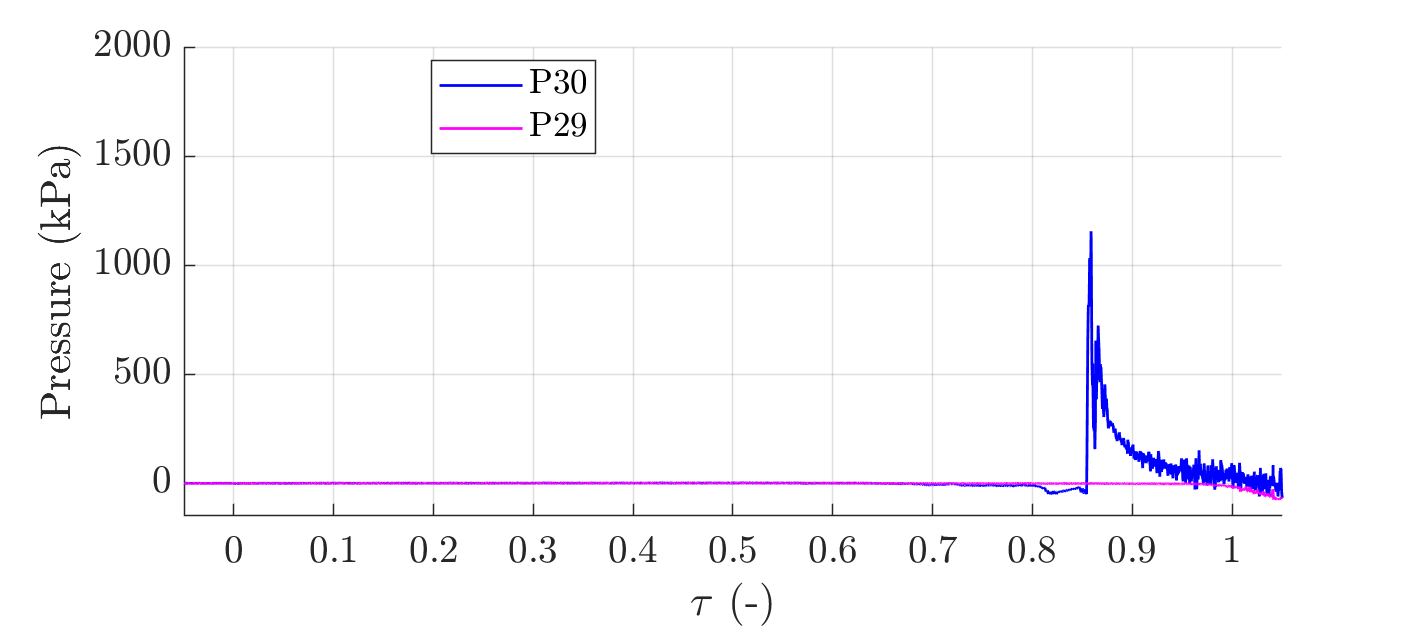}\label{subfig:row8}} \quad 
\subfigure[Row $r7$]{\includegraphics[width=0.46\textwidth]{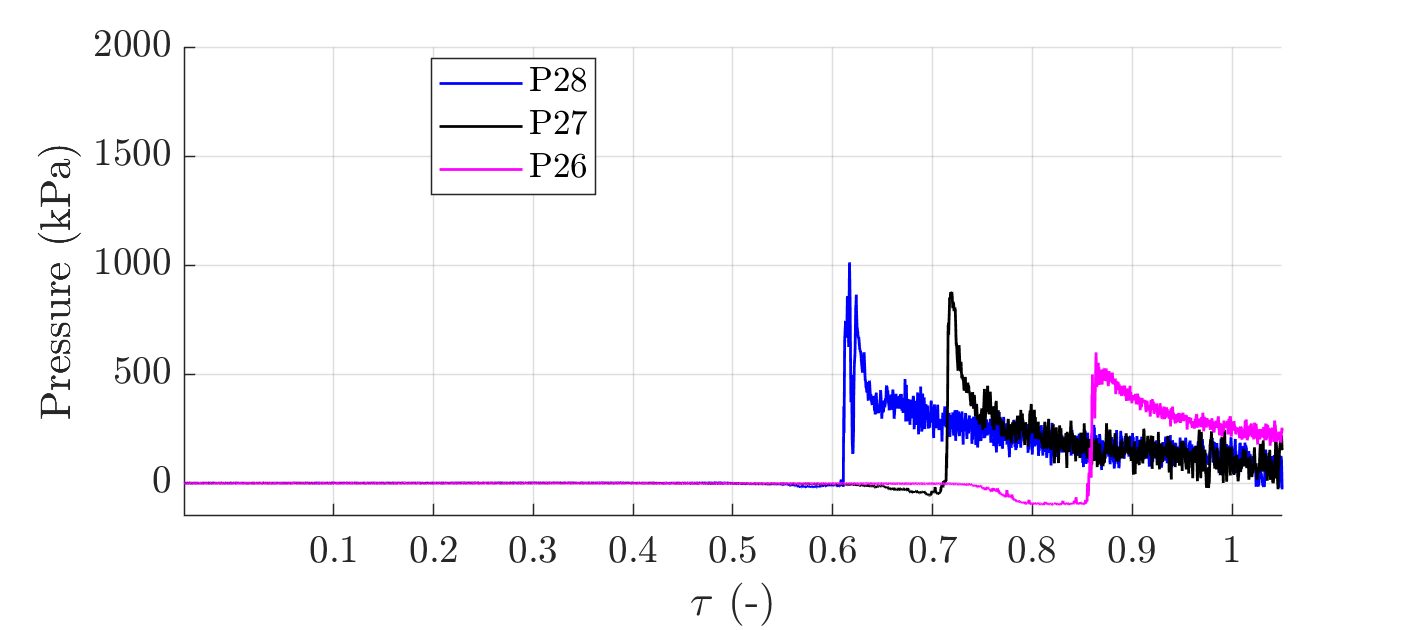}\label{subfig:row7}} \\
\subfigure[Row $r6$]{\includegraphics[width=0.46\textwidth]{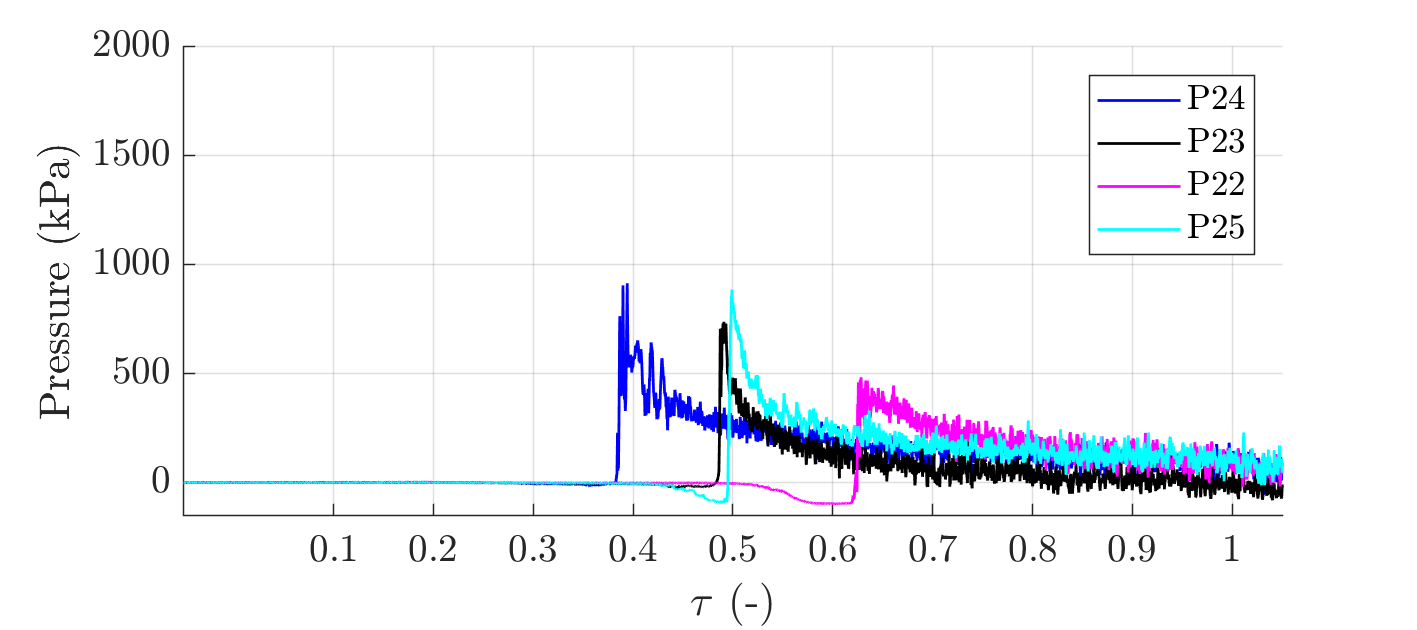}\label{subfig:row6}}
\quad
\subfigure[Row $r5$]{\includegraphics[width=0.46\textwidth]{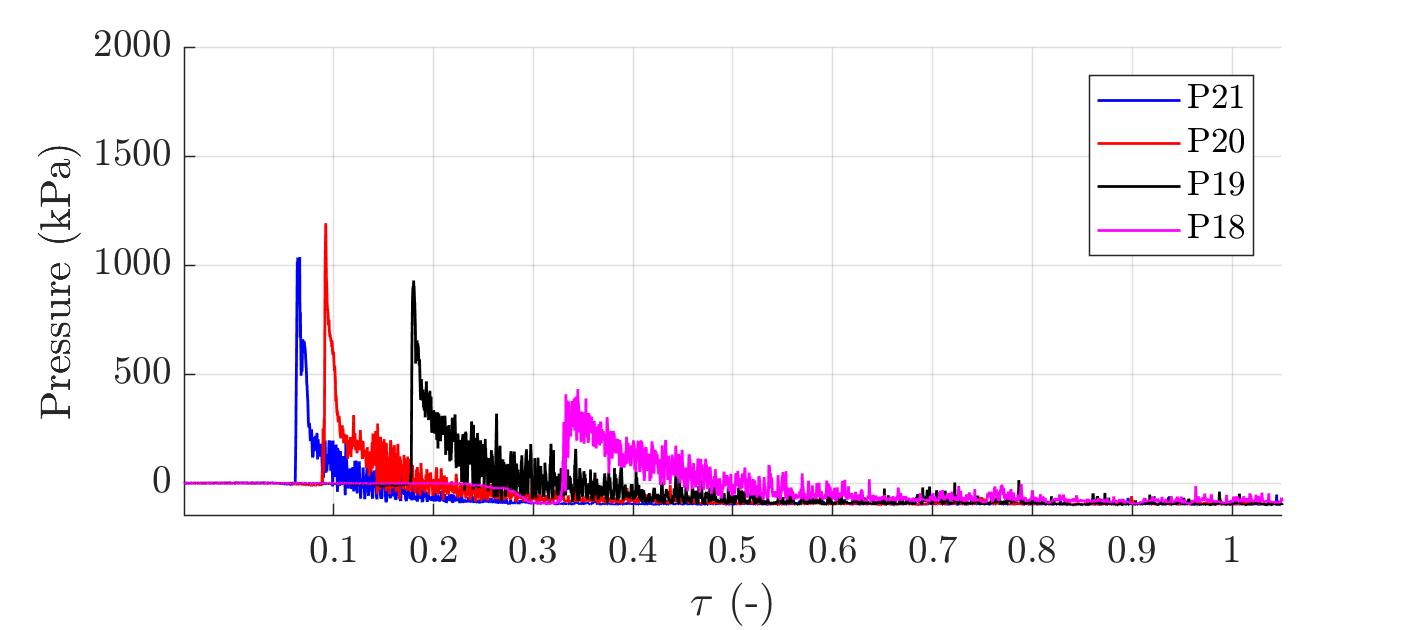}\label{subfig:row5}}
\\
\subfigure[Row $r4$]{\includegraphics[width=0.46\textwidth]{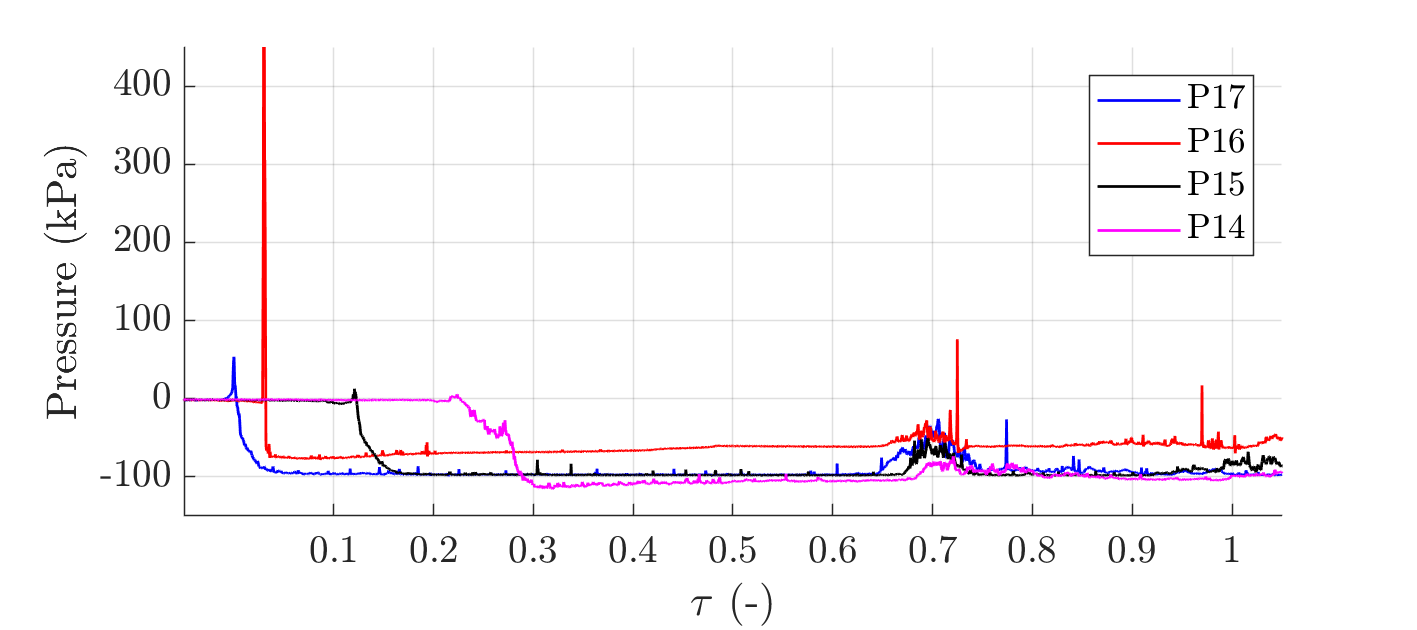}\label{subfig:row4}}
\quad
\subfigure[Row $r3$]{\includegraphics[width=0.46\textwidth]{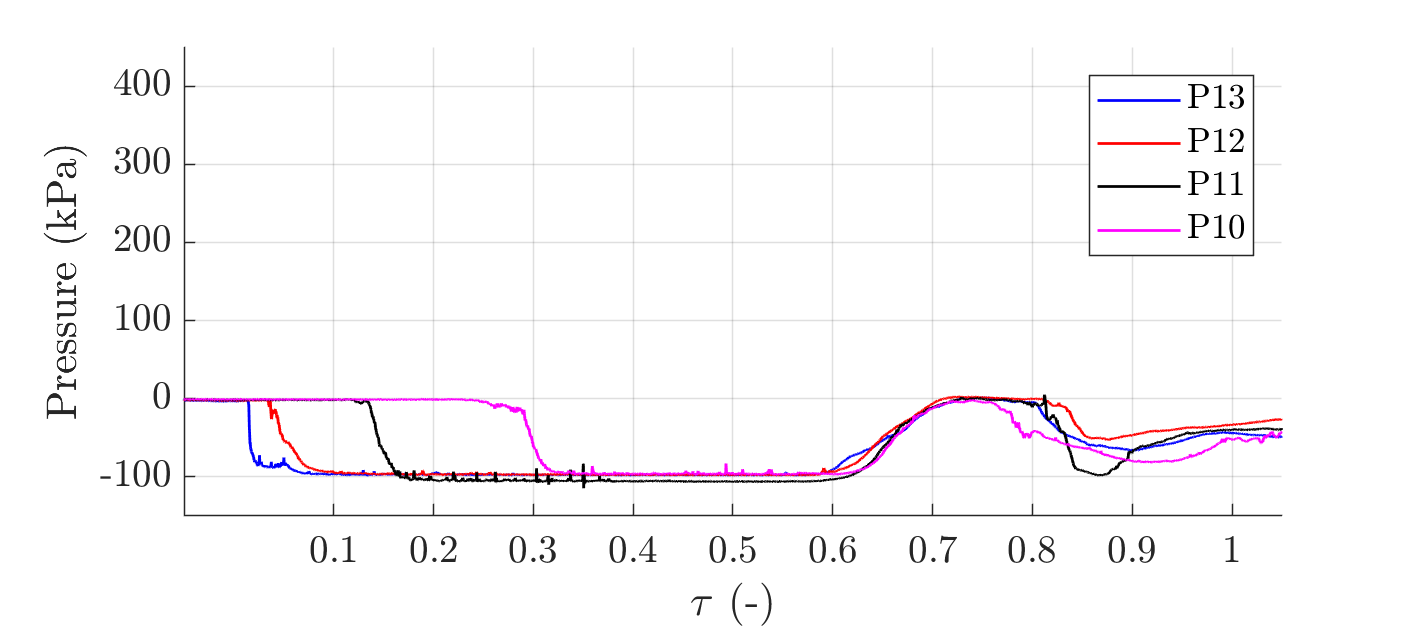}\label{subfig:row3}}
\\
\subfigure[Row $r2$]{\includegraphics[width=0.46\textwidth]{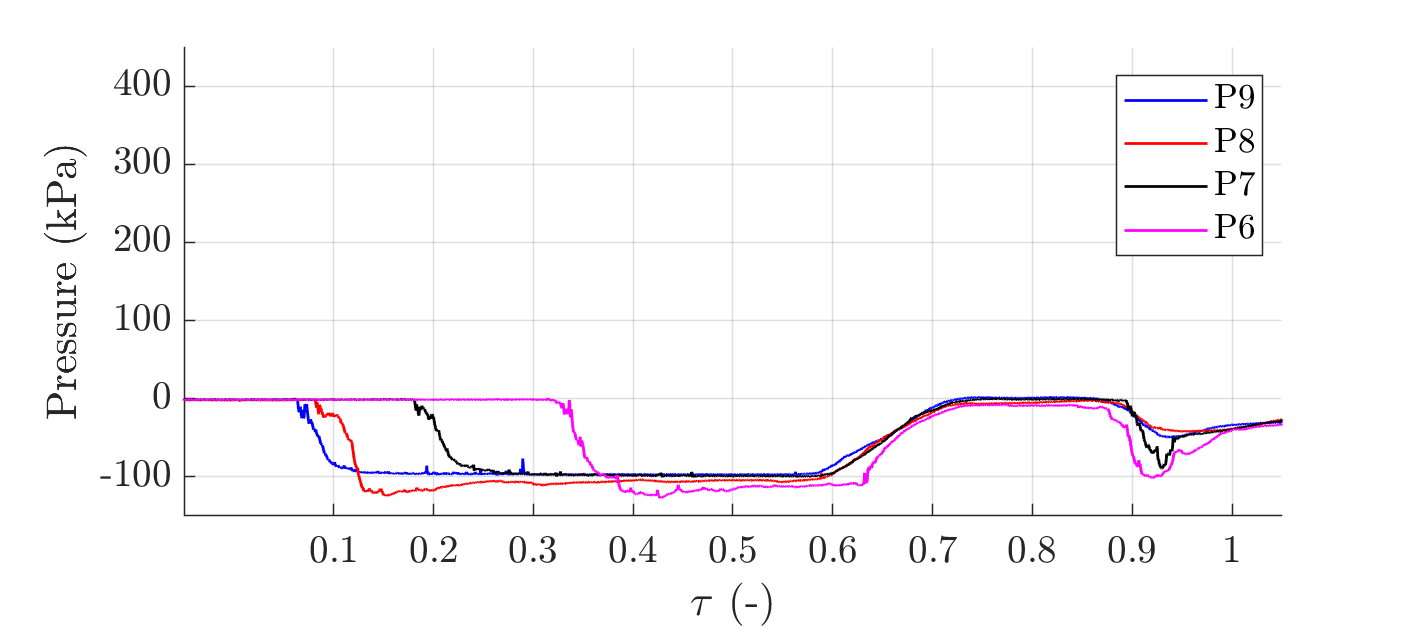}\label{subfig:row2}}
\quad
\subfigure[Row $r1$]{\includegraphics[width=0.46\textwidth]{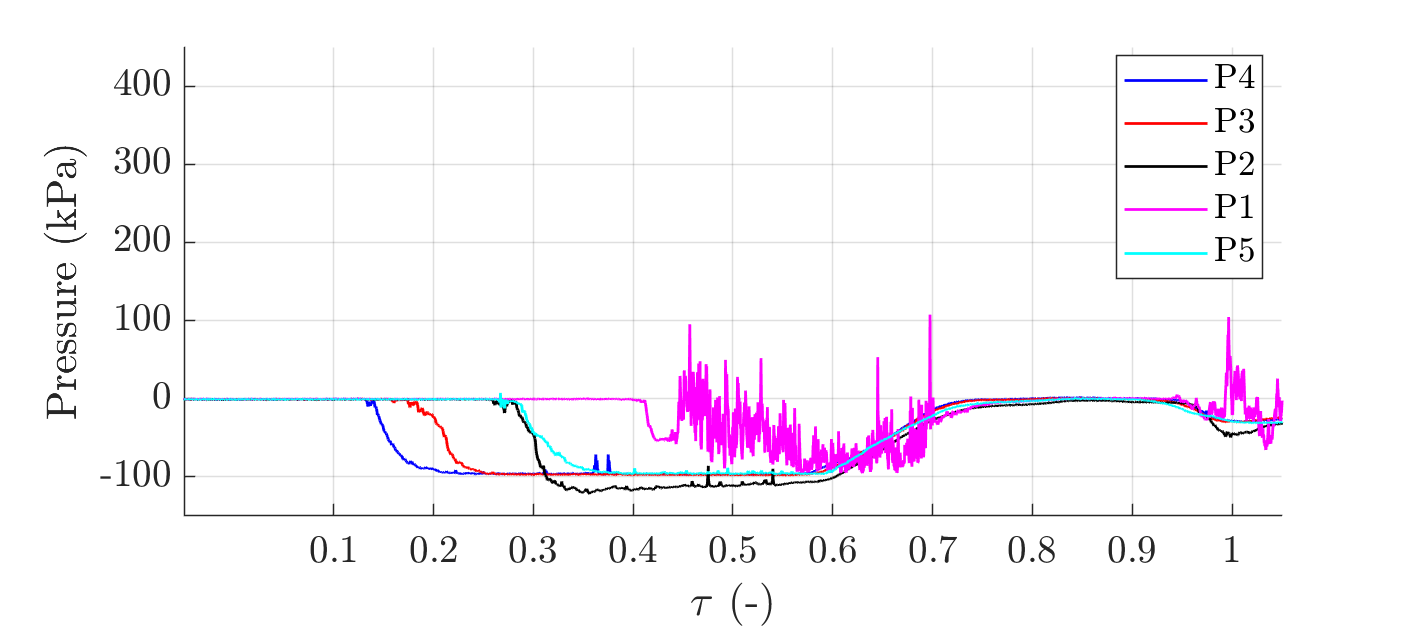}\label{subfig:row1}}
\\
\caption{Pressure probe time histories row by row for the shape \textbf{S2} moving at $U$=40~m/s, at a pitch angle of $\alpha$=6$^{\circ}$.}
\label{pressure_rows_1H2222_17_07_2018_1}
\end{figure}

%
Qualitative considerations on the hydrodynamics of the water entry of the 
double curvature specimens can be made 
through a comparative analysis 
of the pressure time histories and the underwater high-speed videos.
Such an analysis is conducted on 
the test of the shape \textbf{S2}, advancing with a horizontal velocity 
$U$=40~m/s, velocity ratio $V/U$=0.0375 and pitch angle $\alpha$=~6$^{\circ}$.
Video frames taken at different $\tau$, displaying the flow evolution 
on the bottom of the specimen, are shown in Figure 
\ref{fig:UWframes_1H2222_17_07_2018_1}.
The time histories of pressure measured by probes located in the same 
row (see Figure \ref{fig:Pressure_Probe_Positions} for the row definitions) 
are shown in Figure \ref{pressure_rows_1H2222_17_07_2018_1}.
From the underwater frames shown in Figure 
\ref{fig:UWframes_1H2222_17_07_2018_1} it is possible to recognize the wetted
area as the darker region. During the early stage of impact, i.e. 
in Figure \ref{fig:UWframes_1H2222_17_07_2018_1}$(b,c,d)$, the wetted area grows in 
all directions. The brighter zone around the wetted area represents the 
spray region. 

In the \emph{front part of the specimen} the spray propagates in a way 
similar to that observed during the water entry of a curved plate,
as shown in \citet{iafrati2018effect}.
The spray root is characterized by a localized pressure peak that causes a 
sharp rise of the pressure recorded by the probes, as shown in Figure 
\ref{pressure_rows_1H2222_17_07_2018_1} for rows from $r5$ to $r8$. 
After the passage of the peak, the
pressure gradually diminishes, similarly to what found for the flat plates 
\citep{iafrati2016experimental,iafrati2015high} or convex plates 
\citep{iafrati2018effect}. 
As observed in \citet{iafrati2015high},
the pressure drop recorded by some of
the probes (see for example P22, P25 or P26) before the sharp pressure rise is 
associated with the interaction of the flow in the spray tip with the probe itself,
which can also induce cavitating vortices.
For the probes located at the row $r5$, which for the shape \textbf{S2} at 
6$^{\circ}$~pitch is just ahead of the point of first 
contact, all probes approach a constant negative pressure value, 
slightly higher than the vapour pressure.
From both the underwater images it is possible to see that the spray root 
is curved backwards,
due of the combined action of 
three-dimensional effects and of the transverse curvature of the specimen,
as observed in flat plates \citep{iafrati2016experimental}, 
and in convex plates \citep{iafrati2018effect}, respectively. 
This is also confirmed by the time histories of the pressures, from which it
is observed that the spray root region, i.e. the pressure peak, reaches first the 
probes located in the mid-line, such as P21, P24, P28 and P30, and next the 
probes located at the sides, see for instance the recordings of the
probes on $r5$ in \ref{subfig:row5} .
In Figure \ref{subfig:row6}, the time histories of probes P23 and 
P25, are very close to each other, with the pressure peak exhibiting about 
the same delay compared to the probe P24, confirming that a satisfactory level
of symmetry is achieved.

In the \emph{rear part of the specimen},
i.e. the region behind \textbf{C}, the hydrodynamics is 
quite different from that observed in the front part. In the early stage the 
wetted area propagates backwards, although at a lower speed than that observed 
in the front part. This is coherent with the lower the values of the 
pressure peaks measured by the probes located at the rear, such as P17.
Soon after the passage of the spray root, the pressures decrease quite rapidly
and become negative. Such a behaviour is associated to the shape
of the body and with the impact conditions.
A qualitative explanation
can be provided from two different points of view,
i.e. focusing on the flow in the midplane (longitudinal plane) 
or on the flow in the transverse plane.

By focusing on the \emph{longitudinal plane} (midplane)
in the trolley-fixed reference frame,
the results of the two-dimensional case of an asymmetrical wedge
entering water with a horizontal velocity \citep{judge2004initial,xu2008numerical,krastev2018asymmetric}
can provide some elements for the interpretation 
of the flow phenomena in the present case.
In fact, as a consequence of the kinematic asymmetry of the flow
due to the horizontal speed, the pressure on the front part 
of the specimen is higher than that at the rear. 
From a theoretical standpoint,
it is expected that a velocity singularity will occur at the apex
\citep{judge2004initial}. 
In real flows, a strong local velocity acceleration is observed
there, accompanied by a pressure drop, which is similar to what occurs
on the suction side of hydrofoils
\citep{faltinsen2005hydrodynamics,newman}.
The flow field presents also similarities
with that of the vertical water entry of inclined asymmetric wedges
\citep{semenov2006nonlinear,riccardi2004water}.
In \citet{xu2008numerical}, in which a wedge entering
water with both a vertical and a horizontal velocity
is studied using a self-similar approach, even if 
the velocity ratio $U/V$ is 
much lower than the present case, a local pressure drop 
to negative values is observed at the wedge apex.

An alternative interpretation of the pressure behaviour can be obtained 
by focusing on the flow in the \textit{transverse plane}, 
in an earth-fixed reference frame,
by exploiting the so called 2D+$t$ approach 
\citep{fontaine1998prediction,iafrati2010comparisons}.
In the front part the plate always penetrates into
the water, whereas in the rear part, depending on 
the local inclination of the tangent
to the body surface in the longitudinal plane and 
on the $U/V$ velocity ratio,
the fluid flow problem changes from a water entry 
to a water exit one \citep{tassin2013two} and 
when this happens, negative pressures develop.
The development of low and even negative pressures occurs very shortly 
after the first contact with water and, based on the above considerations,
the pressure drop is expected to be higher if the gradient in the slope of the
body surface is more significant, which means in presence of larger longitudinal 
curvature. 

The simulations of the horizontal planing motion
of the same scaled fuselage models at low
speeds and at the same pitch angle, discussed in \citet{spinosa2022hydrodynamic},
may also offer insight on the 
hydrodynamics of the water entry at at high horizontal speed. 
This work also demonstrates the attainment of negative pressure 
behind point \textbf{H}. Furthermore, it is shown 
that the pressure does not become negative abruptly 
at point \textbf{H}. Instead, the low pressure
region extends slightly ahead of this point.

As observed in \citet{iafrati2019cavitation}, the pressure drop can be so high
that the vapour pressure value may be reached. With a water temperature
between 15 and 18~$^{\circ}$~C, the vapour pressure relative
to the atmospheric value, denoted as $p_{\textrm{vap}}$, is in the range 
-99.6 and-99.2 kPa.
This explain the formation of a cavitation 
region \citep{acosta1973hydrofoils,wang2001dynamics,brennen2014cavitation},
as shown in the underwater images of Figure \ref{fig:UWframes_1H2222_17_07_2018_1} 
at the non-dimensional time $\tau \geq 0.2$.
In the trolley-fixed reference frame, 
the cavitation region has a sort
of reverse "D" shape, which follows the pressure distribution on the bottom
of the specimen. As the specimen enters in water, 
the forward front of the cavitation
region maintains rather straight and located 
about the point of curvature change \textbf{C}, 
whereas the rear edge of the
cavitation region spreads backwards towards the trailing edge of the specimen.
Once the cavitation region reaches the trailing edge of the specimen, 
the water vapour inside it, at vapour pressure, comes into contact 
with air at atmospheric pressure. As a consequence, air is suddenly 
entrained into the water vapour cavity, 
forming a \emph{ventilation zone} that expands in the
opposite direction, i.e towards the leading edge.
This behaviour is
typical of the so called \emph{cavitation-ventilation regime},
described in more detail in Section \ref{sec:effect_speed} based
on the analysis of the water entry of the
specimen \textbf{S3} \citep{iafrati2019cavitation,iafrati2020experimental}
and also occurs for \textbf{S2} at the same horizontal speed.

The expansion of the cavitation region can be also 
recognized by observing the time histories of the pressure probes,
see Figure \ref{pressure_rows_1H2222_17_07_2018_1}. 
The passage of the rear edge of the cavitation region corresponds to the abrupt 
drops to the vapour pressure value. 
Moving backwards, i.e. from row $r4$ to row $r1$, the instant at which the 
drops occur for the midline probes (P17, P13, 
P9 and P4) increases. This time delay is 
related to the propagation velocity of the rear edge of the cavitation region. 
Comparing the pressure time histories across the same rows,
it can be noted that pressure drops of the 
midline probes are delayed with respect to those of the probes
at the sides, due to the combined effect of
of the body shape and of the curvature of the rear edge
of the cavitation region.
When ventilation starts and air is entrained
within the cavity, pressures start increasing progressively,
reaching the zero-pressure value. 
It is worth noting that, in contrast to the pressure drop associated
with the propagation of the cavitation region, the rise in pressure 
values during this phase is gradual and occurs at extremely close time instants
for all pressure probes positioned on the same row. 
The increase in pressure values occurs later 
in time, when progressing from row $r1$ to $r4$.
This clearly indicates that the ventilation front is advancing 
at an extremely high speed, as previously observed for the shape
\textbf{S3} in \citet{iafrati2019cavitation}.

\begin{figure}
\centering
\subfigure[$\tau$=0.30]
{\includegraphics[width=0.45\textwidth]{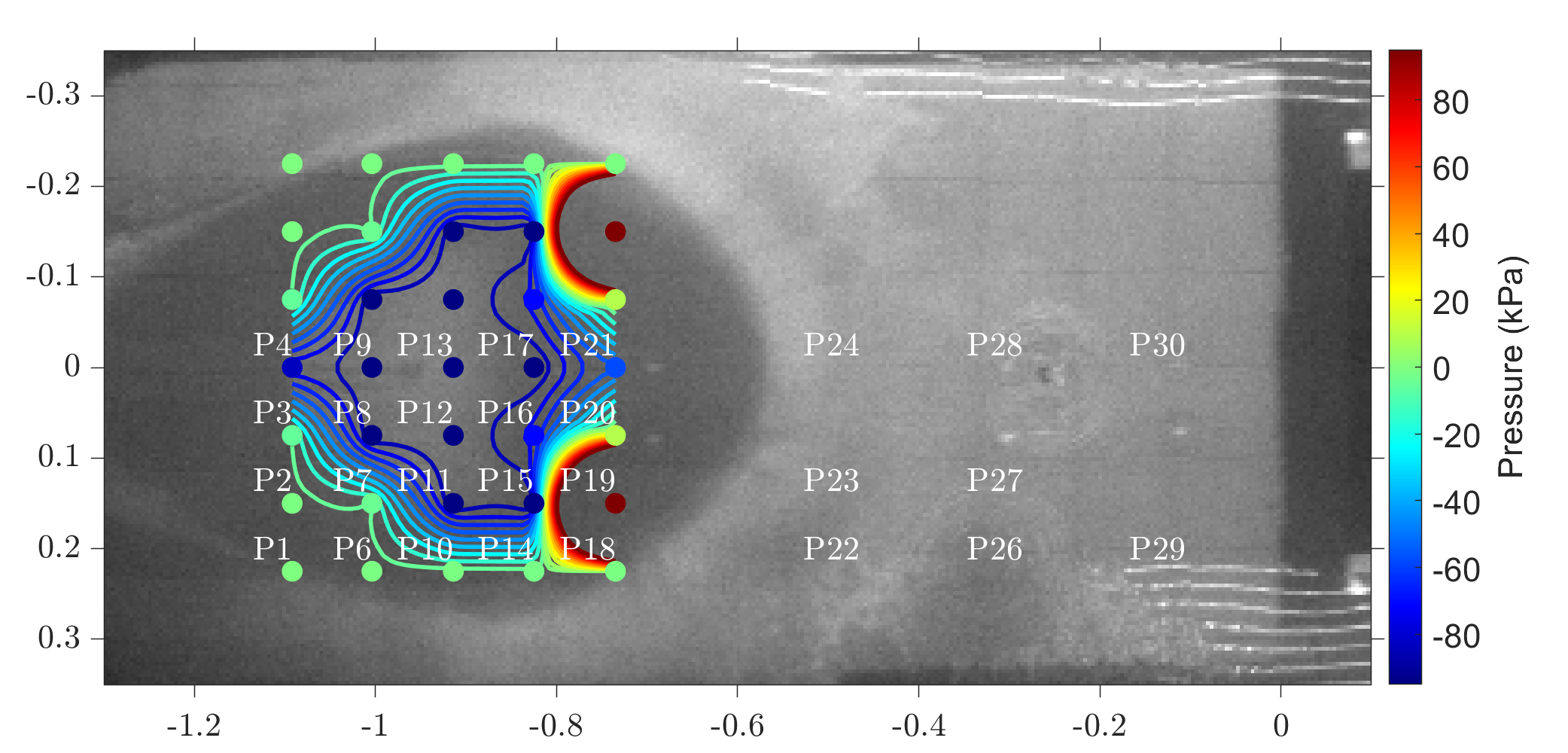}} \quad
\subfigure[$\tau$=0.60]
{\includegraphics[width=0.45\textwidth]{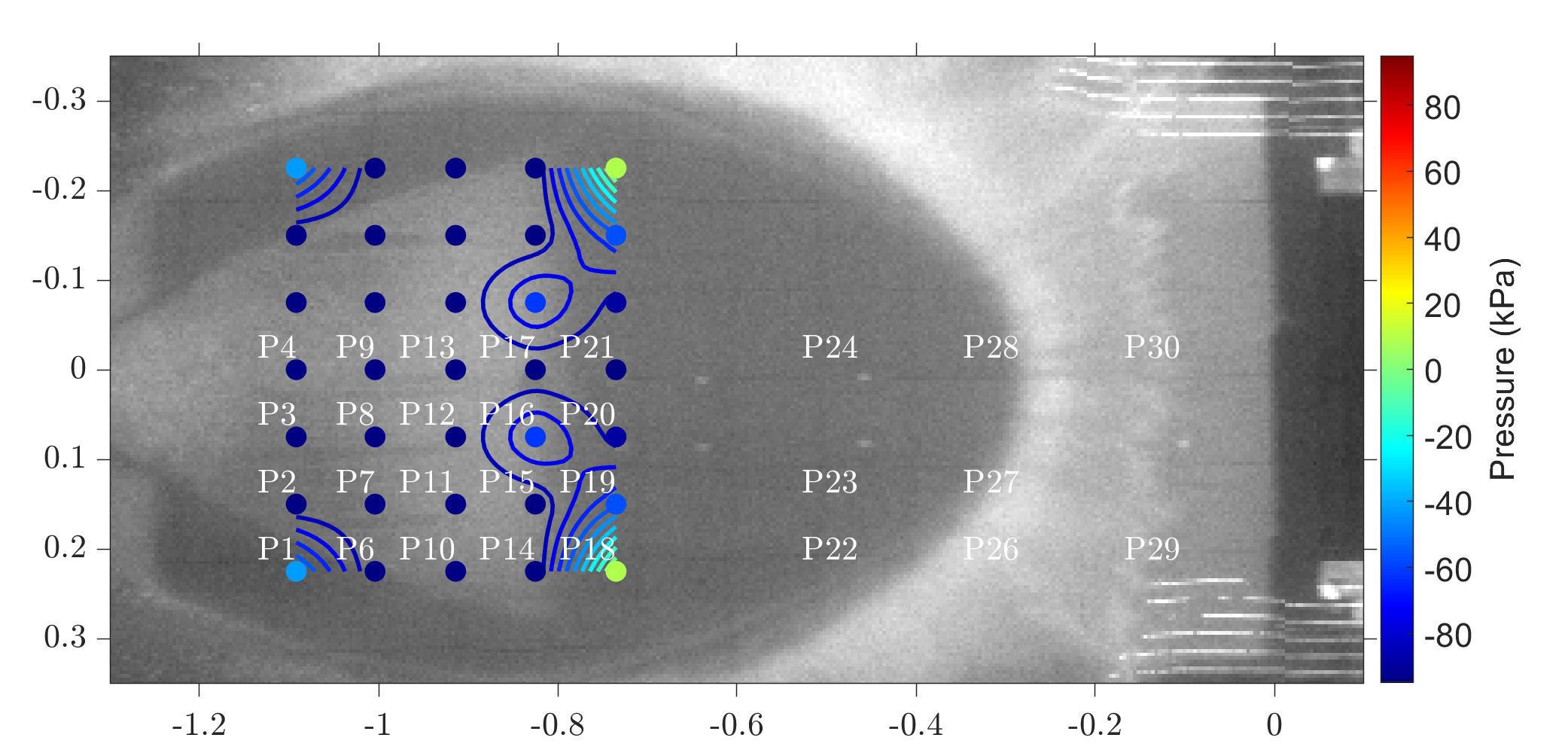}} \\
\subfigure[$\tau$=0.70]
{\includegraphics[width=0.45\textwidth]{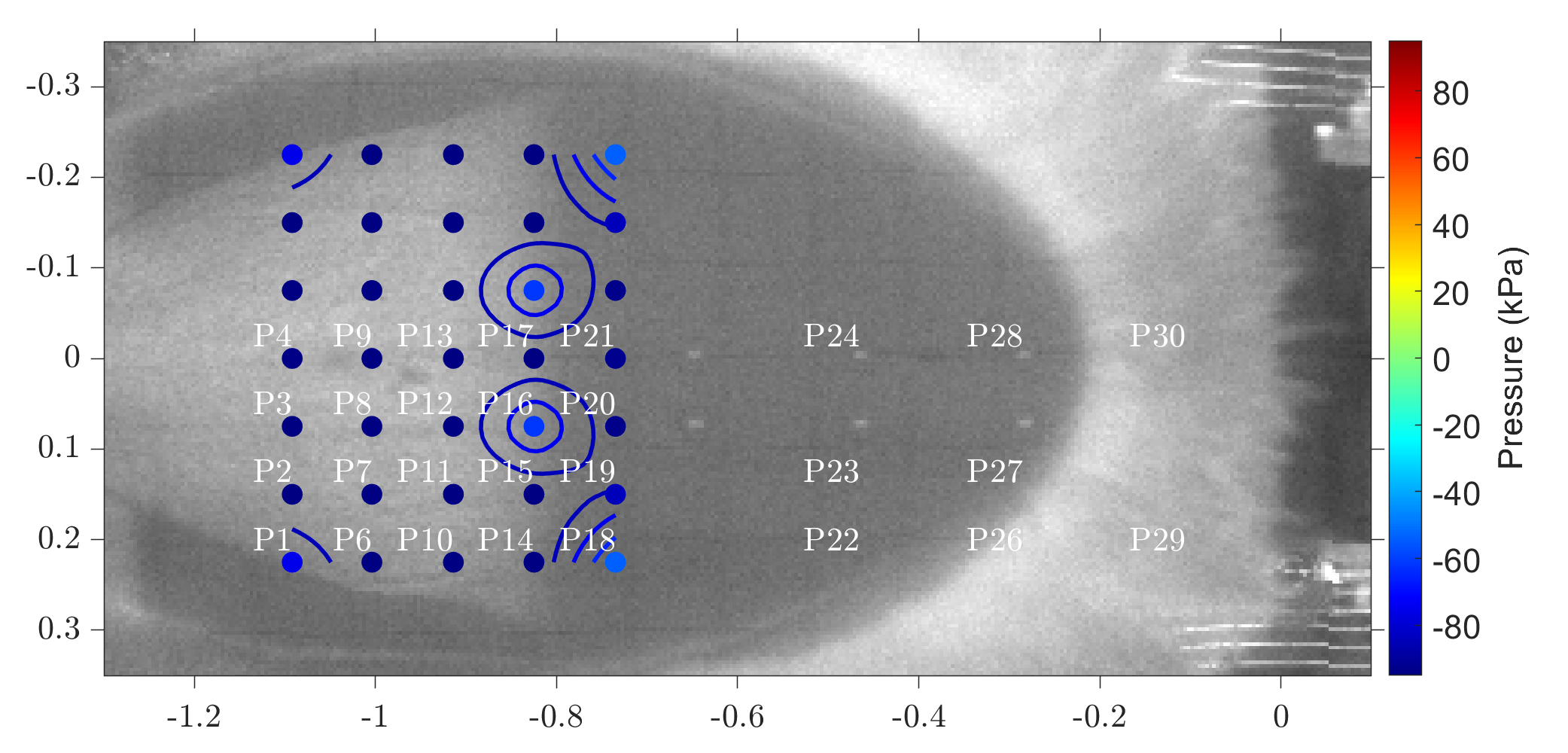}} \quad
\subfigure[$\tau$=0.72]
{\includegraphics[width=0.45\textwidth]{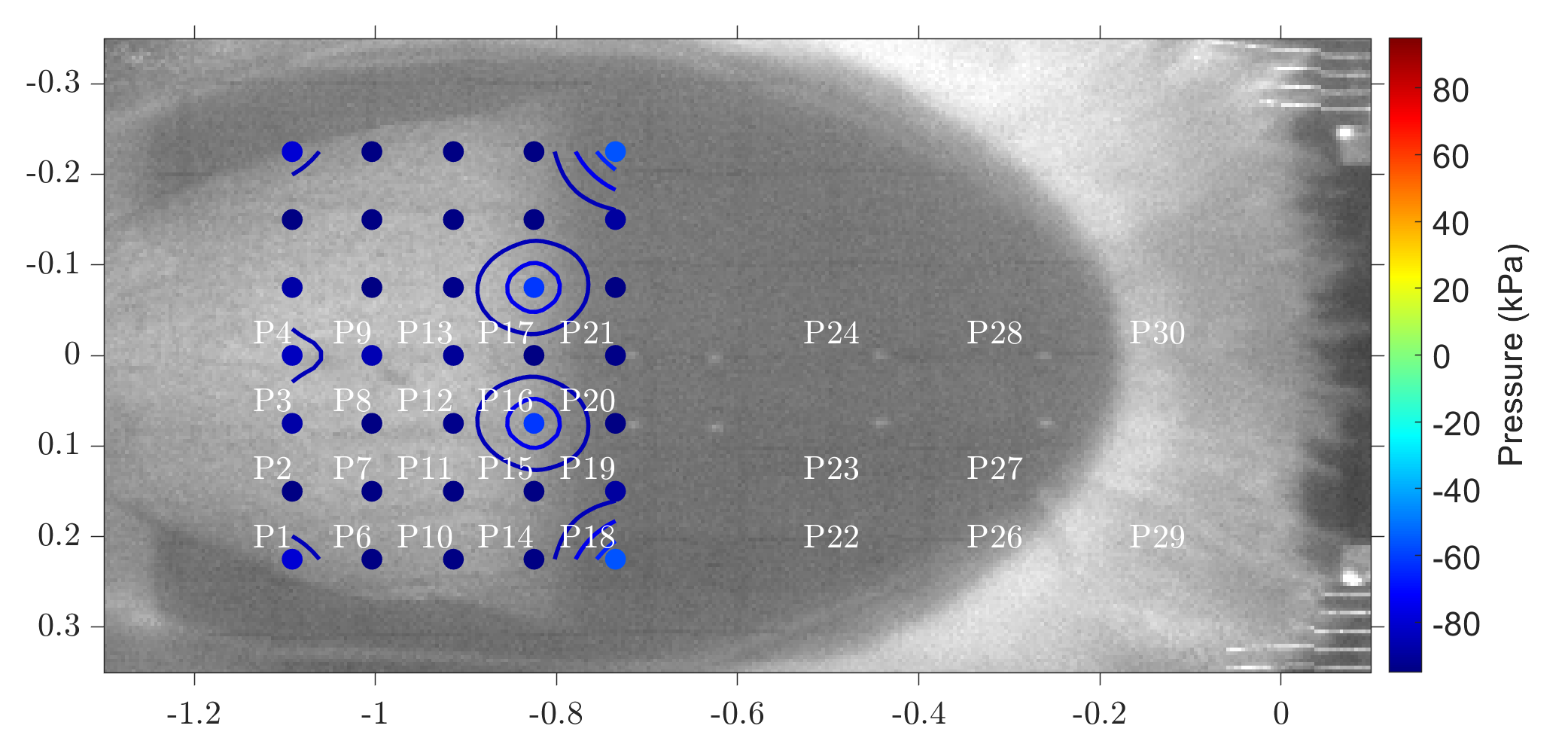}} \\
\subfigure[$\tau$=0.74]
{\includegraphics[width=0.45\textwidth]{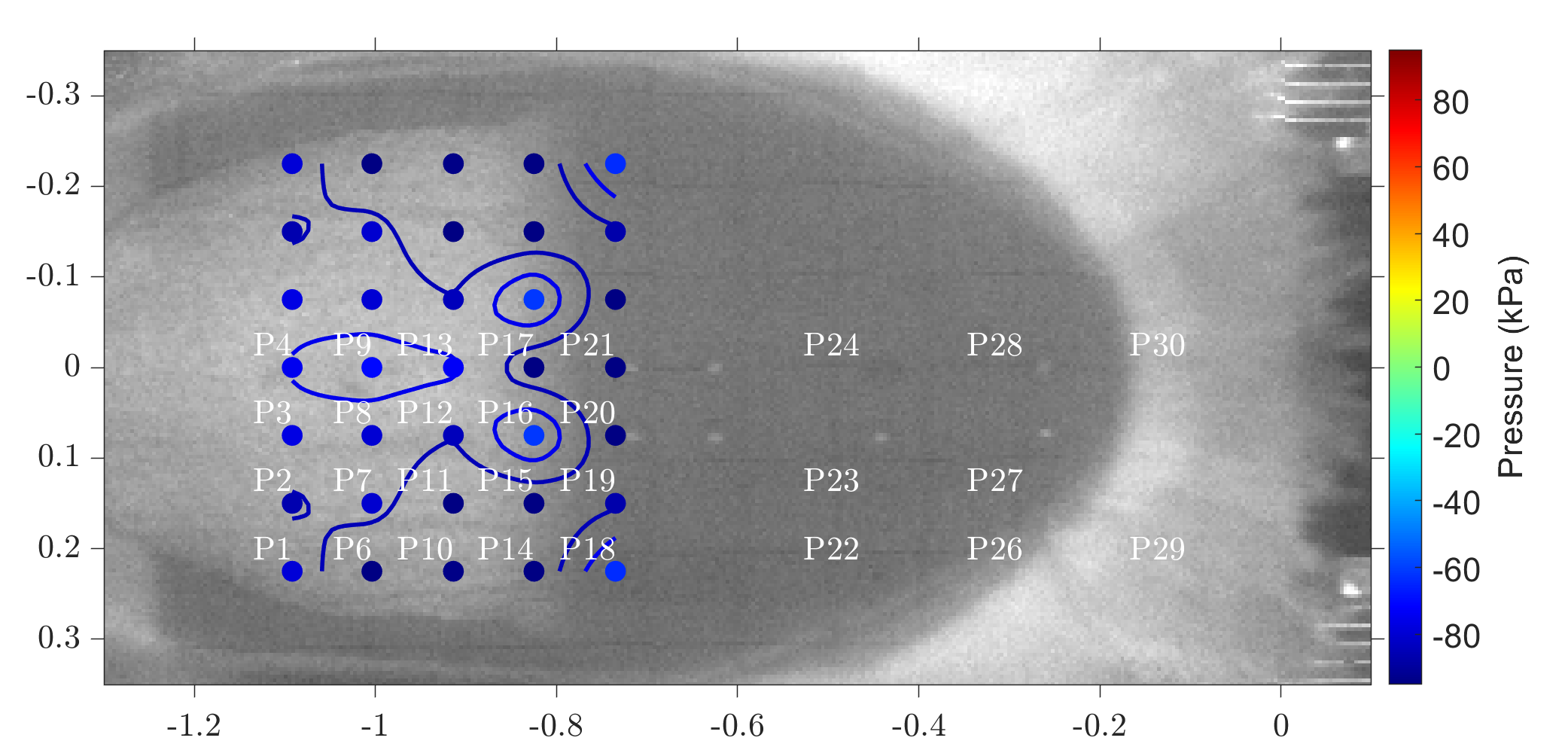}} \quad
\subfigure[$\tau$=0.76]
{\includegraphics[width=0.45\textwidth]{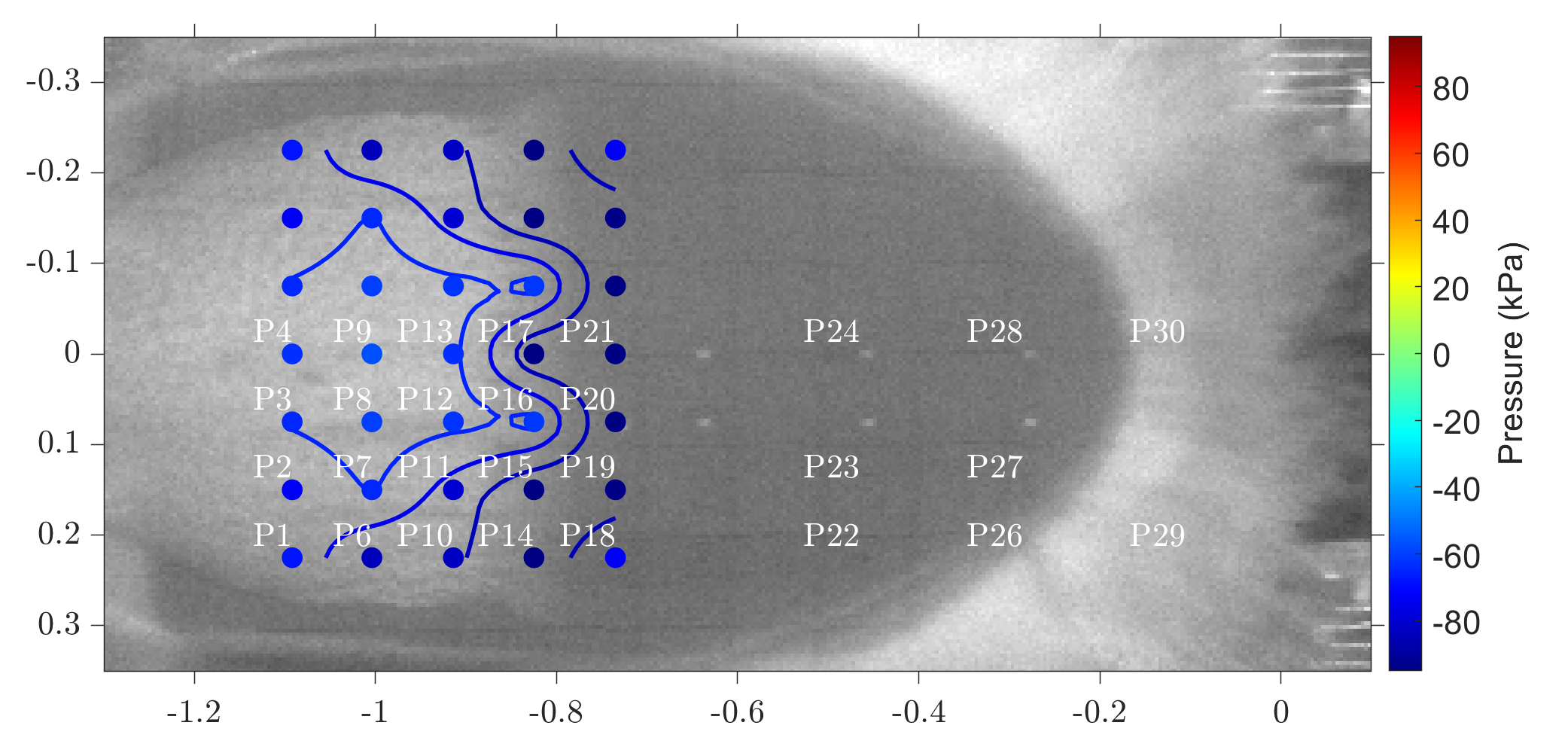}} \\
\subfigure[$\tau$=0.78]
{\includegraphics[width=0.45\textwidth]{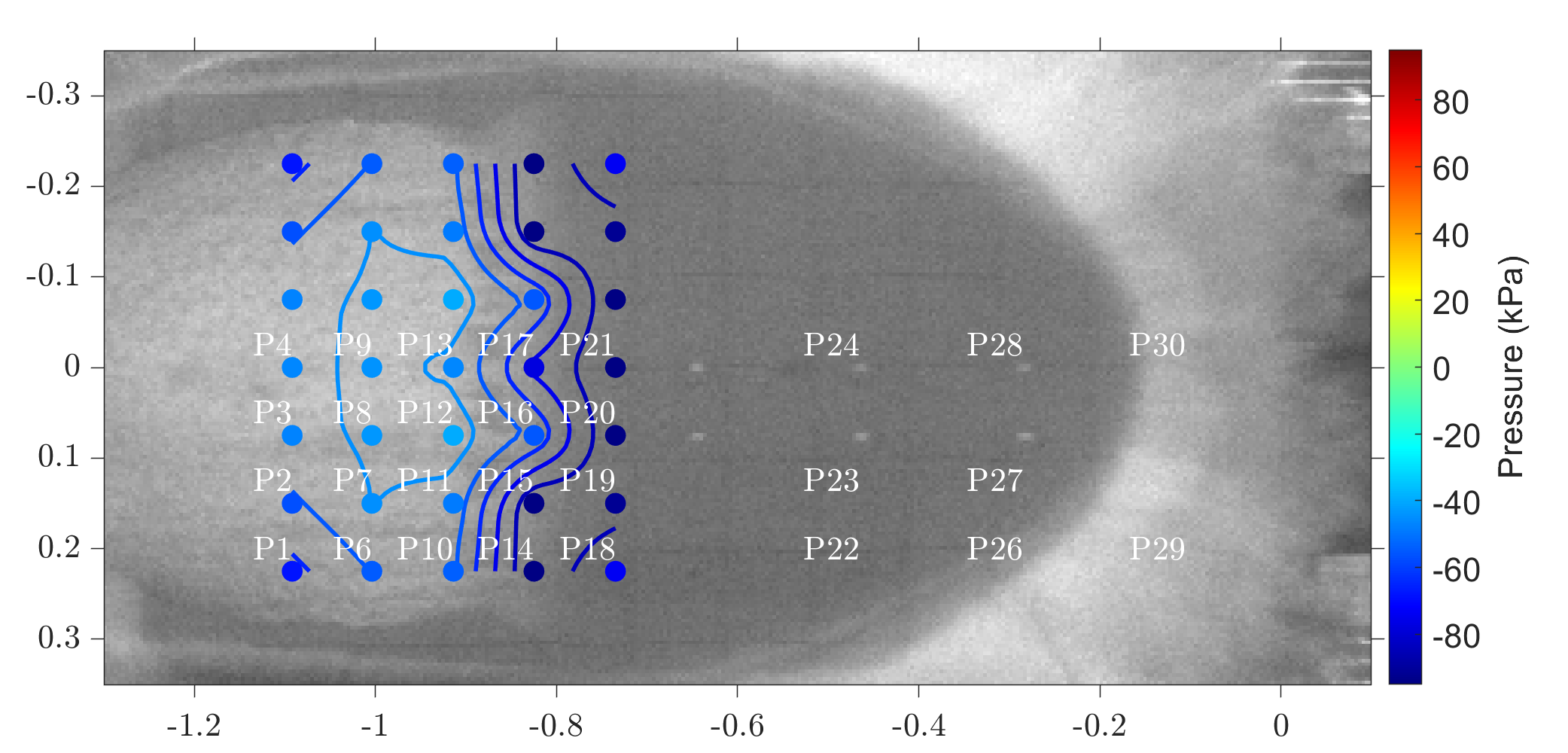}} \quad
\subfigure[$\tau$=0.8]
{\includegraphics[width=0.45\textwidth]{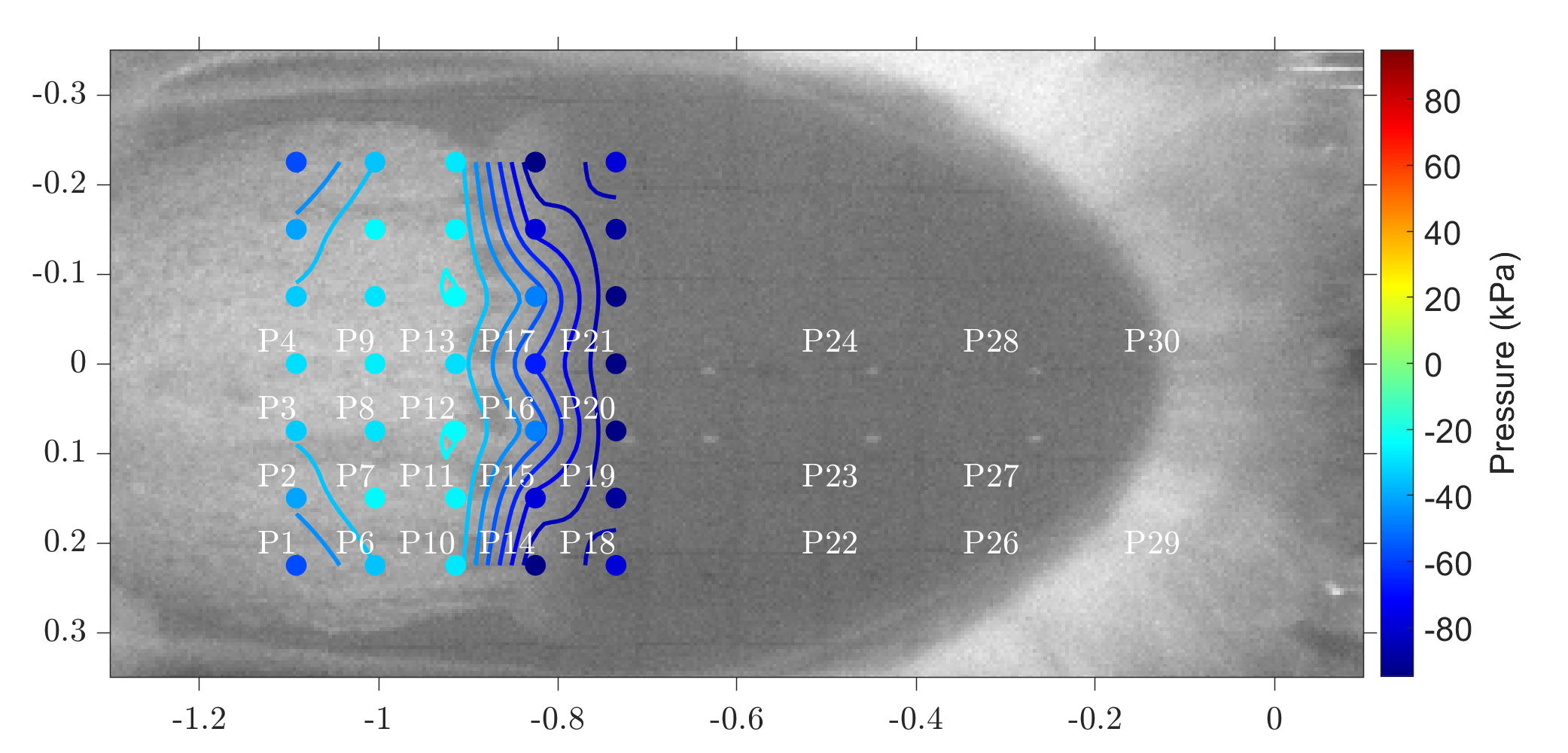}} 
\caption{Pressure iso-lines in the rear part of the specimen 
overlapped to the underwater video frames for the test of 
shape \textbf{S2} at U=40~m/s, at $\alpha = 6^{\circ}$ at
the specfied $\tau$ values. The symmetry about the mid-line is exploited.}
\label{fig:S2_isolines}
\end{figure}

A comprehensive global overview of the shape
and evolution of the cavitation region can be derived
by constructing pressure iso-lines using the available pressure data 
from row $r1$ to $r5$ at different time instants.
In Figure~\ref{fig:S2_isolines}, iso-lines are superimposed 
on the corresponding video frames at $\tau = 0.3$, $\tau = 0.6$, 
and from $\tau = 0.7$ to $\tau = 0.8$ with a finer non-dimensional
time resolution of $\Delta \tau = 0.02$.
Although the probe spacing is not fine enough 
to provide a very accurate interpolation, especially in this case where 
relevant pressure gradients are present, it is evident that 
after the first contact with water, the pressure iso-lines 
also assume the shape of a reverse D and expand in time
in the interval between $\tau$=~0.3 and $\tau$=~0.6. 
The pressure values in time decreases overall and from a certain $\tau$ 
reach the value of $p_{\textrm{vap}}$ within the inner iso-line,
which matches the shape of the cavitation region visible in the video frames.
As the cavitation region reaches the trailing
edge, the ventilation front starts moving
back very fast towards the leading edge. The
ventilation front is visible especially at $\tau$=0.8,
where the iso-lines at p~$\approx$~0 have advanced
towards the leading edge, having almost reached the point \textbf{C}.
The ventilation front at this time instant
appears to be quite straight in the transverse direction.

Observing the pressure time histories, 
it can be observed that ventilation occurs 
within the time interval between $\tau \approx 0.7$
and 0.8. 
By focusing the pressure iso-lines over this time interval,
as shown in Figure \ref{fig:S2_isolines}, it is possible
to gain more insight on the ventilation dynamics.
In fact, it is observed that the pressure begins
to rise back to the atmospheric value first at the centre, 
since the cavitation region reaches the trailing edge first there.
However, the ventilation front develops rapidly 
and at about $\tau$=0.78.
becomes roughly parallel
to the transverse direction, as also observed
for the shape \textbf{S3} in \citet{iafrati2019cavitation}.
%

\begin{figure}
	\centering
	\includegraphics[width=0.75\textwidth]{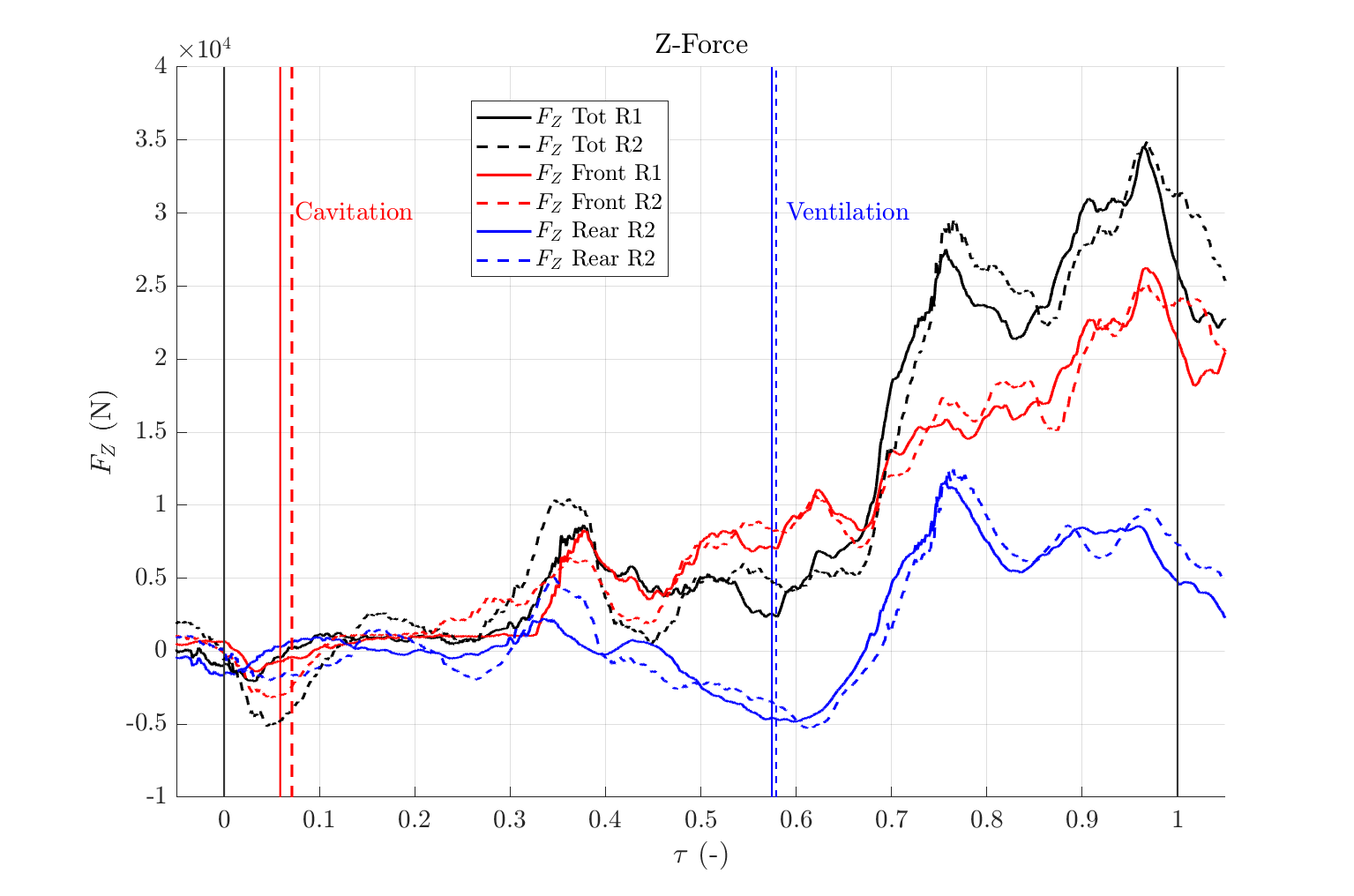}
	\caption{Time histories of the forces recorded by the front (red) 
	and rear (blue) load cells and of the total normal load (black), for shape 
\textbf{S2} at 6$^{\circ}$ pitch angle and $U$=40~m/s. 
The vertical red and blue lines 
indicate the start of cavitation and of ventilation respectively.
The solid and dashed lines refer to the first
and second test repeat respectively (denoted as R1 and R2).}
	\label{fig:1H2222_17_07_2018_1_Zforces}
\end{figure}

In Figure \ref{fig:1H2222_17_07_2018_1_Zforces} the normal loads acting on 
the specimen recorded by the rear and front cells are plotted. The data of
two test repeats, denoted as R1 and R2, performed at the same conditions 
are drawn.
The red and blue vertical lines indicate the non-dimensional time
at which the cavitation and the ventilation phases start,
respectively. These time instants are determined by
using the data of the pressure probes: the cavitation phase is assumed to start 
when the pressure at probe P17 drops to the vapour pressure value,
while the ventilation phase when the pressure at probe P4
begins to rise to atmospheric pressure.
The time histories of the force components exhibit some oscillations, 
which are more evident when the loads are low. These oscillations
are partly due to the vibrations of the trolley, as it moves along the guide,
and partly to the free play at the joints.
In spite of these oscillations, the data of the two repeats display a quite 
similar trend.
The loads measured by the front cells are always positive and grow 
progressively as time elapses, which is consistent with the positive 
pressures recorded by the probes located from $r5$ to $r8$
(see Figure \ref{fig:LoadCell_Ref_Frame}).
The load measured by the cells at the rear is rather low at the beginning, 
which is presumably associated with the development of cavitation, starting  
slightly before $\tau=0.1$ at this specific test condition.
As the cavitation region expands, the loads measured at the
rear undergo a reduction, which is rather evident between 
$\tau \approx 0.4$ and $\tau \approx 0.6$.
As soon as the ventilation starts, the loads at the rear start increasing and 
turn positive. Such growth is consistent with the increase in the 
pressure at the rear of the body, when the pressure passes from the 
vapour pressure value to zero (atmospheric pressure). The load at the rear remains almost
constant after $\tau \approx 0.8$, whereas the total load continues to grow up 
to $\tau=1$, when the spray root has reached the leading edge. At this time
the front face of the trolley starts penetrating into the water, circumstance 
that reduces the horizontal speed and then the loads.

In the next sections, comparisons are established between tests performed at 
different conditions, in order to evaluate the effect of the horizontal speed, 
pitch angle and body curvature on pressures and loads.
%
\subsection{Effect of the ditching speed}
\label{sec:effect_speed}
%
As previously noted in \citet{iafrati2019cavitation,iafrati2020experimental},
significant alterations in the hydrodynamics of the water entry
are observed when the horizontal speed is varied 
for the shape \textbf{S3}.
As shown in Table\ref{tab:test_conditions}, when the horizontal speed is varied 
while keeping the ratio $V/U$ constant, the vertical speed also varies.
In particular, at least four regimes 
were identified: a \emph{no-cavitation regime}
(for $U$ lower than about 27~m/s), in which no cavitation
is observed; an \emph{incipient cavitation} regime 
(between $U \approx$ 27~m/s and 35~m/s) in which a cavitation
region expanding in time only up to a certain $\tau$ can be seen; 
an \emph{intermediate regime}, (between $U \approx$ 35~m/s
and 37~m/s) (denoted in
\citet{iafrati2019cavitation} as \textit{Full Cavitation Bubble}
and \textit{Cavitation Alternate with Ventilation}), in which
the cavitation region reaches the trailing edge, followed
by the formation of a local ventilation region, which 
does not expand in relevant way;
finally a \emph{cavitation-ventilation regime} ($U$ higher than 37~m/s),
in which the ventilation region exhibits a successive
fast expansion towards the leading edge of the specimen.

\begin{figure}
\centering
\subfigure[$U$=30~m/s  - Frame $\tau$~=0.2]
{\includegraphics[width=0.4\textwidth]{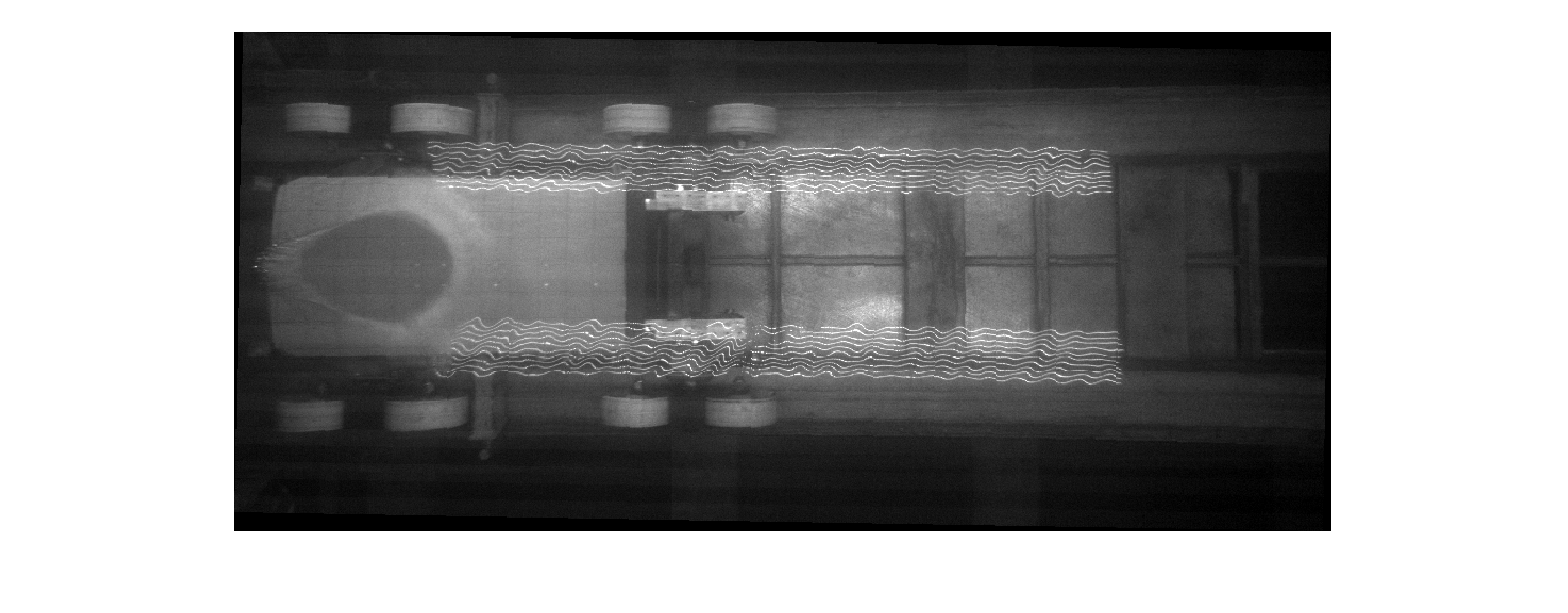}} 
\quad
\subfigure[$U$=35~m/s  - Frame $\tau$~=0.2]
{\includegraphics[width=0.4\textwidth]{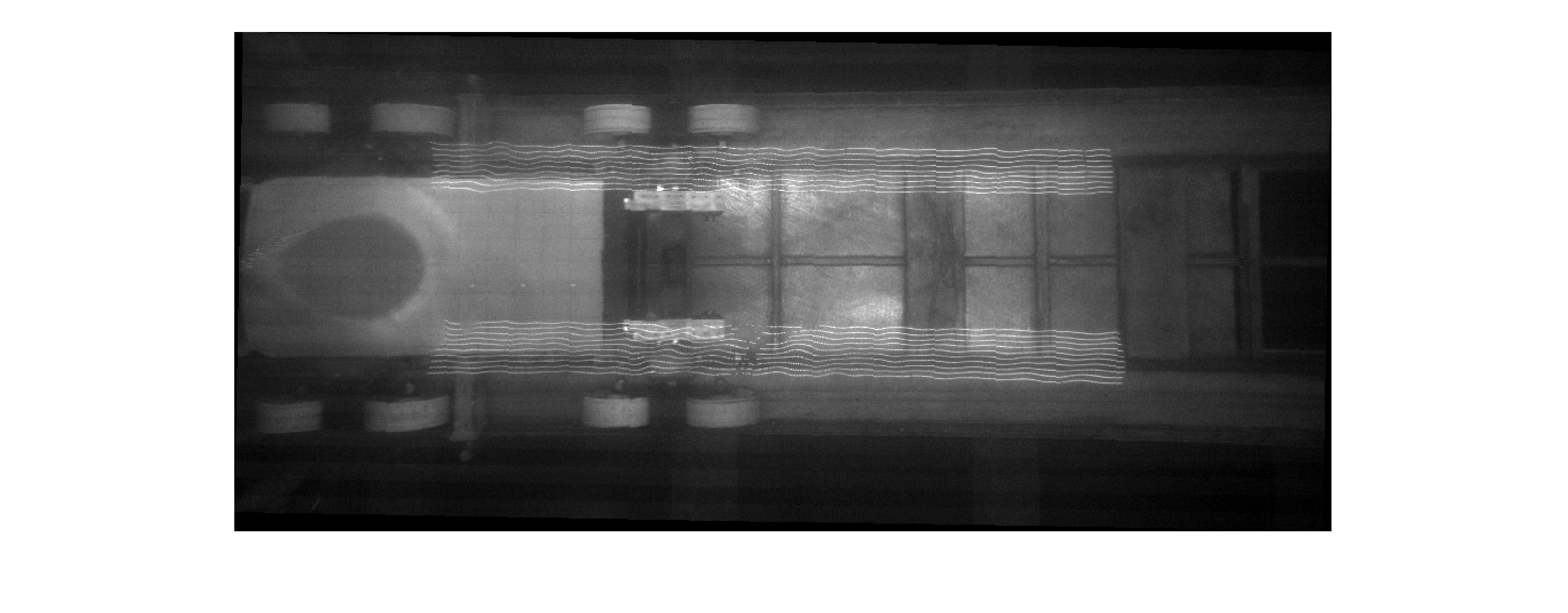}}
\\
\subfigure[$U$=30~m/s  - Frame $\tau$~=0.6]
{\includegraphics[width=0.4\textwidth]{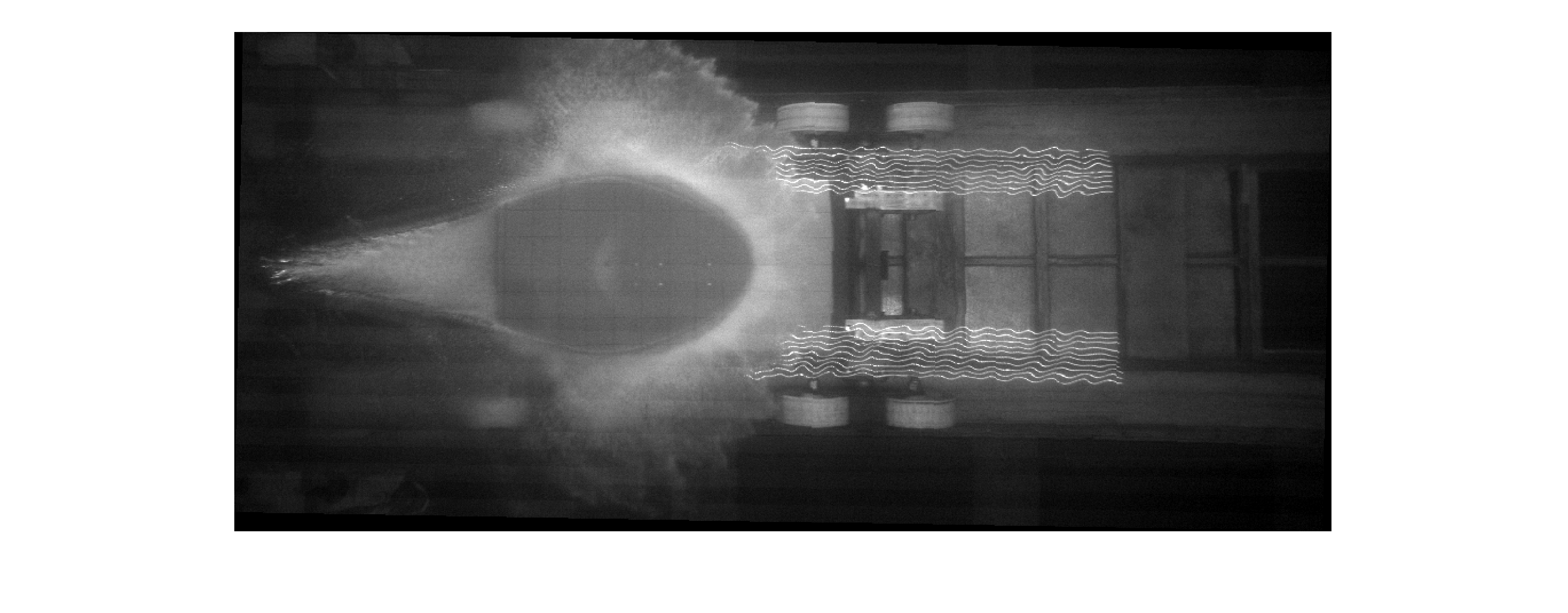}} 
\quad
\subfigure[$U$=35~m/s  - Frame $\tau$~=0.6]
{\includegraphics[width=0.4\textwidth]{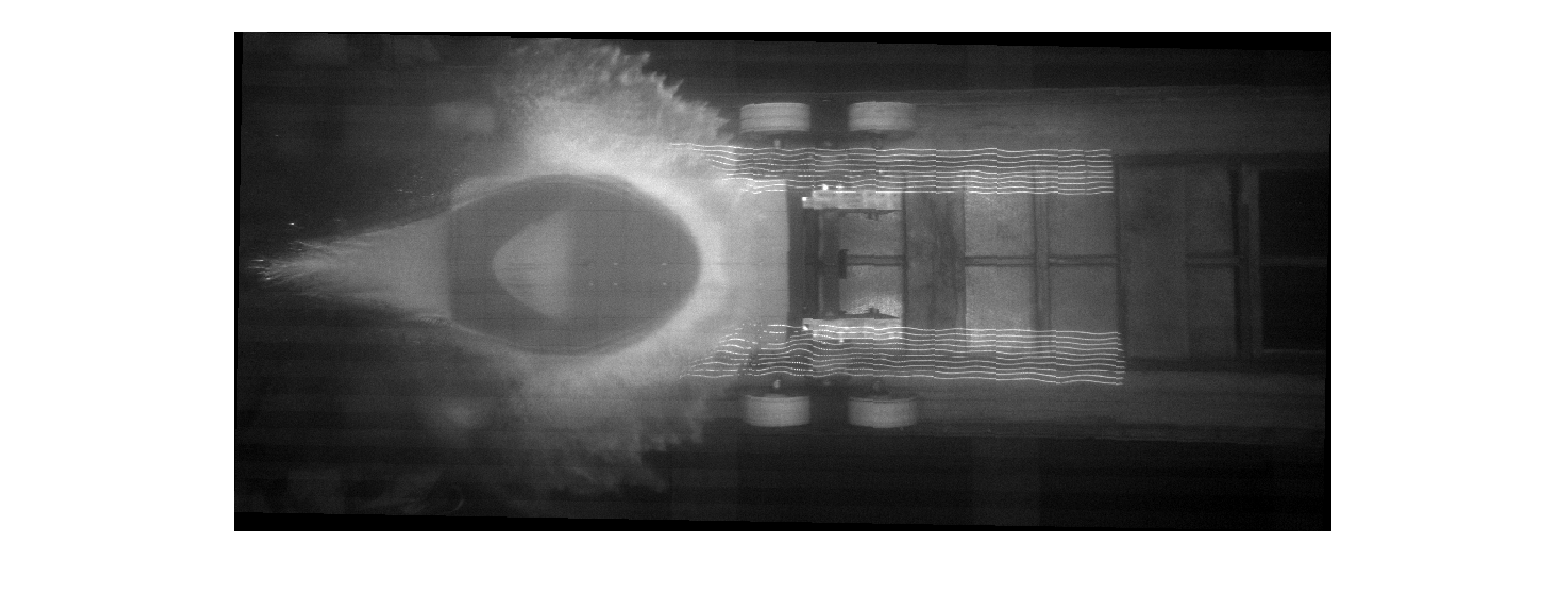}}
\\
\subfigure[$p$ front probes - midline - $U$=~30~m/s]
{\includegraphics[width=0.4\textwidth]{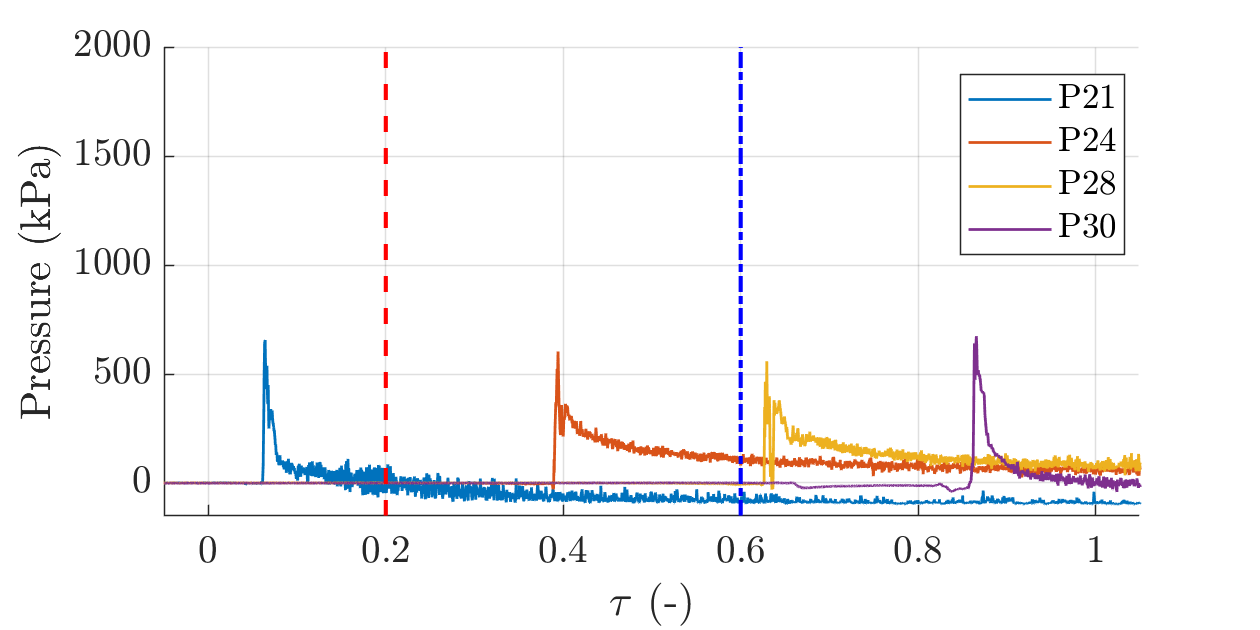}} 
\quad
\subfigure[$p$ front - probes midline -  $U$=~35~m/s]
{\includegraphics[width=0.4\textwidth]{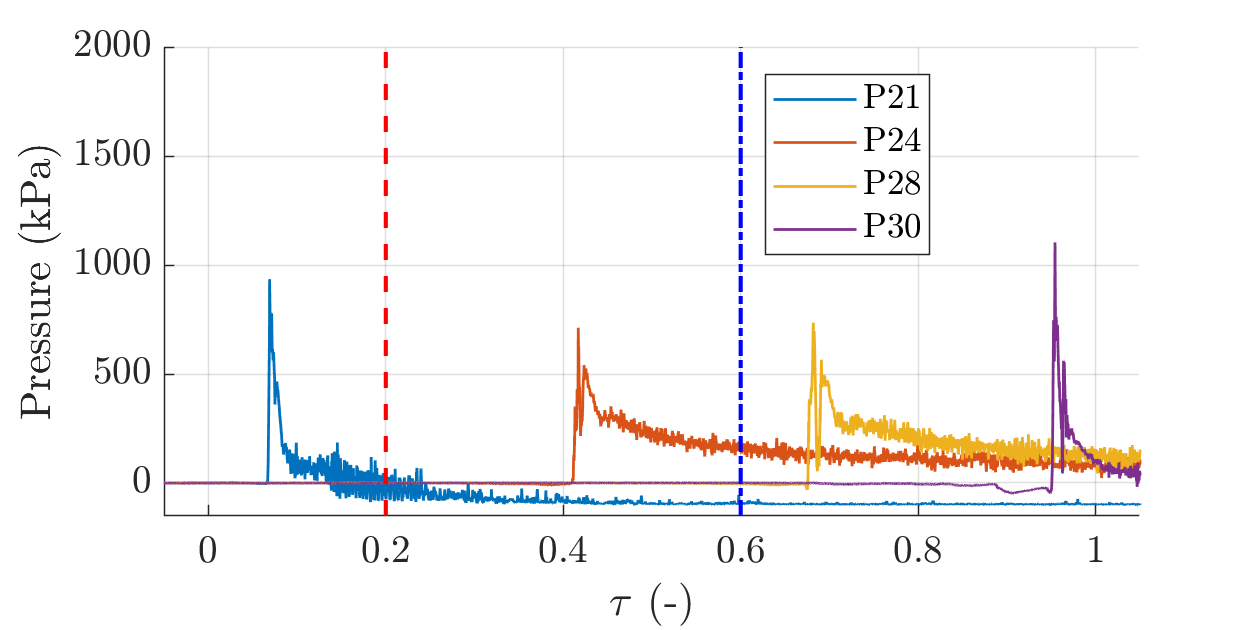}}
\\
\subfigure[$p$ front probes - line l3 $U$=~30~m/s]
{\includegraphics[width=0.4\textwidth]{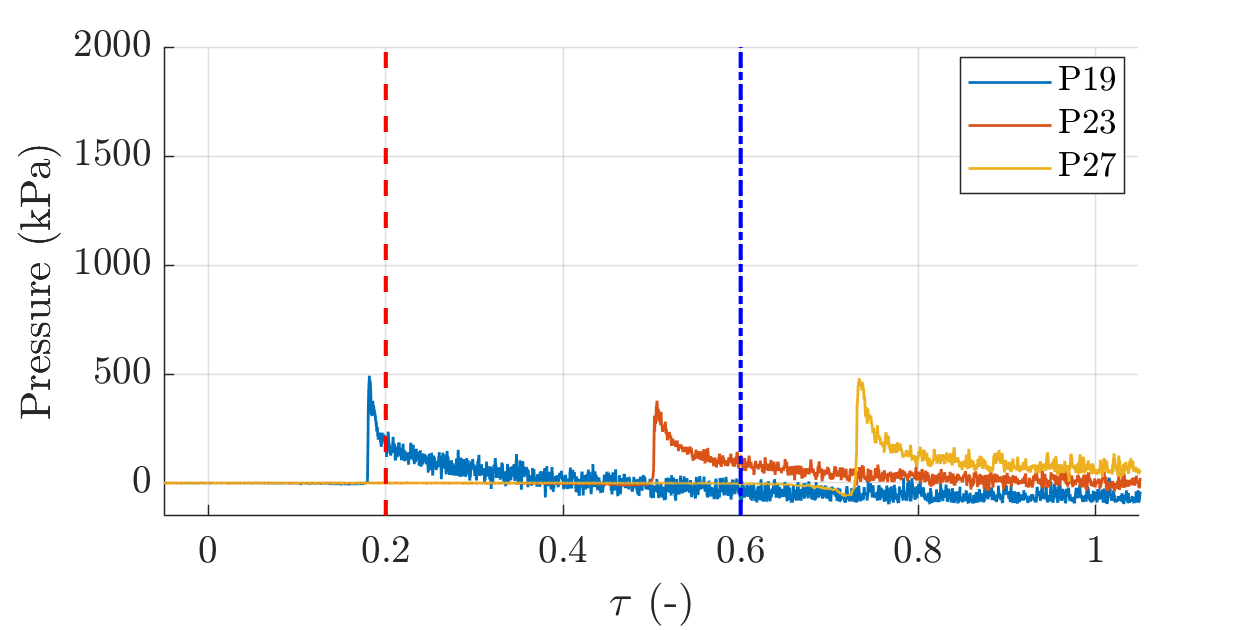}} 
\quad
\subfigure[$p$ front probes - line l3 $U$=~35~m/s]
{\includegraphics[width=0.4\textwidth]{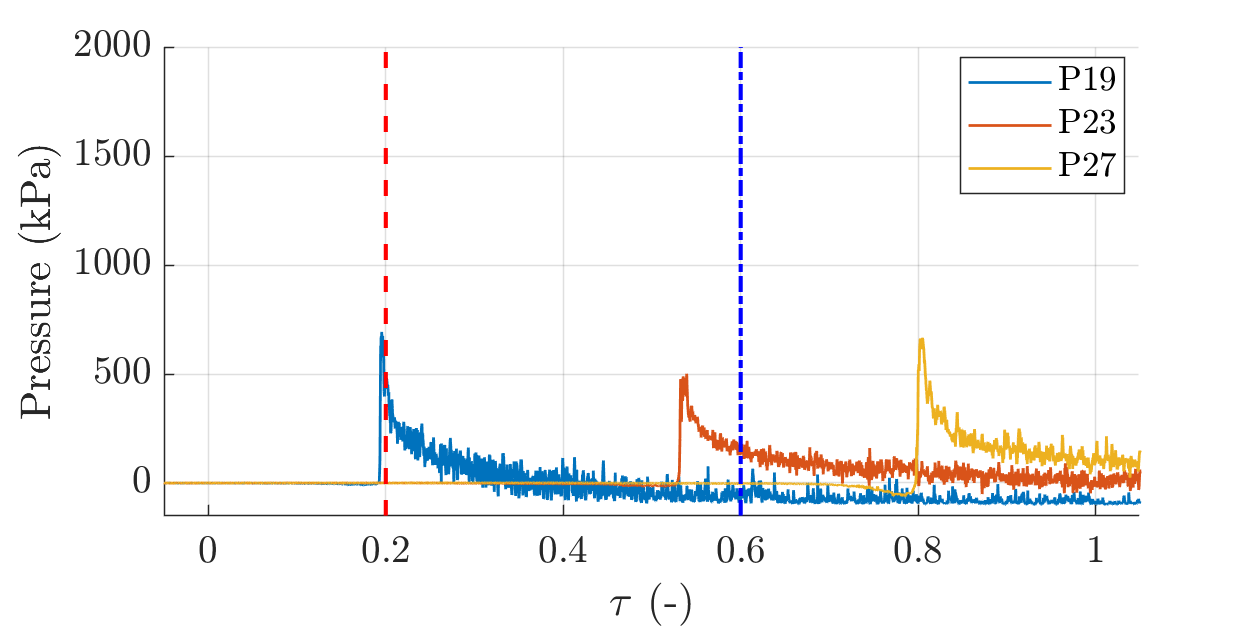}}
\\
\subfigure[$p$ midline - rear probes -  $U$=~30~m/s]
{\includegraphics[width=0.4\textwidth]{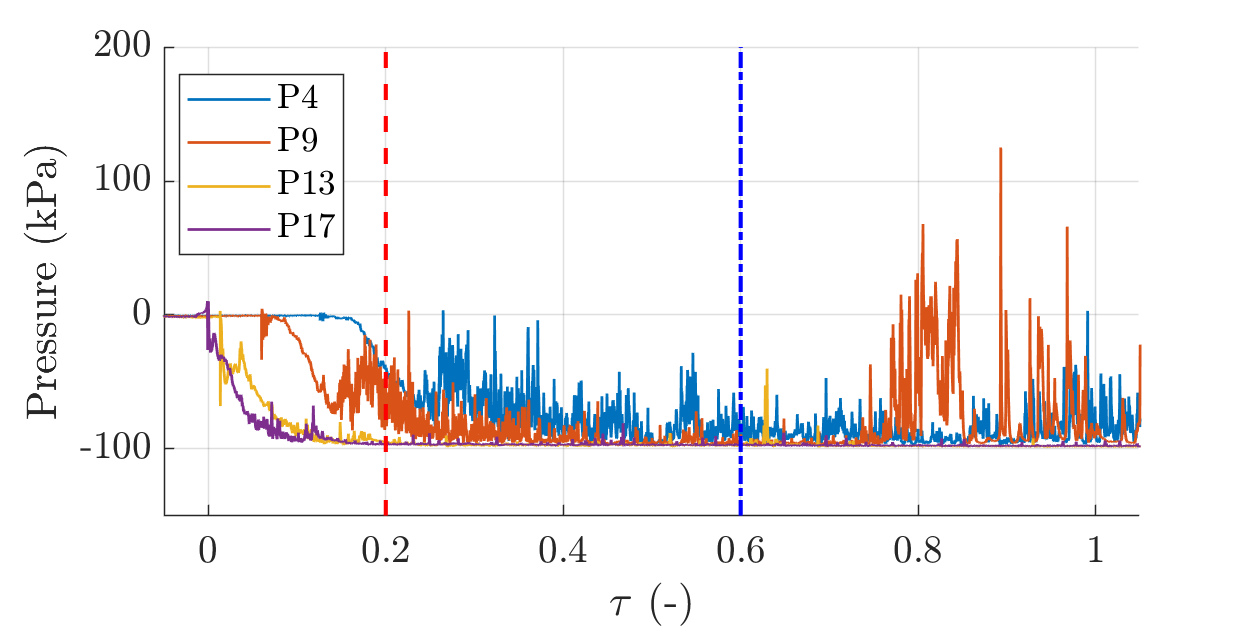}} 
\quad
\subfigure[$p$ midline - rear probes - $U$=~35~m/s]
{\includegraphics[width=0.4\textwidth]{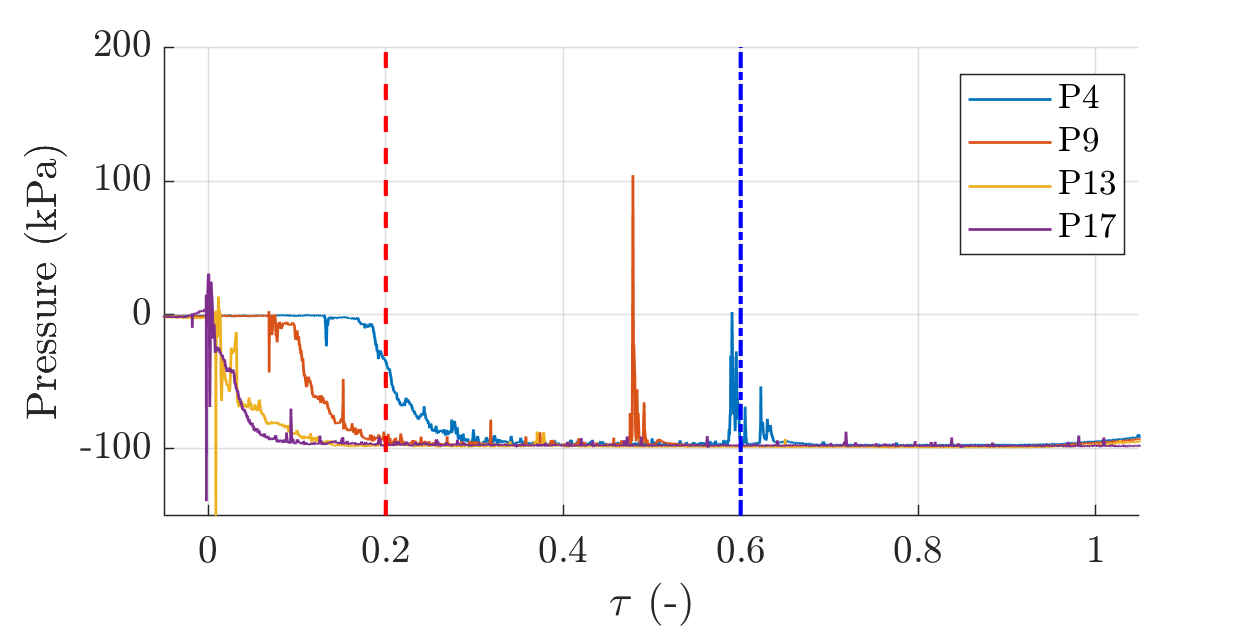}}
\\
\subfigure[$p$ line l3 $U$=~30~m/s]
{\includegraphics[width=0.4\textwidth]{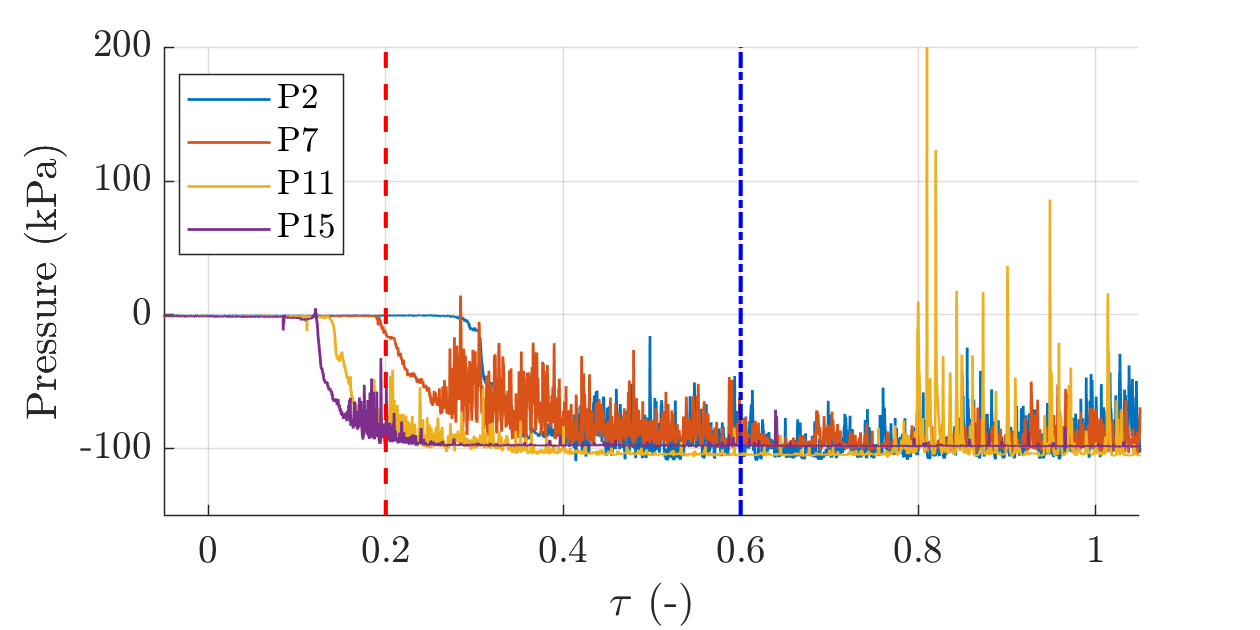}} 
\quad
\subfigure[$p$ line l3 $U$=~35~m/s]
{\includegraphics[width=0.4\textwidth]{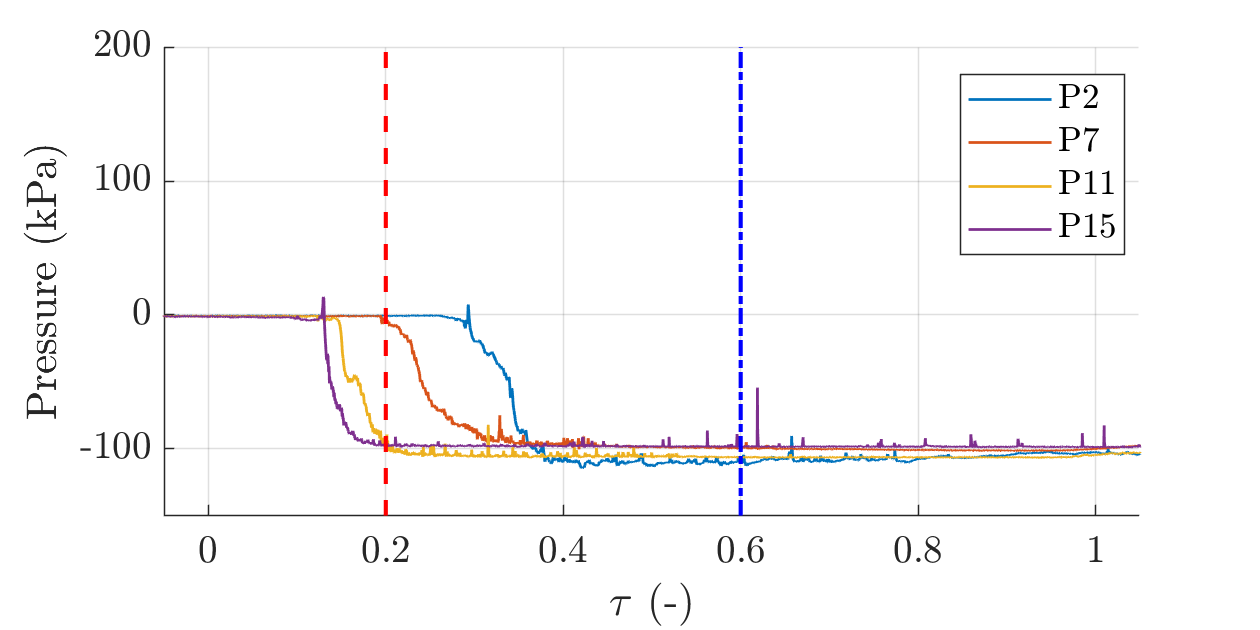}}
\caption{Front and rear pressure probe time histories 
and under-water video frames
comparison for the tests of shape \textbf{S2}, at a
pitch angle 6$^{\circ}$ and $U$=30~m/s and 35~m/s.}
\label{fig:Pressure_UWframes_1H2X22_30_35}
\end{figure}

\begin{figure}
\centering
\subfigure[$U$=40~m/s  - Frame $\tau$~=0.2]
{\includegraphics[width=0.4\textwidth]{1H2222_17_07_2018_1_figures/UW_frame_tau_0.2.png}} 
\quad
\subfigure[$U$=45~m/s  - Frame $\tau$~=0.2]
{\includegraphics[width=0.4\textwidth]{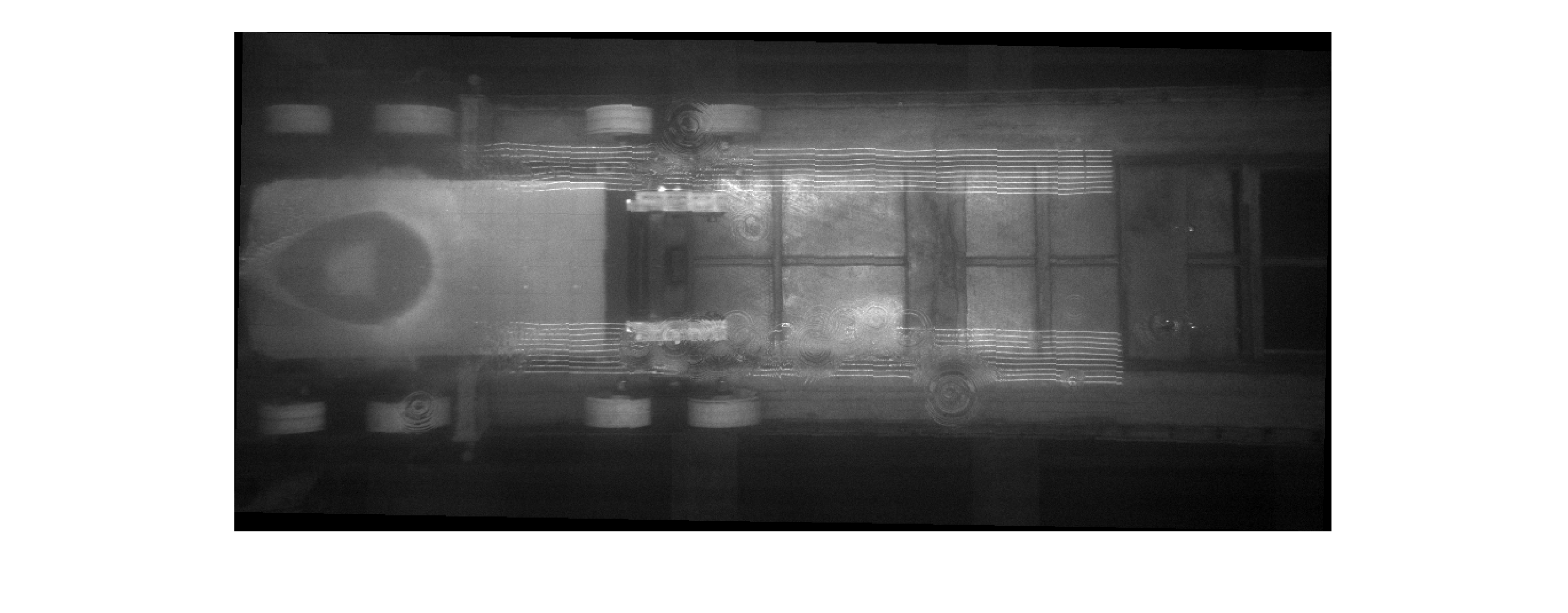}}
\\
\subfigure[$U$=40~m/s  - Frame $\tau$~=0.6]
{\includegraphics[width=0.4\textwidth]{1H2222_17_07_2018_1_figures/UW_frame_tau_0.6.png}} 
\quad
\subfigure[$U$=45~m/s  - Frame $\tau$~=0.6]
{\includegraphics[width=0.4\textwidth]{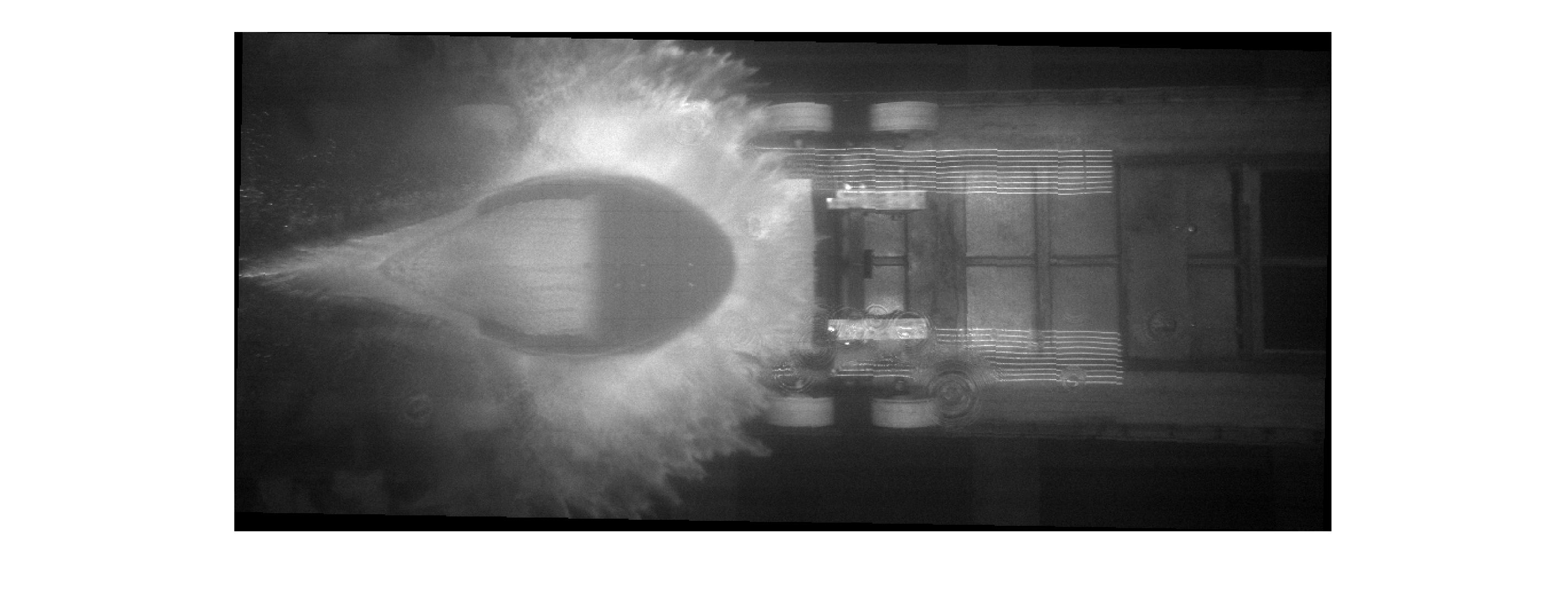}}
\\
\subfigure[$p$ front probes - midline - $U$=~40~m/s]
{\includegraphics[width=0.4\textwidth]{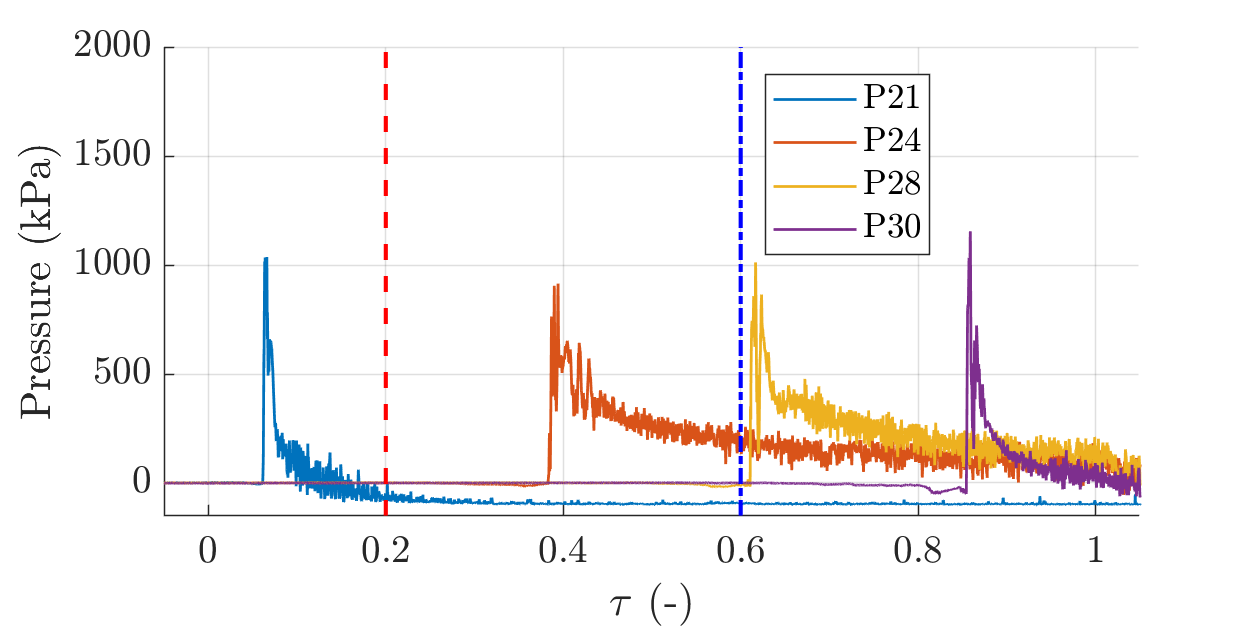}} 
\quad
\subfigure[$p$ front - probes midline -  $U$=~45~m/s]
{\includegraphics[width=0.4\textwidth]{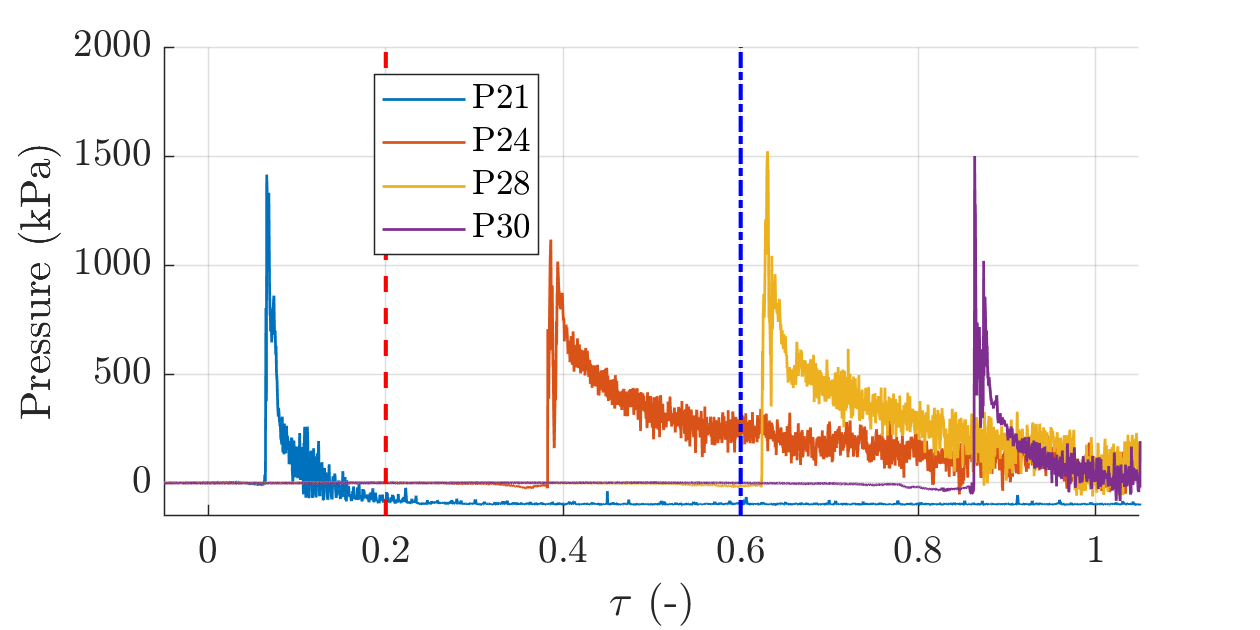}}
\\
\subfigure[$p$ front probes - line l3 $U$=40~m/s]
{\includegraphics[width=0.4\textwidth]{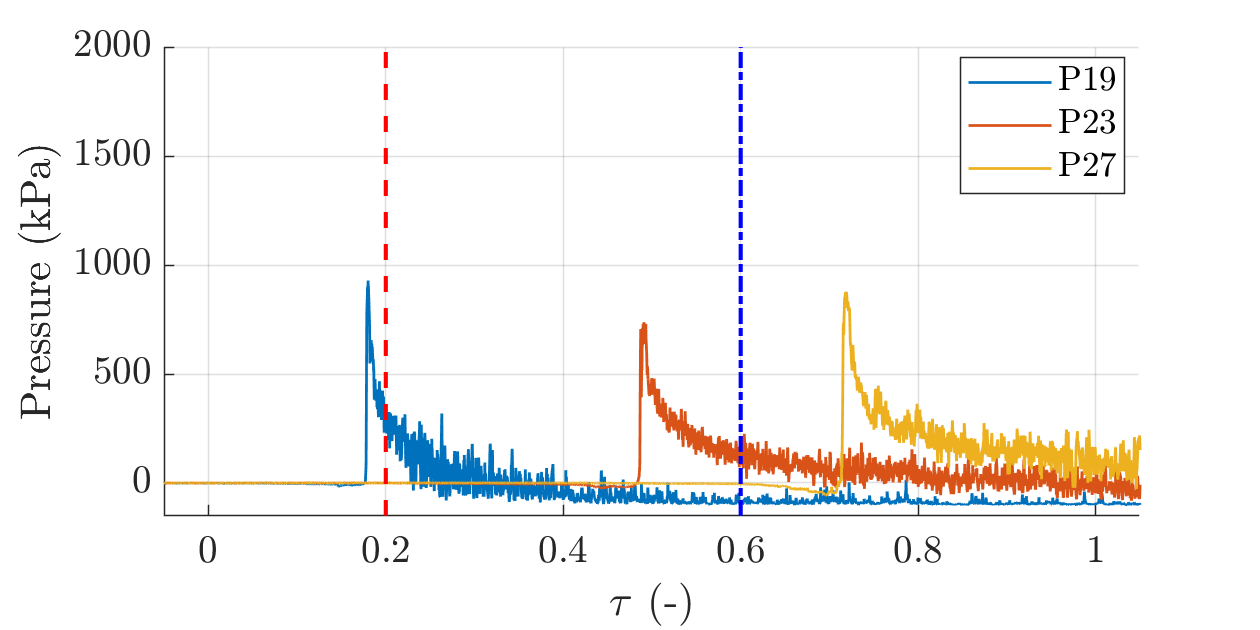}} 
\quad
\subfigure[$p$ front probes - line l3 $U$=45~m/s]
{\includegraphics[width=0.4\textwidth]{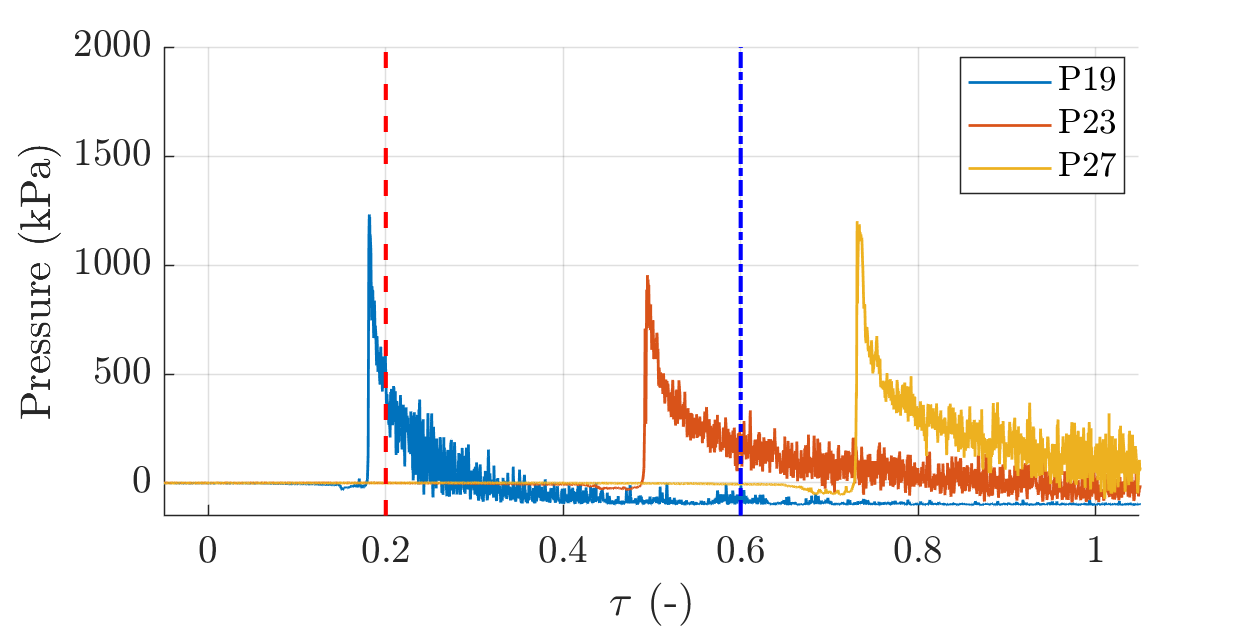}}
\\
\subfigure[$p$ midline - rear probes -  $U$=40~m/s]
{\includegraphics[width=0.4\textwidth]{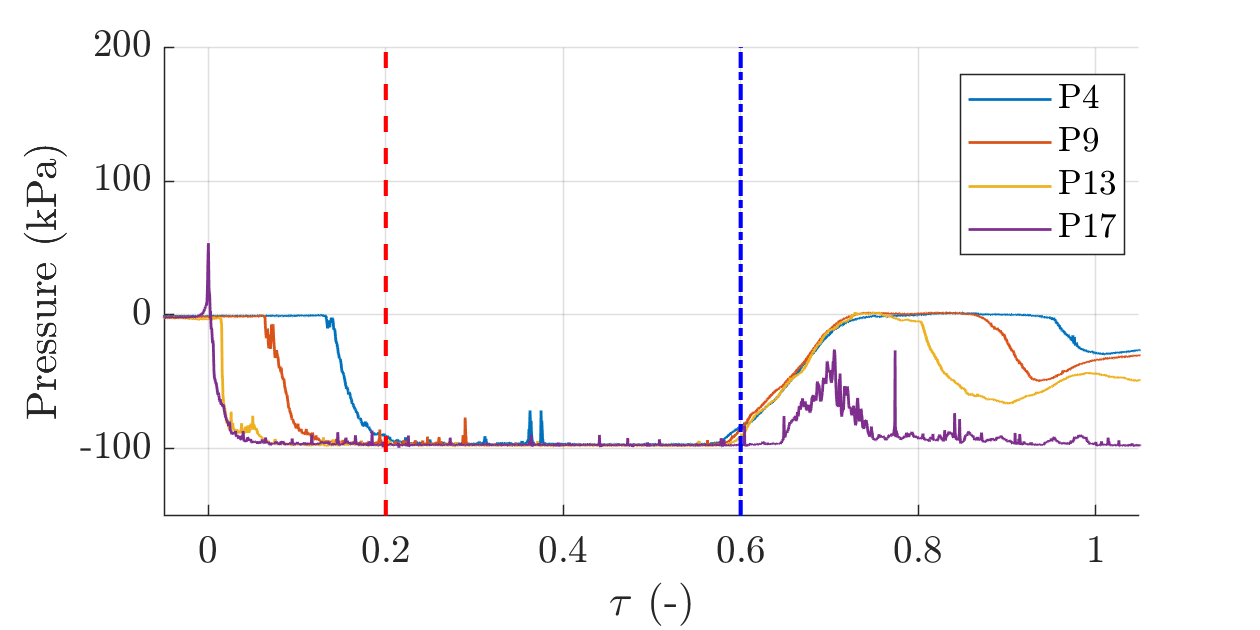}} 
\quad
\subfigure[$p$ midline - rear probes - $U$=45~m/s]
{\includegraphics[width=0.4\textwidth]{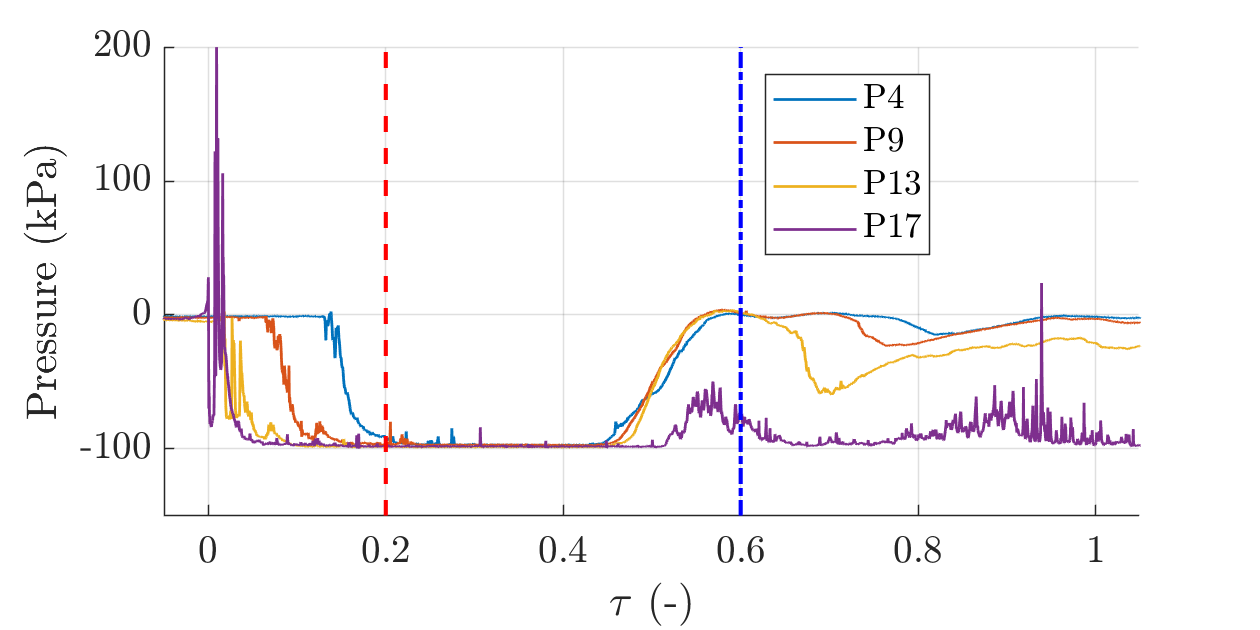}}
\\
\subfigure[$p$ line l3 $U$=40~m/s]
{\includegraphics[width=0.4\textwidth]{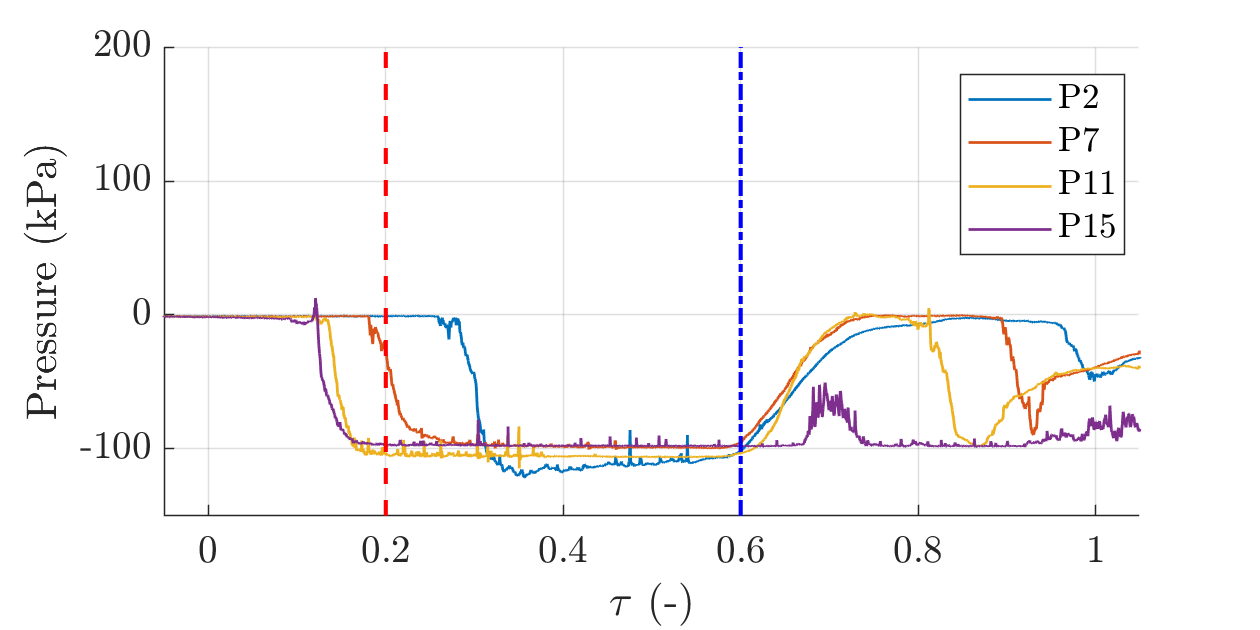}} 
\quad
\subfigure[$p$ line l3 $U$=~45~m/s]
{\includegraphics[width=0.4\textwidth]{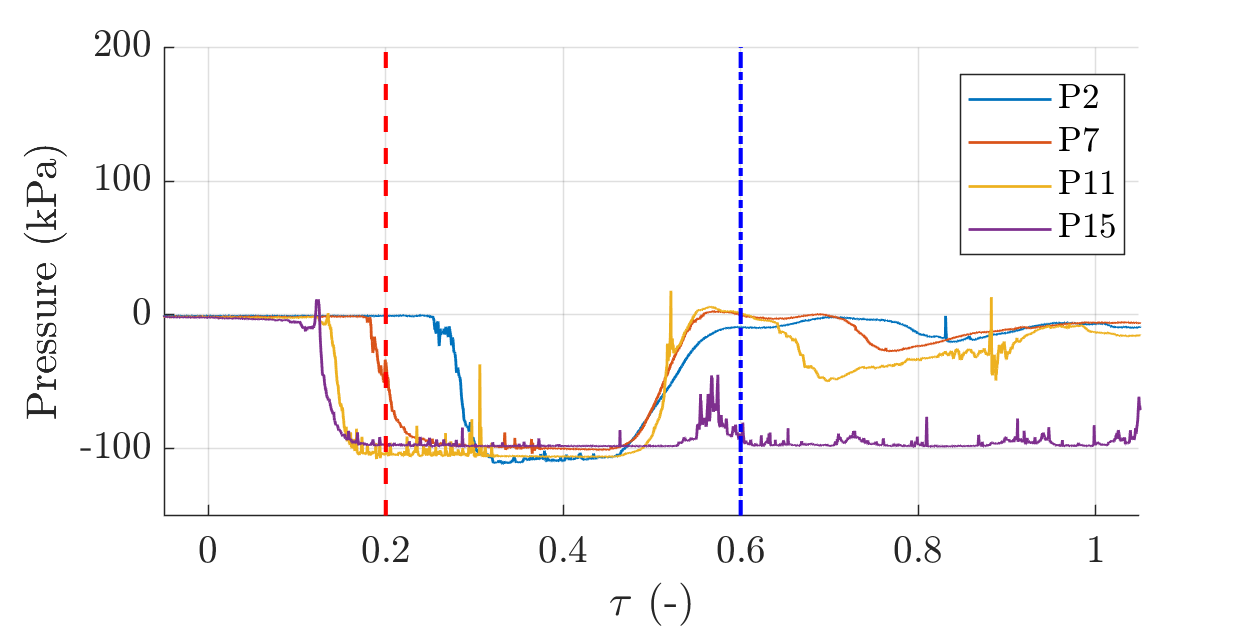}}
\caption{Front and rear pressure probe time histories 
and under-water video frames
comparison for the tests of shape \textbf{S2}, at a
pitch angle 6$^{\circ}$ and $U$=40~m/s and 45~m/s.}
\label{fig:Pressure_UWframes_1H2X22_40_45}
\end{figure}

In the present work, the effect of the 
horizontal speed is analysed for the shape \textbf{S2},
entering water with a pitch angle of 6$^{\circ}$ and at 
velocity ratio of $V/U = 0.0375$.
The analysis is performed in terms of pressures and loads and 
through the observation of the underwater images
taken at two non-dimensional times, i.e. $\tau=0.2$ and $\tau=0.6$. 
A comparison is presented between the time histories of
pressures recorded by the probes located on the midline and on line $l3$.
(Figure \ref{fig:LoadCell_Ref_Frame}). In order to make the comparison easier,
the probes lying along rows $r5-r8$ and $r1-r4$ are 
plotted in separate graphs.
The results for $U$=30~m/s and 35~m/s are shown in 
Figure \ref{fig:Pressure_UWframes_1H2X22_30_35},
whereas those for $U$=40~m/s and 45~m/s in
Figure \ref{fig:Pressure_UWframes_1H2X22_40_45}.
%

As shown in Figures \ref{fig:Pressure_UWframes_1H2X22_30_35} and 
\ref{fig:Pressure_UWframes_1H2X22_40_45} the \emph{probes located in the front}, i.e. 
ahead of the point \textbf{H}, are characterized by the typical sharp rise, followed by
a gentle decay. It is worth noting that, for all probes, the non-dimensional 
time $\tau$ at which the pressure peak is attained is almost independent of the
horizontal speed, whereas the peak value increases with the speed. 
As expected, the peaks recorded by the probes along the line
$l3$ occur at a later time with respect to the peaks recorded by the 
probes at the midline in the same row.
This is a consequence of the combined effect of the transverse curvature 
of the specimen and of the curvature of the spray root, 
which is associated with the possibility 
for the water to escape from the sides \citep{iafrati2016experimental}.

The analysis of the pressure measurements of the \emph{probes located behind the 
point} \textbf{H} is less straightforward and the support of the underwater 
images is needed for a correct interpretation.
In fact, looking at the underwater frames in Figures 
\ref{fig:Pressure_UWframes_1H2X22_30_35} and 
\ref{fig:Pressure_UWframes_1H2X22_40_45} it can be seen that at $\tau$=0.2 
the cavitation region is clearly visible only at $U$=40~m/s and 45~m/s. At 
$\tau$=0.6 the cavitation region is visible at all speeds, 
however at $U$=40~m/s and 45~m/s it has already reached the trailing edge of the
specimen, whereas at $U$=30~m/s and at $U$=35~m/s it has
expanded less significantly.
\begin{figure}
\centering
\subfigure[$U$=30~m/s $\tau$~=1]
{\includegraphics[width=0.48\textwidth]{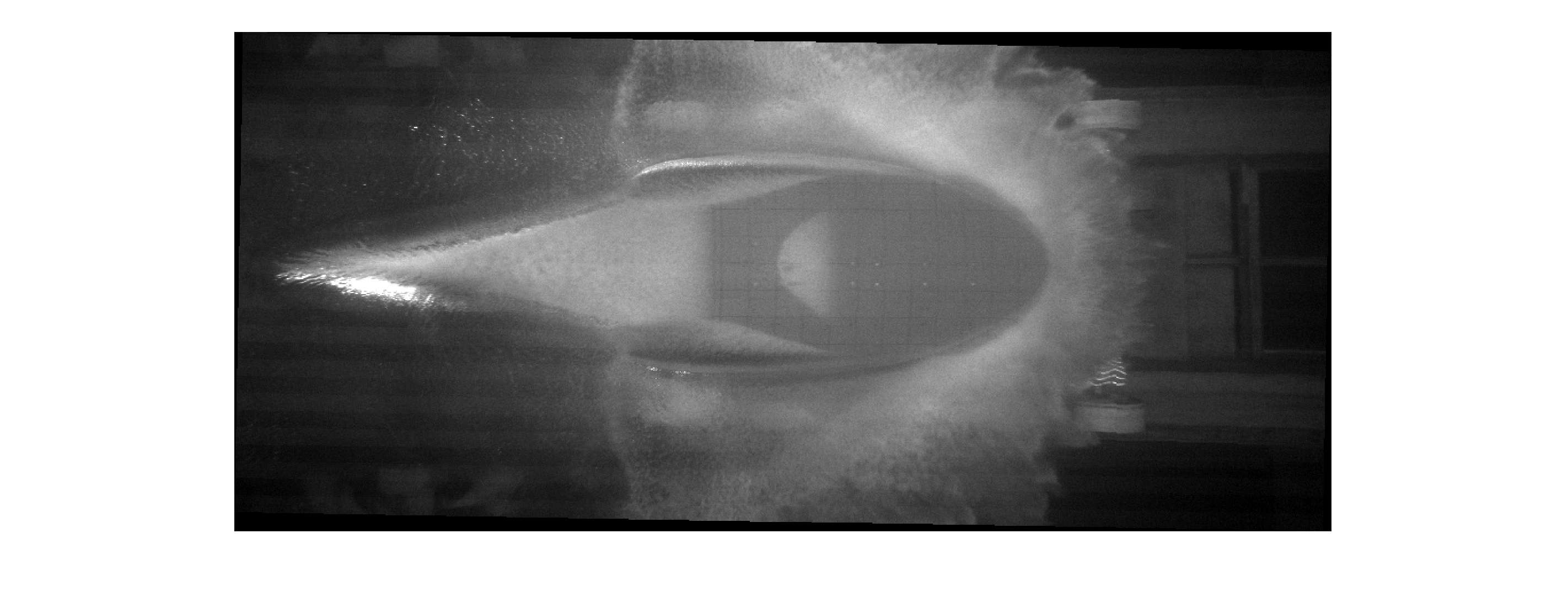}} 
\subfigure[$U$=35~m/s $\tau$~=1]
{\includegraphics[width=0.48\textwidth]{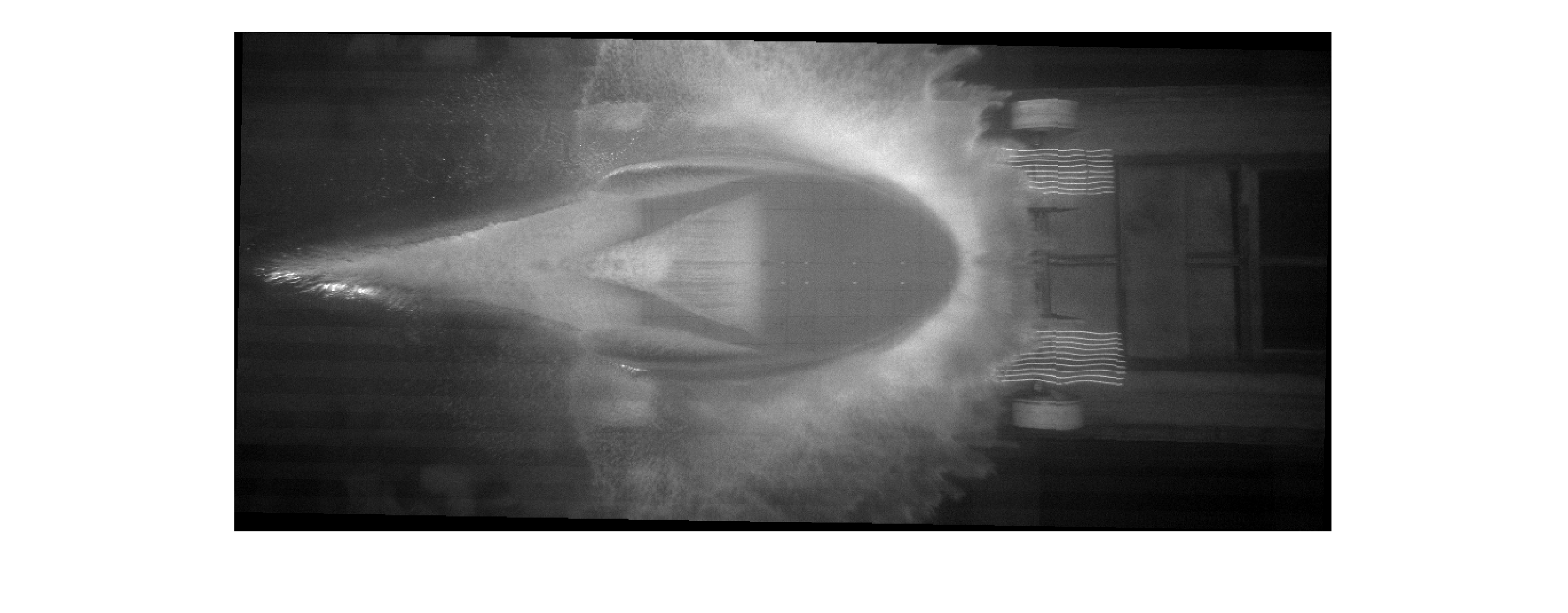}}
\\
\subfigure[$U$=40~m/s $\tau$~=1]
{\includegraphics[width=0.48\textwidth]{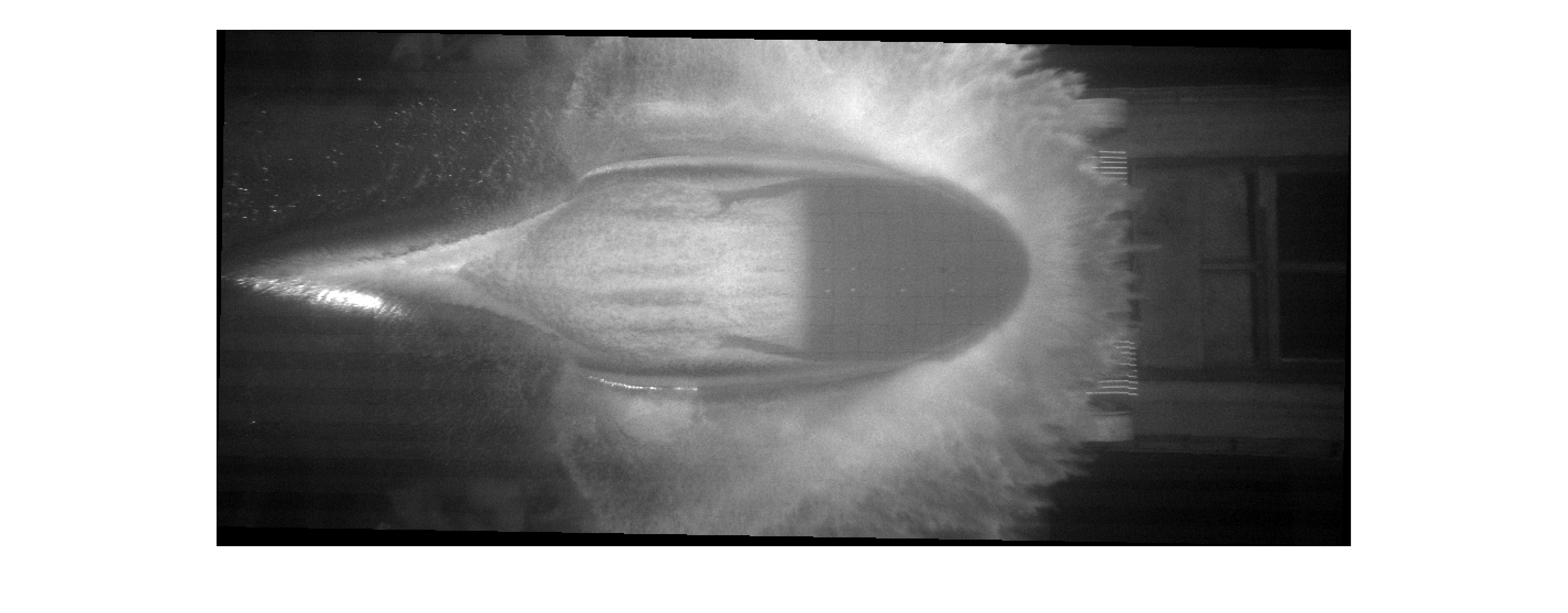}}
\subfigure[$U$=45~m/s $\tau$~=1]
{\includegraphics[width=0.48\textwidth]{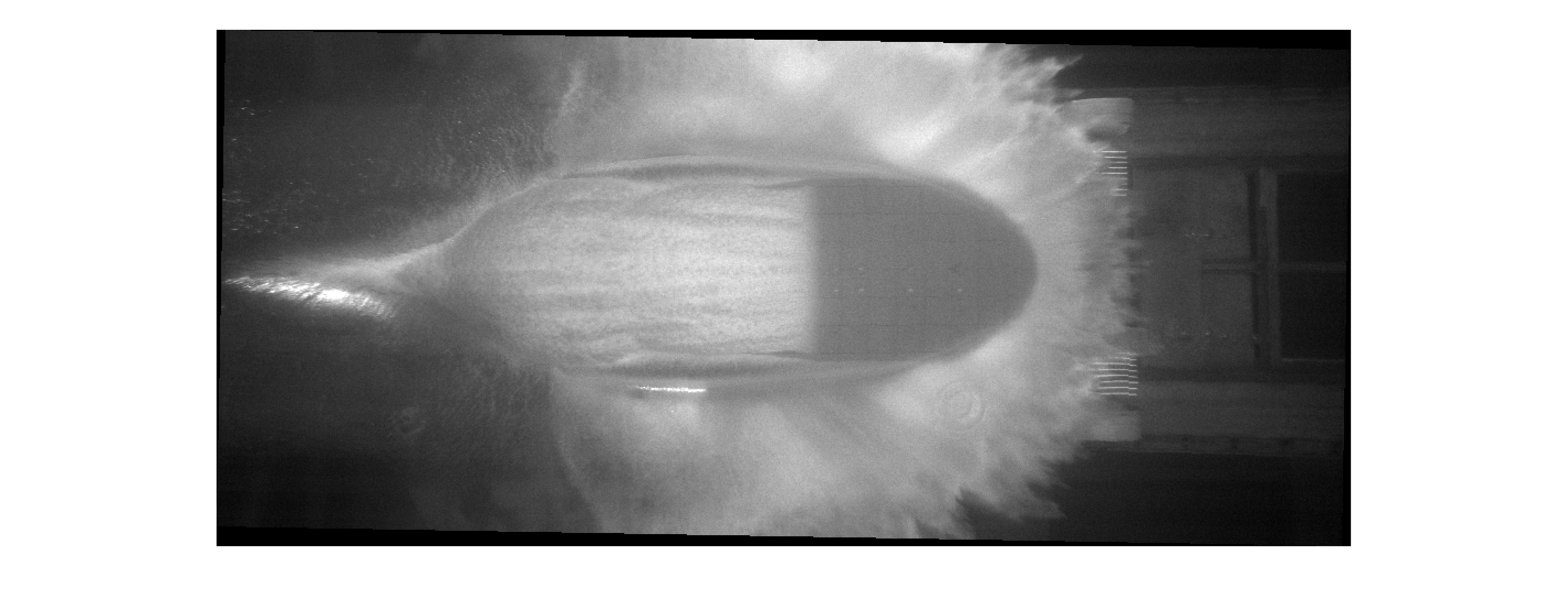}}
\caption{Frames of the underwater video taken at $\tau$=1 for the shape \textbf{S2}, at 
a pitch angle of $\alpha$=6$^{\circ}$ and at the different speeds.}
\label{fig:UWframes_1H2X22_latest_phase}
\end{figure}
%
By looking at the signals recorded by the probes located at the rear, all of
them exhibit a descent to the vapour pressure value. 
It should be noted that at U=30~m/s, possibly due to the incipient 
cavitation regime, signals from rear probes P2, P4, P7, P9, and P11 
exhibit oscillatory behaviour during this descent.
The start time of the descent is consistent across all speeds,
while the descent itself becomes sharper as the horizontal speed increases.
At $U$=40~m/s and 45~m/s, all probes, except for P15 and P17 (located on $r4$), 
display a rise towards the atmospheric value due, to the occurrence of ventilation, 
which is consistent with what shown by the underwater images. The rise starts 
earlier at higher speed. The probes P15 and P17, being located in the region
where the longitudinal curvature is rather high, remains in the cavitation
region and are not reached by the ventilation front.

To gather more insight on the cavitation and ventilation
modalities, especially in the final part of the impact
phase, in Figure \ref{fig:UWframes_1H2X22_latest_phase},
the video frames at $\tau$=1 at all speeds are shown. It is
possible to see that $U$=30~m/s at $\tau$=1 the cavitation
region has expanded with respect to $\tau$=0.6, 
see Figure \ref{fig:Pressure_UWframes_1H2X22_30_35}(c), but its rear edge has not moved 
significantly backwards. Therefore, this case can be ascribed to the 
\emph{incipient cavitation regime}, also discussed in \textbf{S3} \citep{iafrati2019cavitation}.
At $U$=35~m/s the cavitation region has reached the trailing edge and a ventilation
region appears, which however remains confined around the
trailing edge area, similarly to what happens in the \emph{intermediate regime},
discussed above. 
Finally, at $U$=40~m/s and $U$=45~m/s $\tau$~=1 the ventilation
front has clearly moved back towards the leading edge, therefore
in these cases a \emph{cavitation-ventilation} regime is observed. 
As a final comment, it can be observed 
that the threshold values that delineate the 
various regimes in \textbf{S2} are analogous to those of \textbf{S3}.

Starting from this last consideration, this aspect can be related not only 
to the velocity values, but also to the cavitation number, which is defined as follows: 
\begin{equation}
	Ca = \frac{p_{\textrm{atm}}-p_{\textrm{vap}}}{1/2 \, \rho \, U^2}.
\end{equation}
The cavitation number, other relevant nondimensional parameters,
and boundary layer characteristics for the tested shape conditions \textbf{S2} and \textbf{S3} are 
reported in Table \ref{tab:S2_parameters} and Table \ref{tab:S3_parameters}, respectively.
Table \ref{tab:S3_parameters} includes additional test
conditions that are not analysed in detail in this
article but are discussed in \citet{iafrati2019cavitation}.
\begin{table}
\begin{center}
\def~{\hphantom{0}}
\begin{tabular}{ccccccc}
$U$ [m/s] & Ca [-] & Cavitation Modality & $Re$ [-] & $\delta$ [mm] & $C_f$ [-] & $\ell_v$ [m] \\
29.87 & 0.222 & Incipient    & $3.70 \times 10^7$ & 14.1 & 0.00242 & $9.63 \times 10^{-7}$ \\
34.50 & 0.167 & Intermediate & $4.28 \times 10^7$ & 13.7 & 0.00237 & $8.43 \times 10^{-7}$ \\
40.25 & 0.122 & Cavit.+Vent. & $4.99 \times 10^7$ & 13.2 & 0.00231 & $7.31 \times 10^{-7}$ \\
46.19 & 0.093 & Cavit.+Vent. & $5.73 \times 10^7$ & 12.9 & 0.00226 & $6.44 \times 10^{-7}$ \\
\end{tabular}
\caption{Characteristic non-dimensional and boundary layer parameters for Shape \textbf{S2}}
\label{tab:S2_parameters}
\end{center}
\end{table}
\begin{table}
\begin{center}
\def~{\hphantom{0}}
\begin{tabular}{ccccccc}
$U$ [m/s] & Ca [-] & Cavitation Modality & $Re$ [-] & $\delta$ [mm] & $C_f$ [-] & $\ell_v$ [m] \\
21.00 & 0.450 & No Cavit.    & $2.60 \times 10^7$ & 15.1 & 0.00256 & $1.33 \times 10^{-6}$ \\
26.80 & 0.276 & No Cavit.    & $3.32 \times 10^7$ & 14.4 & 0.00246 & $1.06 \times 10^{-6}$ \\
30.60 & 0.212 & Incipient    & $3.79 \times 10^7$ & 14.0 & 0.00241 & $9.42 \times 10^{-7}$ \\
34.50 & 0.167 & Intermediate & $4.28 \times 10^7$ & 13.7 & 0.00237 & $8.43 \times 10^{-7}$ \\
35.70 & 0.156 & Intermediate & $4.43 \times 10^7$ & 13.6 & 0.00235 & $8.17 \times 10^{-7}$ \\
37.20 & 0.143 & Cavit.+Vent. & $4.61 \times 10^7$ & 13.5 & 0.00234 & $7.86 \times 10^{-7}$ \\
40.20 & 0.123 & Cavit.+Vent. & $4.98 \times 10^7$ & 13.2 & 0.00231 & $7.32 \times 10^{-7}$ \\
45.20 & 0.097 & Cavit.+Vent. & $5.60 \times 10^7$ & 12.9 & 0.00227 & $6.57 \times 10^{-7}$ \\
\end{tabular}
\caption{CCharacteristic non-dimensional and boundary layer parameters for Shape \textbf{S3}}
\label{tab:S3_parameters}
\end{center}
\end{table}
For both shapes no cavitation occurs for 
$\textrm{Ca} \geq$~0.25. 
At $\textrm{Ca}$ values around 0.20–0.22, incipient cavitation 
is observed, while for $\textrm{Ca} \leq$~0.2 the cavitation region extends 
downstream, eventually leading to ventilation.
This trend with $\textrm{Ca}$ is consistent 
with the observations of \citet{wang2001dynamics} and 
aligns with the description of the stages of sheet cavitation 
on hydrofoils discussed in \cite{huang2019review}.

As a final note, it should be reminded that 
the effect of Reynolds number in water entry flows is generally 
considered secondary, as supported by \citet{moghisi1981experimental, facci2015numerical}. 
Additionally, the pressure drop at the rear can be explained solely by 
potential theory \citep{faltinsen2005hydrodynamics,newman,semenov2006nonlinear,riccardi2004water},
indicating a negligible influence of viscous effects. 
Furthermore, data in Tables \ref{tab:S2_parameters} and \ref{tab:S3_parameters} 
show that both boundary layer thickness and viscous length, estimated 
approximately using the flat plate analogy do not vary significantly across the tested
range of horizontal velocities. Therefore, boundary layer effects should
remain essentially consistent across the cases and speeds under investigation.

\begin{figure}
\centering
\includegraphics[width=0.95\textwidth]{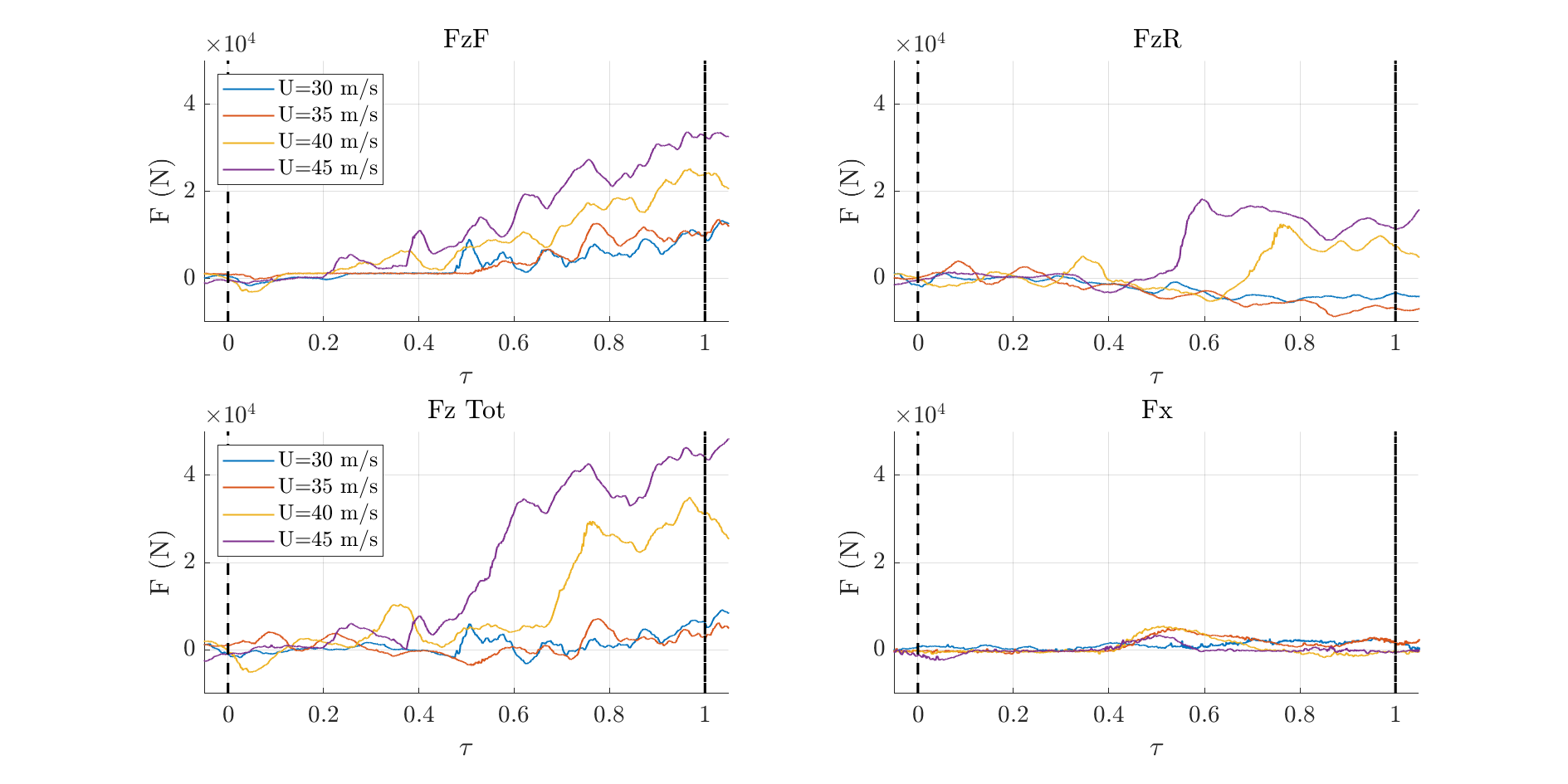}
\caption{Time histories of the forces for the tests of shape \textbf{S2}, 
at pitch angle 6$^{\circ}$ and different horizontal speeds.}
\label{fig:Forces_S2_6deg_RAW}
\end{figure}

The force acting in the $z$-direction and measured by the load cells located
at the front and at the rear are shown on the upper panels of 
Figure \ref{fig:Forces_S2_6deg_RAW}.
%
The normal force recorded by the front load cells $F_{zF}$
is more sensitive to the 
pressures acting on the front portion of the specimen,
and for this reason it displays an increasing trend, which is consistent
with the forward propagation of the spray root.
The force intensity grows with speed, in line with the corresponding 
increase of the pressures.
On the other hand, the rear load cells are more sensitive to the pressures 
acting on the back portion of the specimen and then, 
aside from the very early stage of the impact phase,
they undergo the effects of the 
cavitation and ventilation phenomena. The load at the rear $F_{zR}$ 
exhibit a similar behaviour at the different speeds,
assuming more and more negative values as the cavitation area expands 
up to about $\tau$= 0.5. 
Successively, when the ventilation starts, earlier for the test at $U$=45~m/s and 
then at 40~m/s, the force at the rear undergoes a sharp growth.
It is worth noticing that this sharp variation of the loads
introduces a significant displacement of the centre of loads during the water entry, 
which may have an important effect on the aircraft dynamics in a real ditching scenario.
The total loads acting in the $z$- and $x$-directions are provided in
the lower panels of 
Figure \ref{fig:Forces_S2_6deg_RAW}. 
The data indicate that the
$x$-component is much smaller than the $z$-component and is characterized by
a small hump occurring about $\tau$= 0.5. 
Although already visible in terms of the rear and forward components, the total 
$z$-force is characterized by a large increase with the horizontal speed,
starting from the time at which the ventilation starts. Before that time, 
the positive force in the front part is partly balanced by the negative loads 
at the rear. When ventilation starts, there is a significant increase in the total 
loads due to the jump in the pressure values acting on the rear portion of the 
specimen. A positive contribution in this time interval also derives from 
the increase in the front force, as a consequence of the growth of the wetted area
and to the forward propagation of the spray root, 
with the associated pressure peak.

%
\subsection{Effect of the pitch angle}
\label{sec.effect_pitch}
\begin{figure}
\centering
\includegraphics[width=0.85\textwidth]{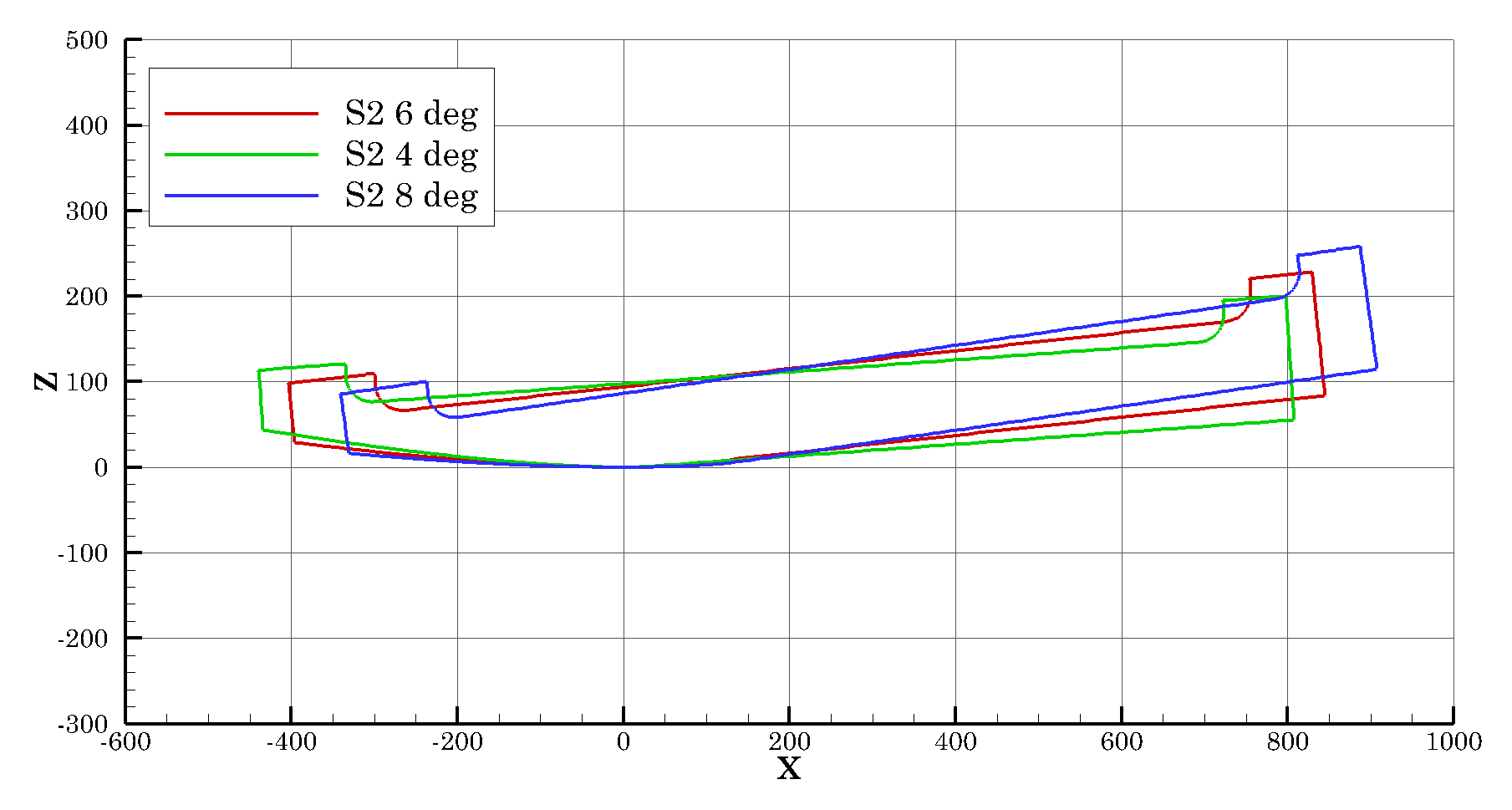}
\caption{Comparison among the midline profiles of the shape \textbf{S2}
 at different pitch angles and aligned with respect 
to the first contact point \textbf{H}.} 
	\label{fig:comparison_S2_pitches_midplane}
\end{figure}

\begin{figure}
	\centering
	\includegraphics[width=0.75\textwidth]{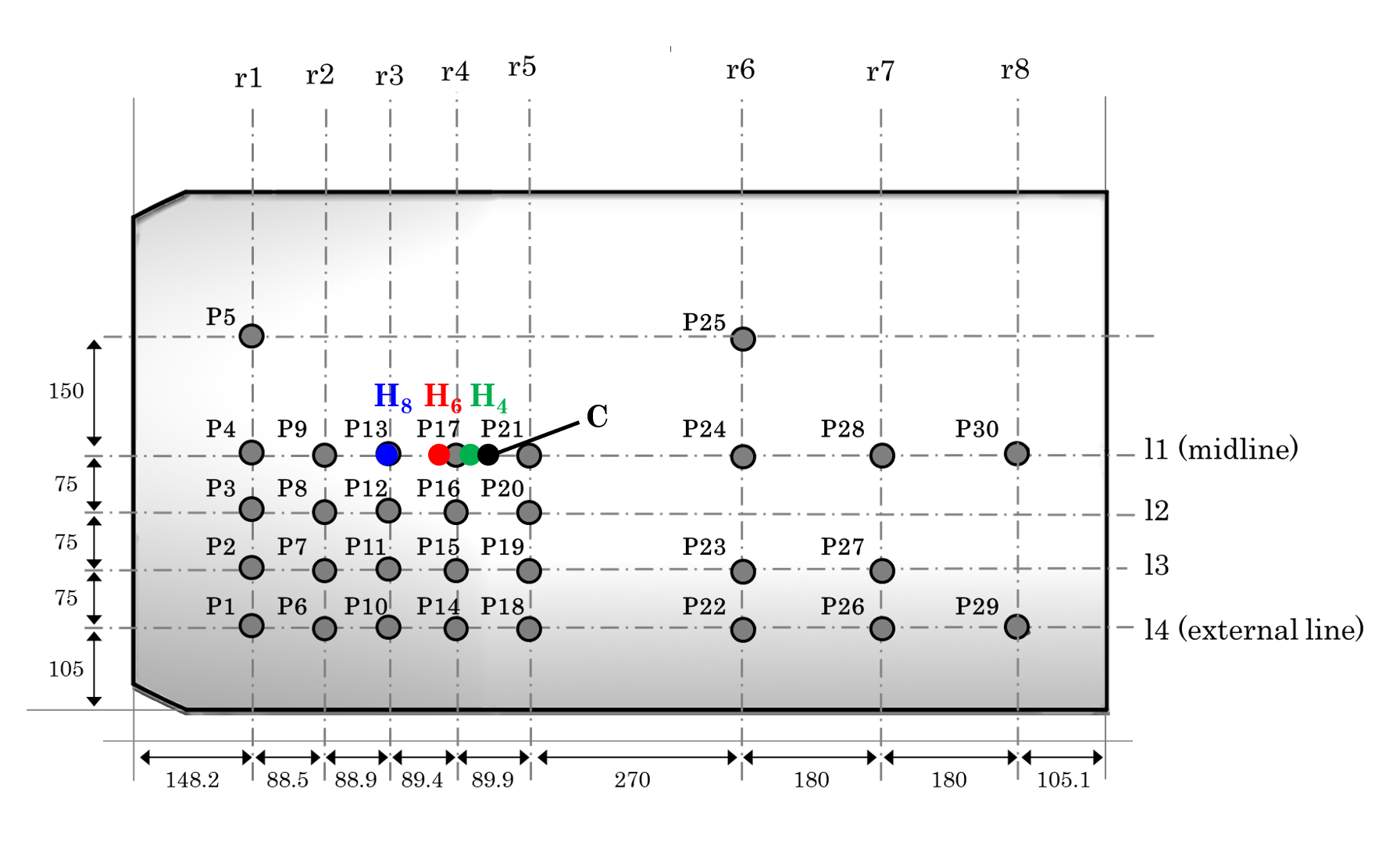}
	\caption{Position of the curvature change point \textbf{C} 
	and of the points of first contact \textbf{H$_4$}, \textbf{H$_6$}
	and \textbf{H$_8$}, which correspond to the pitch angles of 4$^{\circ}$,
	6$^{\circ}$ and 8$^{\circ}$, respectively.}
	\label{fig:S2_4_6_8deg_FirstContactPoints_PressProbes}
\end{figure}

%
The effect of the pitch angle on pressures and loads is somewhat more
complicated by the fact that the first contact point \textbf{H} changes
with it, and therefore the local geometry, in particular the
longitudinal curvature around the first contact point, is different. 
This feature is highlighted in Figure \ref{fig:comparison_S2_pitches_midplane}, 
where the mid-plane sections of the specimen \textbf{S2} at the three
pitch angles of 4$^{\circ}$, 6$^{\circ}$ and 8$^{\circ}$, aligned with
respect to the first contact point, are plotted.
It is worth noticing that the differences among the three profiles are
not so relevant behind \textbf{H} but are very evident in the front part.
As the first contact point \textbf{H} moves backwards as the pitch angle 
increases, the relative positions of the probes with respect to the point of
initial contact are not the same, as shown in Figure 
\ref{fig:S2_4_6_8deg_FirstContactPoints_PressProbes}.
%

\begin{figure}
\centering
\subfigure[Pitch 4$^{\circ} - \tau$=0.6]
{\includegraphics[width=0.31\textwidth]{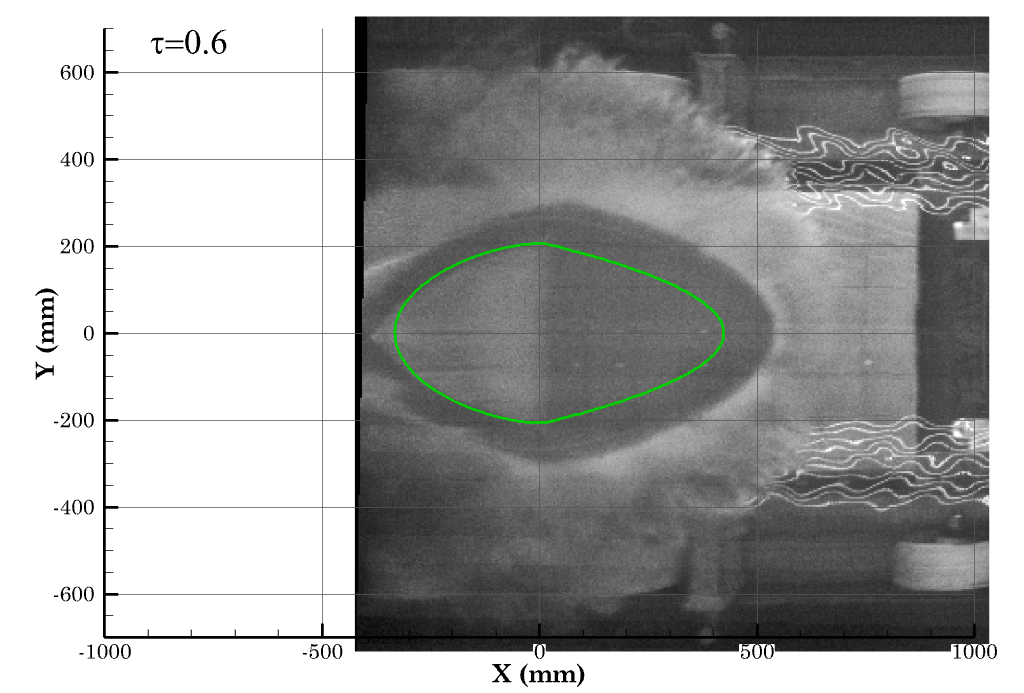}} 
\quad
\subfigure[Pitch 6$^{\circ} - \tau$=0.6]
{\includegraphics[width=0.31\textwidth]{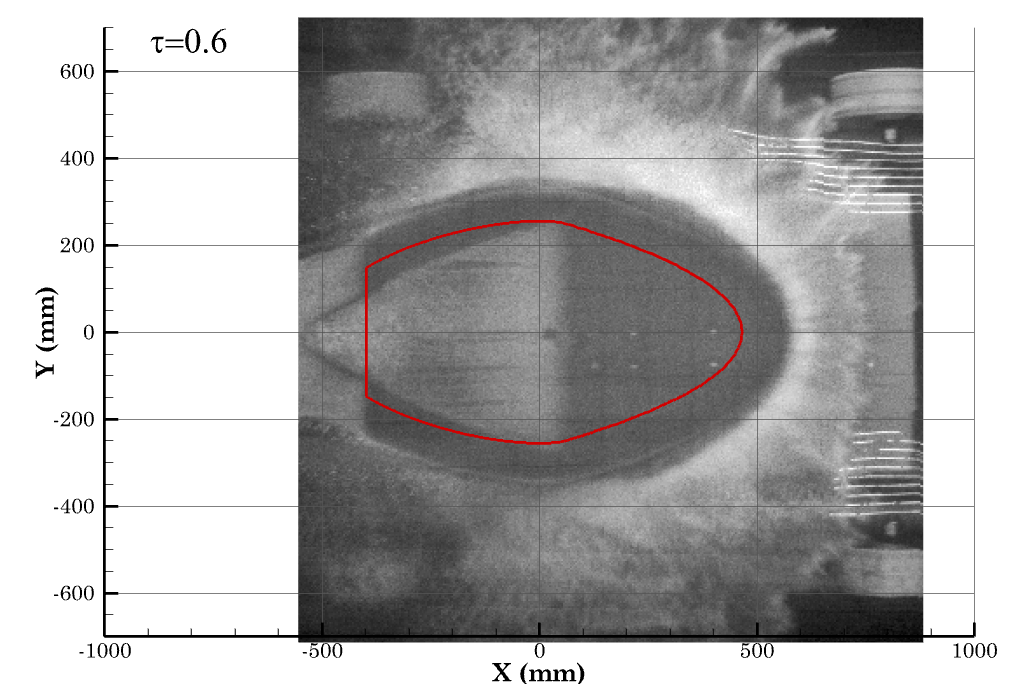}}
\quad 
\subfigure[Pitch 8$^{\circ} - \tau$=0.6]
{\includegraphics[width=0.31\textwidth]{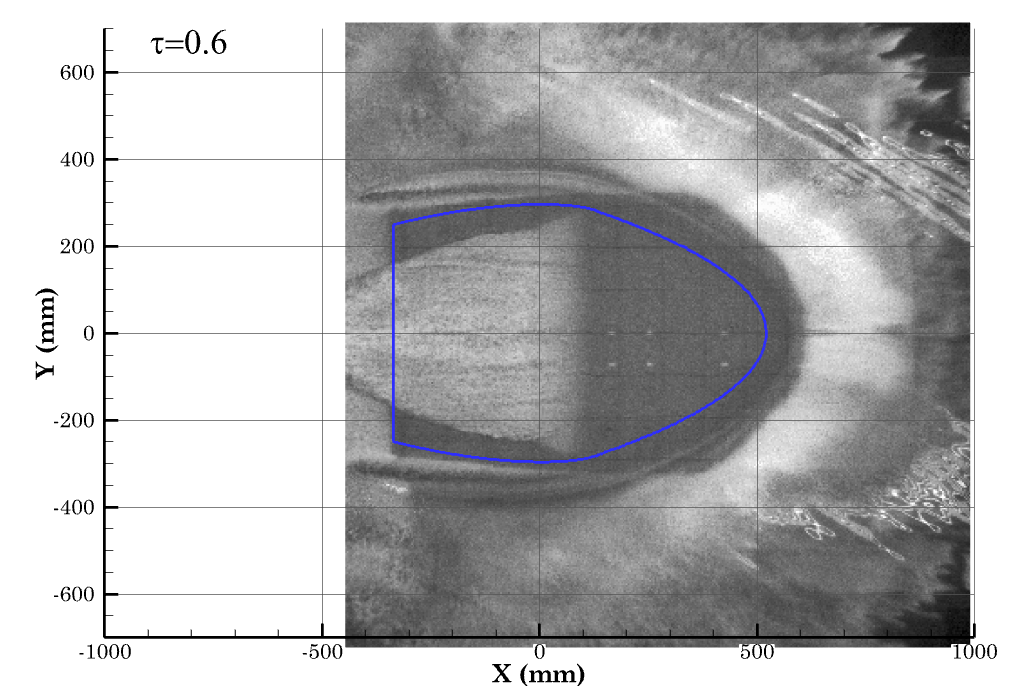}}
\\
\subfigure[Pitch 4$^{\circ} - \tau$=0.9]
{\includegraphics[width=0.31\textwidth]{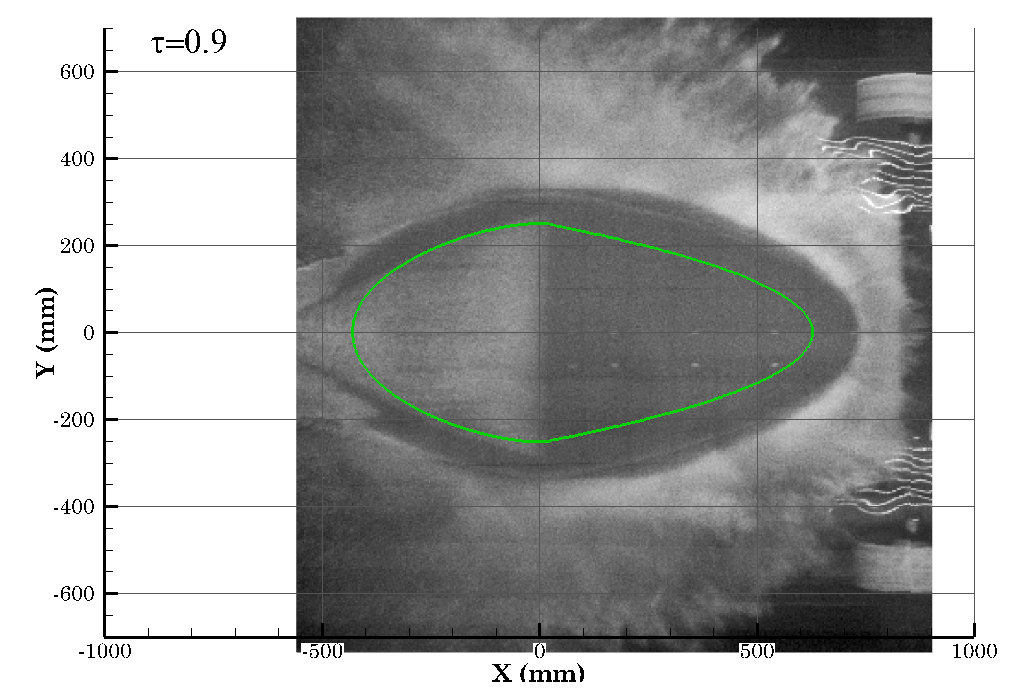}} 
\quad
\subfigure[Pitch 6$^{\circ} - \tau$=0.9]
{\includegraphics[width=0.31\textwidth]{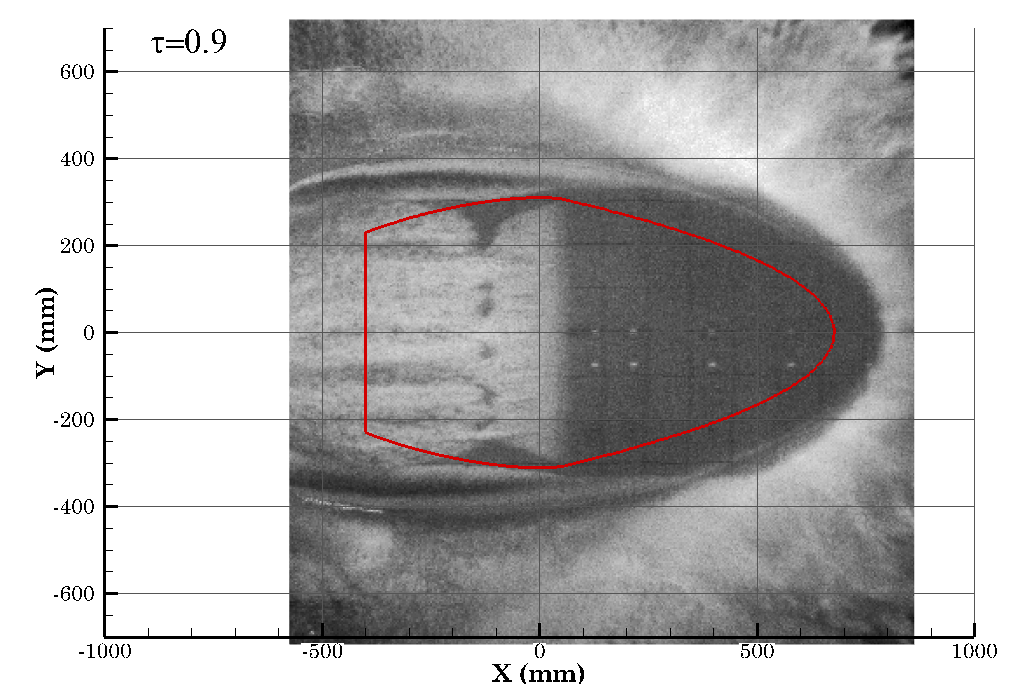}}
\quad 
\subfigure[Pitch 8$^{\circ} - \tau$=0.9]
{\includegraphics[width=0.31\textwidth]{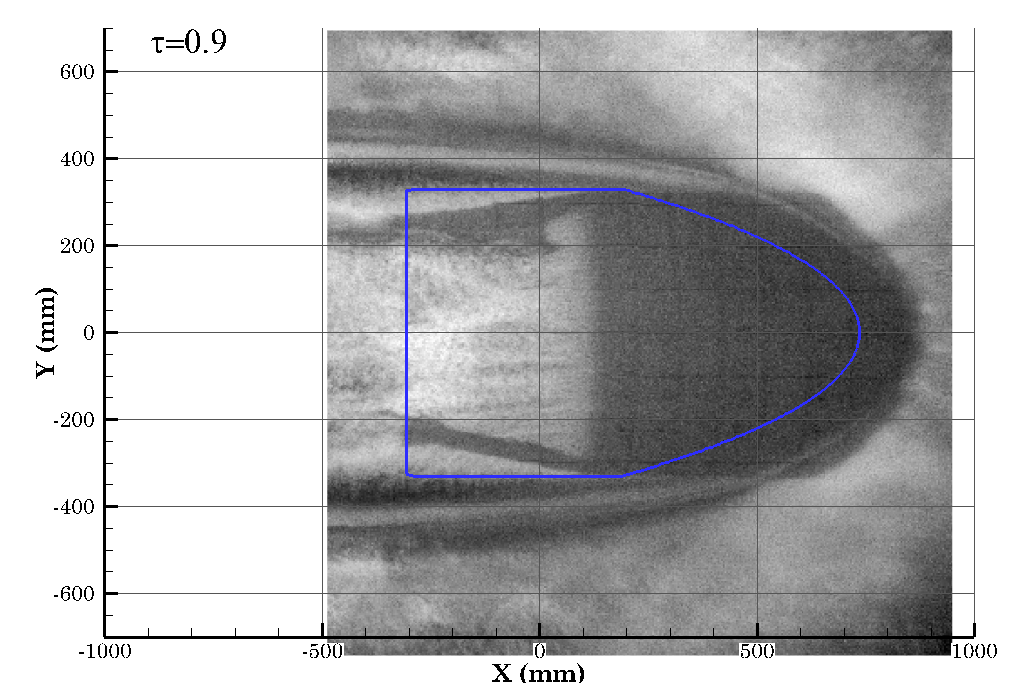}}
\caption{Under-water frames comparison
at $\tau$=~0.6 and 0.9 for the specimen \textbf{S2} at $U$=40~m/s and
at the pitch angles 4$^{\circ}$, 6$^{\circ}$ and 8$^{\circ}$.}
\label{fig:UW_frames_diff_pitch}
\end{figure}

%
In figure \ref{fig:UW_frames_diff_pitch} the underwater frames at 
$\tau$=~0.6 and $\tau$=~0.9 for the
different pitch angles are shown, together with the edges of
geometric intersection between
the specimen inclined at different pitch angles and
the still water level at $\tau$=0.6 and $\tau$=0.9. 
%
%
By comparing the wetted area with the geometric intersection
areas, a large
consistency in found, aside from the presence of the cavitation and
ventilation zones 
at the rear as well as the pile-up and the spray region in front.

The expansion of the cavitation zone 
to the trailing edge is observed to be faster at higher 
pitch angles and consequently the onset of 
ventilation occurs earlier. 
Such a situation is expected. In fact, as clearly shown 
in Figure \ref{fig:comparison_S2_pitches_midplane}
the slope gradient about the point \textbf{H}, while traversing the lower 
specimen profile from the leading edge to the trailing edge, 
is more pronounced in the case of higher pitch angles. 
This results in the development of lower pressures at the 
rear during the impact phase, 
thereby facilitating a faster expansion of the cavitation region. 
In addition, it is observed that
at 6$^{\circ}$ and at 8$^{\circ}$ and at $\tau$=0.6 
the cavitation region has already reached the trailing edge and 
a ventilation front has already started propagating forwards.
Instead, at 4$^{\circ}$ the shape of the cavitation region 
has not changed relevantly between $\tau$=0.6 and $\tau$=0.9.
Such a situation is similar to the case of the 
intermediate cavitation regimes 
discussed in Section \ref{sec:effect_speed}. 

\begin{landscape}
\begin{figure}
\subfigure[$p$ midline - Front probes - Pitch 4$^{\circ}$]
{\includegraphics[width=0.4\textwidth]{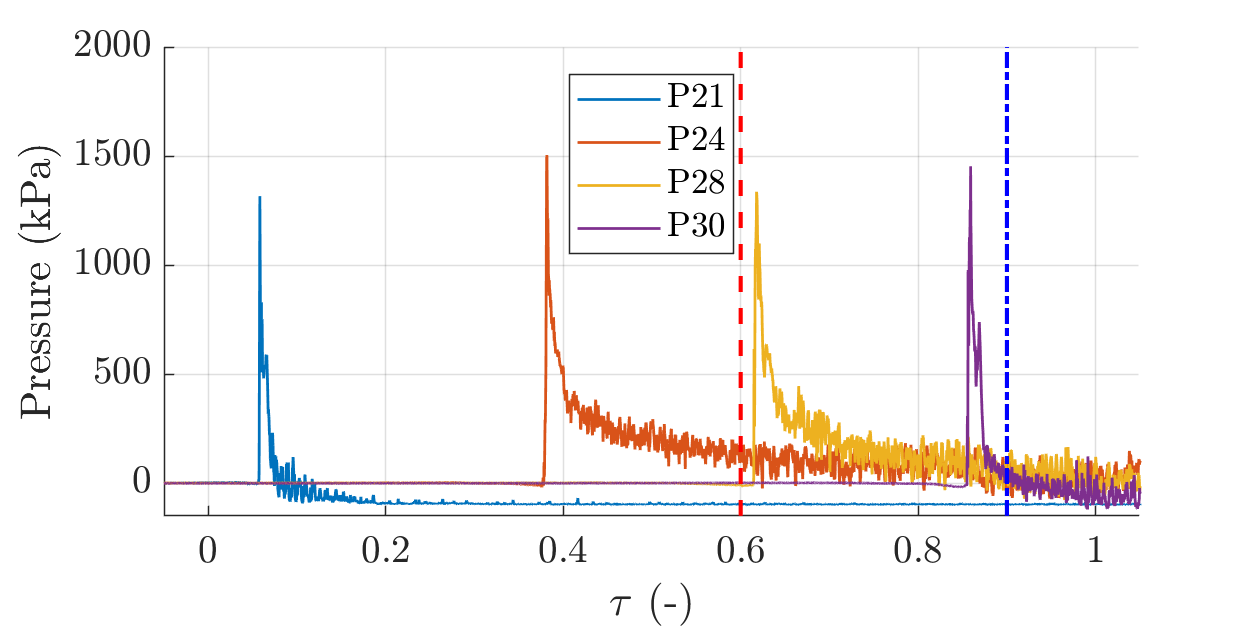}} 
\qquad \qquad
\subfigure[$p$ midline - Front probes - Pitch 6$^{\circ}$]
{\includegraphics[width=0.4\textwidth]{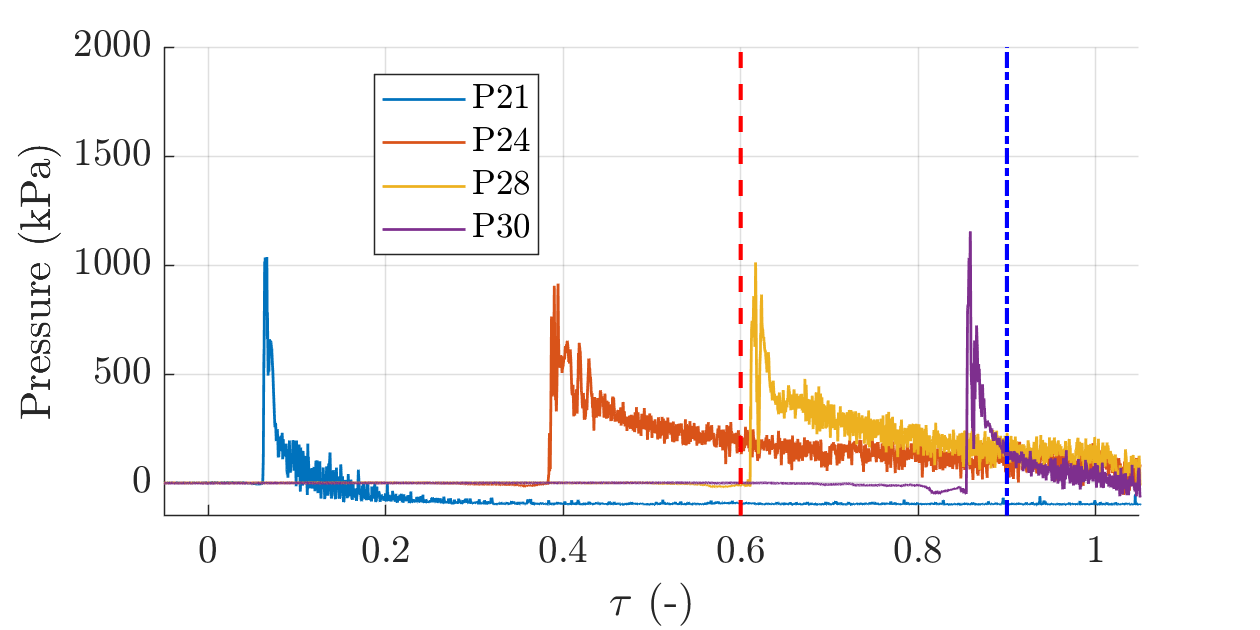}}
\qquad \qquad
\subfigure[$p$ midline - Front probes - Pitch 8$^{\circ}$]
{\includegraphics[width=0.4\textwidth]{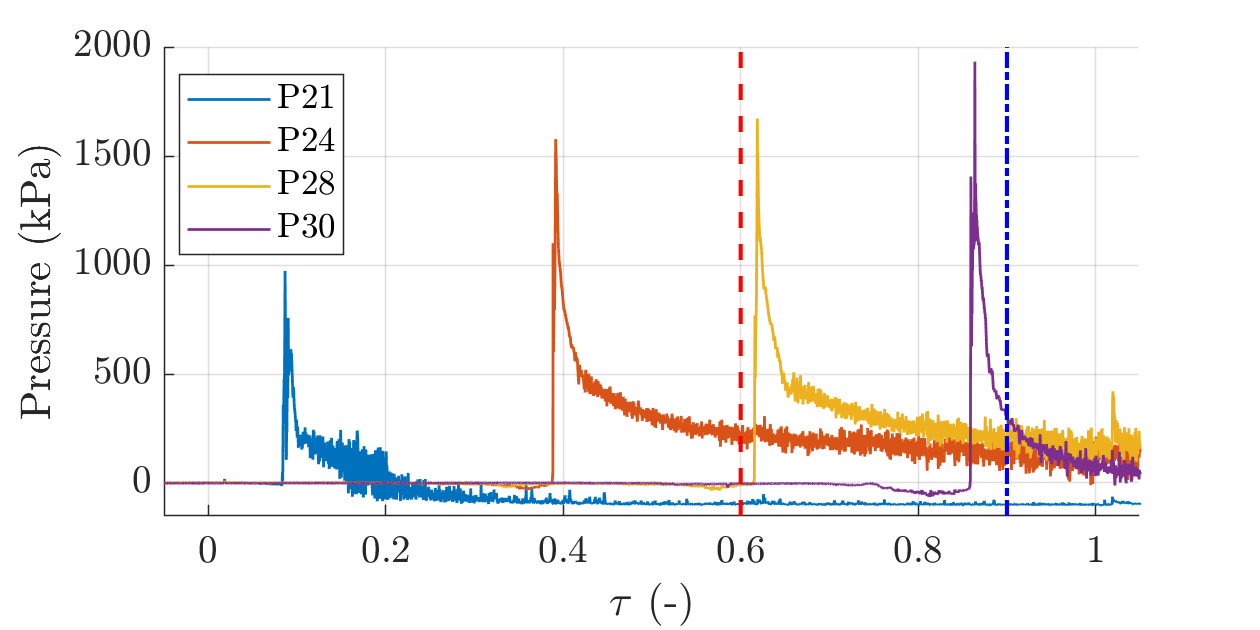}}
\\
\subfigure[$p$ line $l3$ - Front probes - Pitch 4$^{\circ}$]
{\includegraphics[width=0.4\textwidth]{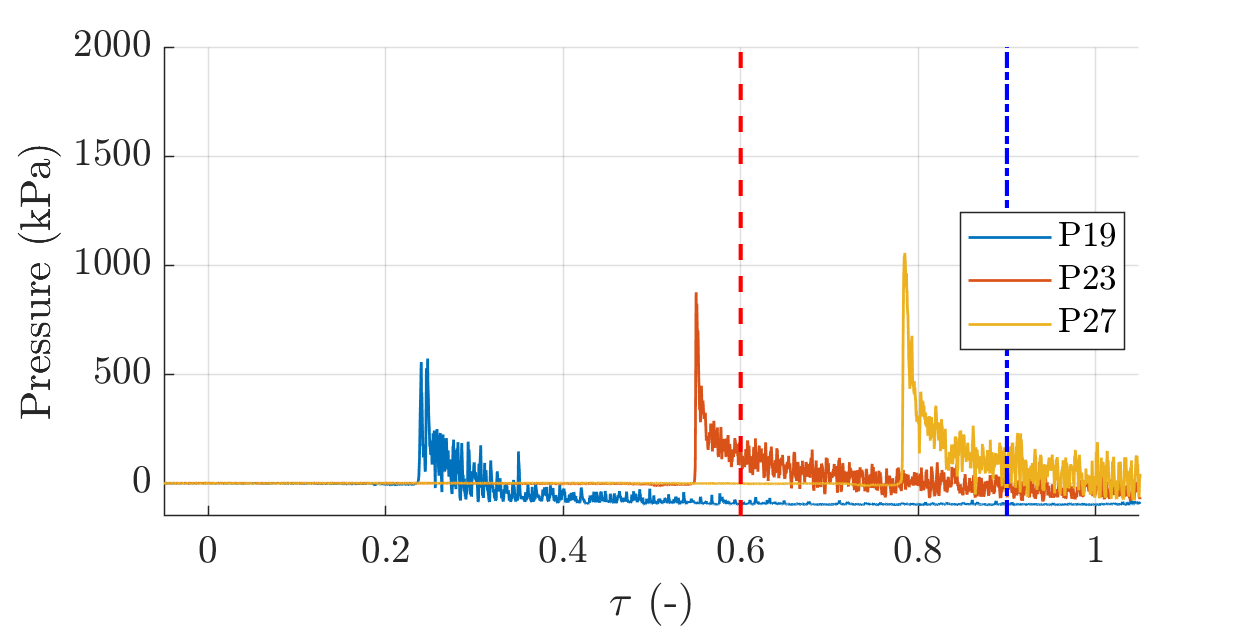}} 
\qquad \qquad
\subfigure[$p$ line $l3$ - Front probes - Pitch 6$^{\circ}$]
{\includegraphics[width=0.4\textwidth]{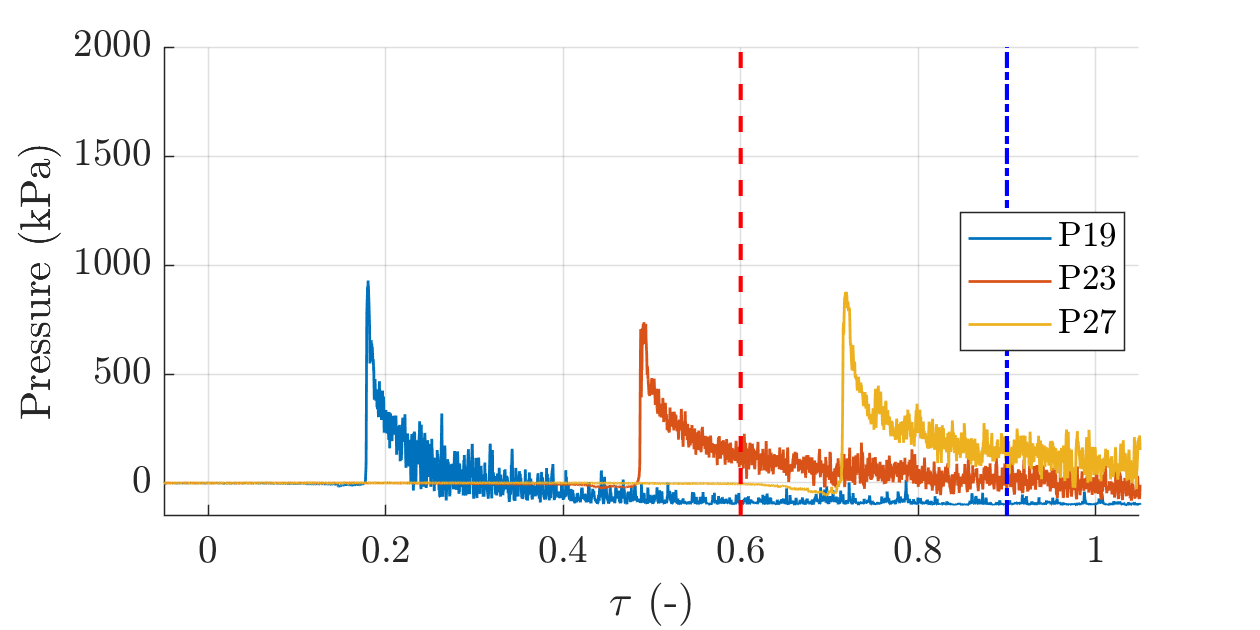}}
\qquad \qquad
\subfigure[$p$ line $l3$ - Front probes - Pitch 8$^{\circ}$]
{\includegraphics[width=0.4\textwidth]{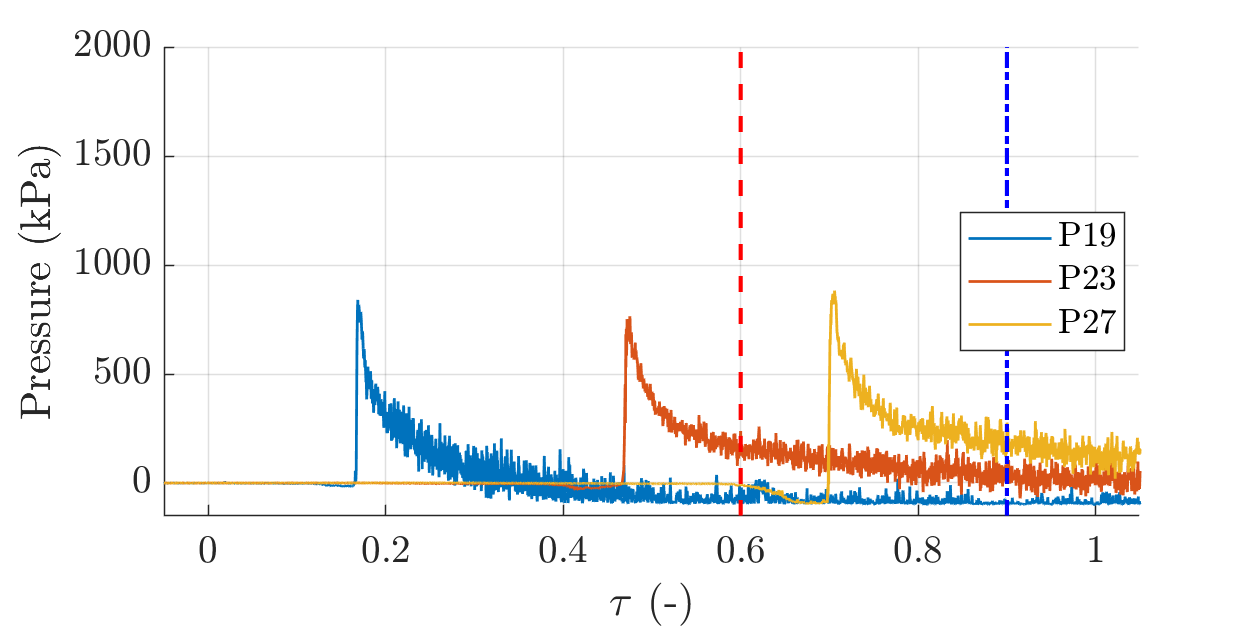}}
\\
\subfigure[$p$ midline - Rear probes - Pitch 4$^{\circ}$]
{\includegraphics[width=0.4\textwidth]{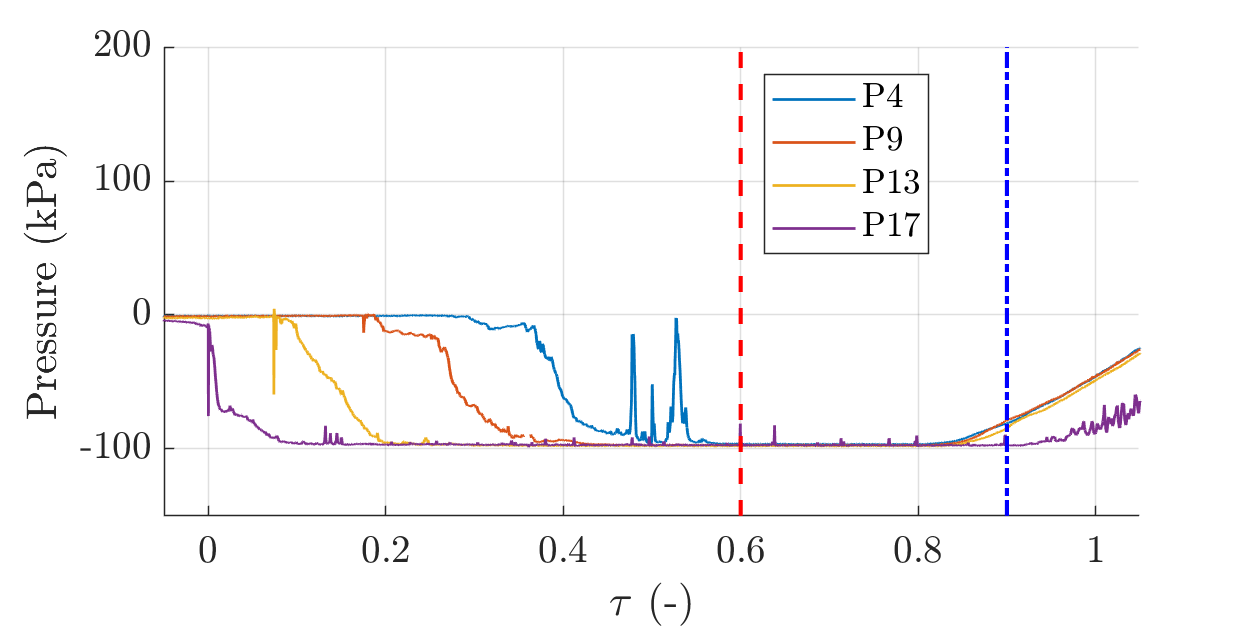}} 
\qquad \qquad
\subfigure[$p$ midline - Rear probes - Pitch 6$^{\circ}$]
{\includegraphics[width=0.4\textwidth]{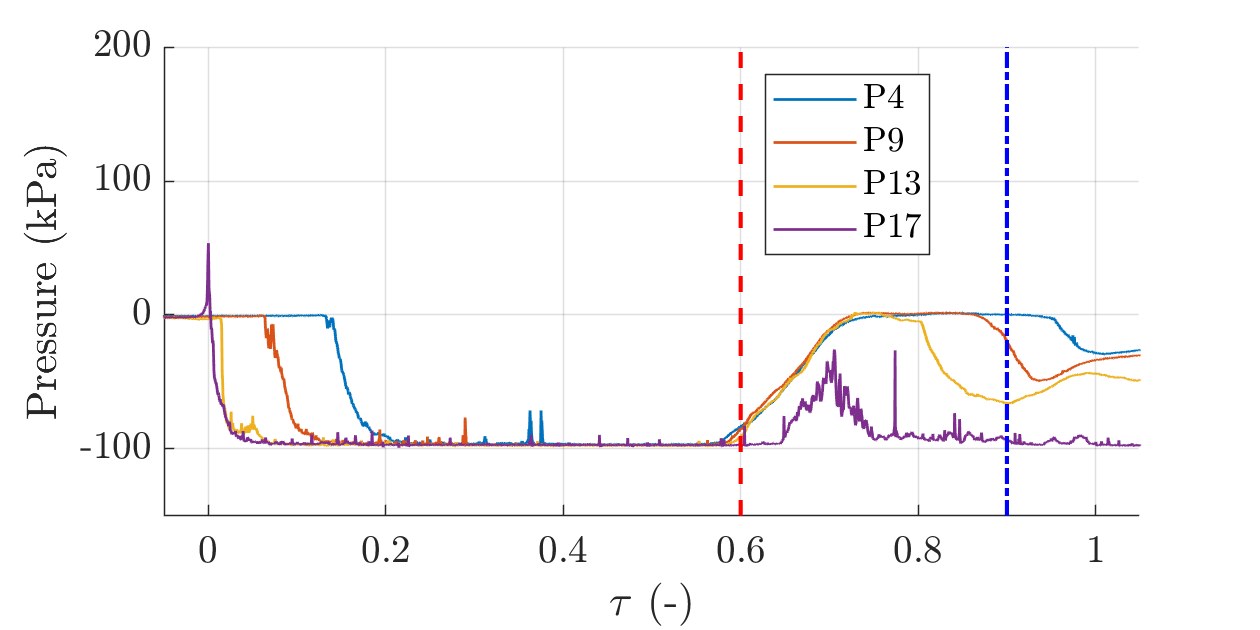}}
\qquad \qquad
\subfigure[$p$ midline - Rear probes - Pitch 8$^{\circ}$]
{\includegraphics[width=0.4\textwidth]{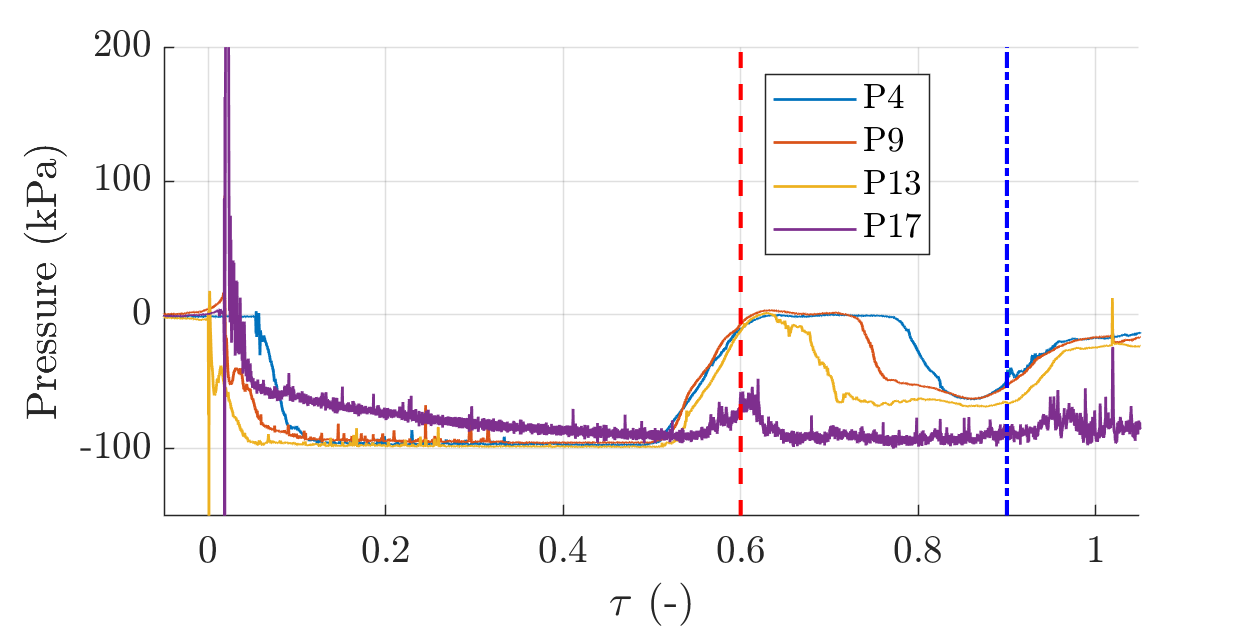}}
\\
\subfigure[$p$ line $l3$ - Rear probes - Pitch 4$^{\circ}$]
{\includegraphics[width=0.4\textwidth]{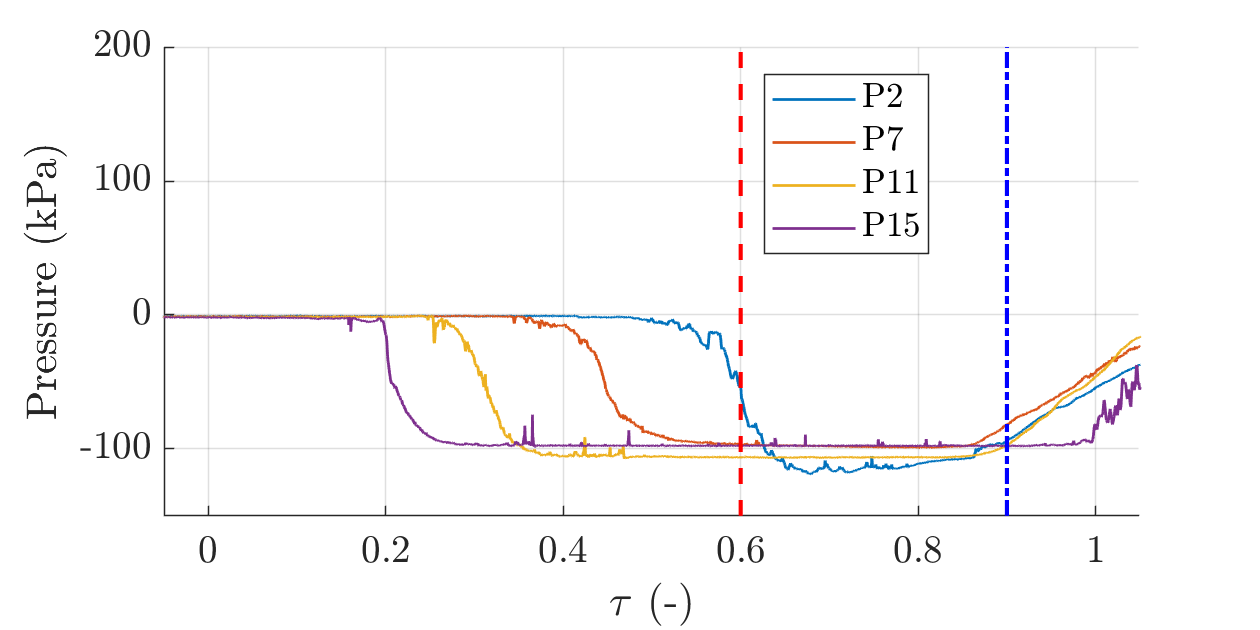}} 
\qquad \qquad
\subfigure[$p$ line $l3$ - Rear probes - Pitch 6$^{\circ}$]
{\includegraphics[width=0.4\textwidth]{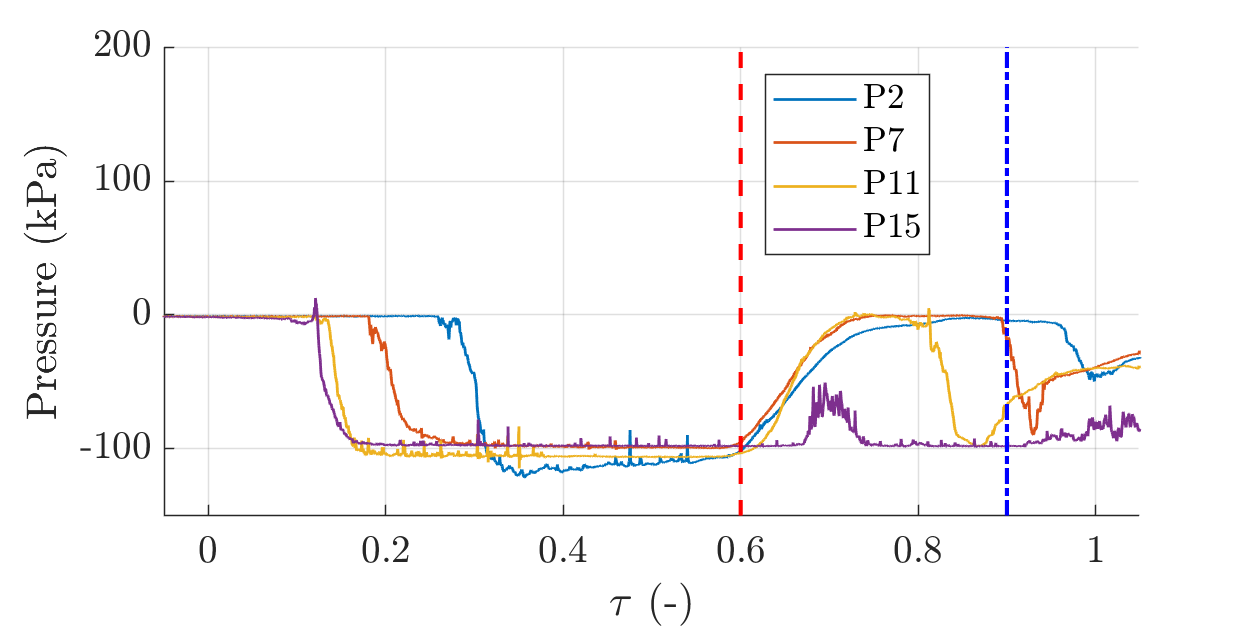}}
\qquad \qquad
\subfigure[$p$ line $l3$ - Rear probes - Pitch 8$^{\circ}$]
{\includegraphics[width=0.4\textwidth]{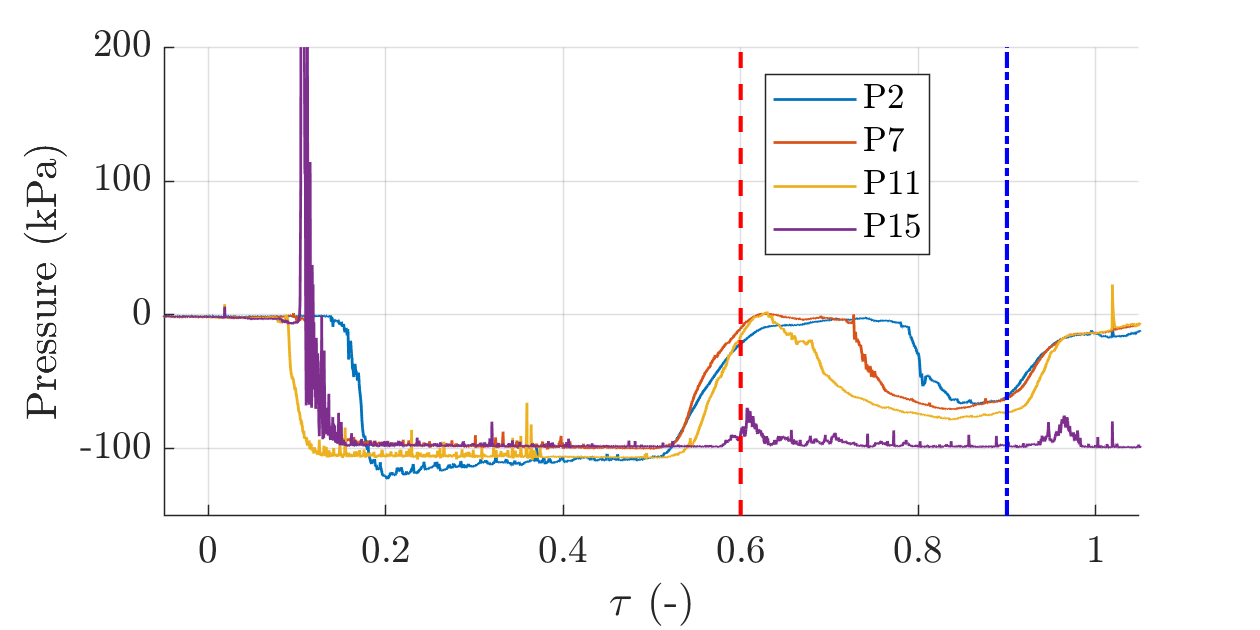}}
\caption{Front and rear pressure probe time histories 
for the specimen \textbf{S2} at $U$=40~m/s and
at the pitch angles 4$^{\circ}$, 6$^{\circ}$ and 8$^{\circ}$.}
\label{fig:Pressure_UWframes_1H222X}
\end{figure}
\end{landscape}

In Figure \ref{fig:Pressure_UWframes_1H222X}(a-f)
and in Figure \ref{fig:Pressure_UWframes_1H222X}(g-l)
the time histories of the pressures recorded by the probes 
located along the midline \textit{l1} and line \textit{l3}
are shown for the groups of rows from $r5$ to $r8$ and from
$r1$ to $r4$, respectively. 
%
%
Looking at the pressure time histories \emph{in the front part of the 
specimen} in Figure \ref{fig:Pressure_UWframes_1H222X}(a)-(f),
it is observed that the pressure peaks are of the same order
of magnitude for all three shapes and are delayed on the line $l3$, 
as a consequence of the curvature of the spray root region induced by
both the body curvature and three-dimensional effects.
The effect of the pitch angle is more 
evident by looking at the pressures values after the peak, which 
approach an almost constant value, higher for larger pitch
angles, in agreement with what found in \cite {iafrati2016experimental}
for the impact of a flat plate. 
Something similar happens for the pressures recorded by the probes
located along the sensor line $l3$. 

As for the time histories recorded by the \emph{rear pressure probes},
shown in Figure \ref{fig:Pressure_UWframes_1H222X}(g)-(l), it
is shown that the drop to the vapour pressure values of all the
probes located along the mid-line occurs over a shorter time period
going from $\alpha$=~4° to $\alpha$=~6° and then $\alpha$=~8°.
Such a behaviour also is noticed at sensor line $l3$, where all pressure
signals are delayed with respect to the midline due to the transverse curvature
of the body.
The pressure time histories also confirm that the
duration of the cavitation phase is longer at 4$^{\circ}$, where 
ventilation starts just at the end of the impact phase, while at
both 6$^{\circ}$ and 8$^{\circ}$ it starts at about $\tau=0.6$.
It also appears that at 6$^{\circ}$ and 8$^{\circ}$,
when ventilation occurs, the pressure rise to the atmospheric 
pressure is quite sharp. In contrast, at 4$^{\circ}$,
the pressure growth is much milder. 
Such a pressure behaviour, in agreement with what discussed above
regarding the video frames, is again similar to the \emph{intermediate
regimes} of cavitation, also observed for the shape \textbf{S3}
in \citet{iafrati2019cavitation}.

As the pitch angle increases, it is observed that 
the peak pressure recorded by the P17 probe also increases.
At 4$^{\circ}$ no evident peak appears.
An explanation for such a circumstance could be
that, as shown in Figure \ref{fig:S2_4_6_8deg_FirstContactPoints_PressProbes},
since the probe P17 is located very close to the first contact, 
some air cushioning effects may enter into play. Air cushioning
may induce a local lowering of the free surface, especially for
water entry at high speed and at low pitch angles, as
discussed for flat plates in \citet{iafrati2015high}. 
Air entrapment may also occur at higher pitch angles during
the initial moments of the water impact, which may explain
why a positive pressure peak is not recorded by the probe P13,
which at 8$^{\circ}$ pitch is located at the first contact point.

\begin{figure}
	\centering
	\includegraphics[width=0.95\textwidth]{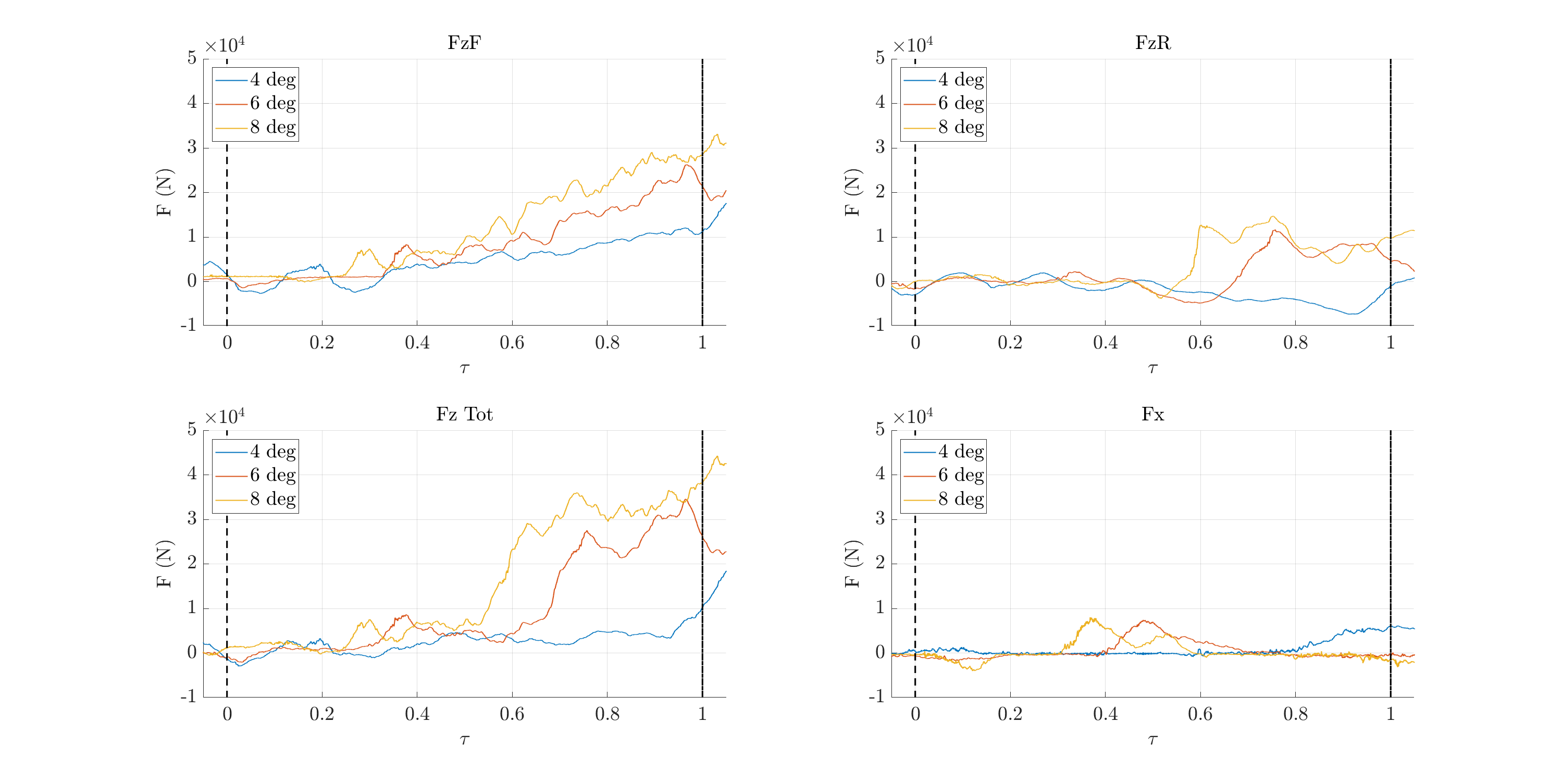}
	\caption{Time histories of the forces at
	the three pitch angles for the shape \textbf{S2} at U=40 m/s.}
	\label{fig:Forces_S2_AllPitches_RAW}
\end{figure}

The time histories of the loads at the three pitch angles are shown in 
Figure \ref{fig:Forces_S2_AllPitches_RAW}.
It can be observed that, as expected, the forces at the front increase
as the pitch angle increases by about a factor 2 or more, as the pitch angle
passes from 4$^{\circ}$ to 8$^{\circ}$.
The forces at the rear remain very low up to about half of the impact
phase. Then, at 4$^{\circ}$ the negative loads are more evident, whereas
at 8$^{\circ}$ a sudden jump to higher values occurs as a consequence of
ventilation.
%
%
\subsection{Effects of the body curvatures}
\label{sec:effect_shape}
%
%
Another aspect worth of investigation is the role played by the body
curvatures. As already detailed in Section \ref{sec:shapes},
three different shapes are considered. The
specimens \textbf{S2} and \textbf{S3} have the same longitudinal profile 
in the mid-plane but different curvatures in the transverse direction,
\textbf{S3} being characterized by a rather ``flat'' elliptical cross-section, 
while \textbf{S2} by a circular cross-section. The specimen \textbf{S1B}
has a somewhat smoother longitudinal profile compared to the others and
a circular cross section, but with a smaller radius than \textbf{S2}.
%

\begin{landscape}
\begin{figure}
\centering
\subfigure[Shape \textbf{S1B}]
{\includegraphics[width=1.45\textwidth]{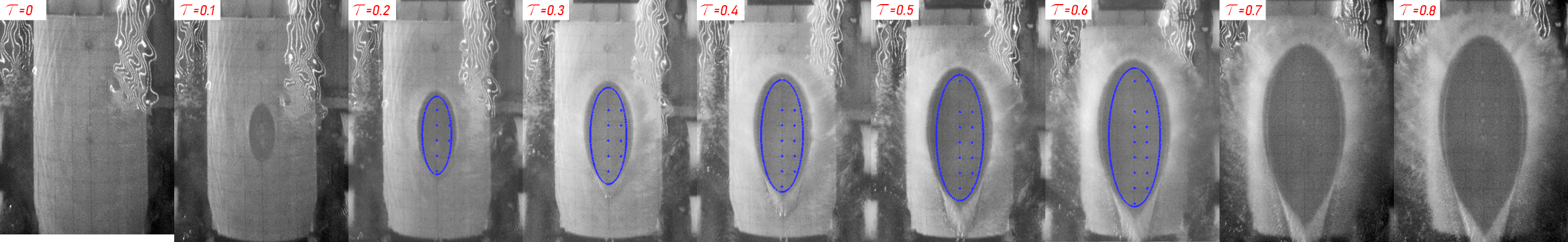}} 
\\
\subfigure[Shape \textbf{S2}]
{\includegraphics[width=1.45\textwidth]{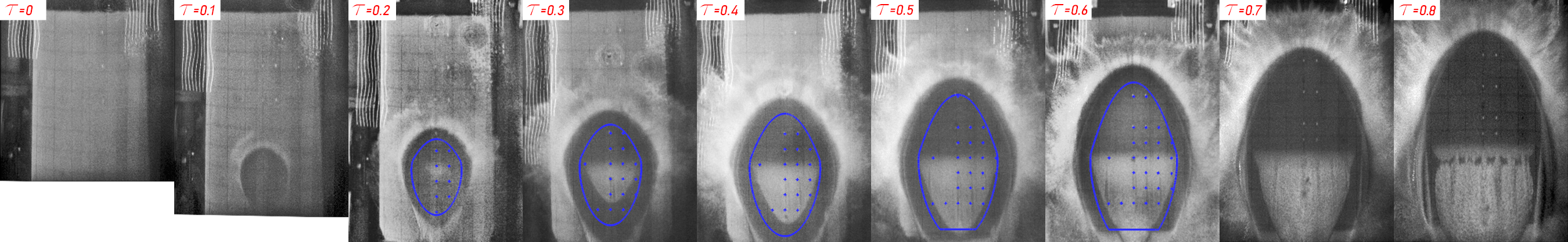}}
\\
\subfigure[Shape \textbf{S3}]
{\includegraphics[width=1.45\textwidth]{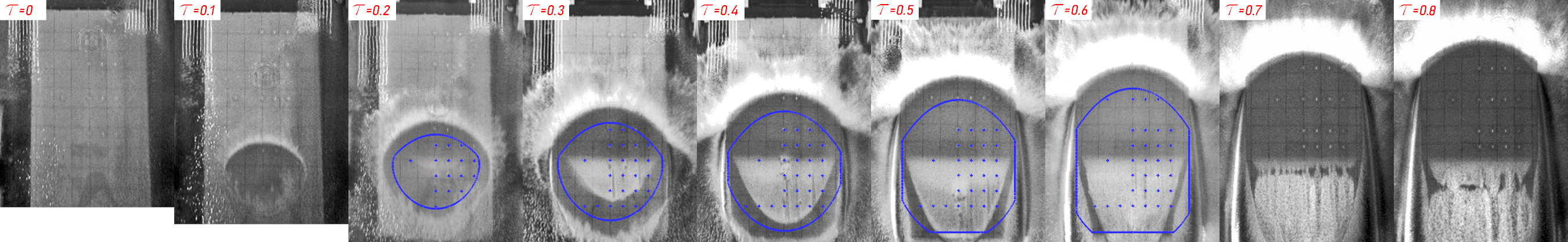}}
\caption{Under-water video frames
of the specimens \textbf{S1B}, \textbf{S2} and \textbf{S3}
during the water entry $U$=40~m/s and pitch angle 6$^{\circ}$
from $\tau$=0 to $\tau=0.8$. For the frames 
from $\tau$=0.2 to $\tau=0.6$ the edges of the
geometric intersection area with the still water
level are also plotted in 
blue. The direction of water entry is from bottom to
the top.}
\label{fig:Silvano_Cavitation_Frames}
\end{figure}
\end{landscape}

In order to gain insight on the shape on the spray in front and
of the cavitation and ventilation phenomena at the rear, the 
underwater video-frames taken for the different shapes from $\tau$=0
to $\tau$=0.8 are shown in Figure \ref{fig:Silvano_Cavitation_Frames}. The
frames are cropped and aligned with respect
to the specimen leading edge, to facilitate the comparison.
The edges of the geometric intersection 
areas with the still water
level from $\tau$=0.2 to 0.6 are superimposed to the
corresponding pictures.
%
%
By looking at the underwater images, it can be seen that
for \textbf{S1B}, due to the smoother longitudinal curvature, 
the first point of contact is in the centre of the specimen. 
(see also Figure \ref{fig:ref_frames_points_ALL})
and, differently from
\textbf{S2} and \textbf{S3}, the wetted area is rather symmetric with 
respect to the transverse line passing through the first contact point.
In the time interval from $\tau= 0.2$ to $\tau=0.6$
it is also evident that for \textbf{S1B} the geometric intersection area
exhibits analogous symmetry and is evenly smaller than 
the wetted area in all directions, which implies a rather uniform pile-up
all around.
Transversally, the width of the wetted area reflects the differences in
the cross sections, therefore \textbf{S1B} has the narrowest
section, due to the smaller radius of curvature,
whereas \textbf{S3} has the broadest one due,
to the elliptical profile.

The difference in the transverse curvatures is also responsible for the
differences in the curvature of the forward propagating spray.

It is also evident that in \textbf{S1B} no cavitation region appears during the
impact phase, unlike in \textbf{S2} and \textbf{S3}. As explained in the
previous sections, in a $2D+t$ approximation, the longitudinal curvature
at the rear of the body transform the water entry problem into a water
exit one. As such, the smoother is the longitudinal curvature, the lower
is the water exit speed and the lower is the reduction in the 
pressure. 

As observed for the shape \textbf{S2} in Section \ref{hydrodynamics_S2},
also for the shape \textbf{S3} the cavitation region expands in the rear part, 
whereas the front edge appears to be 
fixed at approximately the point of curvature change \textbf{C}. This
is evident in the frames shown in Figure \ref{fig:Silvano_Cavitation_Frames}.
As already noted in
Section \ref{sec:effect_speed} the differing transverse profile between 
\textbf{S2} and \textbf{S3} does not alter the fact that the 
same regimes of evolution of the cavitation region occur, and the
threshold values of $U$ that delineate them are similar.
However, in agreement with the findings of \cite{spinosa2022hydrodynamic}
the pressures at the rear in \textbf{S3} are expected to
be averagely lower (more negative),
due to the lower three-dimensionality of the flow. 
This results in the cavitation region 
having a faster expansion, which also leads to an earlier
formation of the ventilation zone for this shape. 
A more detailed analysis of the backwards expansion of the cavitation 
region is provided in Section \ref{sec:cavitation_velocity},
where this assertion is verified in a quantitative manner. 

\begin{figure}
\centering
\includegraphics[width=0.70\textwidth]{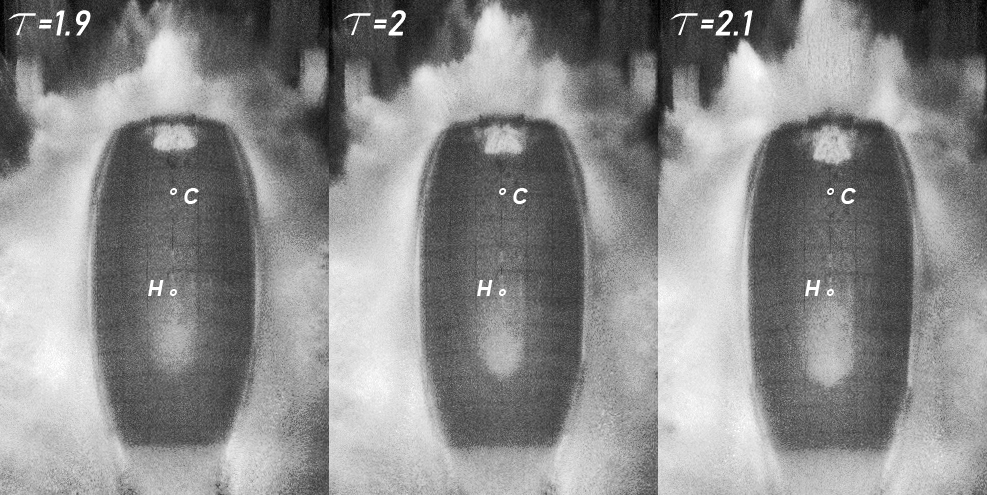}
\caption{Cavitation area observed in the shape \textbf{S1B} after the end of
the impact phase for the test at $U$~=40~m/s and pitch angle 6~$^{\circ}$.
The points \textbf{H} and \textbf{C} are also superimposed.
The direction of water entry is from bottom to the top.}
\label{fig:S1B_CAV}
\end{figure}

Although in a much less evident form, cavitation occurs also
for specimen \textbf{S1B} at the same impact conditions but, as shown in 
Figure \ref{fig:S1B_CAV}, at a much later stage starting 
from $\tau \simeq 1.9$.
In this case, the front edge of 
the cavitation region appears to be
situated in close proximity of the first contact point, whereas
the rear edge does not exhibit a significant backward expansion in time.
This situation is somewhat consistent with the incipient 
cavitation regime described in Section \ref{sec:effect_speed} 
and in \cite{iafrati2019cavitation}.
However, as it can be seen in Figure \ref{fig:S1B_CAV}, at those
times the leading edge of the specimen is already below the still water level,
and a strong jet is formed at the front, which propagates towards the
trailing edge. In this phase the trolley undergoes a sharp deceleration, and 
therefore the flow conditions are rather uncertain. For this reason
the dynamics of the cavitation region for the specimen \textbf{S1B} 
is not discussed further.

\begin{landscape}
\begin{figure}
\centering
\subfigure[$p$ midline - Front probes - Shape \textbf{S1B}]
{\includegraphics[width=0.38\textwidth]{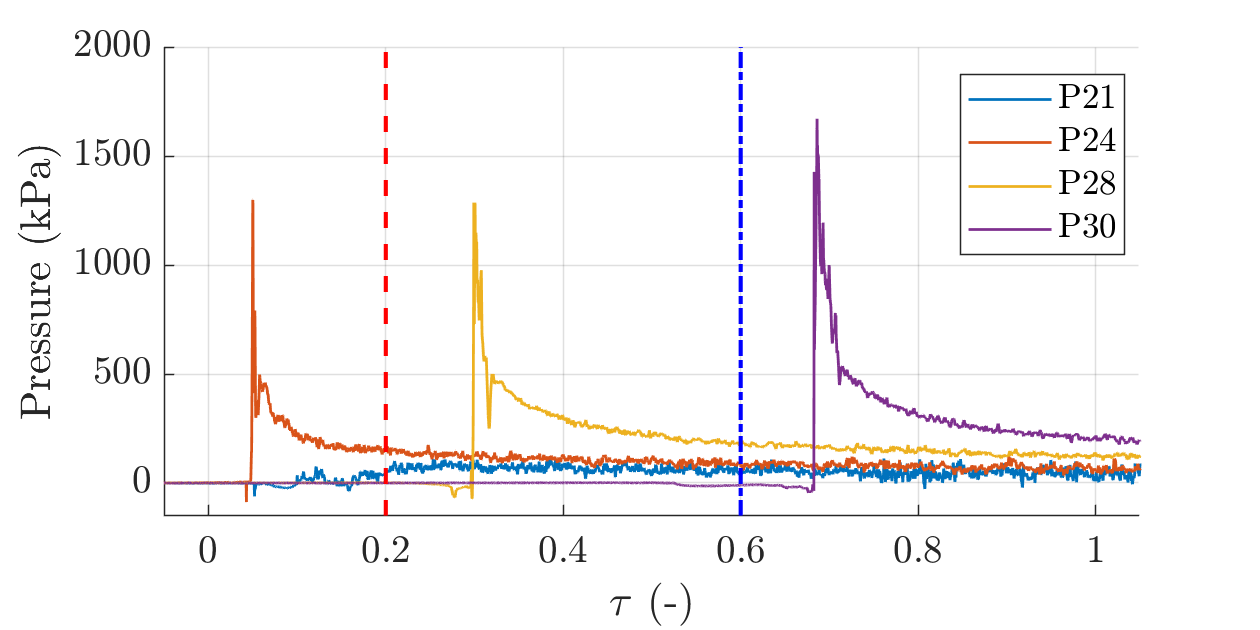}} 
\qquad \qquad
\subfigure[$p$ midline - Front probes -  Shape \textbf{S2}]
{\includegraphics[width=0.38\textwidth]{1H2222_17_07_2018_1_figures/pressure_FRONT_l1_taus_02_06.png}}
\qquad \qquad
\subfigure[$p$ midline -  Front probes - Shape \textbf{S3}]
{\includegraphics[width=0.38\textwidth]{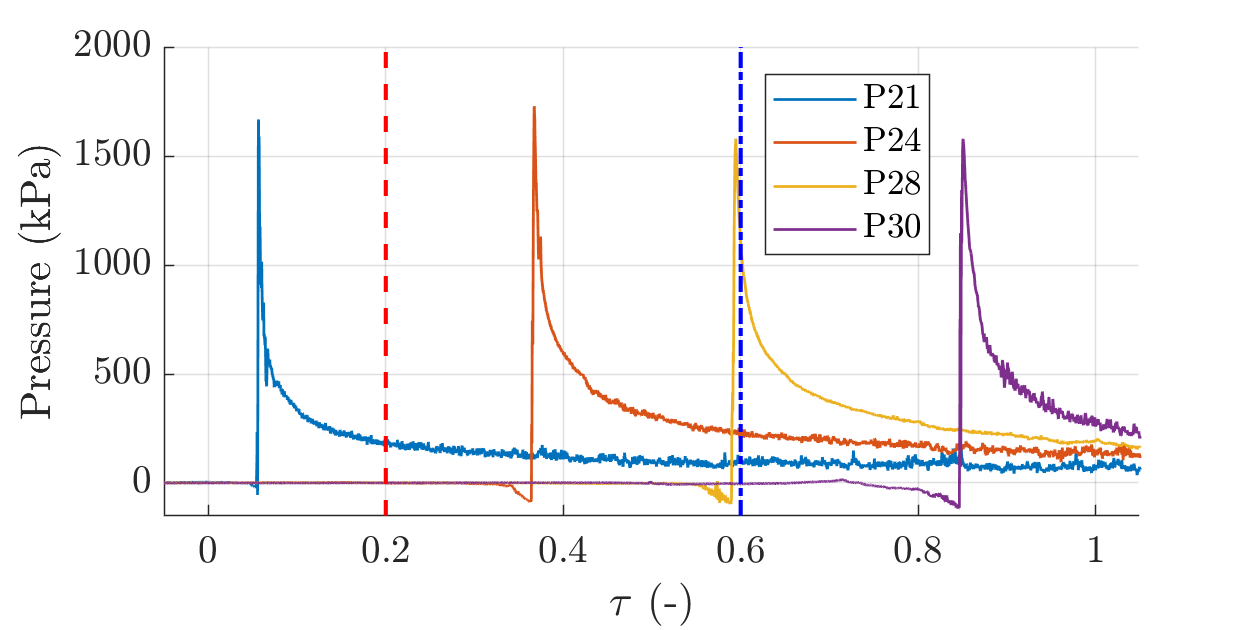}}
\\
\subfigure[$p$ Line l3 -  Front probes - Shape \textbf{S1B}]
{\includegraphics[width=0.38\textwidth]{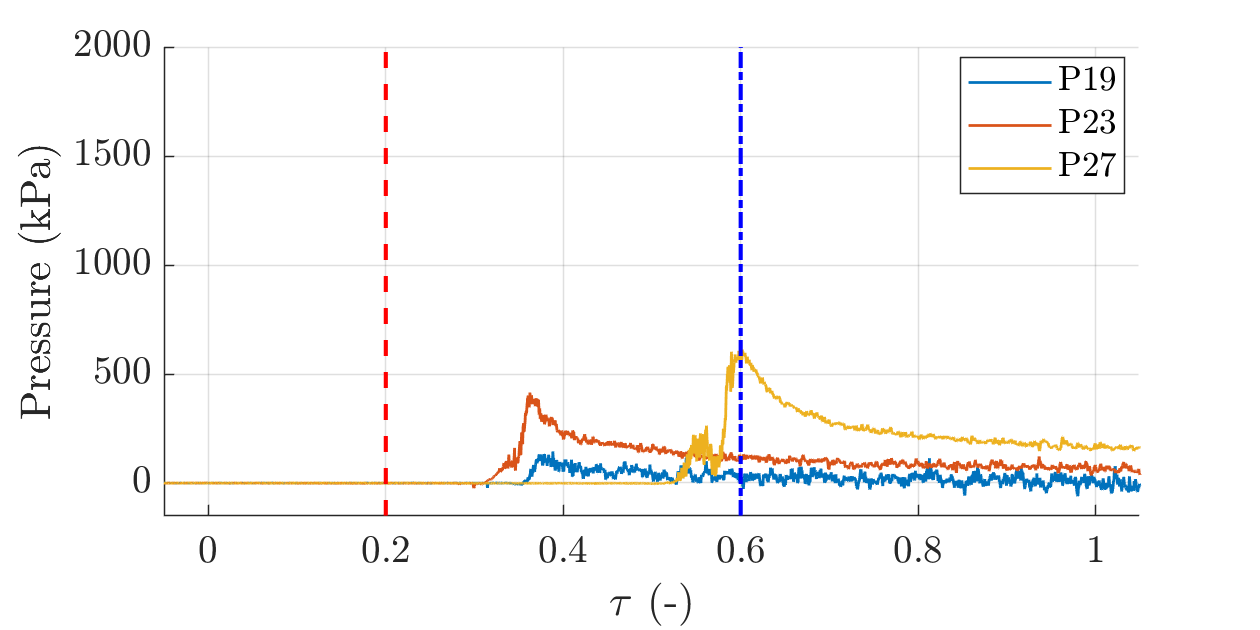}} 
\qquad \qquad
\subfigure[$p$ Line l3 -  Front probes - Shape \textbf{S2}]
{\includegraphics[width=0.38\textwidth]{1H2222_17_07_2018_1_figures/pressure_FRONT_l3_taus_02_06.png}}
\qquad \qquad
\subfigure[$p$ Line l3 -  Front probes - Shape \textbf{S3}]
{\includegraphics[width=0.38\textwidth]{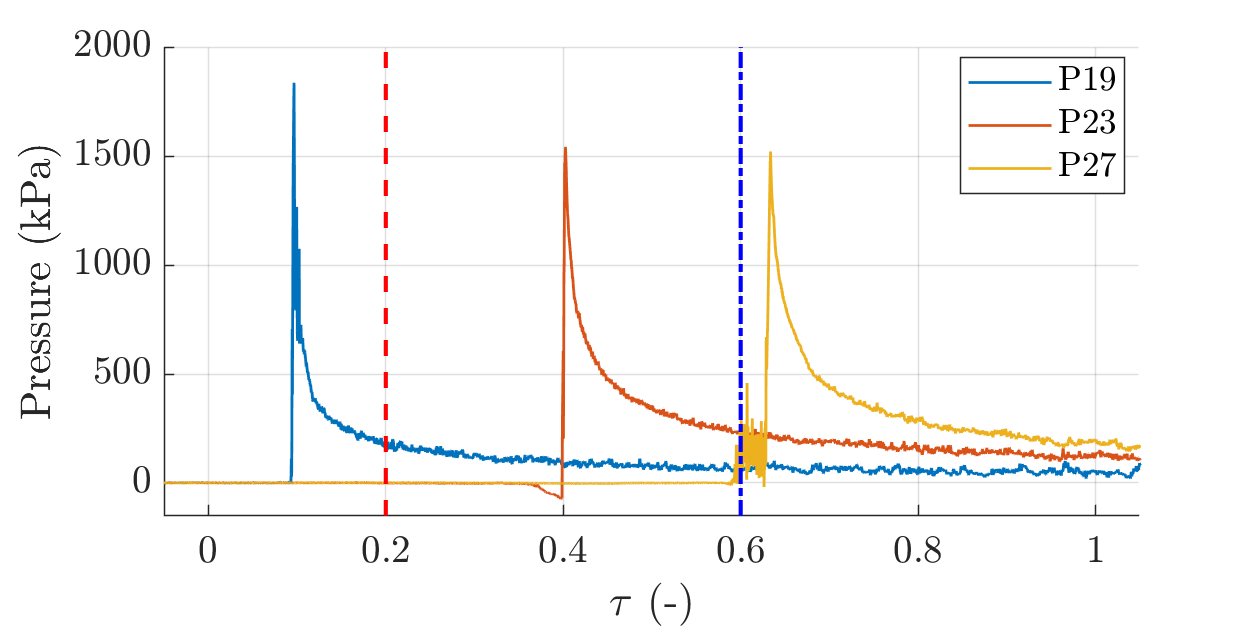}}
\subfigure[$p$ midline - Rear probes - Shape \textbf{S1B}]
{\includegraphics[width=0.38\textwidth]{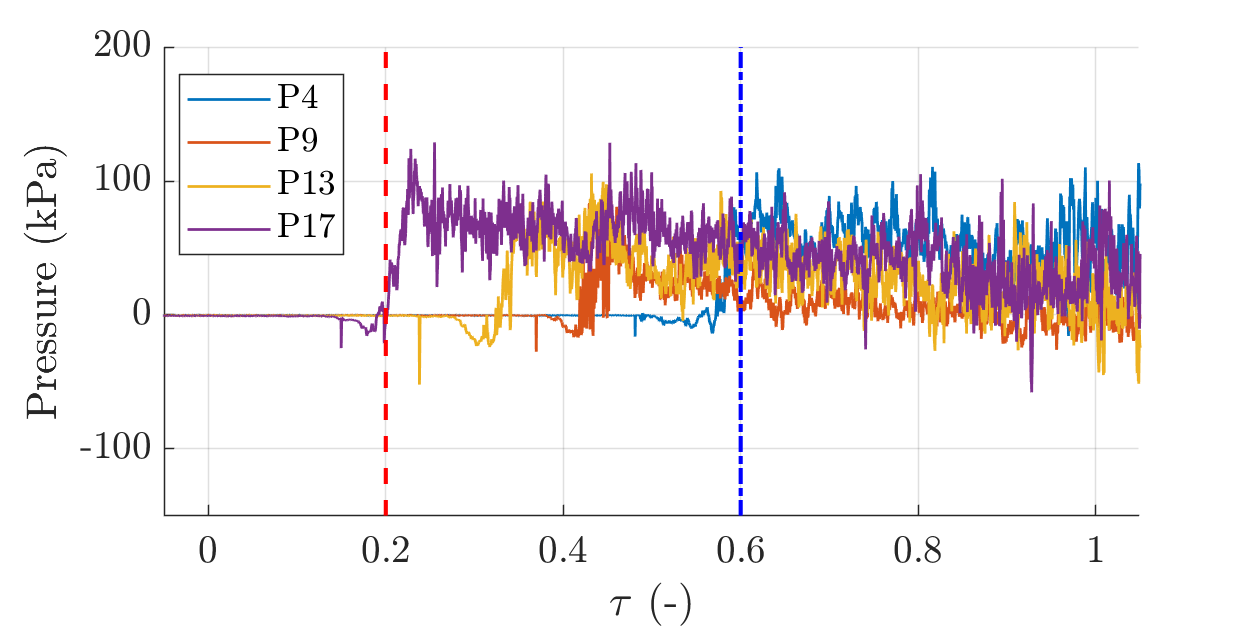}} 
\qquad \qquad
\subfigure[$p$ midline - Rear probes -  Shape \textbf{S2}]
{\includegraphics[width=0.38\textwidth]{1H2222_17_07_2018_1_figures/pressure_REAR_l1_taus_02_06.png}}
\qquad \qquad
\subfigure[$p$ midline -  Rear probes - Shape \textbf{S3}]
{\includegraphics[width=0.38\textwidth]{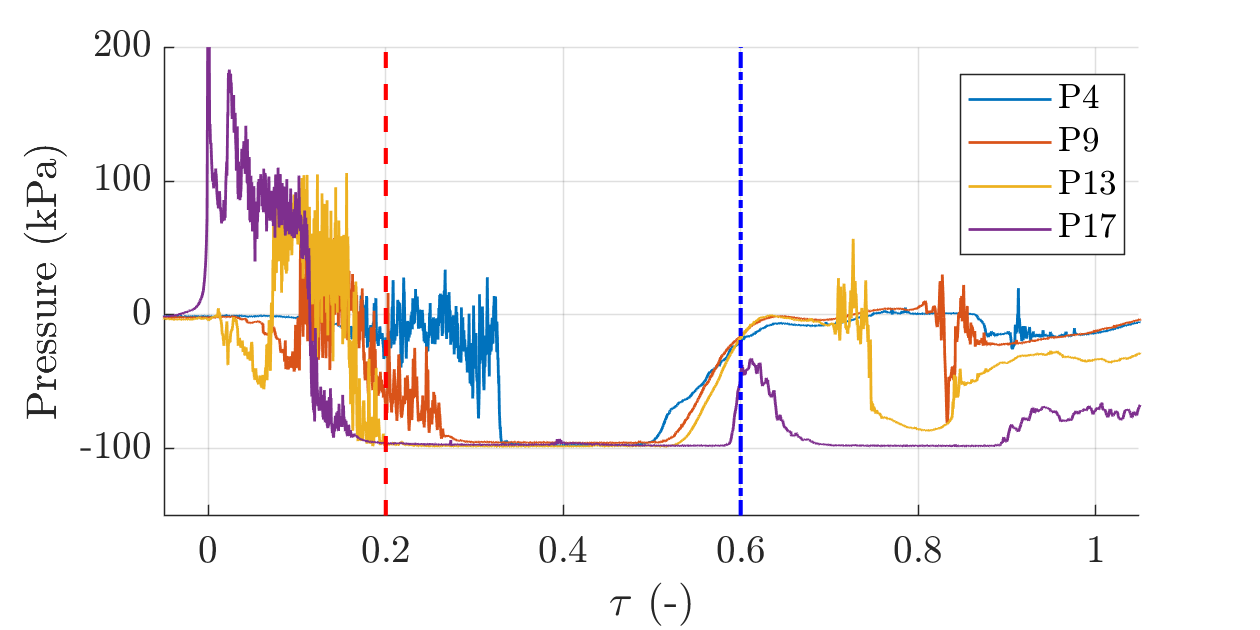}}
\\
\subfigure[$p$ Line l3 -  Rear probes - Shape \textbf{S1B}]
{\includegraphics[width=0.38\textwidth]{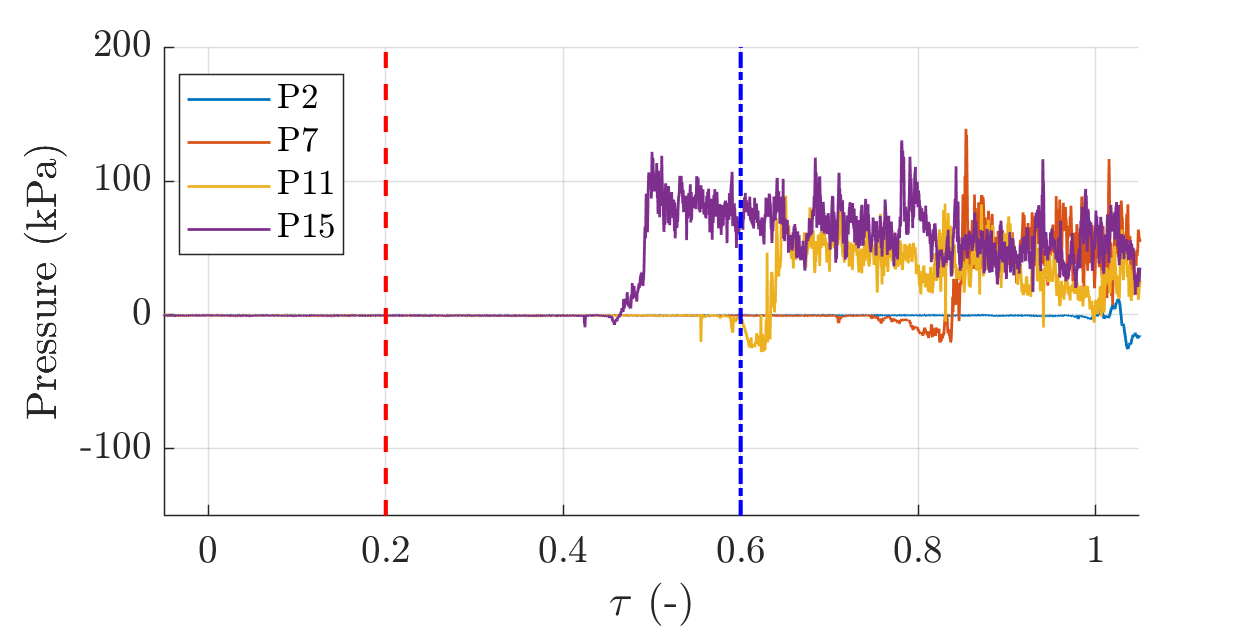}} 
\qquad \qquad
\subfigure[$p$ Line l3 -  Rear probes - Shape \textbf{S2}]
{\includegraphics[width=0.38\textwidth]{1H2222_17_07_2018_1_figures/pressure_REAR_l3_taus_02_06.png}}
\qquad \qquad
\subfigure[$p$ Line l3 -  Rear probes - Shape \textbf{S3}]
{\includegraphics[width=0.38\textwidth]{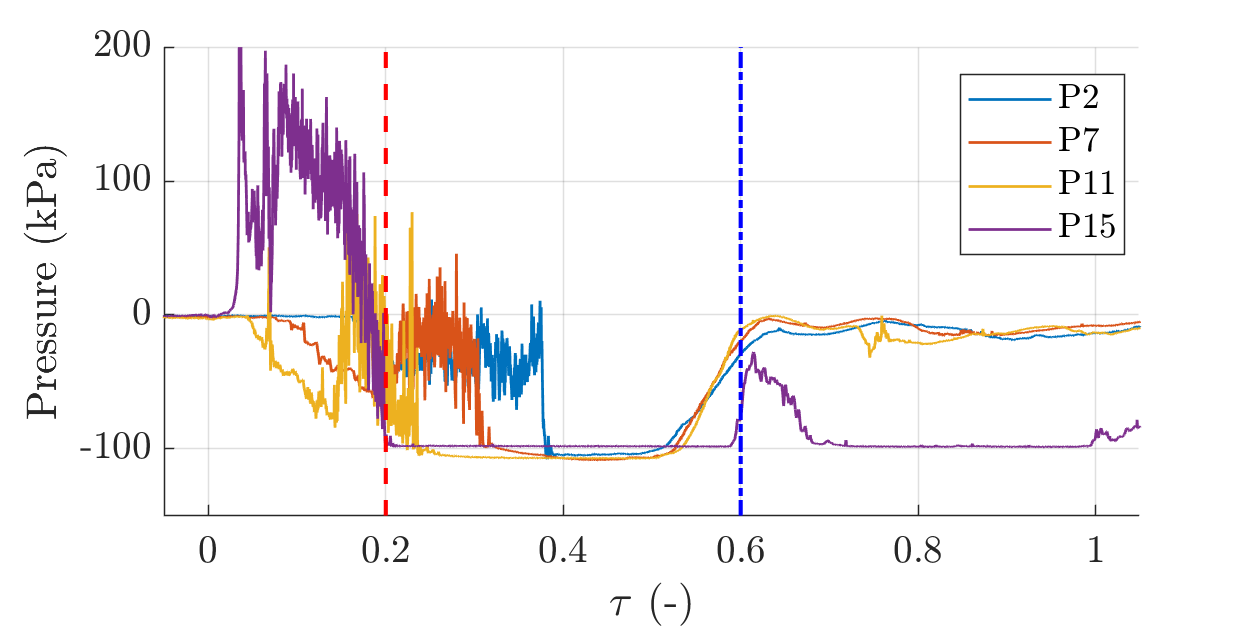}} 
\caption{Front and rear pressure time histories for the three shapes, at
$U$=40~m/s and at a pitch angle of 6$^{\circ}$.}
\label{fig:Pressure_1HX222}
\end{figure}
\end{landscape}

The above considerations are supported by the time histories of pressures
along the mid-line and along line $l3$, 
which are shown in Figure \ref{fig:Pressure_1HX222}.
It is worth noting that in the shape \textbf{S1B}, due to the different
location of the first contact point, the probes located in front portion 
do not include those on the row \emph{r5} (see again Figure 
\ref{fig:S2_4_6_8deg_FirstContactPoints_PressProbes}).
%
%
Looking at the mid-line probes located in the \emph{front part} of the
specimen, apart from some differences in the peak values, the pressures
after the peak display a rather similar behaviour. The effect of the
different curvature in the transverse direction appears evident when comparing 
the time delays between probes located across the same sensor row. 
The largest delays appears in \textbf{S1B}, whereas the 
the smallest one in \textbf{S3}. 
This circumstance can be attributed
to the larger curvature of the spray root
in the specimens with a higher transverse curvature, which allows 
for a greater possibility of the fluid to escape from the sides.
For the same reason, the pressure values recorded by the probes along
the line \emph{l3} are lower for the specimens with higher transverse 
curvature, i.e. \textbf{S1} and \textbf{S2}.

The time histories of the pressure measured by the \emph{rear probes} 
for \textbf{S1B} are generally positive. Differently from
what happens for \textbf{S3}, where the probes display a peak of small
amplitude followed by a rapid descent and a sharp drop to the negative vapour
pressure value, for \textbf{S1B} the pressures display a sharp rise to a
maximum value and then a very gentle decay, with some probes assuming
negative values at about the end of the impact phase. 

Comparing the \textbf{S2} and the \textbf{S3} pressure time histories
at the rear, it can be seen that, presumably due to the
lower transverse curvature and the lower local 
deadrise angle, evident pressure peaks appear only in \textbf{S3}. 
In the next stage, for both \textbf{S2} 
and \textbf{S3} the pressures at the rear exhibit sharp drops to the
vapour pressure value, which is consistent with what shown by 
the underwater images.
The delay between the pressure drops recorded by the probes
placed on the same row between lines $l1$ and $l3$
(for example between P17 and P15 or
between P13 and P11)
is greater in the case of S2, compared to S3.
This is a consequence of the larger curvature
of the rear edge of the cavitation region, 
which appears as a D-shape more elongated
in the longitudinal direction 
(see Figure \ref{fig:Silvano_Cavitation_Frames}(b) and (c)).
The ventilation phase starts earlier in \textbf{S3} than in \textbf{S2},
which confirms that the expansion of the cavitation region is
faster for the former shape, as the cavity reaches 
the trailing edge earlier in time. 

\begin{figure}
\centering
\subfigure[$\tau$=~0.3]
{\includegraphics[width=0.45\textwidth]{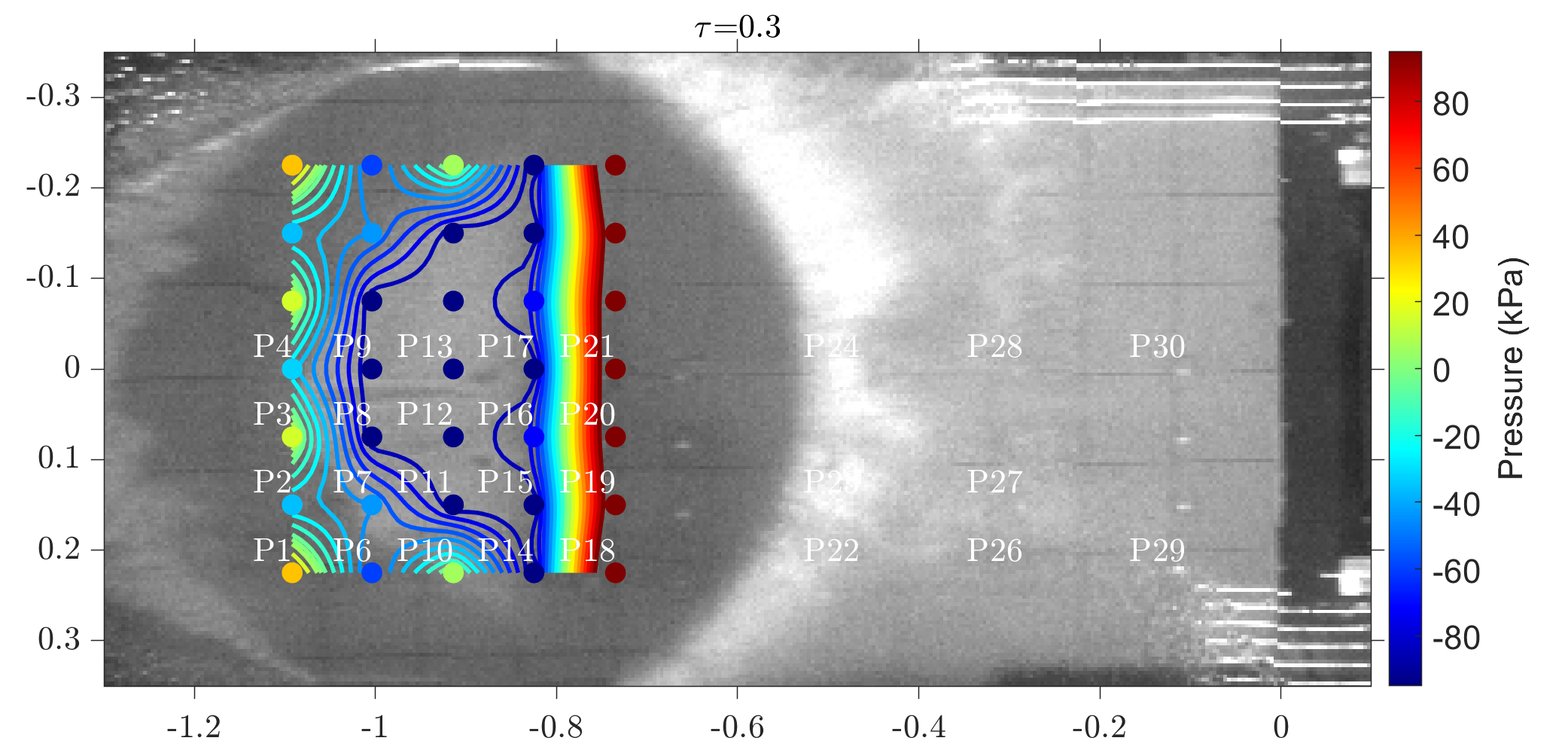}} \qquad
\subfigure[$\tau$=0.50]
{\includegraphics[width=0.45\textwidth]{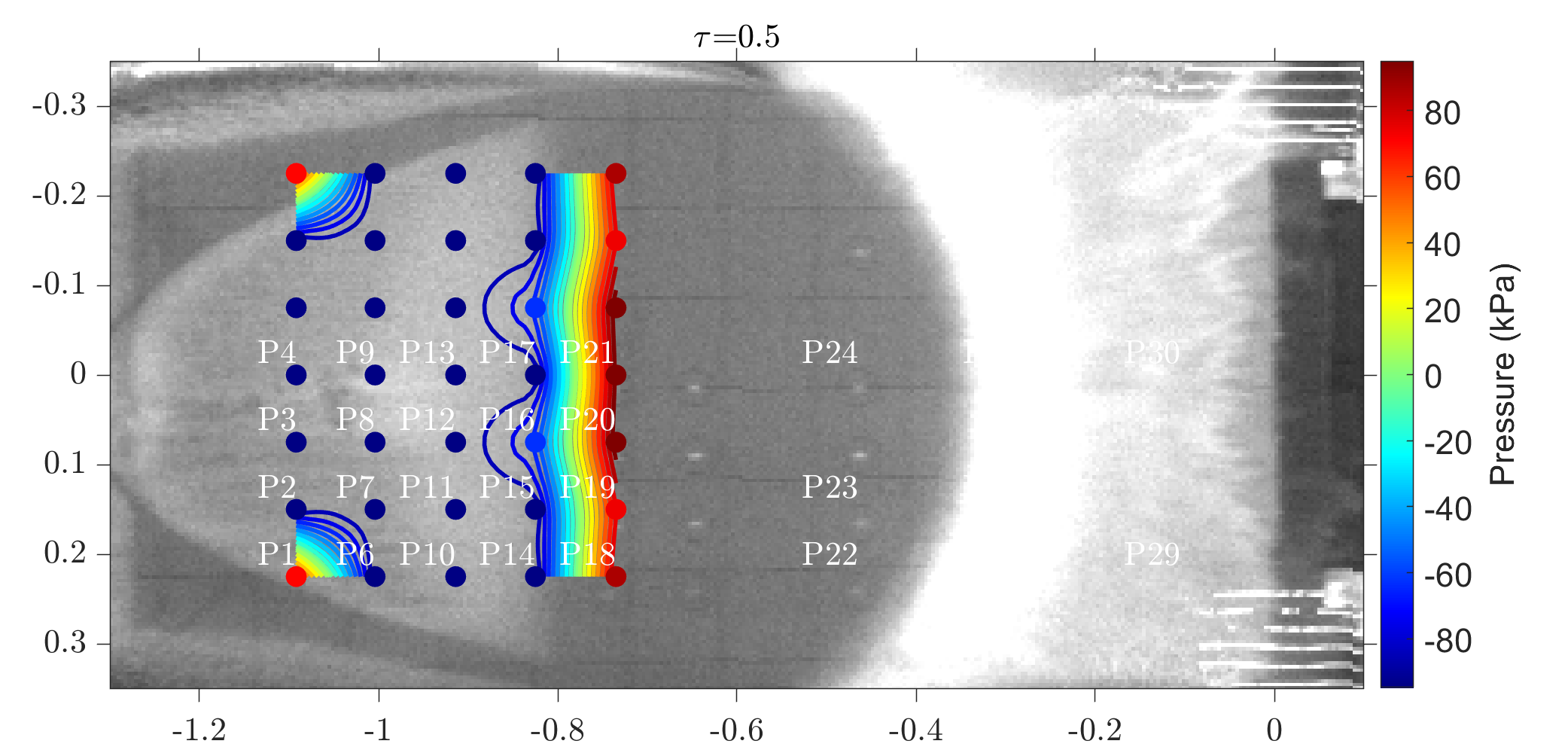}} \\
\subfigure[$\tau$=0.52]
{\includegraphics[width=0.45\textwidth]{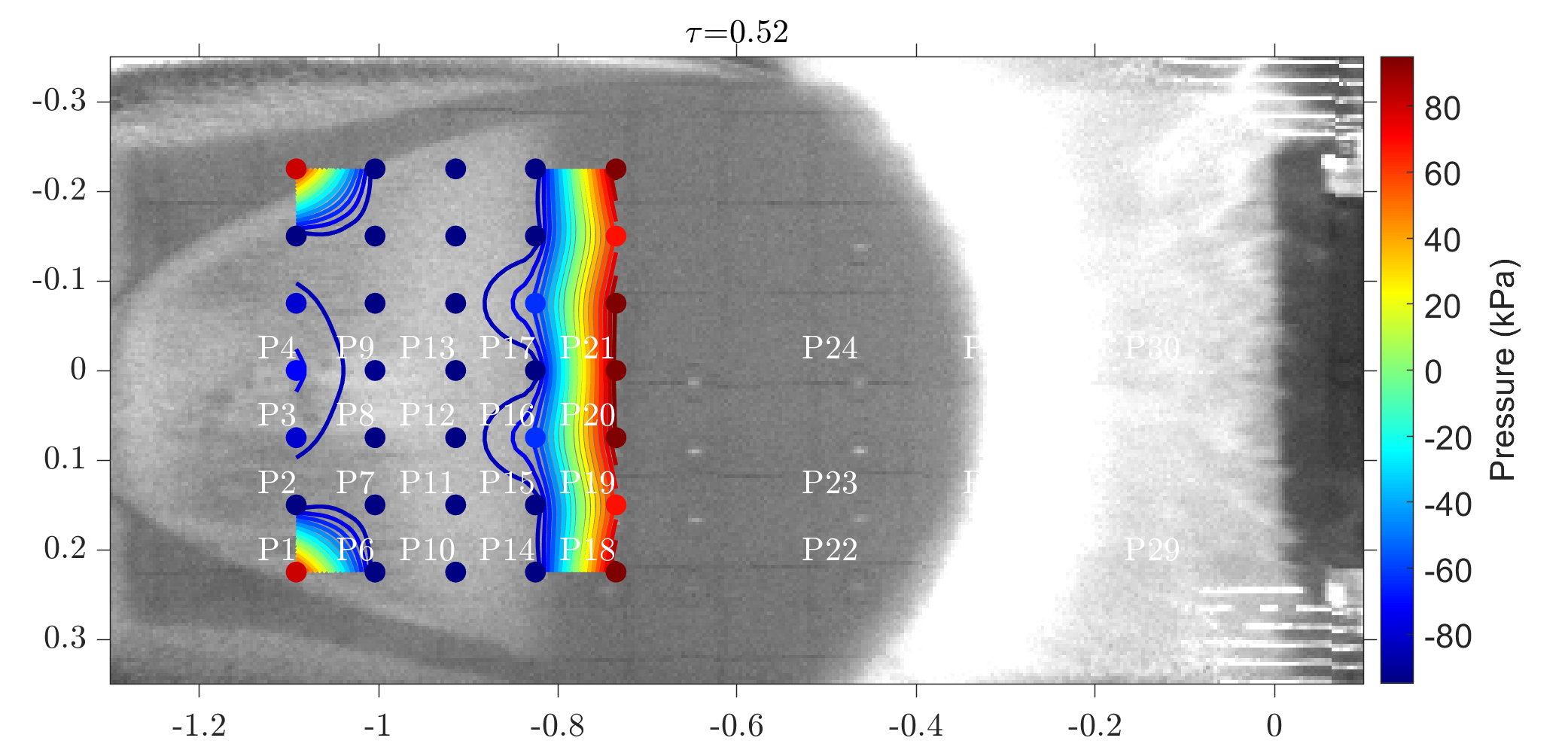}} \qquad
\subfigure[$\tau$=0.54]
{\includegraphics[width=0.45\textwidth]{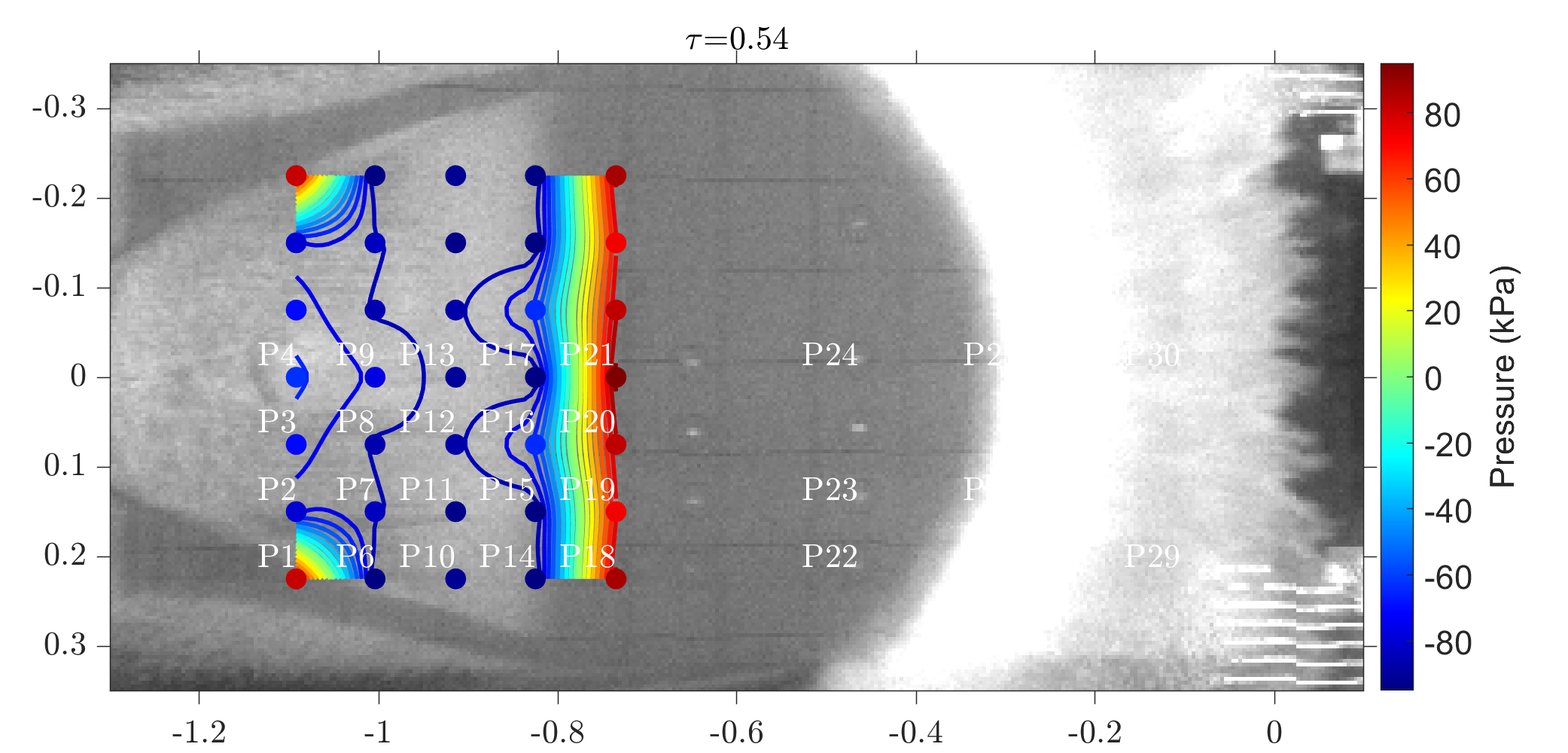}} \\
\subfigure[$\tau$=0.56]
{\includegraphics[width=0.45\textwidth]{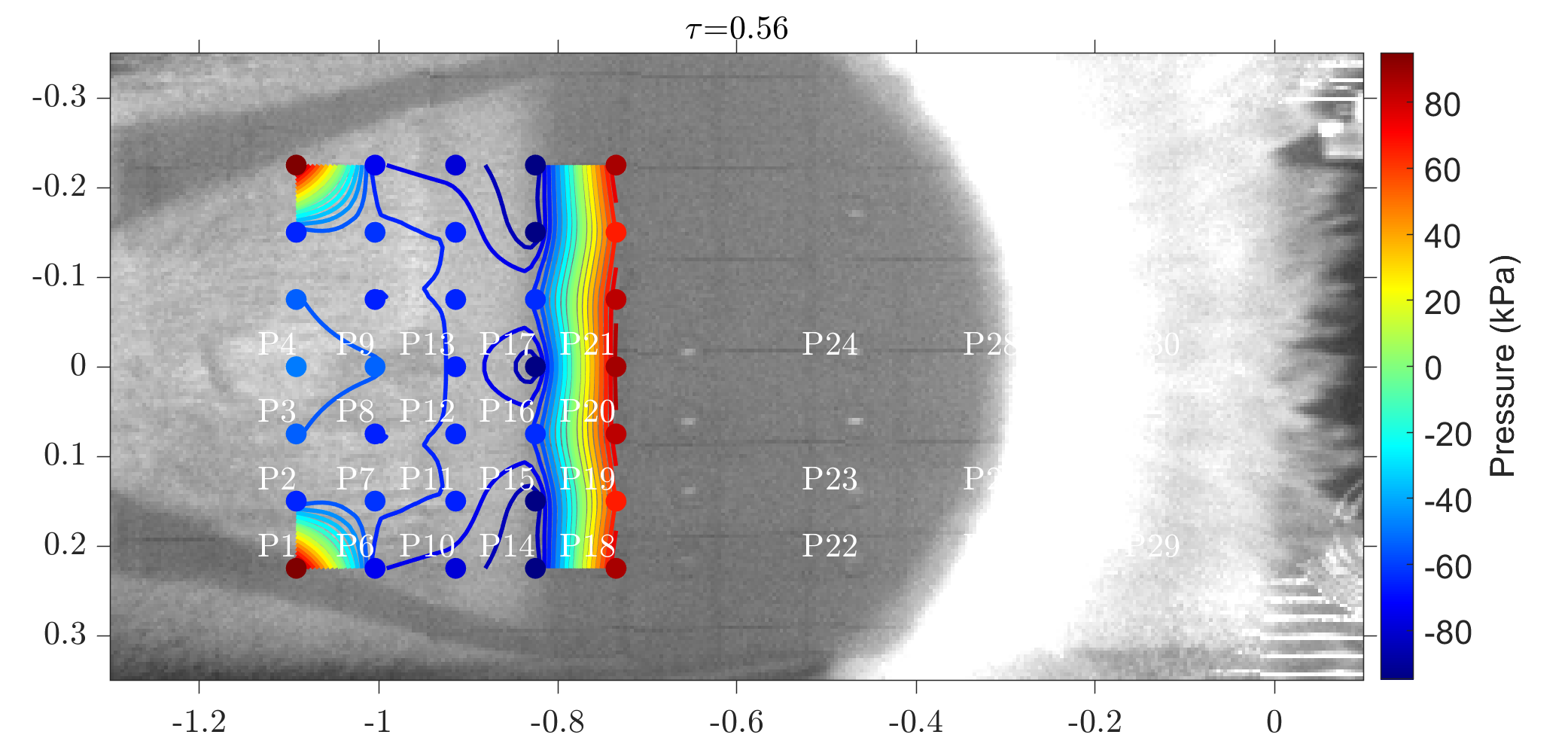}} \qquad
\subfigure[$\tau$=0.58]
{\includegraphics[width=0.45\textwidth]{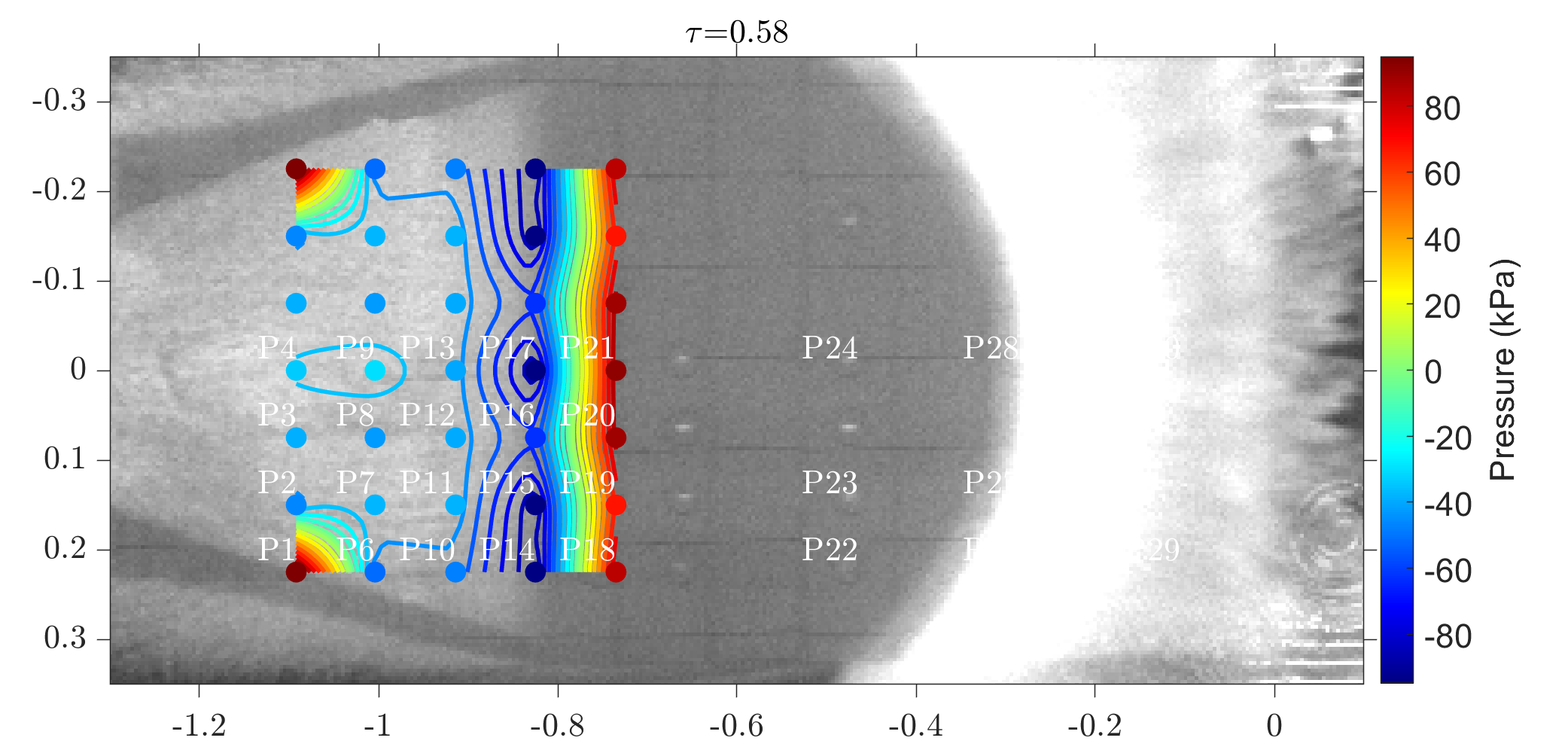}} \\
\subfigure[$\tau$=0.6]
{\includegraphics[width=0.45\textwidth]{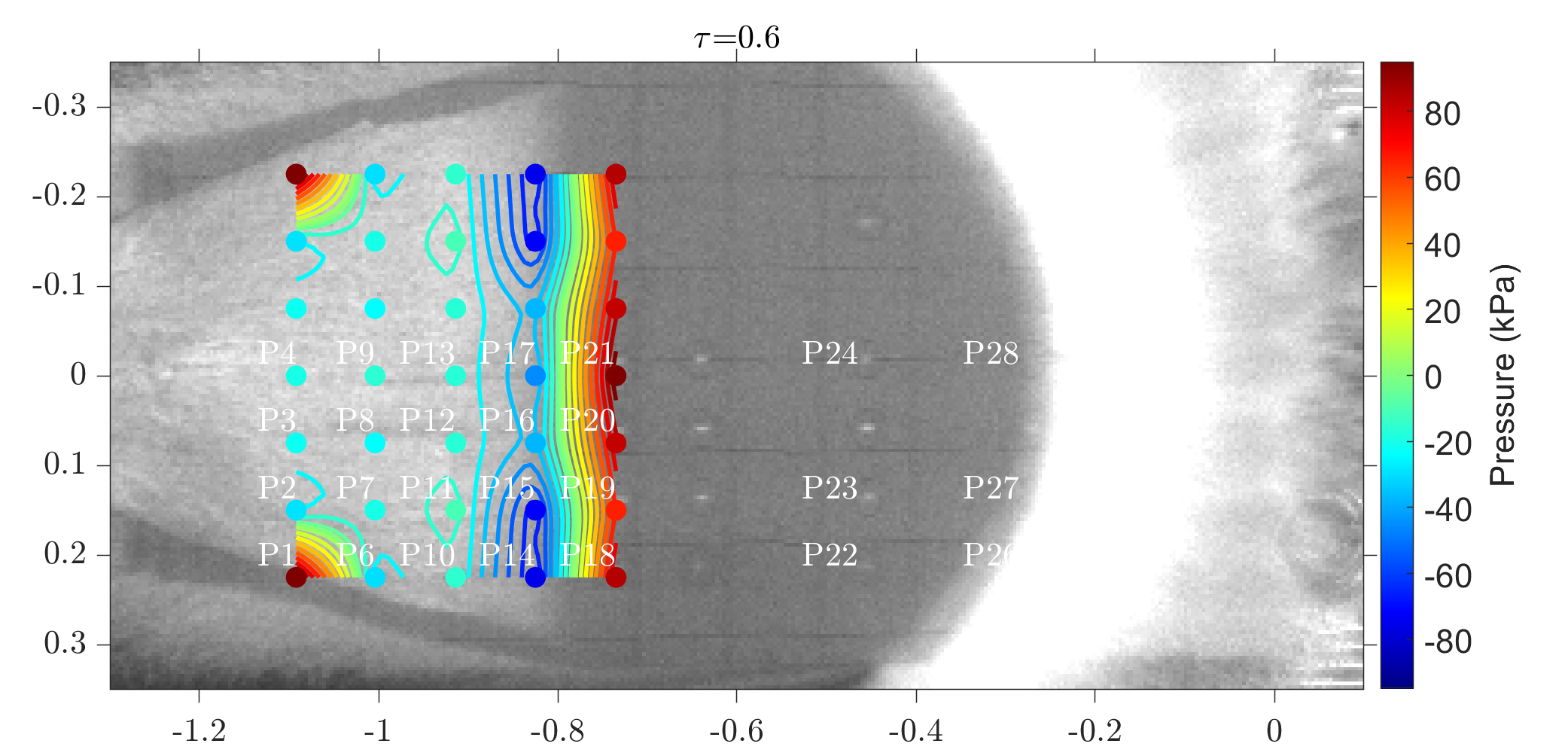}} \qquad
\subfigure[$\tau$=~0.8]
{\includegraphics[width=0.45\textwidth]{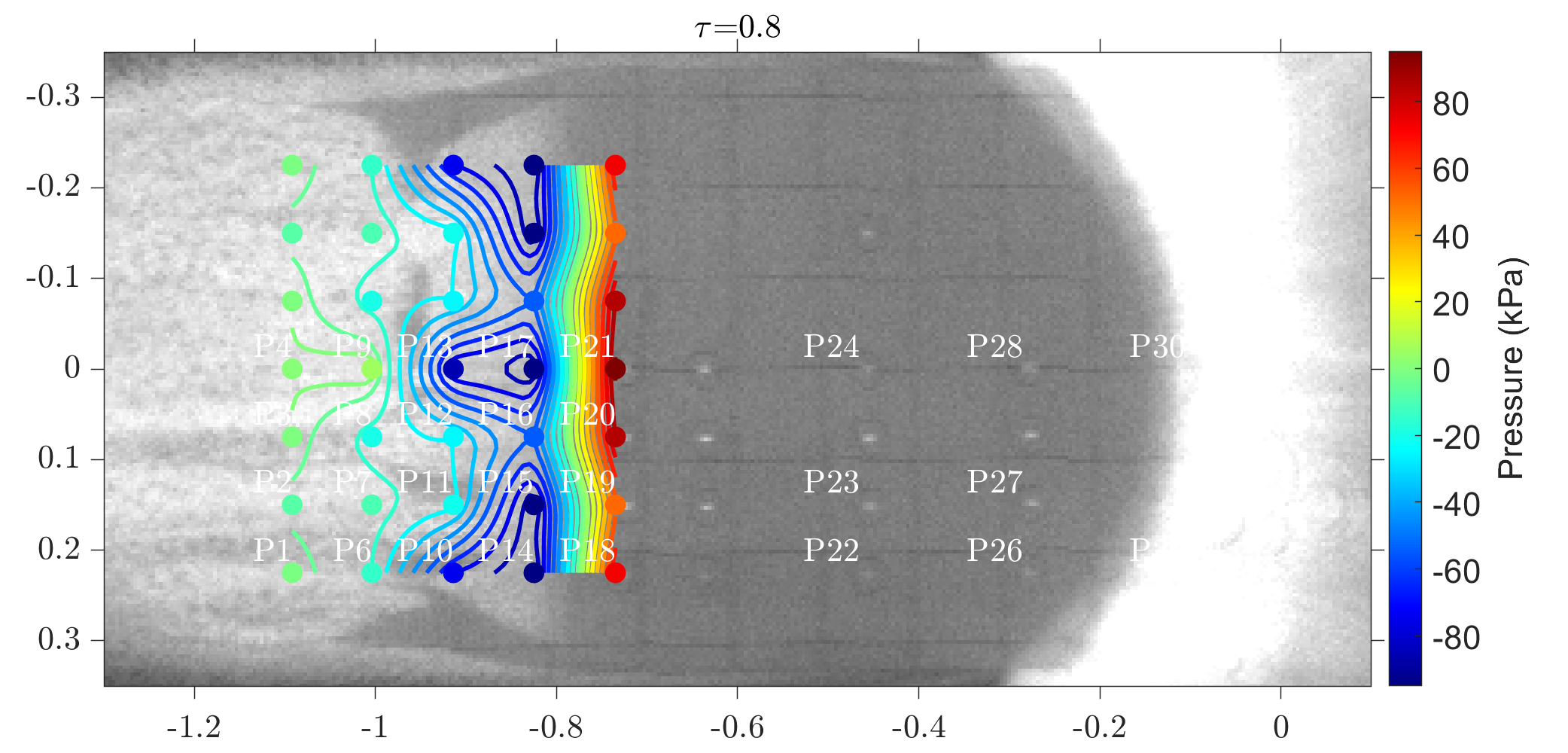}}
\caption{Pressure iso-lines in the rear part of the specimens
\textbf{S3} overlapped to the underwater video frames for the 
water entry at U=40~m/s, at $\alpha = 6^{\circ}$ at the specified
$\tau$ values. The symmetry about the mid-line is exploited.}
\label{fig:S3_isolines}
\end{figure}

The difference between the
the pressure fields of \textbf{S2} and \textbf{S3} on the
bottom can also be visualized when comparing pressure iso-lines
which in Figure~\ref{fig:S3_isolines}, are superimposed 
on the corresponding video frames at $\tau = 0.3$ and
$\tau = 0.8$ and from $\tau = 0.5$ to $\tau = 0.6$
with a finer non-dimensional time resolution of $\Delta \tau = 0.02$.
Firstly, it can be observed that, due to the ``flatter transverse profile''
the pressure at the front edge of the cavitation region 
rises more sharply and uniformly in the transverse direction
for the shape \textbf{S3}, compared to \textbf{S2}.
At $\tau$=0.3 the shape of the 
cavitation area in \textbf{S3} is still
in agreement with the pressure iso-lines, but it has a less elongated
shape compared to \textbf{S2}. However, as the cavitation region 
expands faster in \textbf{S3}, at $\tau$=0.6 the rear edge has already reached 
the trailing edge and the ventilation front has already moved forwards,
up to the point \textit{C}. At $\tau$=0.8, the development of
the ventilation zone has further evolved.

Ventilation in \textbf{S3} occurs in between
$\tau$=0.5 and $\tau$=0.6. In Figure \ref{fig:S3_isolines}
the evolution of the process is visualised at six time instants within this 
non-dimensional time interval. The mechanism of formation 
and development of the ventilation front
in \textbf{S3} is similar to \textbf{S2}, with a curved front
starting from the centre at the trailing edge, which evolves 
very shortly into a straight one and reaches the point \textbf{C}.

\begin{figure}
\centering
\includegraphics[width=0.85\textwidth]{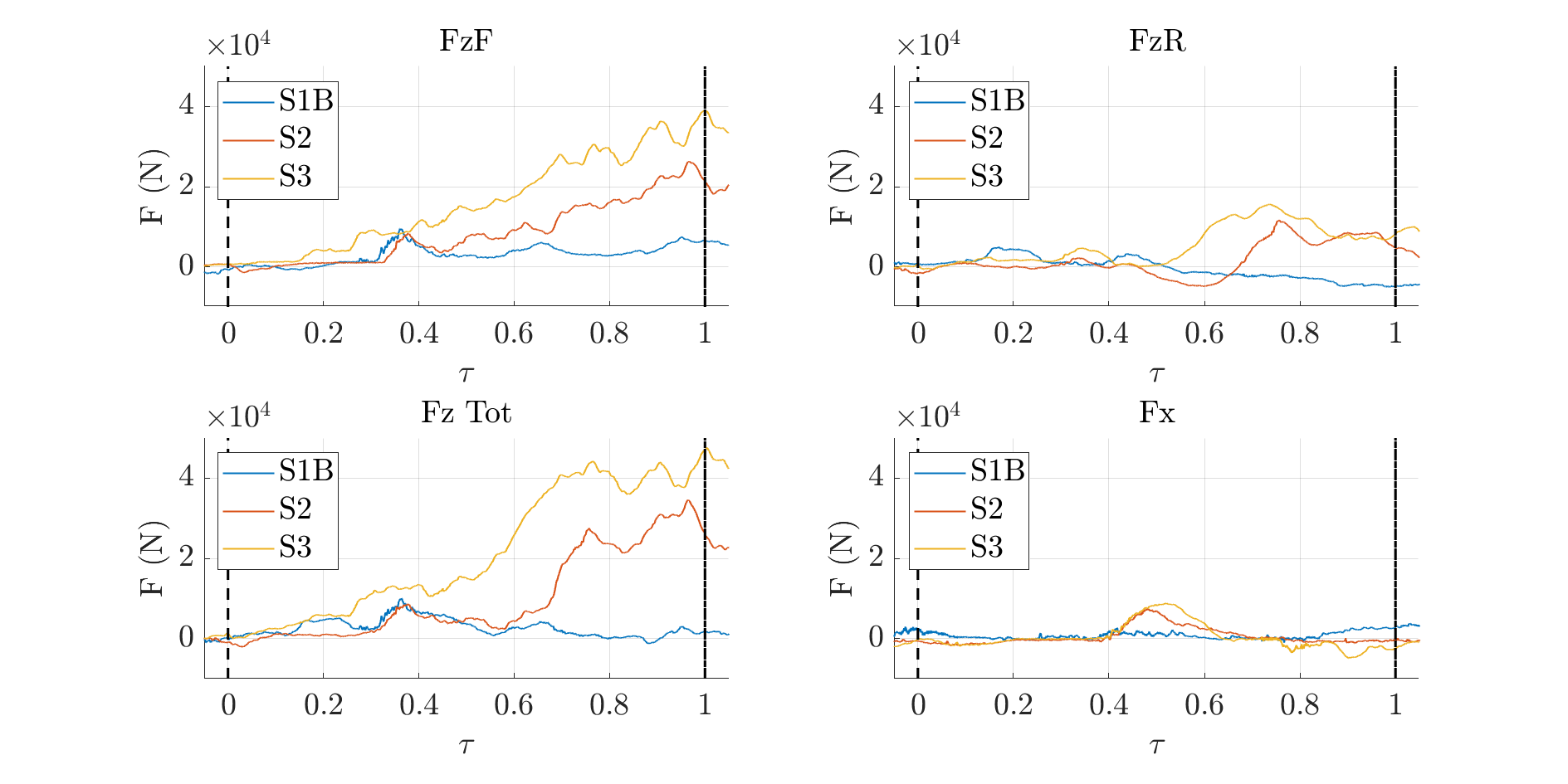}
\caption{Time histories of the forces for the three shapes
at $U$=40~m/s and pitch angle 6$^{\circ}$.}
\label{fig:Forces_AllShapes_U40ms_6deg_RAW}
\end{figure}

The comparison in terms of the $z$-force time histories for the three 
shapes is shown in Figure \ref{fig:Forces_AllShapes_U40ms_6deg_RAW}.
The data indicates that the load measured at the front is 
characterised by an increasing trend for all three shapes. 
The growth rate is higher, as 
the transverse curvature decreases, 
which is consistent with the pressure behaviour discussed above.
The load at the rear exhibit a slightly decreasing trend for
\textbf{S1B}, whereas for both \textbf{S2} and \textbf{S3} 
an initial decreasing trend is observed, associated with the onset of 
cavitation, followed by a rather sharp rise, related to
the ventilation phase. 
As ventilation starts earlier for shape \textbf{S3},
the sharp rise occurs earlier, 
consistently with the pressure time histories.
No substantial differences are found for 
the $x$-component of the force for
the three specimens, which is always very small 
compared to the $z$-component.
%
%
%
\subsection{Cavitation zone propagation velocity}
\label{sec:cavitation_velocity}
%
The underwater videos and pressure data
can be exploited to derive more precise information
on the dynamics of the cavitation region, not only from the qualitative
point of view, but also from the quantitative point of view.
The occurrence of cavitation exerts significant effects 
in terms of both load amplitude and moments, and thus the centre of loads. 
It is therefore of great importance to analyse its evolution in more detail, 
and to evaluate, on a quantitative basis, the influence of the
body curvatures upon it, especially the transverse curvature.

\begin{figure}
\centering
\subfigure[Signal De-Noising]
{\includegraphics[width=0.45\textwidth]{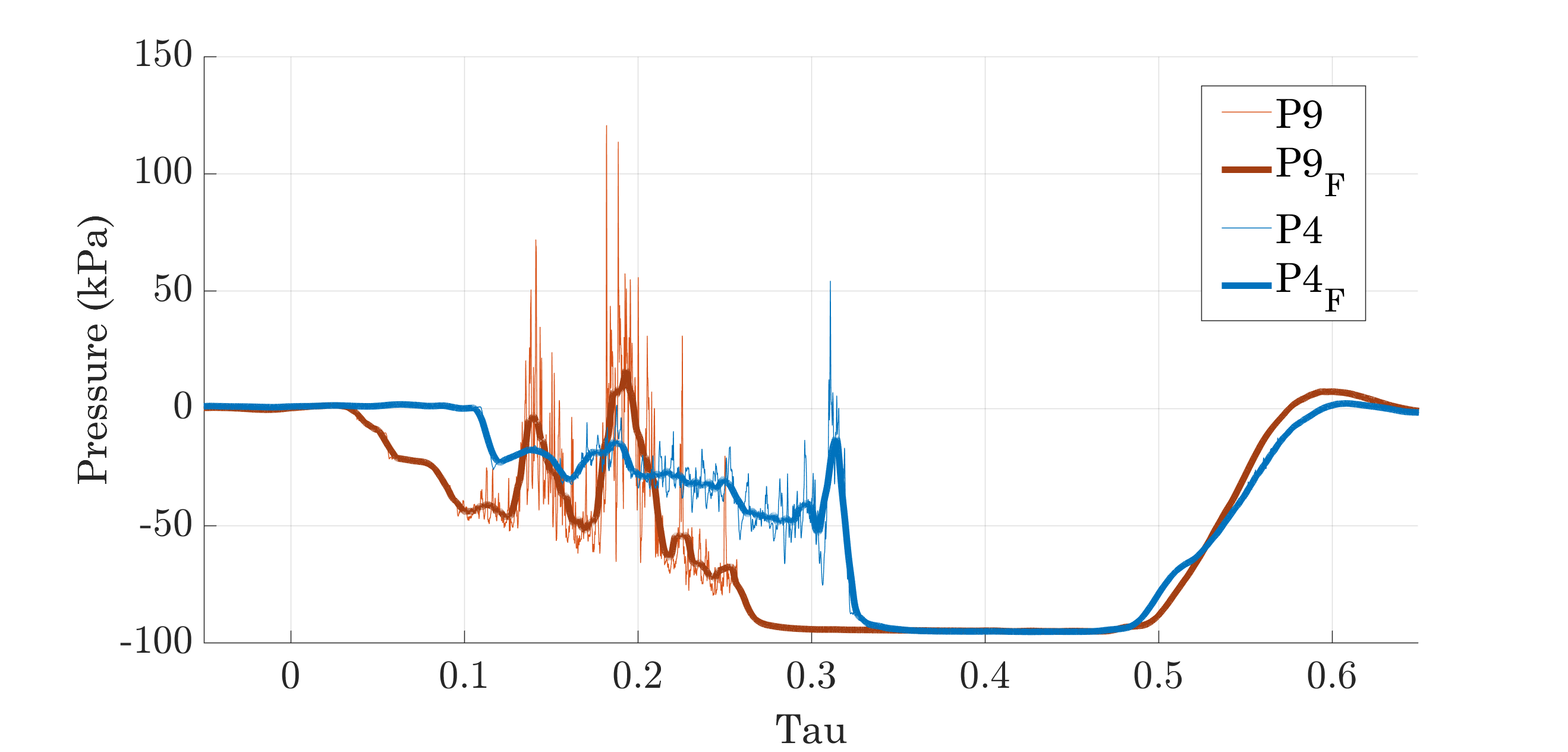}} 
\quad
\subfigure[Delay estimation]
{\includegraphics[width=0.45\textwidth]{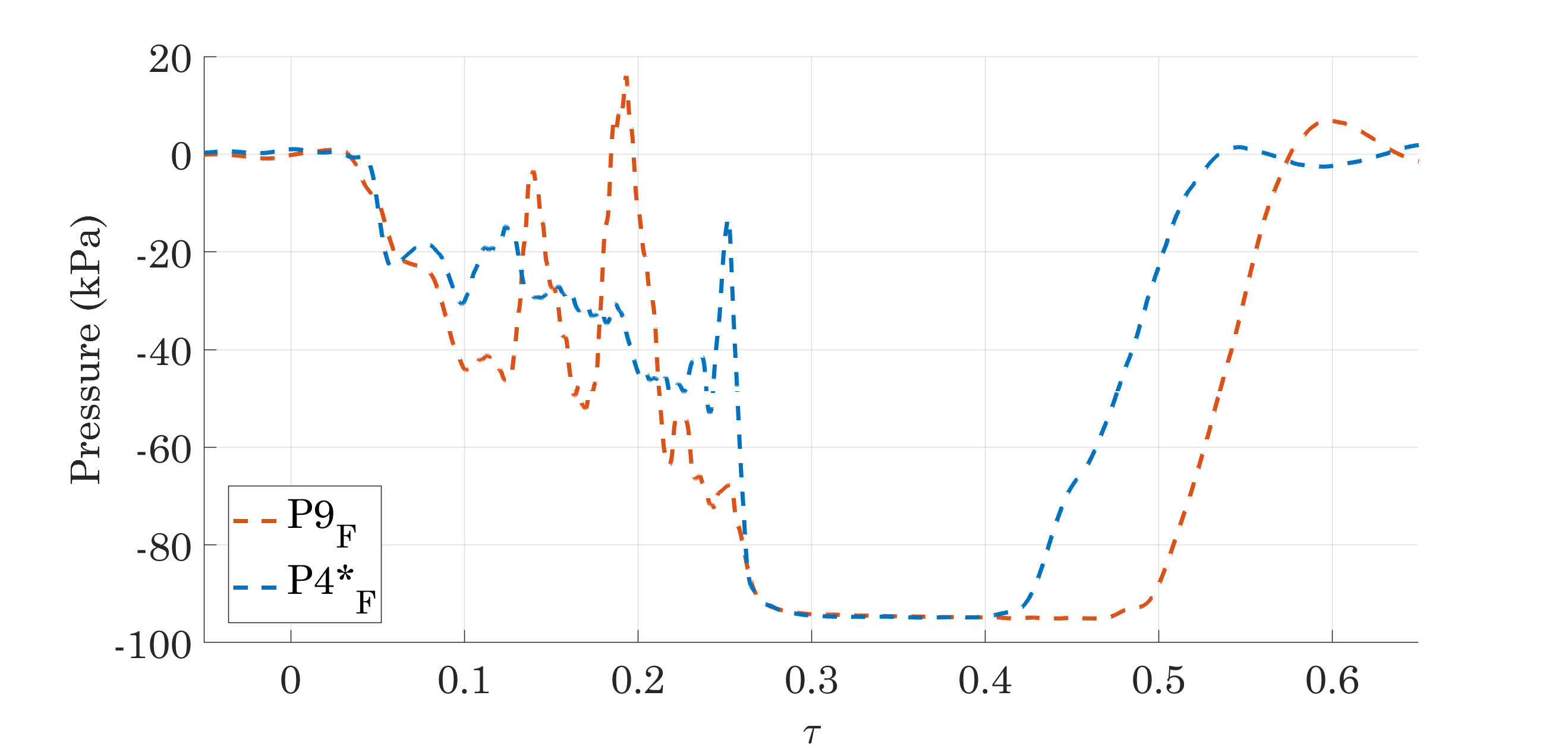}}
\caption{Comparison of (a) two sample signals recorded by
the pressure probes P4 and P9 before and after
the de-noising procedure and (b) overlapping 
of the signal P9 and the signal P4$^*$, which is the one 
recorded by P4 and translated in time for the delay estimation.}
\label{fig:Comparison_Overlap_Signals_Silvano}
\end{figure}

\begin{figure}
\centering
\includegraphics[width=0.8\textwidth]{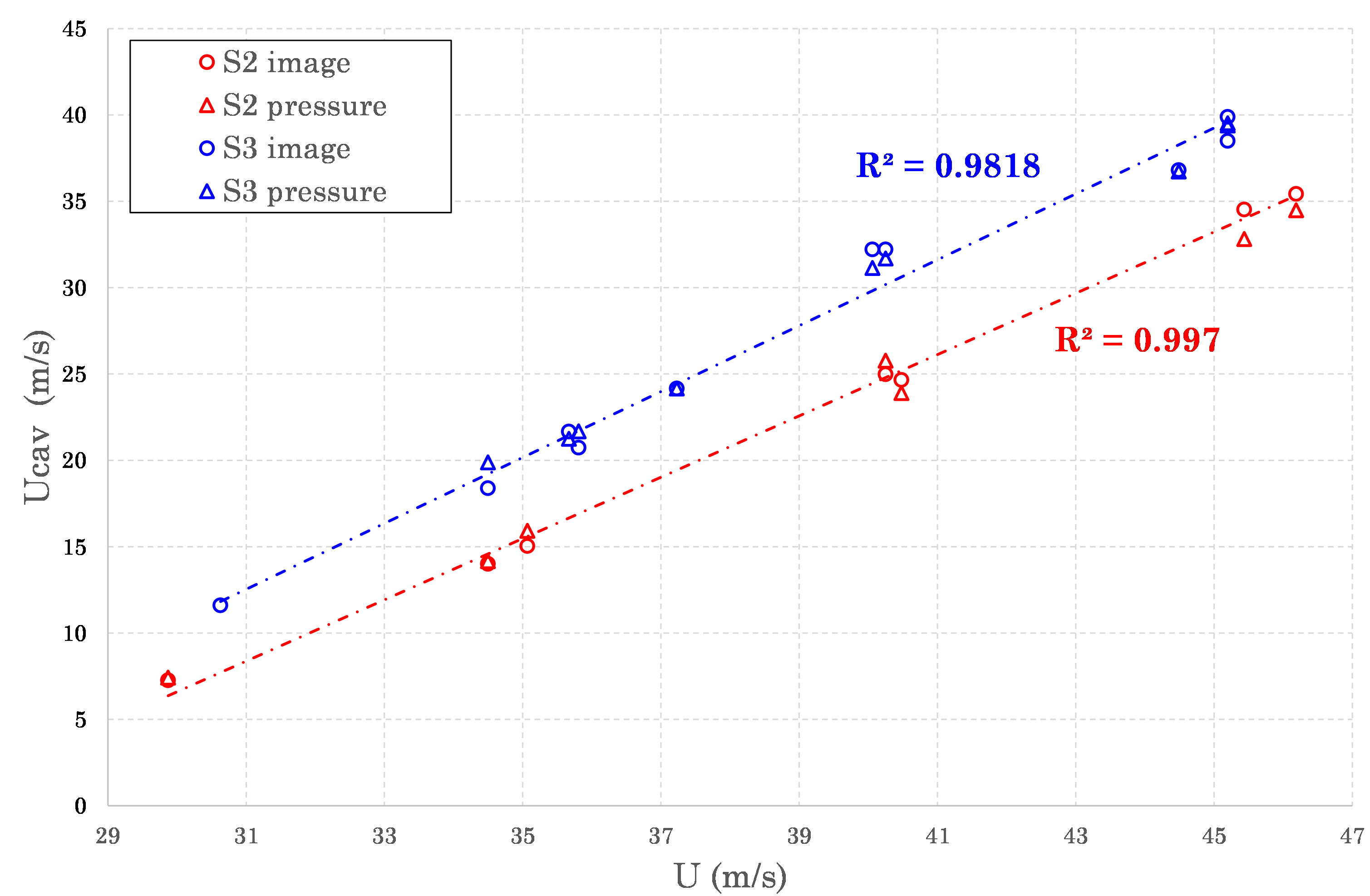}
\caption{Propagation velocities of the cavitation area 
for the three shapes and as a function of $U$,
using the method bases on image analysis (circles) and on
pressure data (triangles). Linear trend lines are added using the 
results of the image analysis method.}
\label{fig:Cavitation_Velocity_Silvano}
\end{figure}

For this purpose a \emph{propagation velocity of the cavitation region},
denoted as $U_{\textrm{cav}}$, is introduced,
which is the velocity at which the rear edge of the cavitation
zone moves backwards in the trolley reference frame.
This propagation velocity can be retrieved either from the analysis of the
underwater video frames or from the pressure signals, as follows.
Starting from the \textit{underwater images} the edge of the cavitation
region is identified and the furthest point at the back is marked for all
frames, which are separated by a constant time interval.
The position of the rear edge in the pictures can be determined within
$\pm$ 2 pixels, which makes the calculation of 
the propagation velocity rather accurate. 
The propagation velocity is computed as an average velocity from the instant
in which the cavitation region clearly appears in the frames up to the
instant at which it reaches its maximum extension, i.e.
when the region reaches a steady size 
in the \emph{incipient cavitation regime}
or as soon as the region reaches the trailing edge in the
\emph{cavitation-ventilation regime}.
The propagation velocity of the cavitation region can also be
determined from the \emph{time histories of the pressure} measured by the
different probes. In fact, the passage of the cavitation front corresponds to 
the sharp drop to the vapour pressure value. Hence, being
the position of the probes on the specimen known, the velocity can be
determined from the time delay between
the pressure drops recorded by different probes.
Since the raw pressure signals, as observed above, are often very noisy and 
display significant spikes, to facilitate the estimation of the
time delays, a de-noising procedure has been developed in 
MATLAB$^{\copyright}$, which includes in sequence a Continuous Wavelet 
Transform - based filter, a conventional low-pass filter and a moving average.  
Hence, the time delay is calculated through a cross-correlation 
of the denoised signals, with the objective of identifying the time shift
that optimises the degree of overlap between the drop portions.
The process is illustrated in detail in Figure
\ref{fig:Comparison_Overlap_Signals_Silvano}.
%
%
The estimated propagation velocities of the cavitation region $U_{\textrm{cav}}$,
based on the image analysis and on the pressure
time histories, are plotted in Figure 
\ref{fig:Cavitation_Velocity_Silvano} for the shapes \textbf{S2}
and \textbf{S3} at 6$^{\circ}$, velocity ratio $V/U$=0.0375, and different 
horizontal speeds. It is worth noticing that in the 
graph supplementary test conditions with different horizontal velocities are 
considered, in addition to those listed in Table \ref{tab:test_conditions}.
$U_{\textrm{cav}}$ is plotted as a function
of the horizontal velocity $U$, and linear fits of the data derived from
the image analysis are also plotted, which both have a coefficient
of determination $R^2$ very close to 1. It is worth
noticing that the propagation velocity computed by using the pressure 
signals are very much in line with those derived from the image
analysis.
The above results indicate that $U_{\textrm{cav}}$
is directly proportional to the horizontal speed U and
that the coefficient of proportionality is between 1.77 and 1.91 
for the shapes \textbf{S2} and \textbf{S3}, respectively.
There is a vertical offset between the two trends, with 
$U_{\textrm{cav}}$ being higher for \textbf{S3} compared to \textbf{S2}, as a
consequence of the lower transverse curvature that reduces the 
three-dimensional effects on the resulting flow, consistently
with the considerations presented in Section \ref{sec:effect_shape}.
%
%
\subsection{Wetted area evolution}
%
In the pioneering works of \citet{von1929impact}
and \cite{wagner1932stoss} on vertical water entry, the growth
of the wetted area is related to the time variation of
the added mass and, therefore, to the impact loads. 
The primary distinction between the two theories is that, 
whereas von Karman's theory considers solely the geometric
intersection of the body with the free, the Wagner method
also accounts for the water pile-up at the spray.
In the work of \cite{iafrati2016experimental} on the water
entry of flat plates with a high horizontal speed, 
it was observed that the value of the pressure peak
and the propagation of the spray root during the impact
phase are related. Consequently, the evolution of the
wetted area and the loads are also  subject to a similar 
relationship. In light of the aforementioned discussion, 
it can be reasonably assumed that a comparable analogy can be 
expected in the front part of double-curvature specimens, 
where the pressures are found to be positive in all areas. 
Conversely, at the rear, the occurrence of negative pressures 
and, consequently, of cavitation and ventilation renders 
such an analogy invalid.
Given the aforementioned considerations, it is considered useful
to conduct a more detailed investigation into the evolution
of the wetted area in relation to the geometric intersection area.

Looking again at Figure \ref{fig:Silvano_Cavitation_Frames},
referring to the water entry of the three shapes
at U=40~m/s, $\alpha$=6$^{\circ}$ and V/U=0.0375, as mentioned already in 
Section~\ref{hydrodynamics_S2}, 
the wetted areas are identifiable as the darker regions at the centre of the
specimen. Inside the wetted area the brighter zone, when present, is
the cavitation region. 
It is observed that, due to the pile-up effect, for \textbf{S1B} and 
\textbf{S2} the actual wetted area is broader than the geometric
intersection area with the 
undisturbed free surface at $\tau$ in between 0.2 and $\tau$=~0.6. 
This is also the case for the shape \textbf{S3} up to $\tau$=~0.3.
However, for \textbf{S3} from a value of $\tau$ between 0.4 and 0.5, 
if the pile-up effect persists at the front, at the rear of the specimen, 
in addition to the cavitation/ventilation zone at the middle, 
some bright zones emerge at the sides.
These zones are most likely caused by ventilation originating
from the sides of the specimen and developing as a result 
of the low pressures in the middle.
Two dark narrow regions of wetted area are still present
between the cavitation/ventilation zones
in the middle and the ventilation zones at the sides.
In particular, it is observed that for \textbf{S3} the wetted area becomes
smaller than the geometrical intersection area at a non-dimensional time
$\tau$ between 0.4 and 0.5. Such an effect becomes more evident
at $\tau$=~0.6. A similar effect was also
observed in \citet{spinosa2022hydrodynamic}, in which a
shrinking of the wetted area with respect to the geometric intersection
area is observed in the planing motion of the
whole fuselages \textbf{S2} and \textbf{S3},
with the effect more enhanced for \textbf{S3}, as a
consequence of the lower pressures at the rear.

\begin{figure}
\centering
\subfigure[Wetted Area Front/Rear $\tau$=0.3]
{\includegraphics[width=0.31\textwidth]{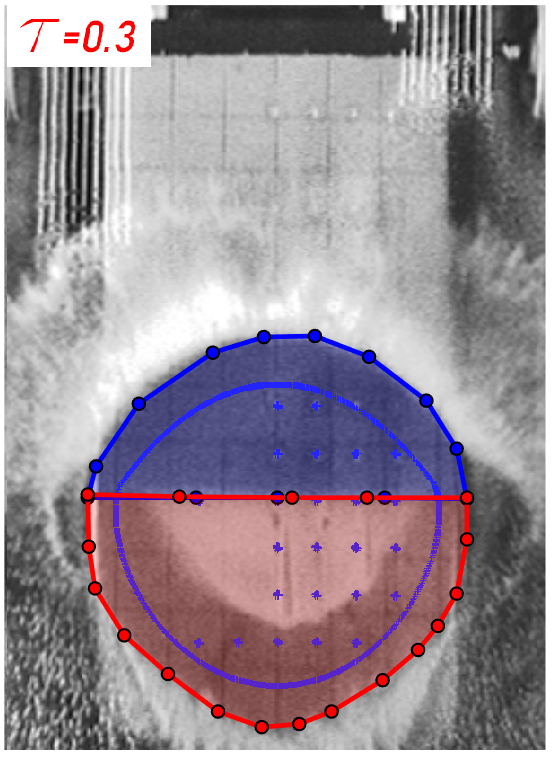}} 
\quad
\subfigure[Geometric Intersection Front/Rear $\tau$=0.3]
{\includegraphics[width=0.31\textwidth]{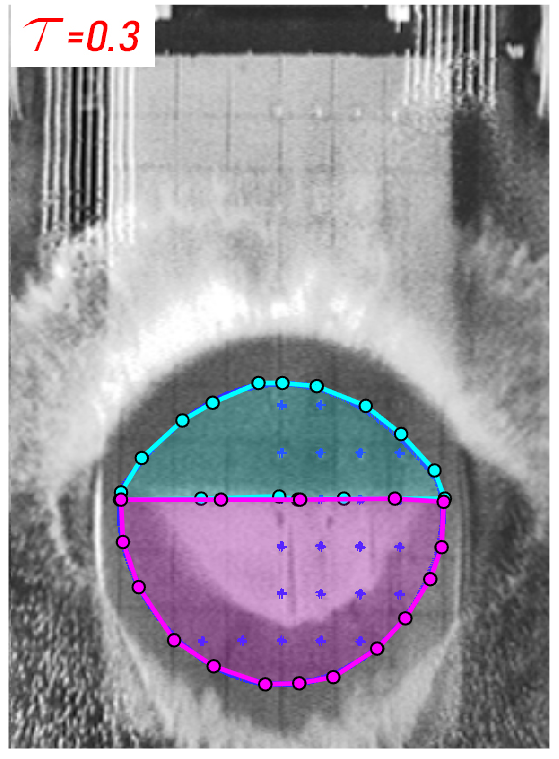}}
\quad 
\subfigure[Cavitation Area $\tau$=0.3]
{\includegraphics[width=0.31\textwidth]{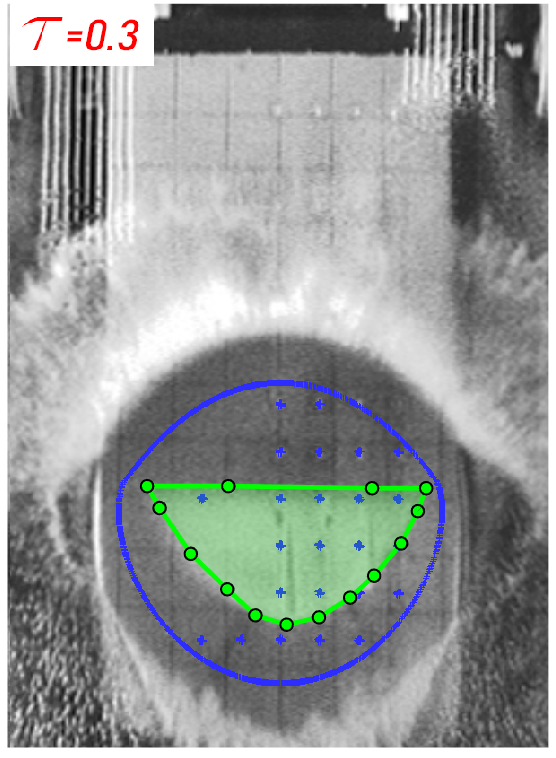}}\\
\subfigure[Wetted Area Front/Rear $\tau$=0.6]
{\includegraphics[width=0.31\textwidth]{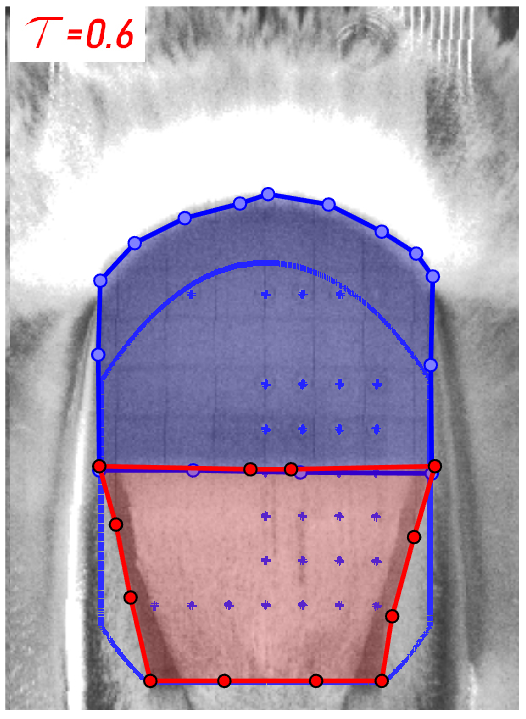}} 
\quad
\subfigure[Geometric Intersection Front/Rear $\tau$=0.6]
{\includegraphics[width=0.31\textwidth]{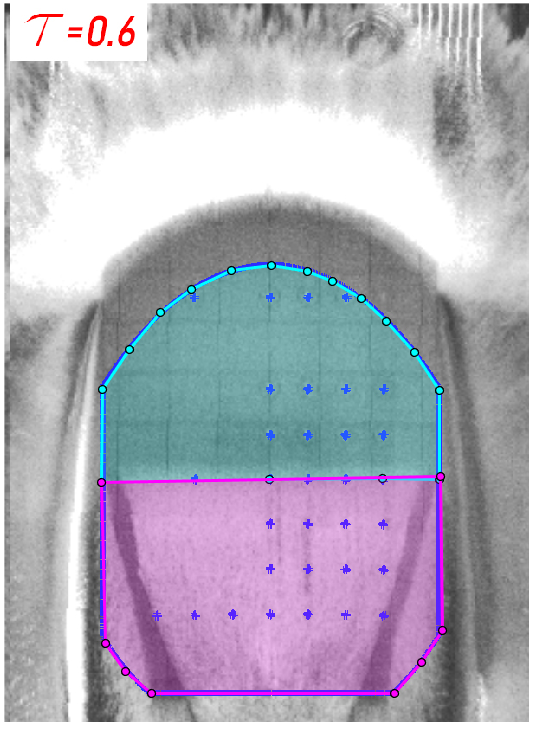}}
\quad 
\subfigure[Cavitation Area $\tau$=0.6]
{\includegraphics[width=0.31\textwidth]{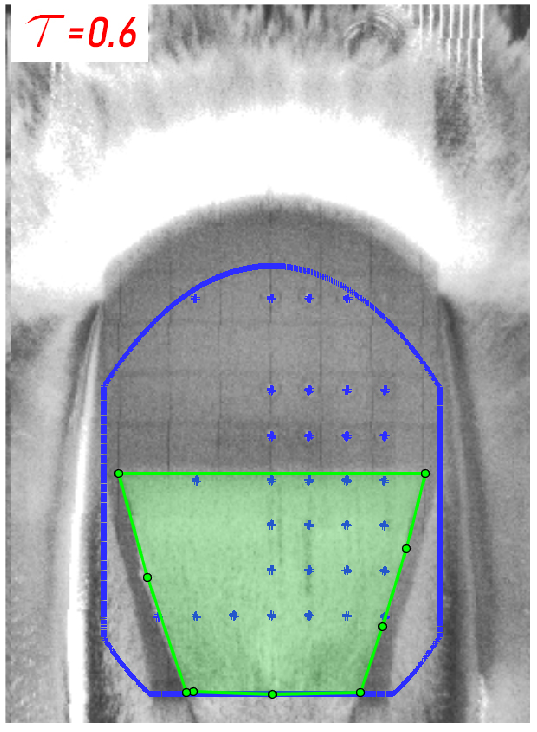}}\\
\caption{Definition of the ROI for (a,d) the wetted
areas in front and at the rear of the specimen; (b,e) geometrical
intersection area at the front and at the rear;
(c,f) cavitation area for the shape \textbf{S3}
at U=40~m/s, pitch angle 6$^{\circ}$ at $\tau$=0.3
(top) and 0.6 (bottom).} 
\label{fig:ROIs_wetted_geom_cav_area_S3_tau03}
\end{figure}

From the quantitative point of view it is possible to estimate
the wetted area from the underwater video frames and to compare
it with the geometrical intersection area. 
This is performed for the three shapes, using
an image processing routine developed in MATLAB$^{\copyright}$.
As shown in 
Figure~\ref{fig:ROIs_wetted_geom_cav_area_S3_tau03},
which refers to shape \textbf{S3} at $\tau$=0.2 and 0.6,
three Region Of Interests (ROIs) are manually
selected and their areas are computed, first as pixel counts
and then as the physical area in m$^2$, by multiplying the pixel count
for the pixel area derived from an image calibration.
In particular, the defined ROIs are the front and 
rear wetted surface, which are delimited by the
$x$-coordinate of the pressure
probe row $r4$, the front and rear part of the geometric
intersection area, still separated by the row $r4$, and finally
the cavitation area.
It is worth noticing that 
using this routine, the wetted areas are not exactly estimated
rather their projection on the plane
$X-Y$, parallel to the ground.

\begin{landscape}
\begin{figure}
\raggedleft
\subfigure[Front part \textbf{S1B}]
{\includegraphics[width=0.4\textwidth]{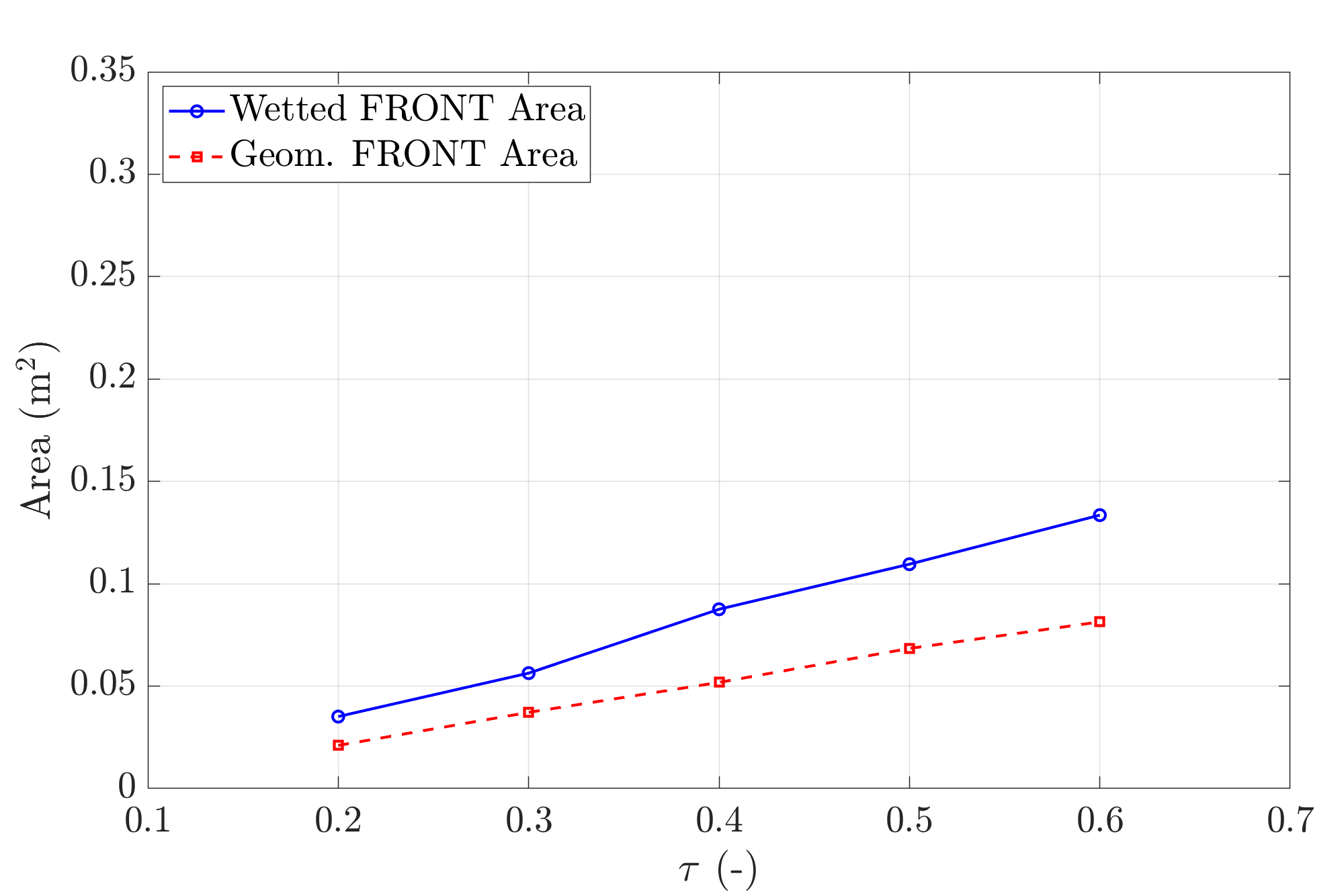}} \quad
\subfigure[Front part \textbf{S2}]
{\includegraphics[width=0.4\textwidth]{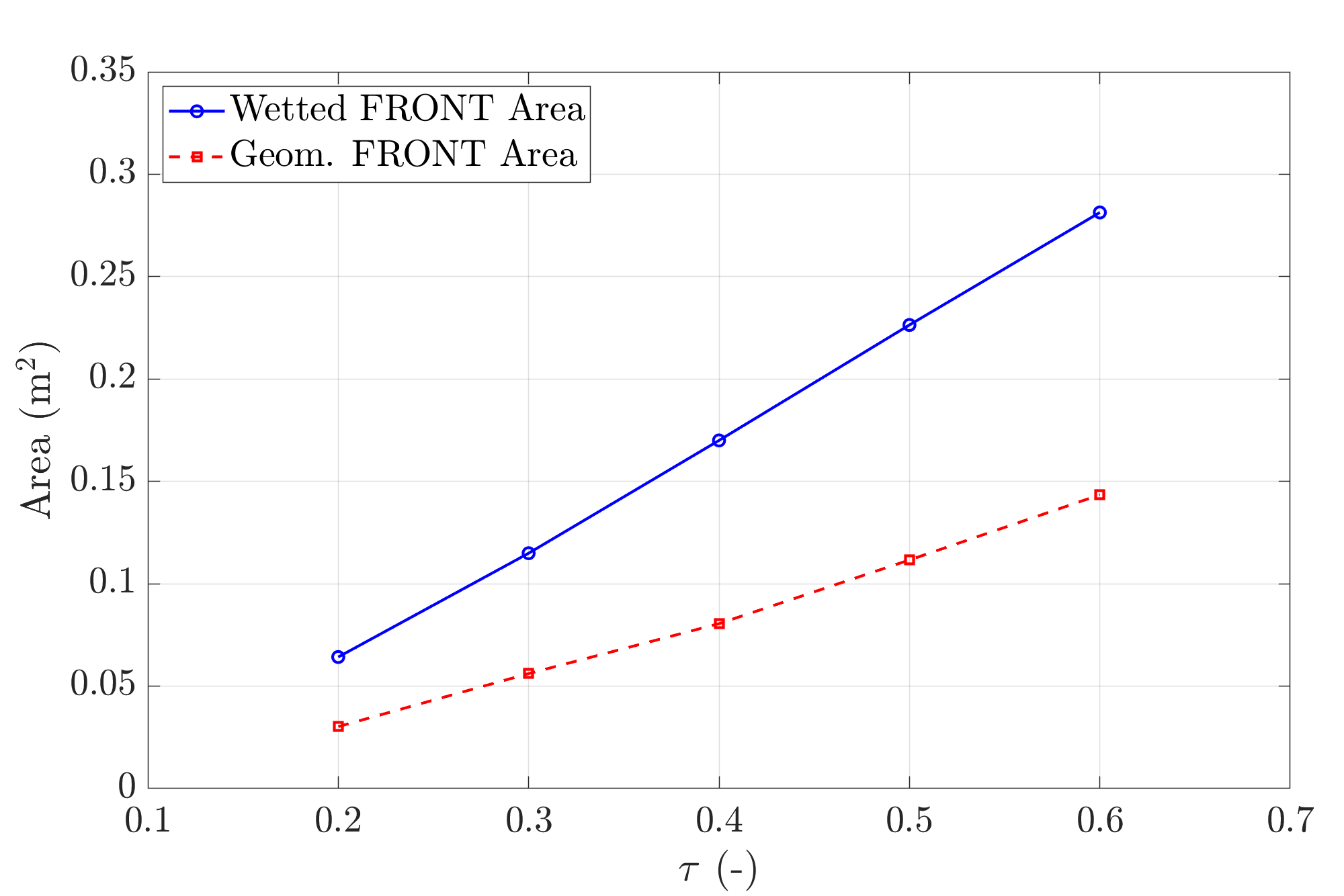}} \quad
\subfigure[Front part \textbf{S3}]
{\includegraphics[width=0.4\textwidth]{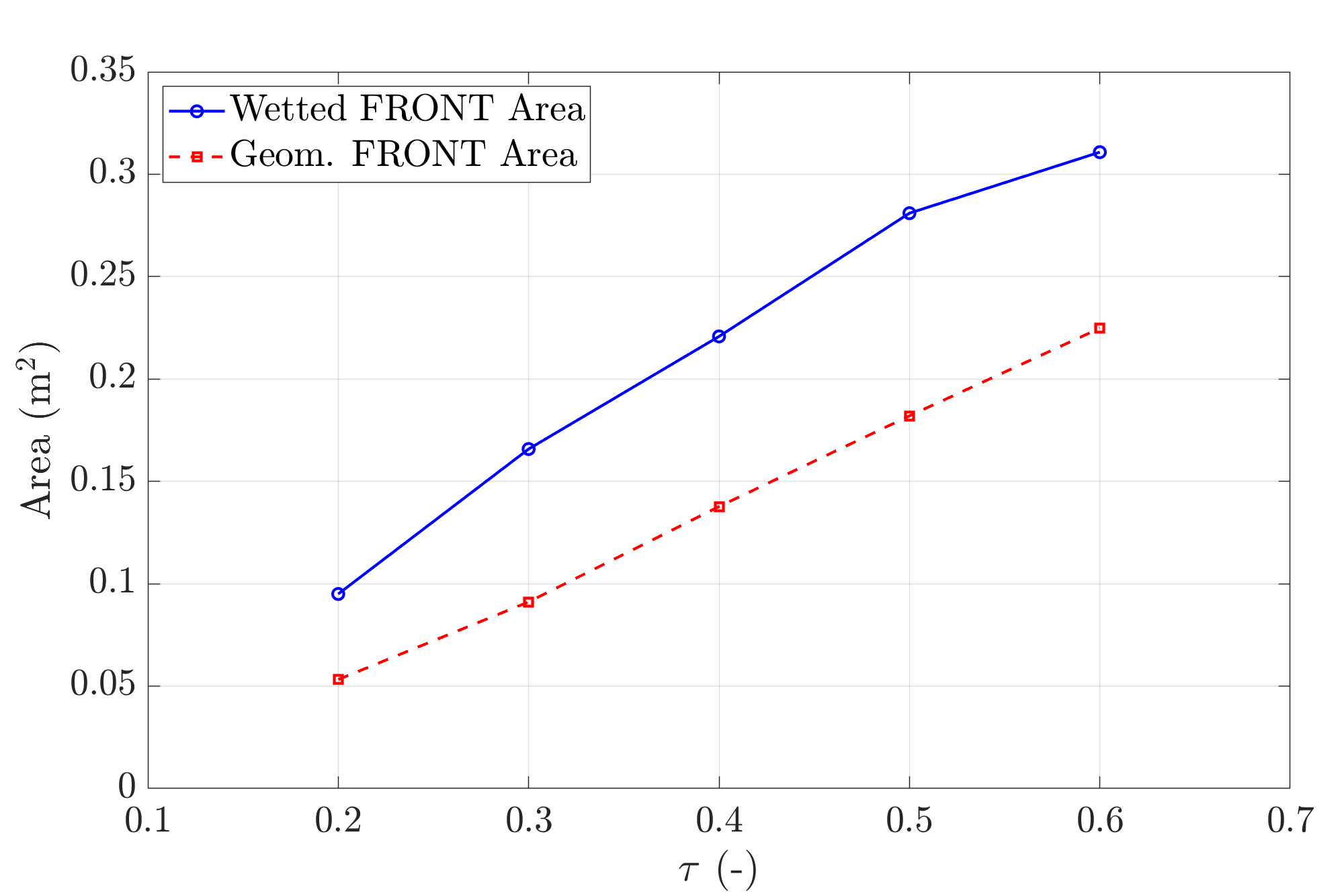}} \\
\subfigure[Rear part \textbf{S1B}]
{\includegraphics[width=0.4\textwidth]{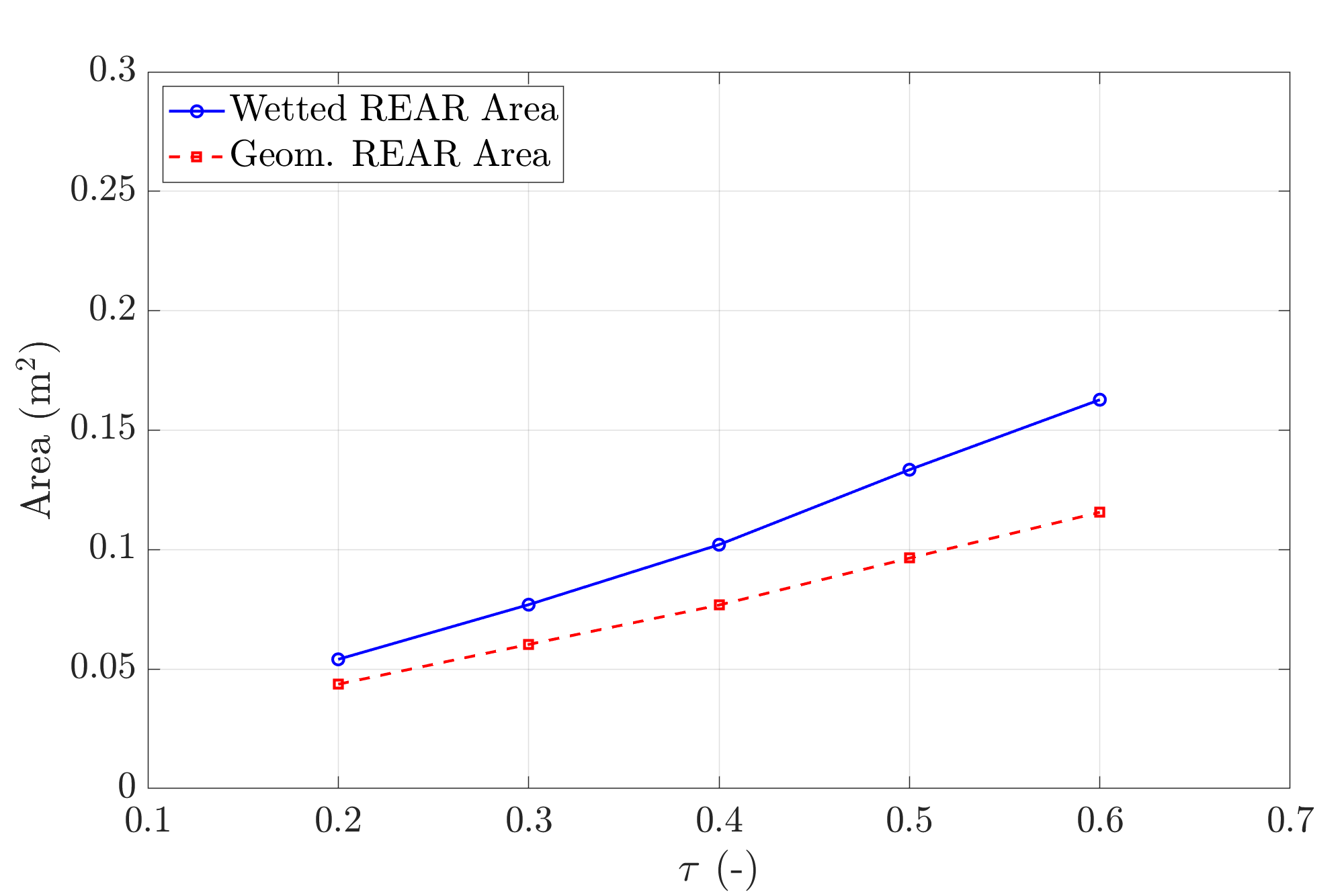}} \quad
\subfigure[Rear part \textbf{S2}]
{\includegraphics[width=0.4\textwidth]{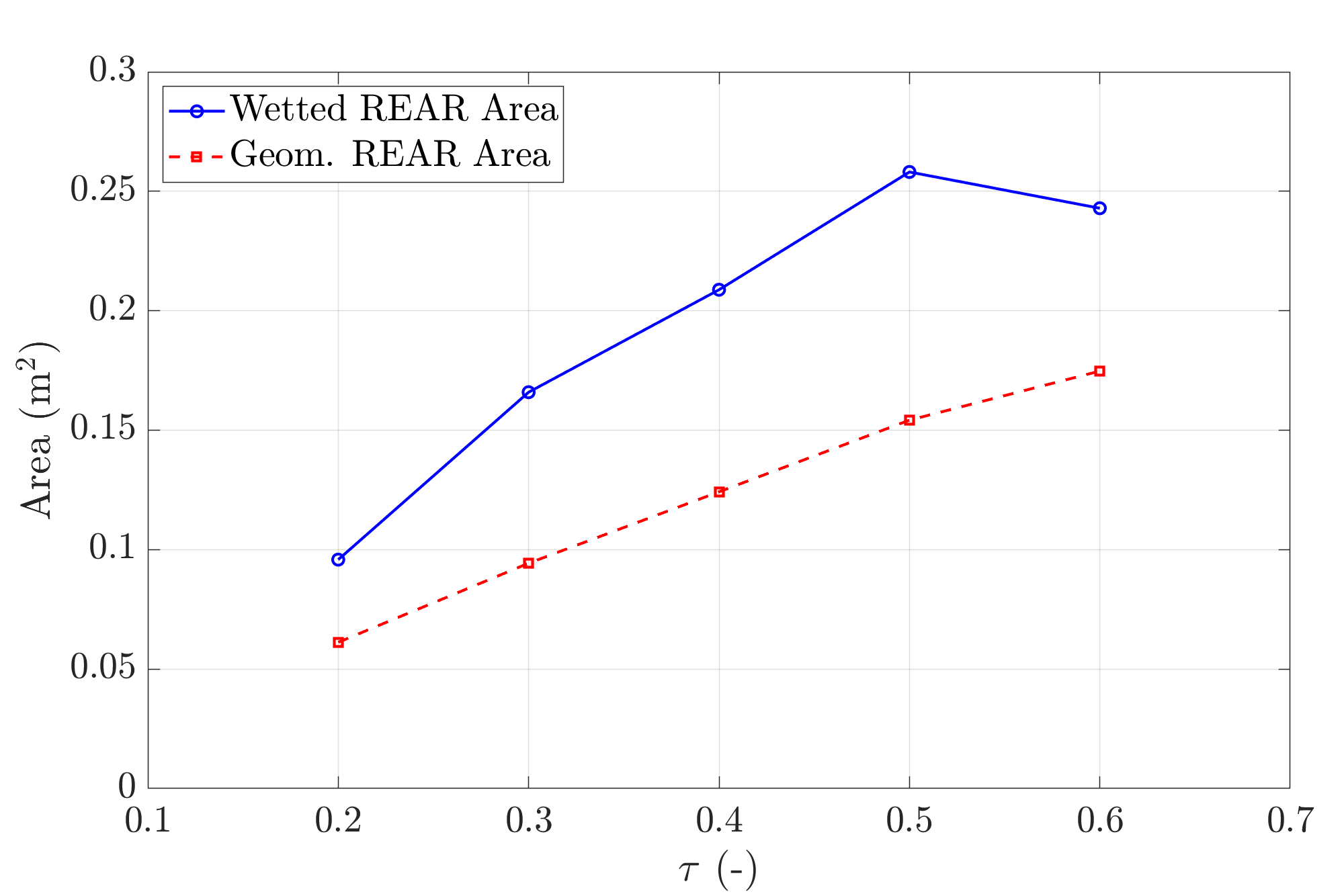}} \quad
\subfigure[Rear part \textbf{S3}]
{\includegraphics[width=0.4\textwidth]{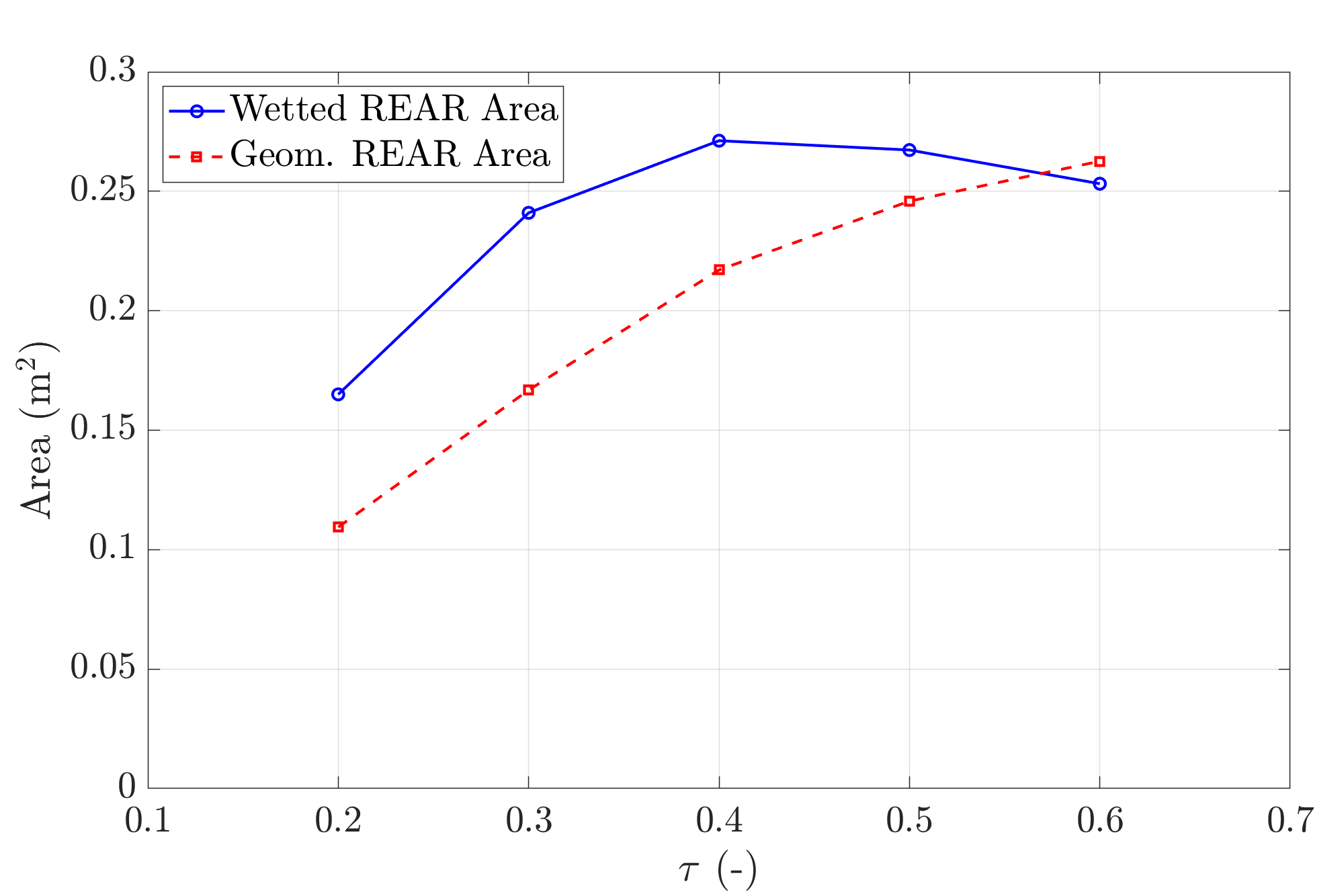}} \\
\subfigure[Cavitation area S2]
{\includegraphics[width=0.4\textwidth]{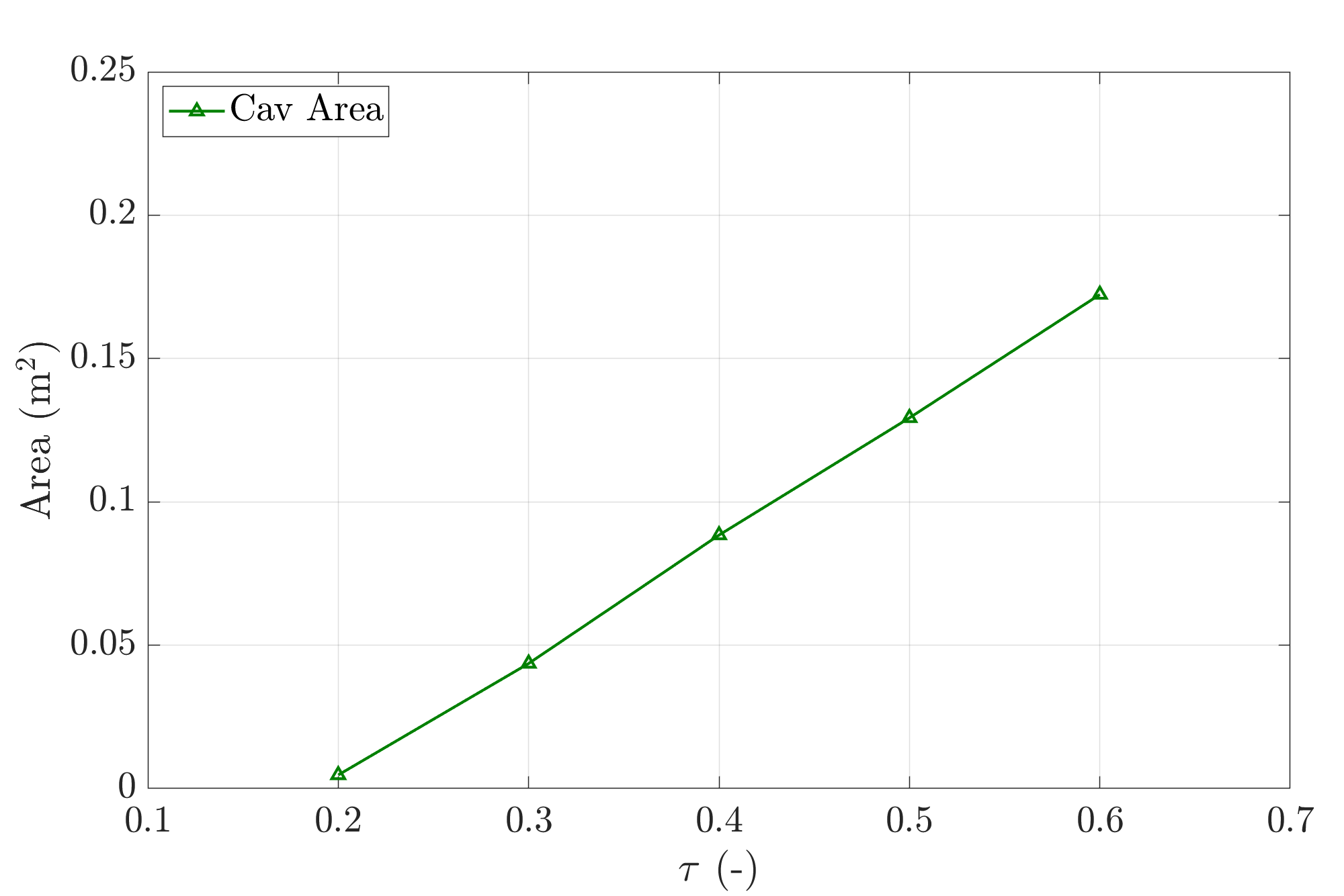}} \quad
\subfigure[Cavitation area S3]
{\includegraphics[width=0.4\textwidth]{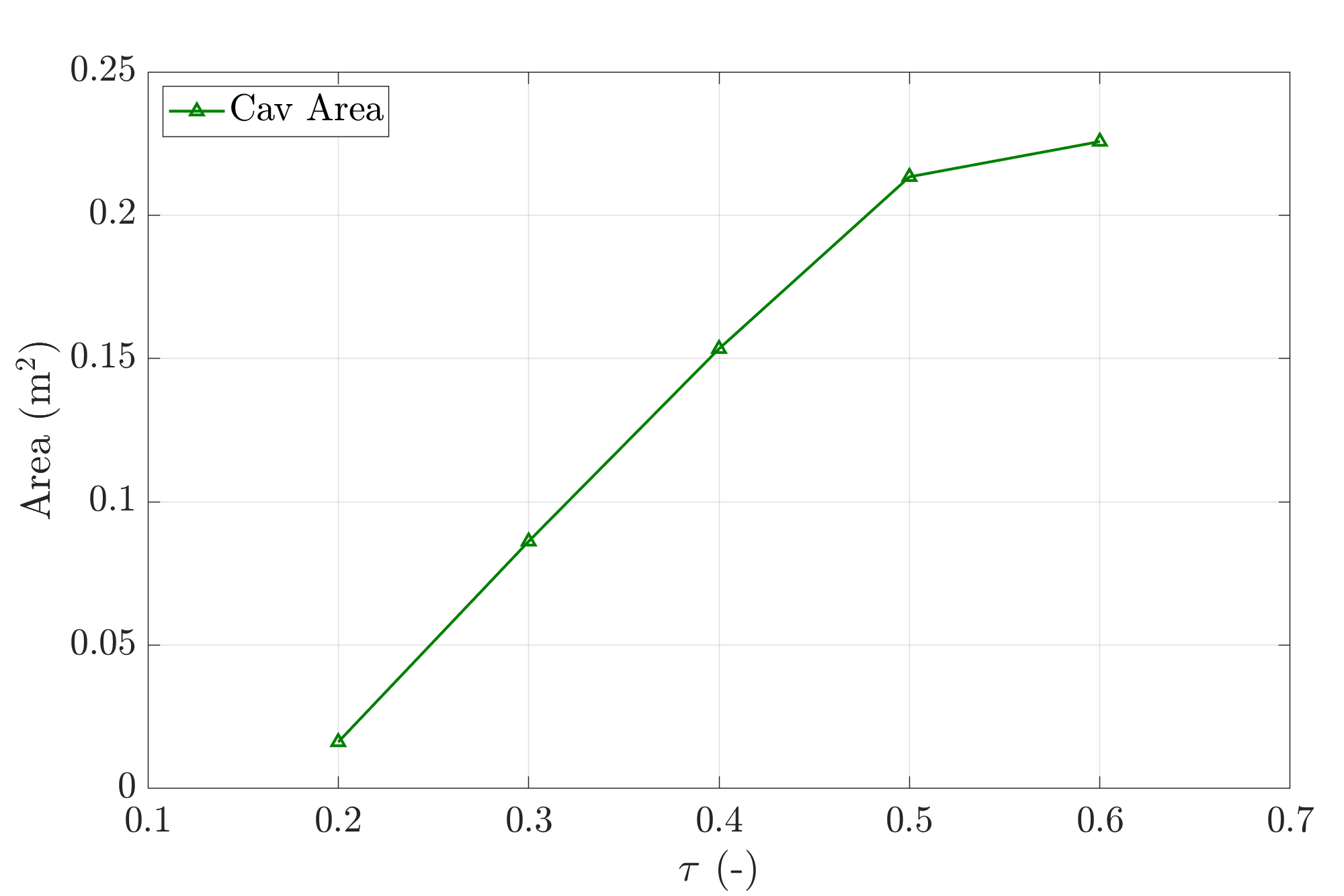}}
\caption{Variation in time wetted areas compared with the geometrical intersection
area for the specimen \textbf{S1B}(a,d), \textbf{S2} (b,e) and \textbf{S3}(c,f)
at pitch angle 6$^{\circ}$ at U=40~m/s from $\tau$=~0.2 
to 0.6. In (g,h) the cavitation
area as a function of $\tau$ is shown for 
the shapes \textbf{S2} and \textbf{S3}.} 
\label{fig:FRONT_REAR_wetted_vs_geom}
\end{figure}
\end{landscape}

A comparison between the values of
front/rear wetted areas and of the geometrical
intersection from $\tau$=0.2 to 0.6 (see also
Figure \ref{fig:Silvano_Cavitation_Frames})
is shown in
Figure \ref{fig:FRONT_REAR_wetted_vs_geom}
for the three shapes
The values of the cavitation areas are plotted in 
Figure \ref{fig:FRONT_REAR_wetted_vs_geom}(g) and (h)
for the shapes \textbf{S2} and \textbf{S3} respectively.
%
It is observed that at the front both the wetted and the geometric 
intersection areas grow between $\tau$=~0.2 and $\tau$=~0.6, in particular
the wetted area grows at a slightly higher rate.
Owing to the much smoother curvature, specimen \textbf{S1B} is
characterized by the lowest growth rates. As to \textbf{S2} and
\textbf{S3}, it can be noticed that in the interval $\tau \in (0.2 \: 0.6)$
the growth of the wetted area is about the same, despite the
fact that \textbf{S3} exhibits a higher growth rate of the geometric
intersection area.

At the rear, in the case of specimen \textbf{S1B} both the geometric
intersection and the wetted areas grow in time, the latter at a
slightly higher rate.
In the case of \textbf{S2} and \textbf{S3}, the values of the 
geometric intersection area display a monotonic trend,
although in the case of \textbf{S3} the growth 
rate exhibits a reduction after $\tau=0.4$. The wetted area grows up
to $\tau=0.5$ and $\tau=0.4$ for \textbf{S2} and \textbf{S3},
respectively. For \textbf{S2} the wetted surface remains almost constant
afterwards, as it reaches the edges of the specimen. In the
case of \textbf{S3} the occurrence of ventilation at the sides
causes a shrinking of the wetted area, which progressively approaches the
value of the cavitation area.
For both \textbf{S2} and \textbf{S3} the cavitation area
grows almost linearly in time, although for \textbf{S3} it approaches an
almost constant value after $\tau$=0.5, once the cavitation zone has
reached the trailing edge of the specimen, and the rear ventilation
starts.
%
%
\section{Conclusions and Future Work}
\label{sec:conclusions}
%
In this paper the water entry at high horizontal speed of rigid specimens
with double curvature has been investigated, based on experimental
data. The effects of the horizontal speed and of the pitch angle are analysed,
keeping constant the vertical-to-horizontal velocity ratio. Three
different shapes with different transverse and longitudinal curvatures 
have been considered.

It has been observed that the presence of a longitudinal curvature plays a key role,
marking an important distinction with respect to 
the water entry of flat plates at similar test conditions \citep{iafrati2016experimental}. 
Indeed, the longitudinal curvature, depending on velocity ratio 
and on the pitch angles, can lead to the occurrence 
of negative pressures (relative to atmospheric pressure) at the rear,
which can reach the vapour pressure value. 
This, in turn, can result in the formation of a cavitation region.
Furthermore, as the region of low pressure
at the rear may expands significantly during the water entry,
it can favour air entrainment either from the trailing 
edge or from the sides, leading to ventilation.
It has been shown that the transverse curvature does not influence much
the cavitation modalities, however
it has an effect on the loads and on the dynamics of the cavitation region,
due to the increased possibility for the fluid to escape from the side,
in case of higher curvature. 

The pitch angle is found to affect the pressures and the loads
in the front part of the body, the loads being higher for
higher pitch angles, consistently with what found for the impact of a
flat plate in \citet{iafrati2016experimental}. 
In addition, given the
change in the gradient slope in the rear part,
the modalities of cavitation may change.
At the higher pitch angle a cavitation-ventilation
regime is observed, whereas at the lower pitch angles
the ventilation region remains confined close to the
trailing edge during the impact phase.

Based on the underwater video frames and on the pressure measurements,
more detailed information on the dynamics of the 
cavitation region and of the wetted area can be retrieved. 
It is observed that the propagation velocity of the
cavitation region increases linearly with the horizontal speed
both for \textbf{S2} and \textbf{S3}. 
Although the slope is approximately equivalent in both case, 
an offset exists between the two trends, being the 
propagation speed higher for \textbf{S3}, as a consequence of the
lower transverse curvature.

Starting from the underwater images, quantitative estimates of the wetted
area, of the the geometric intersection area with the still water
level and of the cavitation area can be retrieved. A distinction is
made between what happens ahead and behind the point of curvature
change. The data indicate
that for the shape \textbf{S1B}, the growth rate of all areas is lower,
and so is the difference between the wetted area and
the geometric intersection area, compared to the specimens 
characterised by higher longitudinal curvatures.
For shape \textbf{S3} a ventilation phenomenon 
from the sides has been also observed,
occurring at the rear, as a consequence of the low pressures developing
beneath. The occurrence of such ventilation zones causes a shrinking of
the wetted area at the rear, which becomes smaller than the corresponding 
geometric intersection area.

In addition to contributing to a deeper understanding 
of the physical phenomena occurring during the water entry
at high horizontal speed and of the effects induced by 
changes in the body shape, the present paper provides a 
rather wide and detailed dataset, which could be used 
for an accurate validation of numerical approaches 
for the simulation of aircraft ditching or other water entry problems. 
Future work will focus on developing routine-based
pattern recognition algorithms to automate the selection
and tracking in time of the wetted area, cavitation areas and the 
following ventilation front in underwater videos, which can be improved
by also exploiting the information coming from the pressure probes.

\section*{Supplementary Material}
Supplementary materials are available for download at: \url{https://www.sciencedirect.com/science/article/pii/S0029801825018931#ecom0001} \\
The graphical abstract can be viewed at this link: \url{https://ars.els-cdn.com/content/image/1-s2.0-S0029801825018931-ga1_lrg.jpg}

\section*{Acknowledgements}
The authors also wish to thank: Flavio Olivieri, the technical supervisor 
of the \textit{High Speed Ditching Facility}, who is also responsible 
for its maintenance. He also participated in the the execution of all the experiments; 
Fabio Carta, the project manager who dealt with the design of the specimens 
and with the installation of the instrumentation on board of them and on the facility; 
Mauro Sale who provided technical support during the execution of the experiments; 
Alessandro Del Buono who gave valuable support in the interpretation of the results.
\section*{Funding}
This project has been funded from the European Union's Horizon 
2020 Research and Innovation Programme under Grant Agreement No. 724139 
(H2020-SARAH: increased SAfety \& Robust certification for ditching of 
Aircrafts \& Helicopters).
\section*{Declaration of interests}
The authors report no conflict of interest.

%





\bibliographystyle{agsm}
\bibliography{sarah_biblio}

\end{document}